\documentclass[acmsmall,screen,dvipsnames]{acmart}

\AtBeginDocument{%
  }

\setcopyright{acmlicensed}
\copyrightyear{xxx}
\acmYear{xxx}
\acmDOI{XXXXXXX.XXXXXXX}
 
\usepackage{amsthm}
\usepackage{hyperref} 
\usepackage{amsfonts} 
\usepackage{bussproofs} 
\usepackage{amsmath}
\usepackage{graphicx}
\usepackage{cmll} 
\usepackage{subcaption} 

\usepackage{titletoc}

  \newcommand\DoToC{%
    \startcontents
    \printcontents{}{1}{\textbf{Contents}\vskip3pt\hrule\vskip5pt}
    \vskip3pt\hrule\vskip5pt
  }

\usepackage{tikz}
\usetikzlibrary{arrows}
\tikzset{
     arrow/.style={-{Stealth[]}}
}
\usetikzlibrary{positioning,arrows.meta}
\usetikzlibrary{decorations.pathmorphing}
\usetikzlibrary{shapes.symbols}
\usetikzlibrary{shapes}
\tikzset{
  invisible/.style={opacity=0},
  visible on/.style={alt={#1{}{invisible}}},
  alt/.code args={<#1>#2#3}{%
    \alt<#1>{\pgfkeysalso{#2}}{\pgfkeysalso{#3}} 
  },
}

\newtheorem{remark}{Remark}
\newtheorem{definition}{Definition}
\newtheorem{proposition}{Proposition}
\newenvironment{sproof}{%
  \proof}{\endproof}

\usepackage{xfrac}
\usepackage{xcolor}

\usepackage[
disable
]{todonotes}
\newcommand\todom[2][]{\todo[color=blue!20,#1]{#2}} 

\usetikzlibrary{calc}
\def\bnf{::=\mbox{ }}
\def\definedas{\mathrel{\stackrel{\makebox[0pt]{\mbox{\normalfont\tiny def}}}{=}}}
\def\orsintax{\mbox{ }|\mbox{ }}
\newcommand{\FV}[2][]{FV^{#1}\!(#2)}
\def\linFV#1{FV^{t}(#1)}
\def\expFV#1{FV(#1)}

\def\tautypJAX{\tau}
\def\sigmatypJAX{\sigma}

\def\typsetJAX{\mathit{Type}}

\def\typrealJAX{\mathbb{R}}
\def\typunitJAX{\typone}
\def\typtensorJAX#1#2{#1\otimes#2}

\def\linvarJAX#1{\dot{#1}}
\def\expvarJAX#1{#1}
\def\literalJAX{\underline{r}}
\def\retJAX#1#2{(#1 ;\linvarJAX{#2} )}
\def\tupleJAX#1#2{\otimes(#1,#2)}
\def\emptytupleJAX{\otimes()}
\def\lintupleJAX#1#2{\dot{\otimes}(\dot{#1},\dot{#2})}
\def\emptylintupleJAX{\dot{\otimes}()}
\def\expfunJAX{\underline{f}(\expvarJAX{x_1},\dots, \expvarJAX{x_n})}
\def\linzeroJAX#1{\dot{0}_{#1}}
\def\linsumJAX#1#2{\linvarJAX{#1}\dot{+}\mbox{ }\linvarJAX{#2}}
\def\linmultJAX#1#2{\expvarJAX{#1}\dot{*}\mbox{ }\linvarJAX{#2}}
\def\dupJAX#1{\text{dup}(\linvarJAX{#1})}
\def\dropJAX#1{\text{drop}(#1)}
\def\letJAX#1#2#3{\text{let }#1=#2\text{ in }#3}

\def\retexpJAX#1#2{(#1;#2)}
 
\def\JAX{\text{JAX}}
\def\JaxA{\text{Linear A}}
\def\JaxB{\text{Linear B}}

\newcommand{\Sem}[1]{[\!\![#1]\!\!]}					
\newcommand{\SemP}[2][]{[\!\![#2]\!\!]_{#1}^{\mathsf p}}	
\newcommand{\SemT}[2][]{[\!\![#2]\!\!]_{#1}^{\mathsf t}}	

\newcommand{\SplitInvJAX}[2][]{\overline{\sigma}^{\text{\tiny{Jax}}}_{{#2;#1}}}

\def\expexprJAX#1{#1^{p}} 

\def\linexprJAX#1{\dot{#1}} 
  
\def\judgmentJAX#1#2#3#4#5{#1 ; #2\vdash^{\!\!\scalebox{.6}{\text{\tiny{Jax}}}} #3\!:\!(#4;#5)}
\def\lincontextJAX#1{\dot{#1}} 

\def\forwardJAX#1#2{\mathcal{F}^{\mathtt{Jax}}_{#1}(#2)}
\def\contextFMJAX{\phi}
\def\unzippingJAX#1{\mathcal{U}^{\mathtt{Jax}}(#1)}
\def\unzippingJAXPre#1{\mathcal{U}^{\mathtt{Jax}\bullet}(#1)}
\def\expunzipJAX#1{\mathcal{U}^{\mathtt{Jax}}(#1) \definedas \retexpJAX{#1}{\emptylintupleJAX}}
\def\linunzipJAX#1{\mathcal{U}^{\mathtt{Jax}}(#1) \definedas \retexpJAX{{\emptytupleJAX}}{#1}}

\def\transpJAX#1#2#3{\mathcal{T}^{\mathtt{Jax}}_{#1; #2}(#3)}

\def\lambdaLL{\mathbf{\lambda} \text{LL}}

\def\PrimalLL{\mathbf{\lambda} \text{LL}^{\mathtt{p}}}
\def\TangentLL{\mathbf{\lambda} \text{LL}^{\mathtt{t}}}
\def\LinearBLL{\mathbf{\lambda} \text{LL}^{\mathtt{A}}}

\def\explambdaLL{\PrimalLL}	
\def\linlambdaLL{\TangentLL}	
\def\lambdaLLFM{\LinearBLL}	

\def\affinebang{\S}
\def\bangterm#1{\oc #1}
\def\derterm#1{\overline{\tiny{\oc}}  #1}
\def\affbangterm#1{\affinebang{#1}}
\def\affderterm#1{#1}
\def\absterm#1#2{\lambda #1.#2}
\def\appterm#1#2{#1\mbox{}#2}
\def\lintupleterm#1#2{\langle #1,#2 \rangle}

\newcommand\Split[2][]{\sigma_{#2}^{#1}} 
\newcommand\SplitInv[2][]{\overline{\sigma}_{#2}^{#1}} 

\def\annabsterm#1#2#3{\lambda #1^{#2}.#3}

\def\linemptytupleterm{\langle \mbox{ } \rangle}
\def\tupleterm#1#2{(#1 , #2)}
\def\letterm#1#2#3{\mathtt{let}\mbox{ }#1=#2\mbox{ }\mathtt{in}\mbox{ }#3}

\def\emptytupleterm{()}
\def\funterm#1{\underline{f}#1}
\def\funtermname#1#2{\underline{#1} #2}
\newcommand\sumterm[2][]{\dot+_{#1}#2}
\newcommand\multterm[2][]{#1 \dot* #2}
\def\zeroterm#1{\underline{0}_{#1}}

\def\realterm{\underline{r}}

\def\nTuple#1{\langle#1\rangle}

\def\SeqA{\underline{\vec r}}	
\def\SeqB{\underline{\vec s}}

\def\PatA{p}		
\def\PatB{q}		
\def\PatExpA{\PatA^\otimes}	
\def\PatExpB{{\PatB^\otimes}}	
\def\PatAddA{\PatA^\with}	
\def\PatAddB{\PatB^\with}	

\def\typaffbang#1{\affinebang #1}
\def\typbang#1{\oc #1}
\def\typwith#1#2{#1 \& #2}
\def\typlollipop#1#2{#1 \multimap #2}
\def\typtensor#1#2{#1 \otimes #2}
\def\typR{\mathbb{R}}
\def\typone{\mathtt{1}}
\def\typtop{\top}

\newcommand{\Iso}[1][]{\mathrm{dual}_{#1}}			
\newcommand{\Osi}[1][]{\overline{\mathrm{dual}}_{#1}}	

\newcommand{\judgment}[4][]{#2 \vdash^{\tiny{#1}} #3 \!:\!\mbox{} #4}
\def\judgmenttransl#1#2#3#4#5{#1 \vdash #2 :\mbox{}#3 \mbox{}\otimes\mbox{} \typaffbang{(#4 \multimap #5)}}

\def\affcontext#1{\affinebang #1}

\def\typrvar{\mathit{v}}

\def\typrbangr{\mathit{\oc_i}}
\def\typrbangL{\mathit{\oc_e}}

\def\typbangW{\mathit{\oc_w}}

\def\typrlollipopL{\mathit{\multimap_e}}
\def\typrlollipopR{\mathit{\multimap_i}}

\def\typrwithL#1{\mathit{\&_{e#1}}}
\def\typrwithR{\mathit{\&}_i}

\def\typrtensorL{\mathit{\otimes_e}}
\def\typrtensorR{\mathit{\otimes_i}}

\def\typroneL{\mathit{1_e}}
\def\typroneR{\mathit{1_i}}
\def\typrfun#1{\mathit{F_#1}}
\def\typrsum{\mathit{S}}
\def\typrmult{\mathit{M}}
\def\typrreal{\mathit{R}}
\def\typrzero{\mathit{Z}}

\newcommand\ContextA[1][\,]{\gamma[#1]}
\newcommand\ExpContextA[2][]{\epsilon_{#1}[#2]}

\def\redbeta{\beta}

\def\SymbTransA{\delta}
\newcommand\TransA[2][]{\SymbTransA_{#1}(#2)}		
\newcommand\jaxdelta[2][]{\TransA[#1]{#2}}	
\def\SymbTransB{\delta^\mathtt{B}}
\newcommand\TransB[2][]{\SymbTransB_{#1}(#2)}

\def\tradexp#1{\tupleterm{#1}{\affbangterm{(\annabsterm{y}{\typtop}{\linemptytupleterm})}}}
\def\tradlin#1#2{\tupleterm{\oc\emptytupleterm}{\affbangterm{(\annabsterm{y}{#1}{#2})}}}

\def\trad#1#2#3{\tupleterm{#1}{\affbangterm{(\annabsterm{y}{#2}{#3})}}}

\def\PrimalT{\mathtt{p}}	
\def\TangentT{\mathtt{t}}	

\def\forward#1#2{\mathcal{F}_{#1}(#2)}
\def\contextFM{\theta}
\def\linearize#1{\TangentT(#1)}
\def\FMtransf#1#2{\tupleterm{#1}{\affbangterm{(\absterm{u}{#2})}}}

\newcommand\unzipping[1]{\mathcal{U}(#1)}
\newcommand\unzippingPre[1]{\mathcal{U}^\bullet(#1)}

\def\transpose#1{\mathcal{T}(#1)}
\newcommand{\lintranspose}[3][]{\mathcal{T}^{#1}_{#2}(#3)}
\def\transptyp#1{\overleftarrow #1}
\def\PIn{
	\PatAddA
	}
\def\POut{
	\PatAddB
	}

\newcommand { \mysizehead } [1] { \footnotesize { \textbf {#1}}}

\newcommand { \mysizetiny } [1] { \scriptsize {#1}}

\newcommand { \mysizebody } [1] { \footnotesize { {#1}}}

\def\Cost#1{\mathcal{W}(#1)}
\def\cost#1{\Cost{#1}}	
\def\CostJAX#1{\mathcal{W}^{\mathtt{Jax}}(#1)}

\def\CostType#1{\mathcal{W}(#1)}
\def\costType#1{\mathcal{W}(#1)}

\def\Rename{\alpha}	
\def\Dom#1{\mathrm{Dom}(#1)}	
\def\Codom#1{\mathrm{Cod}(#1)}	
\newcommand{\RenamePat}[2]{#1\langle#2\rangle}	
\newcommand{\RenameTerm}[2]{#1[#2]}	
\newcommand{\Fusion}[3]{\nu(#1,#2,#3)}		

\def\testSN#1{\mathtt{Test}_{\mbox{}#1}}
\def\RED#1{\mathtt{RED}_{\mbox{}#1}}

\def\decmultimap#1{\overset{\lower.3em\hbox{\scriptsize #1}}{\multimap}}

\def\Backprop{\overleftarrow{\mathbf{D}}}

\newcommand{\myhighlight}[2]{\colorbox{#1}{$\displaystyle #2$}}

\usepackage{enumitem}

\begin{document}

\title{JAX Autodiff from a Linear Logic Perspective\\
(Extended Version)}

\author{Giulia Giusti}
\affiliation{%
  \institution{ENS de Lyon, LIP, CNRS UMR 5668}
  \streetaddress{69342}
  \city{Lyon cedex 07}
  \country{France}}
\email{giulia.giusti@ens-lyon.fr}
 
\author{Michele Pagani}
\affiliation{%
  \institution{ENS de Lyon, LIP, CNRS UMR 5668}
  \streetaddress{69342}
  \city{Lyon cedex 07}
  \country{France}}
\email{michele.pagani@ens-lyon.fr}

\begin{abstract}
    Autodiff refers to the core of the automatic differentiation systems developed in projects like JAX and Dex. 
    Autodiff has recently been formalised in a linear typed calculus by Radul et al in \cite{radul2023you}. 
    Although this formalisation suffices to express the main program transformations of Autodiff, the calculus is very specific to this task, and it is not clear whether the type system yields a substructural logic that has interest on its own.
  
    We propose an encoding of Autodiff into a linear $\lambda$-calculus that enjoys a Curry-Howard correspondence with Girard's linear logic. 
    We prove that the encoding is sound both qualitatively (the encoded terms are extensionally equivalent to the original ones) and quantitatively (the encoding preserves the original work cost as described in \cite{radul2023you}). 
    As a byproduct, we show that unzipping, one of the transformations used to implement backpropagation in Autodiff, is, in fact, optional.
\end{abstract}

 

\received{xxx}
\received[revised]{xxx}
\received[accepted]{xxx}
 
\maketitle


\section{Introduction} \label{sect:introduction}
Consider a program $P$ that computes a real function $\Sem{P}$ from $\mathbb{R}^n$ to $\mathbb{R}$. Automatic differentiation (AD) refers to a family of algorithms for evaluating derivatives and gradients of numerical functions, such as $\Sem{P}$, by leveraging the source code of programs like $P$. This approach contrasts with other methods, such as numerical differentiation, which computes small differences in $\Sem P$, and symbolic differentiation, which manipulates closed forms of $\Sem P$. AD can be presented as a program transformation, similar to a compilation procedure or a domain-specific interpretation applied to $P$.

The directional derivative $\mathrm{D}_{\vec u}\Sem P(x)$ (assuming it exists) intuitively indicates how much a small perturbation at point $x$ along the direction given by the tangent vector $\vec u \in \typR^n$ affects the output of $\Sem P$. The gradient $\nabla\Sem P(x)$, on the other hand, is a vector that points in the direction of the steepest ascent of $\Sem P$ at $x$. AD primarily operates in two modes: the \emph{forward mode}, which efficiently computes the derivative $\mathrm{D}\Sem P$, and the \emph{backward or reverse mode}, which generates a program that evaluates $\nabla\Sem P$. The terminology refers to the execution flow of the computation: the forward mode propagates tangent vectors from the inputs of $P$ to its output, while the backward mode traces back from the output to the inputs.

The literature on AD dates back to the 60s (e.g.~\cite{AD_wengert}), and we can acknowledge three distinct periods or trends. Initially, AD focused on low-level programs with very simple programming primitives. Only narrow fragments of programming languages like FORTRAN or C were considered, encompassing floating-point variables, arrays, branching, goto statements, and while-loops. This approach was a natural choice to ensure the efficiency of the computation while maintaining enough structure to share intermediate results between different subroutines of a program.

A second period or generation of AD has advanced towards comprehensive AD systems for large high-level programming languages such as C\texttt{++} or Python. This approach has surged in the last decade with the development of industrial deep learning libraries like TensorFlow \cite{AbadiBCCDDDGIIK16}, PyTorch \cite{paszke2017automatic}, and JAX \cite{jax2018github,JAX_autodiff}. These libraries apply AD to complex programs which define numerical functions (e.g., neural networks) dynamically and incorporate increasingly complex programming features such as procedure calls, recursive functions, user-defined types, classes, and more.

Returning to a more academic line of research, a third phase or trend is characterised by efforts to formalise these techniques within an idealised framework\footnote{The term "formalisation" may be misunderstood as providing a mechanised proof in a proof assistant. This is too narrow in scope here: by formalisation, we refer to a general theoretical analysis of an algorithm or program transformation—providing definitions and precise statements that can be proven or refuted by counterexamples. This stands in contrast to more "experimental methods" based on testing and real-world runtime evaluations, which offer a different yet complementary approach to program analysis. We do not delve into mechanised proofs of AD using proof assistants in this paper; this remains an ultimate goal and such theoretical modelling is a preliminary step.}. The aim is to develop a formal system that models the core principles of modern AD implementations, abstracting from specific programming language details and other features such as parallel computation and floating-point arithmetic. The goals of this theoretical approach are manifold: to establish soundness proofs, which become less straightforward as program complexity increases; to elucidate the assumptions underlying such proofs, such as program termination, smoothness, and data persistence; to analyse asymptotic complexity (as opposed to performance evaluation in practical testing); and to decompose AD while drawing connections with other concepts in the theory of programming languages.

Our contribution fits into this third line of research and it starts from the paper \cite{radul2023you}, which formalises how AD, specifically its backward mode, is implemented in libraries like JAX and Dex. Recall the notation introduced earlier: a program $P$ computes a real function $\Sem P$ from $\typR^n$ to $\typR$, and there are forward and backward modes for computing the directional derivative $\mathrm{D}\Sem P$ and the gradient $\nabla\Sem P$, respectively. It is well-known that these two notions are dual to each other, in the sense that $\mathrm{D}_{\vec u}\Sem P(x) = \nabla\Sem P(x) \cdot \vec u$ for any point $x$ and tangent vector $\vec u \in \mathbb{R}^n$.

The peculiarity of JAX is to start from this fact and to implement the backward mode as a composition of three intermediate program transformations: the forward mode, denoted here as $\mathcal{F}$, the unzipping $\mathcal{U}$, and the linear transpose $\mathcal{T}$:
\begin{equation}
    \label{eq:soundness_JAX}
    \nabla\Sem P \approx \Sem{\mathcal{T}(\mathcal{U}(\mathcal{F}(P)))}
\end{equation}

The implementation of $\mathcal{F}$ adopts the concept of dual numbers: each numeric variable $x$ in $P$ is paired with a sibling variable $\dot{x}$, where $x$ is termed the \emph{primal} and $\dot{x}$ the \emph{tangent}. Tangent variables $\dot{x}$ store the differential information relative to $x$, which arises from small perturbations in the inputs.

The unzipping transformation $\mathcal U$ divides the program $\mathcal{F}(P)$ into two subroutines: $\mathcal{F}(P)^p$, which computes all primal outputs of $\mathcal{F}(P)$, and $\mathcal{F}(P)^t$, which computes all tangent outputs. Specifically, the primal computation is independent of the tangent values, whereas the tangent computation generally depends on the primal values. Therefore, $\mathcal{F}(P)^t$ is defined as a program that takes as input the tangent variables associated with the inputs of $P$, along with a sequence of primal variables that store the values computed by $\mathcal{F}(P)$ affecting certain tangent variables. This sequence of variables corresponds to the \emph{tape} in some AD literature.

The program $\mathcal{F}(P)^t$ indeed performs only linear algebraic operations, such as vector addition and scalar multiplication. Formally, the transpose of a linear map $X\mapsto Y$ is a linear map $Y^* \mapsto X^*$ where $X^*$ (resp. $Y^*$) is the algebraic dual of $X$ (resp. $Y$). Hence, the final transformation $\mathcal{T}$ transposes $\mathcal{F}(P)^t$, resulting in a program giving the adjoint of $\mathrm{D}\Sem{P}$, i.e.~the gradient of $\Sem{P}$.

The paper \cite{radul2023you} formalises these three transformations using a simply typed calculus called Linear A (Figure~\ref{fig:jaxtyprules}). The typing discipline integrates a form of linearity: tangent variables are subject to specific constructors for copying or erasure, denoted $\mathtt{dup}$ and $\mathtt{drop}$, respectively. Some typing rules are intricate, particularly the one governing primal/tangent compositions, such as $\letterm{(x,\dot{y})}{e_1}{e_2}$. Additionally, rules manipulating purely primal or purely tangent tuples are constrained to tuples of variables.

Linear A is a ``domain-specific'' calculus designed precisely for formalising JAX AD, and in this regard, it achieves its objective of proving soundness, which essentially corresponds to Equation \eqref{eq:soundness_JAX}. However, it remains unclear whether the grammar expression can be embedded into a more general calculus that has independent interests. Additionally, it is uncertain whether the typing system has a logical interpretation, particularly whether the linearity constraints correspond to a substructural logic that controls the contraction or weakening of hypotheses (which are the logical equivalents to data copying or erasure).

We bridge this gap in this paper by encoding Linear A into a linear $\lambda$-calculus, denoted $\lambdaLL$, which establishes a Curry-Howard correspondence with a fragment of linear logic (LL) \cite{girard1987linear}: types correspond with formulas, programs with proofs and the operational semantics is associated with cut-elimination, a crucial procedure in proof-theory proving consistency.  Linear logic is a substructural logic featuring two families of conjunctions: the multiplicative conjunction $\otimes$ and the additive conjunction $\&$. It also includes an exponential modality that relates these two families through the exponential isomorphism $\oc A \otimes \oc B = \oc(A \& B)$, and governs the structural rules of weakening and contraction: only hypotheses of type $\oc A$ can be used multiple times in a proof. The dependency of the conclusion on the hypotheses in a proof (or equivalently, of the output on the inputs in a $\lambdaLL$ term) is governed by the LL implication $A \multimap B$. In our encoding, primal tuples from Linear A are associated with a multiplicative conjunction of exponentiated types $\oc A \otimes \oc B$, while tangent tuples correspond to an additive conjunction $A \& B$ (Section~\ref{sect:JAXtoLambdaLL}). This setup allows the exponential isomorphism to establish a precise link between these two data types (Remark~\ref{rk:primal_bang_tangent}). 

In this context, $\mathtt{dup}$ represents the additive diagonal ($A \multimap A \& A$), and $\mathtt{drop}$ applied to a primal-tangent pair is the weakening rule on the primal part ($\oc A \multimap \typone$, where $\typone$ is the neutral element of $\otimes$) and the terminal rule on the tangent part ($A \multimap \typtop$, where $\typtop$ is the neutral element of $\&$).

\paragraph*{Benefits of this work}
Embedding a domain-specific calculus like Linear A into a $\lambda$-calculus such as $\lambdaLL$, which enjoys a Curry-Howard correspondence with LL, offers several advantages. It allows us to import the cut-elimination rewriting from LL, giving a well-behaving $\beta$-reduction (Figure~\ref{fig:beta_rules} and Theorem~\ref{th:SN} and~\ref{th:confluence}). This step provides an operational semantics for $\lambdaLL$, which was lacking for Linear A, and facilitates the proof of various term equivalences. In fact, we introduce a logical relation equivalence (Section~\ref{subsect:extensional_equivalence}) which guarantees the soundness of our encoding of Linear A (Theorem~\ref{prop:sound_transA}) and verify different translations. Furthermore, we can restate the cost-preservation of Linear A transformations by referring to the number of flops evaluated along a specific rewriting strategy, essentially implementing a call-by-value evaluation strategy (Section~\ref{subsect:cost_model}).

Furthermore, by encoding Linear A into a linear $\lambda$-calculus as $\lambdaLL$ we have placed it within the same theoretical framework as many other formalizations, such as \cite{POPL2020,Vakar2021CHADCH}, enabling a formal comparison with them. A detailed comparison with the system in \cite{POPL2020} is provided in Section~\ref{sect:comparison}.

Another benefit of $\lambdaLL$ lies in the modularity of our AD system. This benefits is tied to $\lambdaLL$'s ability to define the transpose transformation directly within the image set of the $\mathcal{F}$ transformation, without requiring the unzipping transformation. This is a notable byproduct, as the unzipping process imposes an order on the different phases of the back-propagation algorithm. This order typically involves a forward phase (computing $\mathcal{F}(P)^p$ following the input-output execution flow of $P$) followed by a backward phase (computing $\mathcal{T}(\mathcal{F}(P)^t)$, reversing the execution flow of $P$).
In $\lambdaLL$, various combinations of these two phases can be represented: applying the $\mathcal{U}$ transformation, introduced in Equation~\ref{eq:soundness_JAX}, results in a complete separation of the forward and backward phases, similar to Linear A. However, if $\mathcal{U}$ is not applied (or applied selectively to some sub-terms), terms can represent intermediate computations that blend aspects of both the forward and backward passes. This improves the modularity of the backward algorithm and might be particularly beneficial for programs with independent subroutines.

\paragraph*{Contents of the paper} Section~\ref{sect:JAX} briefly introduces Linear A, recalling the main definitions and adapting the notations from \cite{radul2023you}. This serves as a compendium to ensure the paper is self-contained, although for a more comprehensive understanding, we refer the reader to the original paper. 

Section~\ref{sect:lambdaLL} introduces the linear $\lambda$-calculus $\lambdaLL$, its typing system (Figure~\ref{fig:typing_LL}), $\beta$-reduction (Figure~\ref{fig:beta_rules}), and the main properties of this calculus: subject reduction (Theorem~\ref{th:subjred}), strong normalisation (Theorem~\ref{th:SN}), and confluence (Theorem~\ref{th:confluence}). This section follows a standard pattern, so their proofs are moved to the Appendix. 
Subsection~\ref{subsect:extensional_equivalence} defines the logical relation $\sim$ comparing $\lambdaLL$ terms with respect to their extensional behaviour at ground types.  We give an adaptation (Subsection~\ref{subsect:cost_model}) of the work cost notion presented in \cite[Section 4.3]{radul2023you}, where we establish bounds on the number of flops evaluated in a $\beta$-reduction sequence for a specific (yet complete) family of reduction strategies dubbed \emph{safe-reduction}, applicable to a set of terms including those representing Linear A.

Our main original contributions begin in Section~\ref{sect:JAXtoLambdaLL}, where we introduce a translation $\SymbTransA$ from Linear A to $\lambdaLL$ (Figure~\ref{fig:JAXtoLL}). We prove the soundness of this translation with respect to the extensional semantics of Linear A (Proposition~\ref{prop:sound_transA}).  
Sections~\ref{sect:forward}, \ref{sect:unzipping}, and \ref{sect:transpose} respectively define forward, unzipping, and transpose transformations on $\lambdaLL$ terms and establish their commutativity with the $\SymbTransA$ translation modulo $\sim$-equivalence (Theorem~\ref{th:forward_soundness}, Theorem~\ref{th:unzipping_soundness}, and Theorem~\ref{th:transpose_JAX_LL}).  We also demonstrate how the unzipping transformation can be skipped. Section~\ref{sect:comparison} then discusses a comparison with related work, in particular \cite{POPL2020}. Finally, Section~\ref{sect:conclusion} summarises the main results in Figure~\ref{fig:final_picture} and offers some perspectives.

\paragraph*{Related work}
Let us mention \cite{ADSurvey} as a smooth and modern introduction to AD. 
The literature is abounding in this last decade even if we restrict to the third period described above.
Apart from the already cited \cite{radul2023you}, let us mention some approaches, even if because of space limitations we must unfortunately seriously limit our survey.

A pioneering paper is \cite{AbadiP20} defining an operational semantics of a while-language with a reverse derivative expressing backward AD.  The paper proves a soundness property via a denotational model. Soundness of forward AD has been established in variants of the simply typed $\lambda$-calculus, for instance by \cite{DalLago_Logic}, using an open logical relation, and by \cite{Hout_correctness}, additionally employing diffeologies.

A more abstract approach has been developed in the setting of categorical semantics, leading to a series of papers that provide an axiomatisation of the backward mode starting with a notion of \emph{reverse derivative category} \cite{cockett_et_al:LIPIcs.CSL.2020.18,Cruttwell_Gallagher_Lemay_Pronk_2022,cruttwell_et_al:LIPIcs.CSL.2024.21,Cruttwell2021}. The main benefit of this approach is to offer an abstract framework expressing the notions of AD, or, more generally, gradient-based optimizations (e.g.~\cite{Zanasi_et_al}) in more general spaces then simple real or complex vector spaces. In fact, AD is usually restricted to programs handling tuples of real or complex numbers. Let us note however that this body of works emphasises semantic soundness, while our primary concern is efficiency. Our transformations are designed to preserve the flops workload of the original AD algorithms, enabling gradient computation with a numerical cost comparable to that of computing partial derivatives. This perspective marks a significant shift in focus, as discussed at the beginning of Section~\ref{sect:transpose}.

Many of the papers mentioned above describe AD as syntactic operators that compute derivatives and gradients of the numerical programs to which they are applied. Let us mention to a series of papers \cite{Vakar2021CHADCH,LucatelliCHAD} that present the two modes of AD as program transformations within a system called \emph{CHAD}. The distinctive feature of this approach is that it provides a precise categorical framework, which unambiguously (once the categorical structure is fixed) defines both forward and backward AD as homomorphic functors. A somewhat related system is presented in \cite{kerjeanPedrotDelta}, derived from the so-called \emph{G\"odel's Dialectica interpretation} of intuitionistic logic.
These systems are based on linear typed $\lambda$-calculi that can be naturally represented in $\lambdaLL$. In particular, their forward and transpose transformations share the same types as our $\mathcal F$ and $\mathcal T$. Our approach, however, differs as we focus on JAX AD, defining in particular the backward mode directly in terms of the forward mode (see \eqref{eq:soundness_JAX}), whereas the aforementioned systems define the two modes independently and are grounded in a categorical framework. Furthermore, \cite{Vakar2021CHADCH,LucatelliCHAD,kerjeanPedrotDelta} do not discuss the efficiency of their program transformations --- a major concern which require a subtle definition of our transformation $\mathcal T$ (see the discussion in Section~\ref{sect:transpose}).
Let us mention however that the most recent work \cite{Vakar2024CHAD} does tackle this issue for CHAD too, describing an efficient implementation based on a careful representation of tangent data types. A more detailed comparison between our approach and theirs is then left for future work.

Another approach based on a linear $\lambda$-calculus is presented in \cite{POPL2020}, which likewise formalises forward and backward AD as separate program transformations. The main difference with \cite{POPL2020} is that this latter relies on the notion of a \emph{back-propagator}, achieving asymptotic efficiency by assigning a dedicated operational semantics to this construct. We refer to Section~\ref{sect:comparison} for a more detailed comparison between the back-propagator approach and ours. We also mention \cite{Vakar_linear}, which presents a Haskell implementation of \cite{POPL2020} incorporating several optimisations, and \cite{MazzaP21}, which extends \cite{POPL2020} to recursive programs and achieves a result of almost everywhere correctness.

All these papers present AD as program transformations or as syntactic constructs, however an alternative approach is also quite popular, consisting in looking at AD as a kind of domain specific interpreter. The two approaches go under the names of \emph{define-then-run} and \emph{define-by-run}. Let us mention \cite{Poittier_AD} as a pedagogical and modern account to this latter, providing a (machine-checked) proof of soundness by means of a separation logic.

\section{Brief summary of Autodiff} \label{sect:JAX}
\subsection{Syntax and Semantics}  
Linear A is an idealised language formalising the core of JAX Autodiff -- an implementation of AD in projects like JAX \cite{jax2018github}. The main feature is that the syntax marks which variables store primal values and which variables carry tangent values.

We will recall here the core of Linear A as presented in~\cite{radul2023you}, with some minor notational variations. We refer 
to the Appendix~\ref{app:JAX} and 
to \cite{radul2023you} for a more comprehensive presentation. In the original paper, Linear A is a first order language because it includes definitions of functions at top level, we will not consider this feature of the language as it is not interesting for our purposes. 

\emph{JAX types} are nested tuples of the ground type of reals:
\begin{equation}
    \tag{JAX Types}
    \label{eq:JAXtypes}
    \tautypJAX, \sigmatypJAX \bnf \typrealJAX \orsintax \typunitJAX
        \orsintax \typtensorJAX{\tautypJAX}{\sigmatypJAX}
\end{equation}

JAX considers two disjoint copies of this set: $\typsetJAX \uplus \{\cdot\}\times\typsetJAX$. The elements from the first copy are called \emph{primal types} and the ones from the second copy are called \emph{tangent types}. 

We adopt Church-style typing: the type of each variable is fixed, once and for all. Variables then inherit the primal/tangent terminology and we denote by $\linvarJAX{x}$ a tangent variable, i.e. a variable supposed to have a tangent type $(\cdot, \tautypJAX)$. This latter notation allows for omitting the tag $\cdot$ on the tangent types, so simply writing $\linvarJAX{x}:\tautypJAX$ instead of $\linvarJAX{x}:(\cdot, \tautypJAX)$.

\begin{figure*}
    \scalebox{.85}{\parbox{\textwidth+3cm}{
        \begin{center} 
            \AxiomC{\phantom{$\lincontextJAX{\Gamma}$}}
            \UnaryInfC{$\judgmentJAX{\expvarJAX{x}:\tautypJAX}{\linvarJAX{y}:\sigmatypJAX}
                {\retJAX{x}{y}}{\tautypJAX}{\sigmatypJAX}$}
            \DisplayProof
            \quad
            \AxiomC{$\judgmentJAX{\Gamma_1}{\lincontextJAX{\Gamma}_1}{e_1}
            {\tautypJAX_1}{\sigmatypJAX_1}$}
            \AxiomC{$\judgmentJAX{\Gamma_2,\expvarJAX{x}:\tautypJAX_1}
            {\lincontextJAX{\Gamma}_2,\linvarJAX{y}:\sigmatypJAX_1}{e_2}{\tautypJAX}
            {\sigmatypJAX}$}
            \BinaryInfC{$\judgmentJAX{\Gamma_1 \cup \Gamma_2}
            {\lincontextJAX{\Gamma}_1, \lincontextJAX{\Gamma}_2}
            {\letterm{\retJAX{x}{y}}{e_1}{e_2}}{\tautypJAX}{\sigmatypJAX}$}
            \DisplayProof
        \end{center}
    
        \smallskip
        
        \begin{center}
            \AxiomC{\phantom{$\dot \Gamma$}}
            \UnaryInfC{$\judgmentJAX{}{}{{\emptytupleJAX}}{\typunitJAX}{\typunitJAX}$}
            \DisplayProof
            \;
            \AxiomC{$\judgmentJAX{\Gamma}{\lincontextJAX{\Gamma}}
            {e}{\tautypJAX}{\sigmatypJAX}$}
            \UnaryInfC{$\judgmentJAX{\Gamma,\expvarJAX{z}:\typunitJAX}{\lincontextJAX{\Gamma}}{\letterm{\!\otimes\!\!()\!}{\expvarJAX{z}}{e}}{\tautypJAX}{\sigmatypJAX}$}
            \DisplayProof
            \;              
            \AxiomC{\phantom{$\dot \Gamma$}}
            \UnaryInfC{$\judgmentJAX{}{}
            {\emptylintupleJAX}{\typunitJAX}{\typunitJAX}$}
            \DisplayProof
            \;
            \AxiomC{$\judgmentJAX{\Gamma}{\lincontextJAX{\Gamma}}
            {e}{\tautypJAX}{\sigmatypJAX}$}
            \UnaryInfC{$\judgmentJAX{\Gamma}{\lincontextJAX{\Gamma},\linvarJAX{z}:\typunitJAX}
            {\letterm{\emptylintupleJAX}{\linvarJAX{z}}{e}}{\tautypJAX}{\sigmatypJAX}$}
            \DisplayProof               
        \end{center}

        \smallskip
        \begin{center} 
            \AxiomC{\phantom{$\dot \Gamma$}}
            \UnaryInfC{$\judgmentJAX{}{\linvarJAX{x}_1:\tautypJAX_1,\linvarJAX{x}_2:\tautypJAX_2}
            {\lintupleJAX{x_1}{x_2}}{\typunitJAX}{\typtensorJAX{\tautypJAX_1}{\tautypJAX_2}}$}
            \DisplayProof
            \quad
            \AxiomC{$\judgmentJAX{\Gamma}{\lincontextJAX{\Gamma},\linvarJAX{x}_1:\tautypJAX_1,
            \linvarJAX{x}_2:\tautypJAX_2}
            {e}{\tautypJAX}{\sigmatypJAX}$}
            \UnaryInfC{$\judgmentJAX{\Gamma}{\lincontextJAX{\Gamma},
            \linvarJAX{z}:\typtensorJAX{\tautypJAX_1}{\tautypJAX_2}}
            {\letterm{\lintupleJAX{x_1}{x_2}}{\linvarJAX{z}}{e}}
            {\tautypJAX}{\sigmatypJAX}$}
            \DisplayProof
        \end{center} 
        
        \smallskip  
        \begin{center}
            \AxiomC{\phantom{A}}
            \UnaryInfC{$\judgmentJAX{}{}{\linzeroJAX{\tautypJAX}}{\typunitJAX}{\tautypJAX}$} 
            \DisplayProof
            \;
            \AxiomC{\phantom{A}}
            \UnaryInfC{$\judgmentJAX{}{\linvarJAX{x}:\tautypJAX,
            \linvarJAX{y}:\tautypJAX}{\linsumJAX{x}{y}}{\typunitJAX}{\tautypJAX}$}
            \DisplayProof
            \;
            \AxiomC{\phantom{A}}
            \UnaryInfC{$\judgmentJAX{\expvarJAX{x}:\typrealJAX}
            {\linvarJAX{y}:\tautypJAX}{\linmultJAX{x}{y}}{\typunitJAX}{\tautypJAX}$}
            \DisplayProof
            \;
            \AxiomC{\phantom{A}}
            \UnaryInfC{$\judgmentJAX{}{}
            {\literalJAX}{\typrealJAX}{\typunitJAX}$}
            \DisplayProof
        \end{center}
    
        \smallskip
        \begin{center}
            \AxiomC{\phantom{$\dot\Gamma$}}
            \UnaryInfC{$\judgmentJAX{\expvarJAX{x_1}\!:\!\typrealJAX, \expvarJAX{x_2}\!:\!\typrealJAX}{}
            {\!\underline{f}(x_1,x_2)}{\typrealJAX}{\typunitJAX}$}
            \DisplayProof        
            \,
            \AxiomC{\phantom{$\dot\Gamma$}}
            \UnaryInfC{$\judgmentJAX{}{\linvarJAX{x}:\tautypJAX}{\!\dupJAX{x}}
            {\typunitJAX}{\typtensorJAX{\tautypJAX}{\tautypJAX}}$}
            \DisplayProof
            \,
            \AxiomC{$\judgmentJAX{\Gamma}{\lincontextJAX{\Gamma}}{e}
            {\tautypJAX}{\sigmatypJAX}$}
            \UnaryInfC{$\judgmentJAX{\Gamma}{\lincontextJAX{\Gamma}}{\!\dropJAX{e}}
            {\typunitJAX}{\typunitJAX}$}
            \DisplayProof
        \end{center}
        }
        }
        \caption{Linear A Typing Rules. For short, we detail only the case of 
        $\underline{f}$ binary, the general case being immediate.}
        \label{fig:jaxtyprules}
        \todom[inline]{Notice that this system has no weakening, but it can be in somehow produced by using drop. It is difficult however to define it categorically as a weakening as the output type is not just the tensor unit, but a pair of units of two monomial products.}
\end{figure*}    

Figure~\ref{fig:jaxtyprules} shows the grammar of Linear A expressions together with their typing rules. A \emph{judgment} is defined as $\judgmentJAX{\Gamma}{\lincontextJAX{\Gamma}}{e}{\tautypJAX}{\sigmatypJAX}$, where $e$ is the typed expression,
$\Gamma=\{\expvarJAX{x_1}:\tautypJAX_1, \ldots,\expvarJAX{x_n}:\tautypJAX_n\}$ is a set of primal variables, $\lincontextJAX{\Gamma}=\{\linvarJAX{y}_1:\sigmatypJAX_1, \ldots,\linvarJAX{y}_m:
 \sigmatypJAX_m\}$ is a set of tangent variables, and the type $(\tautypJAX;\sigmatypJAX)$ of the expression $e$ is a pair, giving respectively the type of the primal and the tangent result of $e$. 
 
We write $\Gamma_1 \cup \Gamma_2$ for the union of two primal contexts. We use commas to denote disjoint unions, so when we write $\lincontextJAX{\Gamma}_1, \lincontextJAX{\Gamma}_2$ we suppose that $\lincontextJAX{\Gamma}_1$ and $\lincontextJAX{\Gamma}_2$ have no variable in common, otherwise the rule does not hold. Similarly for $\Gamma, x:\tau$. 

Variables are introduced by pairs $(x; \dot y)$ of a primal and a tangent variable (notice the semi-colon separator). In accordance, expressions compose by a primal/tangent $\mathtt{let}$ which is the most peculiar operator of Linear A. The original paper~\cite{radul2023you} considers $n$-ary introduction and elimination rules for both primal and tangent tuples. For a question of space, we consider here only zeroary primal and tangent tuples and binary tangent tuples as these constructions are essential for the JAX Autodiff transformations. The extension to  binary primal tuples is simple (see Appendix~\ref{app:JAX}) 
 as well as their $n$-ary variants, but notational more heavy and not essential for our results.

Finally, we suppose numeric constants $\literalJAX$ and $\underline{f}$ for, resp., real numbers and $n$-ary numeric functions, e.g. $\underline{f}\in \{\underline{exp},\underline{*},\underline{+},\dots\}$. We suppose also a bound $b$ to the possible arity $n$ of the numeric functions. In fact, for short, Figure~\ref{fig:jaxtyprules} details only binary $\underline{f}$, the general case being trivial. We suppose that all functions are differentiable and come together with their partial derivatives $\underline{\partial_i f}$. 

Numeric functions act over primal variables. We have in addition the sum $\dot{+}$ over tangent variables  and the product $\dot{*}$ between a primal variable and a tangent one. Note that primal variables can be duplicated or erased in the environments but they cannot depend on tangent variables. In the contrast, tangent variables can be modified only by linear operators, but may depend on primal variables through scaling $\dot *$.
Finally, Linear A has an explicit copying operator $\text{dup}$ over tangents and a $\text{drop}$ operator erasing both primal and tangent results.

The set of primal (resp.~tangent) free variables of an expression $\expFV e$  (resp. $\linFV e$) is defined as usual by induction on $e$, with the $\mathtt{let}$ operators as binders. 


The \emph{semantics of an expression} $\judgmentJAX{\Gamma}{\lincontextJAX{\Gamma}}{e}{\tautypJAX}{\sigmatypJAX}$ is defined as a pair of two functions $\SemP e$ and $\SemT e$: the former maps real vectors $\vec{\underline{r}}$ associated with $\Gamma$ to a real vector $\SemP[\vec{\underline{r}}] e$ for $\tau$ giving the primal result of $e$; the second map $\SemT e$ takes in input both a real vector $\vec{\underline{r}}$ for $\Gamma$ and a real vector $\vec{\underline{s}}$ for $\dot\Gamma$ and returns a real vector $\SemT[\vec{\underline{r}};\vec{\underline{s}}] e$ for $\sigma$, giving the tangent result of $e$. 

A vector for a typing environment $\Gamma$ (or $\dot\Gamma$) is a map $\vec{\underline r}$ associating each $x:\tau\in\Gamma$ with a vector $\vec{\underline r}(x)$ for $\tau$. Given a subset $\mathcal X\subseteq \Gamma$, we write by $\vec{\underline r}\vert \mathcal X$ the restriction of $\vec{\underline r}$ to the variables in $\mathcal X$.
The semantics $\SemP e$ and $\SemT e$ are then defined inductively on $e$ in the obvious way. 
For example, the definition of	
$\SemT[
		\vec{\underline r};\vec{\underline s}
]{\letterm{(x;\dot y)}{e_1}{e_2}}$ first computes both $\SemP{e_1}$ and $\SemT{e_1}$ by taking into account the values of the primals and tangents free in $e_1$ and then computes $\SemT{e_2}$ by affecting the values $\SemP{e_1}$ and $\SemT{e_1}$ to the variables $x$ and $\dot y$ bounded by the $\mathtt{let}$. More formally:
\[
    \SemT[
        \vec{\underline r};\vec{\underline s}
    ]{\letterm{(x;\dot y)}{e_1}{e_2}}
    \definedas
    \SemT[
        \vec{\underline r}\vert\FV{e_2}, 
        x\mapsto {\SemP[\vec{\underline r}\vert\FV{e_1}]{e_1}}
        ; 
        \vec{\underline s}\vert\linFV{e_2}, 
        \dot y\mapsto{\SemT[\vec{\underline r},\vec{\underline s}\vert\linFV{e_1}]{e_1}}
    ]{e_2}
\]  
where $\SemP[\vec{\underline r}]{\letterm{(x;\dot y)}{e_1}{e_2}}\definedas \SemP[\vec{\underline r}\vert\FV{e_2}, x\mapsto {\SemP[\vec{\underline r}\vert\FV{e_1}]{e_1}}]{e_2}$.

A notion of workload $\cost{e}$ is introduced in~\cite[Section 4.3]{radul2023you}, which basically estimates a bound to the number of flops performed in the computation of $\SemT{e}$. In particular, every non-linear primitive costs $1$, linear addition $\dot +$ and linear multiplication $\dot *$ cost $1$ per scalar $\typR$ type present in the result, and $\mathrm{drop}(e)$ costs $\cost{e}$ plus $1$ for every scalar type $\typR$ in the output of $e$.

\paragraph*{Notational conventions}
The syntax of Linear A is restrictive and some syntactic sugar is convenient for manipulating the ``purely primal'' or ``purely tangent'' parts of an expression. 
We write
`$\letterm{\expvarJAX{x} }{e_1}{e_2}$' for `$\letterm{\retJAX{x}{y}}{e_1}{\letterm{\emptylintupleJAX}{\linvarJAX y}{e_2}}$' and `$\expvarJAX{x}$' for `$\letterm{\linvarJAX{y} }{\emptylintupleJAX}{\retJAX{x}{y}}$'. Similarly: `$\letterm{\linvarJAX{y} }{e_1}{e_2}$' stands for `$\letterm{\retJAX{x}{y}}{e_1}{\letterm{{\emptytupleJAX}}{\expvarJAX x}{e_2}}$' and `$\linvarJAX{y}$' for `$\letterm{\expvarJAX{x} }{{\emptytupleJAX}}{\retJAX{x}{y}}$'. 
We then consider pairs of expressions of different kind and tensors of expressions of equal kind: $\retexpJAX{e_1}{e_2}\approx\mbox{} \letJAX{\expvarJAX{x}}{e_1} {\letJAX{\linvarJAX{y}}{e_2}{(\expvarJAX{x};\linvarJAX{y})}}$, and finally $\dot\otimes(e_1,e_2)\approx\mbox{} \letJAX{\linvarJAX{x}}{e_1} {\letJAX{\linvarJAX{y}}{e_2}{\lintupleJAX{x}{y}}}$. 
See Figure~\ref{fig:jaxDerTypRules} of Appendix~\ref{app:JAX}.
Given a sequence $\theta = (\tau_1,\dots, \tau_n)$ of types, we denote by $\otimes\theta$ the $n$-fold product $\tau_1\otimes(\dots \otimes\tau_n)$. Given two subsequences $\theta_1$ and $\theta_2$ partitioning the variables in $\theta$, we can define the expression $\judgmentJAX{}{\dot{y}_1:\otimes\theta_1,\dot{y}_2:\otimes\theta_2}
{\SplitInvJAX[\theta]{\dot{y}_1,\dot{y}_2}}{\typone}{\otimes\theta}$ which gather together all components of the two tangent tuples.\todom{maybe explicit, if space. RIGHT PARTITION?}
 
One crucial step of JAX Autodiff is to split the primal part from the tangent part of an expression before performing the transpose transformation. The following fragment of Linear A, called \emph{Linear B} in \cite{radul2023you}, uses the conventions introduced above in order to define a $3$-sorted grammar, giving purely primal ($\expexprJAX{e}$) and purely tangent ($\linexprJAX{e}$) expressions and pairs ($d$) of each of them possibly prefixed by a stack of primal let-definitions:
\allowdisplaybreaks
\begin{align*}
    \tag{Linear B}\label{linear_B}
    d \bnf& 
    \retexpJAX{\expexprJAX{e}}{\linexprJAX{e}}
    \orsintax \letterm{\expvarJAX{x}}{\expexprJAX{e}}{d}
    \orsintax \letterm{{\emptytupleJAX}}{\expvarJAX{z}}{d}
    \\
    \tag{Primal}\label{linear_B:primal}
    \expexprJAX{e} \bnf
    & 
      \expvarJAX{x} 
    \orsintax \letterm{\expvarJAX{x}}{\expexprJAX{e_1}}{\expexprJAX{e_2}} 
    \orsintax\literalJAX 
    \orsintax \underline{f}(\expvarJAX{x}_1, \expvarJAX{x}_2) 
     \orsintax \dropJAX{\expexprJAX{e}}
     \orsintax \emptytupleJAX 
    \orsintax \letterm{{\emptytupleJAX}}{\expvarJAX{z}}{\expexprJAX{e}}
    \\
    \tag{Tangent}\label{linear_B:tanget}
    \linexprJAX{e} \bnf& 
    \linvarJAX{x} 
    \orsintax \letterm{\linvarJAX{x}}{\linexprJAX{e_1}}{\linexprJAX{e_2}} 
    \orsintax \dupJAX{x}
    \orsintax \linzeroJAX{\tautypJAX} 
    \orsintax  \linmultJAX{x}{y} 
    \orsintax \dropJAX{\linexprJAX{e}}
    \orsintax \emptylintupleJAX 
    \orsintax \lintupleJAX{e_1}{e_2} \\
    &\orsintax \letterm{\emptylintupleJAX}{\linvarJAX{z}}{\linexprJAX{e}}
    \orsintax \letterm{\tupleJAX{\linvarJAX{x}_1}{\linvarJAX{x}_2}}{\linvarJAX{z}}{\linexprJAX{e}}
\label{eq:linearBsyntax}
\end{align*}    

Notice that a primal $\expexprJAX{e}$ (resp.~tangent $\linexprJAX{e}$) is typed as $\judgmentJAX{\Gamma}{}{\expexprJAX{e}}{\tautypJAX}{\typunitJAX}$ (resp.~$\judgmentJAX{\Gamma}{\lincontextJAX{\Gamma}}{\linexprJAX{e}}{\typunitJAX}{\tautypJAX}$). 

Let us consider the function $g(x,y)=(sin(x)*y)+cos(x)$. This function will serve as the running example throughout the remainder of the paper.
The purely primal expression in Linear B computing $g$ is the following expression $\expexprJAX{e}$ in Figure \ref{fig:toyEx_linB}
which is well-typed as $\judgmentJAX{x:\typR,y:\typR}{}{\expexprJAX{e}}{\typR}{\typunitJAX}$.
It is important to note that $\underline{*}$ and $\underline{+}$ represent primal operations, not the tangent product and sum, as the latter are indicated using dot notation.
\subsection{JAX Autodiff Transformations}
\begin{figure*}
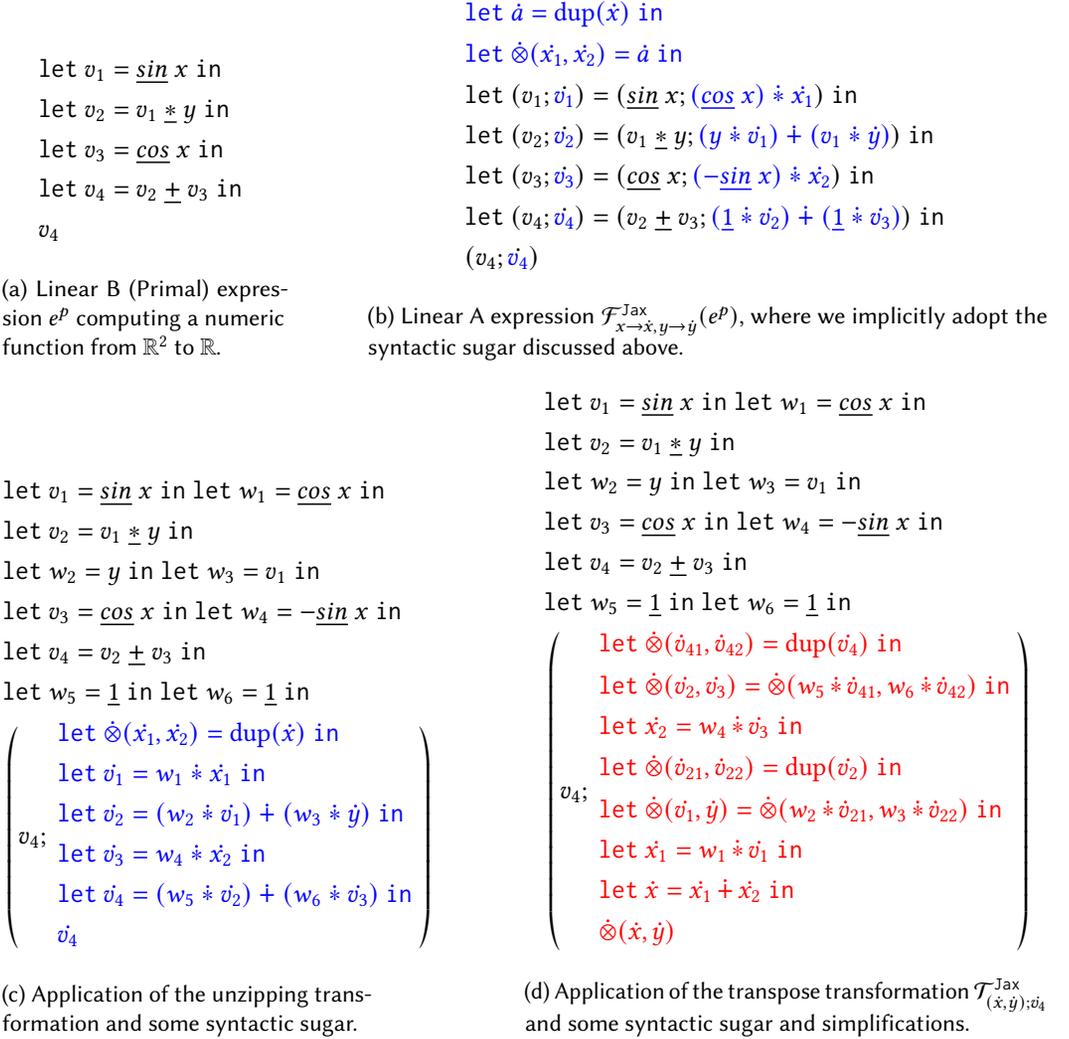

    \vspace{-0.45cm}
    \centering
    \begin{subfigure}[b]{0.27\textwidth}
        \centering
        \begin{align*}
            &\letterm{v_1}{\underline{sin}\mbox{ }x}{}\\
            &\letterm{v_2}{v_1 \mbox{ }\underline{*}\mbox{ } y}{}\\
            &\letterm{v_3}{\underline{cos}\mbox{ }x}{}\\
            &\letterm{v_4}{v_2 \mbox{ }\underline{+}\mbox{ } v_3}{}\\
            &v_4
        \end{align*}
        \caption{Linear B (Primal) expression $\expexprJAX{e}$ computing a numeric function from $\mathbb R^2$ to $\mathbb R$.}
        \label{fig:toyEx_linB}
    \end{subfigure}
    \hfill
    \begin{subfigure}[b]{0.65\textwidth}
        \centering
        \begin{align*} 
            &\color{blue} \letterm{\linvarJAX{a}}{\dupJAX{x}}{}\\
            &\color{blue} \letterm{\lintupleJAX{x_1}{x_2}}{\linvarJAX{a}}{}\\
            \color{black}
            &\letterm{\retJAX{v_1}{\color{blue}v_1}}
            {(\underline{sin}\mbox{ }x; 
            \color{blue}
                (\underline{cos}\mbox{ }x) \mbox{ }\dot{*}\mbox{ } \dot{x_1}
            \color{black})}\\
            &\letterm{\retJAX{v_2}{\color{blue} v_2}}
            {(v_1 \mbox{ }\underline{*}\mbox{ } y; 
            \color{blue}
            (y\mbox{ }\dot{*}\mbox{ }\dot{v_1})\mbox{ }\dot{+}\mbox{ }(v_1\mbox{ }\dot{*}\mbox{ }\dot{y})
            \color{black})}{}\\
            &\letterm{\retJAX{v_3}{\color{blue} v_3}}{(\underline{cos}\mbox{ }x;
            \color{blue}
            (-\underline{sin}\mbox{ }x) \mbox{ }\dot{*}\mbox{ }\dot{x_2}
            \color{black})}{}\\
            &\letterm{\retJAX{v_4}{\color{blue}v_4}}
            {(v_2 \mbox{ }\underline{+}\mbox{ } v_3;
            \color{blue}
            (\underline{1} \mbox{ }\dot *\mbox{ } \dot{v_2})\mbox{ }\dot{+}\mbox{ } (\underline{1} \mbox{ }\dot *\mbox{ }\dot{v_3})
            \color{black})}{}\\
            &\retJAX{v_4}{\color{blue}v_4}
        \end{align*}
        \caption{Linear A expression $\forwardJAX{x\to\dot x, y\to\dot y}{\expexprJAX{e}}$, where
        we implicitly adopt the syntactic sugar discussed above.}
        \label{fig:ex_forward_JAX}
    \end{subfigure} 
    \hfill 
    \begin{subfigure}[b]{0.35\textwidth}
        \centering
        \begin{align*} 
        &\letterm{v_1}{\underline{sin}\mbox{ }x}{\letterm{w_1}{\underline{cos}\mbox{ }x}{}}\\
        &\letterm{v_2}{v_1 \mbox{ }\underline{*}\mbox{ } y}{}\\
        &\letterm{w_2}{y}{\letterm{w_3}{v_1}{}}\\ 
        &\letterm{v_3}{\underline{cos}\mbox{ }x}{\letterm{w_4}{-\underline{sin}\mbox{ }x}{}}\\
        &\letterm{v_4}{v_2 \mbox{ }\underline{+}\mbox{ } v_3}{}\\
        &\letterm{w_5}{\underline{1}}{\letterm{w_6}{\underline{1}}{}}\\ 
        &\left(
            \begin{aligned}
                v_4;\mbox{ }
                \color{blue}
                \begin{aligned}
                    &\letterm{\lintupleJAX{x_1}{x_2}}{\dupJAX{x}}{}\\ 
                    &\letterm{\dot{v_1}}{w_1 \mbox{ }\dot{*}\mbox{ }\dot{x_1}}{}\\
                    &\letterm{\dot{v_2}}{(w_2 \mbox{ }\dot{*}\mbox{ } \dot{v_1}) \mbox{ }\dot{+}\mbox{ }
                    (w_3\mbox{ }\dot{*}\mbox{ } \dot{y})}{}\\ 
                    &\letterm{\dot{v_3}}{w_4 \mbox{ }\dot{*}\mbox{ }\dot{x_2}}{}\\
                    &\letterm{\dot{v_4}}{(w_5\mbox{ } \dot*\mbox{ } \dot{v_2})\mbox{ }\dot{+}\mbox{ }(w_6\mbox{ }\dot*\mbox{ }\dot{v_3})}{}\\
                    &\dot{v_4}
                \end{aligned}
            \end{aligned}
        \right)
        \end{align*}
        \caption{Application of the unzipping transformation and some syntactic sugar.}
        \label{fig:ex_unzipping_JAX}
    \end{subfigure}
    \hfill  
    \begin{subfigure}[b]{0.5\textwidth}
        \centering
        \begin{align*} 
        &\letterm{v_1}{\underline{sin}\mbox{ }x}{\letterm{w_1}{\underline{cos}\mbox{ }x}{}}\\
        &\letterm{v_2}{v_1 \mbox{ }\underline{*}\mbox{ } y}{}\\
        &\letterm{w_2}{y}{\letterm{w_3}{v_1}{}}\\ 
        &\letterm{v_3}{\underline{cos}\mbox{ }x}{\letterm{w_4}{-\underline{sin}\mbox{ }x}{}}\\
        &\letterm{v_4}{v_2 \mbox{ }\underline{+}\mbox{ } v_3}{}\\
        &\letterm{w_5}{\underline{1}}{\letterm{w_6}{\underline{1}}{}}\\ 
        &\left(
            \begin{aligned}
                v_4;\mbox{ }
                \color{red}
                \begin{aligned} 
                    &\letterm{\dot\otimes(\dot{v}_{41},\dot{v}_{42})}{\dupJAX{v_4}}{} \\
                    &\letterm{\lintupleJAX{v_{2}}{v_{3}}}{\dot\otimes(w_5\, \dot{*}\, \dot{v}_{41}, w_6\,\dot{*}\,\dot{v}_{42})}{}\\ 
                    &\letterm{\dot{x_2}}{w_4\,\dot*\,\dot{v_3}}{} \\
                    &\letterm{\dot\otimes(\dot{v}_{21},\dot{v}_{22})}{\dupJAX{v_2}}{} \\
                    &\letterm{\lintupleJAX{v_{1}}{y}}{\dot\otimes(w_2\, \dot{*}\, \dot{v}_{21}, w_3\,\dot{*}\,\dot{v}_{22})}{}\\                     
                    &\letterm{\dot{x_1}}{w_1\,\dot*\,\dot{v_1}}{} \\
                    &\letterm{\dot{x}}{\dot{x_1}\,\dot+\,\dot{x_2}}{} \\
                    &\dot{\otimes}(\dot{x},\dot{y})
                \end{aligned}
            \end{aligned}
        \right) 
        \end{align*}
        \caption{Application of the transpose transformation $\mathcal{T}^{\mathtt{Jax}}_{(\dot x, \dot y); \dot{v_4}}$ and some syntactic sugar and simplifications.}
        \label{fig:ex_transpose_JAX}
    \end{subfigure}
    \caption{Application of JAX Autodiff to the Linear B expression computing the map $(sin(x)*y)+cos(x)$. Color {\color{blue}blue} (resp.~{\color{red}red}) underlines the tangent (resp. tangent transpose) subexpressions introduced by JAX Autodiff.}
    \vspace{-0.38cm}
    \label{fig:exampleJAX} 
\end{figure*}

JAX Autodiff transformations are shown through the example in Figure~\ref{fig:exampleJAX}.

\paragraph*{Forward}
The transformation $\mathcal{F}^{\mathtt{Jax}}$ takes a purely primal expression $\expexprJAX{e}$ in Linear B and a mapping
$\contextFMJAX=\{\expvarJAX{x}_i \rightarrow \linvarJAX{y}_i\}^n_{i=1}$ which associates each primal $\expvarJAX{x}_i:\tau$ free in $\expexprJAX{e}$ with a corresponding tangent variable $\linvarJAX{y}_i:\tau$ and returns an expression in Linear A, which is a pair of a primal (computing the same value as $\expexprJAX{e}$) and a tangent.

\begin{figure*}
    \centering
    \vspace{-.3cm}
    \scalebox{.9}{\parbox{\textwidth}{
    \begin{align*}
        \forwardJAX{\expvarJAX{x} \to \linvarJAX{y}}{\expvarJAX{x}}
        &\definedas
        \retJAX{x}{y}\,,
    \\[3pt]
         \forwardJAX{\contextFMJAX,\{\expvarJAX{x}_i \to \linvarJAX{y}_i\}^2_{i=1}}{\underline{f}(\expvarJAX{x}_1, \expvarJAX{x}_2)}
         &\definedas\mbox{}
        \letterm{\expvarJAX{w_1}}{\underline{\partial_1 f}(\expvarJAX{x}_1,\expvarJAX{x}_2)}{} 
        \letterm{\expvarJAX{w_2}}{\underline{\partial_2 f}(\expvarJAX{x}_1, \expvarJAX{x}_2)}{}
        \\&\hspace{0.4cm}
        \retexpJAX{\underline{f}(\expvarJAX{x}_1, \expvarJAX{x}_2)}{\linmultJAX{w_1}{y_1} \dot{+}\linmultJAX{w_2}{y_2}}\,,
    \\[3pt]
        \forwardJAX{\contextFMJAX_1,\contextFMJAX_2,\{\expvarJAX{z}_i \to 
        \linvarJAX{u}_i\}^k_{i=1}}{\letterm{\expvarJAX{x}}{\expexprJAX{e}_1}{\expexprJAX{e_2}}}& \definedas\mbox{}\letterm{\linvarJAX{a}_1}{\dupJAX{u_1}}{}
         \letterm{\lintupleJAX{w_1}{v_1}}{\linvarJAX{a}_1}{}\cdots\\ 
        &\hspace{0.4cm}\letterm{\linvarJAX{a}_k}{\dupJAX{u_k}}{}
        \letterm{\lintupleJAX{w_k}{v_k}}{\linvarJAX{a}_k}{}\\
        &\hspace{0.4cm}\letterm{\retJAX{x}{y}}{\forwardJAX{\contextFMJAX_1,\{\expvarJAX{z}_i \to \linvarJAX{w}_i\}^k_{i=1}}{e_1}}{}\forwardJAX{\contextFMJAX_2,\{\expvarJAX{z}_i \to \linvarJAX{v}_i\}^k_{i=1}, x\to\dot y}{e_2}
    \end{align*}
    }}
    \vspace{-.3cm}
    \caption{Forward of some purely primal Linear B expressions. In the case of a $\mathtt{let}$ expression we decompose the set its free variables into 
    $dom(\contextFMJAX_i) = \FV{\expexprJAX{e}_i} \setminus (\FV{\expexprJAX{e}_1} \cap \FV{\expexprJAX{e}_2})$ and 
    $\{\expvarJAX{z}_i\}^k_{i=1} = \FV{\expexprJAX{e}_1} \cap \FV{\expexprJAX{e}_2}$.
    }
    \label{fig:FMJAXs}
\end{figure*}    

The definition is by induction on the grammar of $e^p$ above. Figure~\ref{fig:FMJAXs} presents the main cases, and we refer to Figure~\ref{fig:FMJAX} in the Appendix~\ref{app:JAX} for the other cases
; the remaining ones can be easily inferred. The main base case is the transformation of a numeric functional constant $\underline f$ (which we detail for $\underline f$  binary), implementing the chain rule $(f\circ g)' = (f'\circ g)\cdot g'$, where the primals $x_1,x_2$ are the outputs of $g$ and the tangents $\dot y_1,\dot y_2$ give the derivative $g'$ (under the form of a vector of partial derivatives). 
Then $\forwardJAX{}{\underline f(x_1,x_2)}$ returns a Linear A expression having in the primal position the image of the inputs along $f$, and in the tangent position the product of the derivative of $f$ at $(x_1,x_2)$ with the tangent variables $\dot y_1,\dot y_2$.

The definition of $\forwardJAX{}{\letterm{\expvarJAX{x}}{\expexprJAX{e_1}}{\expexprJAX{e_2}}}$ composes $\forwardJAX{}{e_1^p}$ with $\forwardJAX{}{e_2^p}$ by using the primal/tangent $\letterm{(x;\dot y)}{...}{...}$ of Linear A. The typing requires disjoint sets of tangent variables in the environments of the $\mathtt{let}$ and the $\mathtt{in}$ expressions. Then the composition is post-processed by a bunch of $\mathtt{dup}$ terms in order unify these environments. This construction seems ad-hoc, but our encoding in Section~\ref{sect:JAXtoLambdaLL}  will show it as a simple instance of the contraction of the environment of the linear logic additive conjunction $\with$. 

An easy induction gives that for any $\Gamma;\vdash\expexprJAX{e}:(\tau,\mathtt 1)$ and renaming $\contextFMJAX$,
we have that $\Gamma;\contextFMJAX(\Gamma)\vdash\forwardJAX{\contextFMJAX}{\expexprJAX{e}}:(\tau,\tau)$
is a well-typed judgment of Linear A (see Theorem 5.1 in \cite{radul2023you}). 

Figure \ref{fig:ex_forward_JAX} gives the result of applying the $\mathcal{F}^{\mathtt{Jax}}_{x\to\dot x, y\to\dot y}$ transformation to the expression $e^p$ in Figure~\ref{fig:toyEx_linB}. This gives a Linear A expression of type $\judgmentJAX{\expvarJAX{x}:\typrealJAX, \expvarJAX{y}:\typrealJAX}{\linvarJAX{x}:\typrealJAX, \linvarJAX{y}:\typrealJAX}
            {\forwardJAX{x\to\dot x, y\to\dot y}{\expexprJAX{e}}}{\typrealJAX}{\typrealJAX}$.
 Moreover, observe that the part in black in Figure~\ref{fig:ex_forward_JAX} is computing out the function $g(x,y)=(sin(x)*y)+cos(x)$ and the part in {\color{blue} blue} is computing the directional derivative of $g$ which is $\mathrm{D}_{(x',y')}(g) (x,y)=(cos(x)*y)*x'+sin(x)*y'-sin(x)*x'$.          

\paragraph*{Unzipping}
The  transformation $\mathcal{U}^{\mathtt{Jax}}$ disentangles primal and tangent values mapping Linear A into Linear B. The definition is by induction on Linear A and consists in splitting each primal/tangent $\mathtt{let}$ into a pure primal $\mathtt{let}$ and a pure tangent $\mathtt{let}$, moving the former towards the root of the syntactic tree and the latter towards the tangent leaves. The final result $\mathcal{U}^{\mathtt{Jax}}(e)$ will be a stack of pure primal $\mathtt{let}$'s $\letterm{x_1}{e_1^p}{\dots\letterm{x_n}{e_n^p}{\dots}}$ followed by a pair of a primal value and a purely tangent expression. 
Figure $\ref{fig:UnzipJAXs}$ gives the main cases of the definition of $\mathcal{U}^{\mathtt{Jax}}$
(see Figure~\ref{fig:UnzipJAX} in Appendix~\ref{app:JAX} for all cases)
, denoting by a metavariable $E$ such a stack of pure primal $\mathtt{let}$'s. 
E.g., Figure~\ref{fig:ex_unzipping_JAX} gives the unzipping of Figure~\ref{fig:ex_forward_JAX}.

For any Linear A expression $e$, $\unzippingJAX{e}$ is a well-typed expression of Linear B of the same type as $e$ and the same semantics (see Theorem 6.2 in \cite{radul2023you}).

\begin{figure*}
	\vspace{-.3cm}
    \scalebox{.9}{\parbox{\textwidth+2cm}{
        \centering
        \begin{align*}
            \unzippingJAX{\retJAX{x}{y}}\definedas\retJAX{x}{y}\,,
	        \quad & \quad
	    \unzippingJAX{\underline f(x_1,x_2)}\definedas(\underline f(x_1,x_2);\dot\otimes(\,))\,,
            \\
            \unzippingJAX{\letterm{\retJAX{x}{y}}{e_1}{e_2}}\definedas & \mbox{ }E_1 \texttt{ in } \letterm{x}{\expexprJAX{e_1}} E_2 \texttt{ in } \retexpJAX{\expexprJAX{e_2}}{\letterm{\linvarJAX{y}}{\linexprJAX{e_1}}{\linexprJAX{e_2}}}
        \end{align*}
        }}
    \vspace{-.3cm}
    \caption{Some cases of unzipping. In the case of a \texttt{let} expression, we suppose $\unzippingJAX{e_i}= E_i \texttt{ in } \retexpJAX{\expexprJAX{e_i}}{\linexprJAX{e_i}}$.}
    \label{fig:UnzipJAXs} 
\end{figure*}

\paragraph*{Transpose}
The transformation $\mathcal{T}^{\mathtt{Jax}}$ is an endotransformation of Linear B transposing the tangent part of an expression and keeping invariant the primal part.

The core of the definition of $\mathcal{T}^{\mathtt{Jax}}$ is on purely tangent expressions $\judgmentJAX{\Gamma}{\lincontextJAX{\Gamma}}{\linexprJAX{e}}{\typunitJAX}{\tautypJAX}$ giving a $\judgmentJAX{\Gamma}{\linvarJAX{u}:\tautypJAX}{\transpJAX{\theta}{\linvarJAX{u}:\tautypJAX}{\linexprJAX{e}}}{\typunitJAX}{\otimes \theta}$ depending on an enumeration $\theta$ of $\lincontextJAX{\Gamma}$ and a free tangent variable $\dot u$ associated with output of $\linexprJAX{e}$. The output type of $\transpJAX{\theta}{\linvarJAX{u}:\tautypJAX}{\linexprJAX{e}}$ is $(\typunitJAX,\otimes\theta)$, where $\otimes\theta$ represents the nested product $\tau_1\otimes(\tau_2\otimes\dots(\tau_{n-1}\otimes\tau_n)\dots)$ for $\theta=(\dot y_1:\tau_1,\dots, \dot y_n:\tau_n)$. Figure~\ref{fig:TlinJAXs} gives the main cases of this definition, for the other cases see Figure~\ref{fig:TlinJAX} in the Appendix~\ref{app:JAX}. 
\allowdisplaybreaks
\begin{figure*}
    \vspace{-.3cm}
    \scalebox{.88}{\parbox{\textwidth+1.5cm}{
    \centering 
    \begin{align*}
        \transpJAX{\linvarJAX{x}:\tautypJAX}{\linvarJAX{u}:\tautypJAX}{\linvarJAX{x}}\definedas \mbox{}
        \linvarJAX{u}
        \quad
        \transpJAX{\{\linvarJAX{x}:\tautypJAX,\linvarJAX{y}:\tautypJAX\}}{\linvarJAX{u}:\tautypJAX}{\linvarJAX{x}\dot{+}\linvarJAX{y}}\definedas &\; \mbox{}
            \dupJAX{u}
        \quad
        \transpJAX{\emptyset}{\linvarJAX{u}:\tautypJAX}{\linzeroJAX{\tautypJAX}}\definedas  \mbox{}
            \dropJAX{\linvarJAX{u}}
        \quad
        \transpJAX{\theta}{\linvarJAX{u}:\typunitJAX}{\dropJAX{\linexprJAX{e}}}\definedas  \mbox{}
        \linzeroJAX{\otimes \theta}
        \quad
        \transpJAX{\linvarJAX{y}:\tautypJAX}{\linvarJAX{u}:\tautypJAX}{x\dot{*}\linvarJAX{y}}\definedas  \mbox{}
        x\dot{*}\linvarJAX{u}  
        \\
        \transpJAX{\theta}{\linvarJAX{u}:\tautypJAX}{\letterm{\linvarJAX{x}}{\linexprJAX{e_1}}{\linexprJAX{e_2}}}\definedas &\;\mbox{}
        \letterm{\lintupleJAX{x}{u_2}}
        {\transpJAX{\linvarJAX{x}:\sigmatypJAX,\theta\cap \linFV{\linexprJAX{e_2}}}{\linvarJAX{u}:\tautypJAX}{\linexprJAX{e_2}}}
        {\letterm{\linvarJAX{u}_1}
        {\transpJAX{\theta\cap \linFV{\linexprJAX{e_1}}}{\linvarJAX{x}:\sigmatypJAX}{\linexprJAX{e_1}}}
        {\SplitInvJAX[\theta]{\linvarJAX{u}_1,\linvarJAX{u}_2}}}
    \end{align*}
    }}
    \vspace{-.3cm}
    \caption{Transpose of some purely tangent Linear B expressions.}
    \label{fig:TlinJAXs}
\end{figure*}

The definition of $\mathcal{T}^{\mathtt{Jax}}(\letterm{\linvarJAX{x}}{\linexprJAX{e_1}}{\linexprJAX{e_2}})$ reverses the order of the composition and the dependence between $\dot e_1$ and $\dot e_2$. The transpose first computes $\mathcal{T}^{\mathtt{Jax}}(\linexprJAX{e_2})$ storing its result in a pair $(\dot u_1, \dot y)$ and then performs $\mathcal{T}^{\mathtt{Jax}}(\linexprJAX{e_1})$ by using the result $\dot u_1$ which is associated to the $\dot x$ dependence of $\dot e_2$ from $\dot e_1$. Notice also the duality between $\mathrm{dup}$ and $\dot +$ and between $\mathrm{drop}$ and $\dot 0$. In our encoding into $\lambdaLL$, $\mathrm{dup}$ will be the diagonal in the additive conjunction $\with$ and $\mathrm{drop}$ the introduction of the neutral element of this conjunction $\top$.

The transpose transformation is then lifted to the primal constructs of Linear B by a simple commutation: $\transpJAX{\theta}{\linvarJAX{u}:\tautypJAX}{\retexpJAX{\expexprJAX{e}}{\linexprJAX{e}}}\definedas  \retexpJAX{\expexprJAX{e}}{\transpJAX{\theta}{\linvarJAX{u}:\tautypJAX}{\linexprJAX{e}}}$ and $\transpJAX{\theta}{\linvarJAX{u}:\tautypJAX}{\letterm{\expvarJAX{x}}{\expexprJAX{e}}{d}}\definedas  
\letterm{\expvarJAX{x}}{\expexprJAX{e}}{\transpJAX{\theta}{\linvarJAX{u}:\tautypJAX}{d}}$, see Figure~\ref{fig:TJAX} in the Appendix~\ref{app:JAX}. 

Given a Linear B expression $\Gamma;\dot\Gamma\vdash d:(\tau,\sigma)$ and an enumeration $\theta$ of $\dot\Gamma$ in $d$, we have that $\Gamma;\linvarJAX{u}:\sigma\vdash\transpJAX{\theta}{\linvarJAX{u}:\sigma}{d}:(\tau,\otimes\theta)$
is a well-typed judgement of Linear B
(see Theorem 7.1 in \cite{radul2023you}). 

Figure~\ref{fig:ex_transpose_JAX} shows the application of $\mathcal{T}^{\mathtt{Jax}}_{(\dot x, \dot y); \dot{v_4}}$ to the expression in Figure~\ref{fig:ex_unzipping_JAX}. Precisely, Figure~\ref{fig:ex_unzipping_JAX} slightly simplifies the result of $\mathcal{T}^{\mathtt{Jax}}_{(\dot x, \dot y); \dot{v_4}}$ without changing its semantics, for the sake of readability. For instance, the first two red lines in Figure~\ref{fig:ex_transpose_JAX} corresponds to the transpose of the let definition of ${\color{blue}{\dot v_4}}$ in Figure~\ref{fig:ex_unzipping_JAX}, the third red line corresponds to the transpose of the let definition of ${\color{blue}\dot{v_3}}$ and so forth. This yields an expression of type $(\mathbb{R}; \mathbb{R} \otimes \mathbb{R})$ with free variables $\expvarJAX{x}:\typrealJAX$, $\expvarJAX{y}:\typrealJAX$, $\linvarJAX{v_4}:\typrealJAX$.

\section{$\lambdaLL$} \label{sect:lambdaLL} 
We introduce $\lambdaLL$ as an extension of the linear logic $\lambda$-calculus (see e.g.~\cite{ABRAMSKY19933,HYLAND1993273,MARAIST1995370,Wadler_linear,Barber1996DualIL,Ehrhard_cbpv,Zdancewic_linear}) to the ground type of the real numbers $\typR$ and a set of functional symbols which are associated with differentiable functions.
The presentation of $\lambdaLL$ follows a standard pattern and the acquainted reader may want to jump to subsections~\ref{subsect:extensional_equivalence},~\ref{subsect:examples} and~\ref{subsect:cost_model} to have an immediate preview of the special features that will be used to study JAX Autodiff. 

\subsection{Syntax and Type System}
\label{subsect:syntax} 
The grammar of types is defined as  follows:
\begin{equation*} \tag{Types}
    A,B,C \bnf 
        \typR 
        \orsintax \typlollipop{A}{B}
        \orsintax \typone 
        \orsintax A\otimes B
        \orsintax \typtop
        \orsintax A\& B
        \orsintax \typbang{A}  
    \label{eq:types}
\end{equation*}

Linear types distinguish between a resource of a given type $A$ that is used exactly once from a resource of  \emph{exponential modality} $\typbang{A}$ which can be used at will (zero, one, or many times). 

Linear A data-types are nested tuples, representing multidimensional numeric arrays. We express them with two different families mirroring the distinction between primal and tangent data.  
\begin{align*}
	\tag{$\otimes\oc$-sequence Types}
    D,E &\bnf \typR \orsintax \typone \orsintax \typtensor{\oc D}{\oc E}\label{eq:mult_types}\\
    \tag{$\with$-sequence Types}
    L,H &\bnf \typR \orsintax \typtop \orsintax H\with L 
    \label{eq:mult_add_types}
\end{align*}

These two notions of tuples appear in most of the literature on formalisations of AD discussed in the introduction. For example, following the terminology of \cite{Vakar2021CHADCH,LucatelliCHAD}, $\with$-sequences correspond to products of ``linear types'', whereas $\otimes\oc$-sequences correspond to products of ``cartesian types''. See also Remark~\ref{rk:primal_bang_tangent}, which shows how the latter can be viewed, in a sense, as the exponential promotion of the former.

\begin{figure*} 
    \scalebox{.95}{\parbox{\textwidth+1cm}{
        \begin{center}
            \RightLabel{\!\footnotesize{$\typrvar$}}
            \AxiomC{\phantom{$A$ linear}}
            \UnaryInfC{$\judgment{x:A}{x}{A}$}
            \DisplayProof
            \qquad
            \RightLabel{\!\footnotesize{$\typrbangL$}}
            \AxiomC{\phantom{$\judgment{x:A, \Delta}{M}{B}$}}
            \UnaryInfC{$\judgment{\oc x:\oc A}
            {x}{A}$}
            \DisplayProof
           \qquad
            \RightLabel{\!\footnotesize{$\typrbangr$}}
            \AxiomC{$\judgment{\oc\Gamma}{M}{A}$}
            \UnaryInfC{$\judgment{\oc\Gamma}{\bangterm{M}}
            {\typbang{A}}$}
            \DisplayProof
            \qquad
            \RightLabel{\!\footnotesize{$\typbangW$}}
            \AxiomC{$\judgment{\Delta}{M}{B}$}
            \UnaryInfC{$\judgment{\oc x:\oc A, \Delta}{M}{B}$}
            \DisplayProof 
        \end{center}
        
        \medskip
        \begin{center}
            \RightLabel{\!\footnotesize{$\typrlollipopR$}}
            \AxiomC{$\judgment{p:A,\Delta}{M}{B}$}
            \UnaryInfC{$\judgment{\Delta}{\absterm{p}{M}}{\typlollipop{A}{B}}$}
            \DisplayProof
            \quad
            \RightLabel{\!\footnotesize{$\typrlollipopL$}}
            \AxiomC{$\judgment{\oc\Gamma_1,\Delta_1}{M}
            {\typlollipop{A}{B}}$}
            \AxiomC{$\judgment{\oc\Gamma_2,\Delta_2}{N}{A}$}
            \BinaryInfC{$\judgment{\oc\Gamma_1\cup\oc\Gamma_2,\Delta_1,\Delta_2}{\appterm{M}{N}}{B}$}
            \DisplayProof
        \end{center}
        
        \medskip
        \begin{center}
            \RightLabel{\!\footnotesize{$\typroneR$}}
            \AxiomC{\phantom{A}}
            \UnaryInfC{$\judgment{}{\emptytupleterm}
            {\typone}$}
            \DisplayProof
            \;
            \RightLabel{\!\footnotesize{$\typroneL$}}
            \AxiomC{$\judgment{\Delta}{M}{B}$}
            \UnaryInfC{$\judgment{\emptytupleterm:\typone, \Delta}
            {M}{B}$}
            \DisplayProof
           \;
            \def\defaultHypSeparation{\hskip .01in}
            \RightLabel{\!\footnotesize{$\typrtensorR$}}
            \AxiomC{$\judgment{\oc\Gamma_1,\Delta_1}{M}{A}$}
            \AxiomC{$\judgment{\oc\Gamma_2,\Delta_2}{N}{B}$}
            \BinaryInfC{$\judgment{\oc\Gamma_1\cup\oc\Gamma_2,\Delta_1, \Delta_2}
            {\tupleterm{M}{N}}{\typtensor{A}{B}}$}
            \DisplayProof
            \;
            \def\defaultHypSeparation{\hskip .01in}
            \RightLabel{\!\footnotesize{$\typrtensorL$}}
            \AxiomC{$\judgment{\PatA:A,\PatB:B,\Delta}{M}{C}$}
            \UnaryInfC{$\judgment{\tupleterm{\PatA}{\PatB}:A\otimes B, \Delta}
            {M}{C}$}
            \DisplayProof
        \end{center}
        
        \medskip
        \begin{center}
            \RightLabel{\!\footnotesize{$\typrwithR$}}
            \AxiomC{$\judgment{\Delta}{M_1}{A_1}$}
            \AxiomC{$\judgment{\Delta}{M_2}{A_2}$}
            \BinaryInfC{$\judgment{\Delta}{\nTuple{M_1, M_2}}{A_1\with A_2}$}
            \DisplayProof
            \quad
            \RightLabel{\!\footnotesize{$\typrwithL{i},\,\tiny i\in\{1,2\}$}}
            \AxiomC{$\judgment{\PatA_i:A_i,\Delta}{M}{B}$}
            \UnaryInfC{$\judgment{\nTuple{\PatA_1,\PatA_2}:A_1\with A_2,\Delta}{M}{B}$}
            \DisplayProof
            \quad
            \RightLabel{\!\footnotesize{$\typtop$}}
            \AxiomC{\phantom{$i\in\{1,2\}$}}
            \UnaryInfC{$\judgment{\Delta}{\nTuple{\,}}{\typtop}$}
            \DisplayProof
        \end{center}
        
        \medskip
        \begin{center}
            \RightLabel{\!\!\footnotesize{$\typrreal$}}
            \AxiomC{$r$ real number}
            \UnaryInfC{$\judgment{}{\realterm}{\typR}$}
            \DisplayProof
            \;
            \RightLabel{\!\!\footnotesize{$\typrfun{2}$}}
            \AxiomC{$f$ binary map}
            \UnaryInfC{$\vdash \underline f :  \oc\typR\otimes \oc\typR \multimap \oc\typR$}
            \DisplayProof
%
	   \;
            \RightLabel{\!\!\footnotesize{$\typrzero$}}
            \AxiomC{\phantom{A}}
            \UnaryInfC{$\judgment{\Delta}{\underline 0}{\typR}$}
            \DisplayProof
            \;
            \RightLabel{\!\footnotesize{$\typrsum$}}
            \AxiomC{\phantom{A}}
            \UnaryInfC{$\judgment{}{\dot +}{\typR\with\typR\multimap\typR}$}
            \DisplayProof
            \;
            \RightLabel{\!\!\footnotesize{$\typrmult$}}
            \AxiomC{\phantom{A}}
            \UnaryInfC{
            	$\judgment{}{\dot*}{\typR\multimap\typR\multimap\typR}$
	    }
            \DisplayProof
        \end{center}
        }}
        \caption{$\lambdaLL$ Typing Rules. 
        For short, we consider only the case of $\underline{f}$ binary, the general case being immediate.}\label{fig:typing_LL}
\end{figure*}

The generation rules for the syntax of well-typed terms of $\lambdaLL$ are given in Figure~\ref{fig:typing_LL}, together with the typing rules. 
As for Linear A, we adopt a Church style typing: each variable has its type fixed once and for all. It is convenient to handle destructors as \emph{patterns} binders, these latter being constructors of pairwise distinct variables:
\begin{align*}
	\tag{Patterns}
    \PatA, \PatB
    	&\bnf x 
	\orsintax \oc x
	\orsintax \emptytupleterm
	\orsintax  \tupleterm{\PatA}{\PatB} 
	\orsintax \nTuple{\PatA,\PatB} 
\end{align*} 
where we suppose $\FV{p}\cap\FV{q}=\emptyset$, so a variable occurs at most once in a pattern.
We say that a pattern is \emph{exponential} whenever it is of the form $\oc x$.
We use meta-variables $\PatAddA, \PatAddB$ for denoting \emph{patterns of $\with$-sequence types}.

A typing environment $\Gamma$ is a finite set of patterns. We write $\typbang{\Gamma}$ whenever all patterns in the environment are exponential. 

As for Linear A, $\realterm$ and $\underline{f}$ are meta-variables varying over, respectively, numerals for real numbers, 
and $n$-ary numeric functions: Figure~\ref{fig:typing_LL} details only the case of $n=2$.
All numerical functions are differentiable and are equipped with their partial derivatives $\underline{\partial_i f}$.
We have dedicated symbols for the specialised sum $\dot+$ and product $\dot*$ with scalars which have a different typing with respect to the typing of their sibling numerical functions.

We adapt the same conventions as for Linear A, in particular commas stand for ``disjoint unions''.  A difference with Linear A is that now typing environments are sets of patterns, not simply variables: so $\Delta$ is disjoint from $\Delta'$ means that no variable appears in both a pattern of $\Delta$ and a pattern of $\Delta'$.  
Namely, the rule $\typrlollipopL$  in Fig.~\ref{fig:typing_LL} is asking that the free variables in common between $M$ and $N$ belong to an exponential pattern in the environment.  
Notice that a variable $x:\oc A$ of exponential type is not an exponential pattern and so cannot be duplicated or erased. This restriction is known to be necessary to guarantee the subject reduction in a linear type system\footnote{\label{footnote:subject_reduction}In fact, $f:A\multimap \oc B, x:A\vdash (\lambda y^{\oc B}.(y,y))(fx): \oc B\otimes\oc B$ would be derivable if it were possible to copy variables of exponential type, but $f:A\multimap \oc B, x:A\vdash  (fx,fx): \oc B\otimes\oc B$ would not. Similar examples are known in the literature, see e.g.~\cite{Wadler_no_subst}. The solution adopted here is in the spirit of Barber's dual intuitionistic linear logic \cite{Barber1996DualIL}, based on dual environments. }. 

\begin{remark}
All rules except those in the last line are standard in linear logic \cite{girard1987linear}. 
Notice that $\dot+$ takes an additive  pair $\typR\with\typR$ and returns $\typR$, while $\dot*$ morally takes a multiplicative pair $\typR\otimes\typR$ and returns $\typR$, reflecting the difference in linear algebra between addition, which is a linear operation, and scalar multiplication, which is a bilinear operation.
Notice here the crucial difference of the two linear logic conjunctions: the multiplicative conjunction $\otimes$ corresponds to the tensor product of vector spaces, transforming any bilinear map into a unique linear map, while the additive conjunction $\with$ is the direct product of vector spaces, so $\typR^2\with\typR^4$ is isomorphic to $\typR^6$, while $\typR^2\otimes\typR^4$ is isomorphic to $\typR^8$.
\end{remark}

The set of free variables $\FV M$ of a term $M$ is defined as usual, in particular $\FV{\lambda \PatA.M}= \FV{M}\setminus\FV{\PatA}$.  Giving $\Gamma=\PatA_1:A_1,\dots, \PatA_n:A_n$, we define $\FV{\Gamma}\definedas\bigcup_i\FV{\PatA_i}$.  
We may write $\lambda\PatA^A.M$ if we wish to explicit the type of $\PatA$.

\paragraph*{Notational conventions}
We may use the let notation for the application to an abstraction, i.e.~$\appterm{(\absterm{\PatA}{M})}{N}$ can be written:
$
	\letterm{\PatA}{N}{M}
$.
It is known that the formula $\typone\with A$ expresses in linear logic the affine resource of type $A$: a value of type $A$ that can be used at most once. 
This modality will be used in our encoding of JAX Autodiff, so we introduce the following notation: $ \affinebang A\definedas \typone\with A $ and $\affinebang M\definedas \nTuple{(\,), M}$.

In particular, $\affinebang\PatA$ is a pattern of type $\affinebang A$, whenever $\PatA$ is a pattern of type $A$.
The following typing rules are then derivable:
\begin{center} 
        \AxiomC{$\judgment{\PatA:A,\Delta}{\!M\!}{B}$}
        \UnaryInfC{$\judgment{\affinebang \PatA: \affinebang A,\Delta}{\!M\!}{B}$}
     	\DisplayProof
	\,
         \AxiomC{$\judgment{\Delta}{\!M\!}{B}$}
         \UnaryInfC{$\judgment{\affinebang \PatA: \affinebang A,\Delta}{\!M\!}{B}$}
	\DisplayProof
	\,
        \AxiomC{$\judgment{\oc\Delta,\affinebang\Sigma}{\!M\!}{B}$}
        \UnaryInfC{$\judgment{\oc\Delta,\affinebang\Sigma}{\!\affinebang M\!}{\affinebang B}$}
    \DisplayProof 
\end{center}
Figure~\ref{fig:mainfigure} gives examples of $\lambdaLL$ terms adopting these conventions. 

Some additional notation will be useful for the additive tuples. First, we may denote the $n$-fold additive product $\nTuple{M_1,\nTuple{M_{2},\dots M_n} \dots}$ as an $n$-ary tuple $\nTuple{M_1, M_2, \dots, M_n}$. We can use shortcut like $\nTuple{M_i}_{i=1}^n$, or even $\nTuple{M_i}_{i}$ if $1$ and $n$ are clear from the context or irrelevant. We adopt similarly notation for the types: $\with_{i=1}^nA_i$ or $\with_iA_i$. 

We may use set-theoretical notation to manipulate sequences. Namely, if $\theta$ is a sequence of variables, $x\in\theta$ (resp.~$X\subseteq\theta$) means that $x$ (resp.~$X$) varies over all elements (resp.~sets of elements) in $\theta$. Moreover if $\theta= (x_1,x_2,x_3,x_4)$, then $\theta\setminus\{x_2,x_3\}=(x_1,x_4)$ and $\theta\cap\{x_2,x_3\}=(x_2,x_3)$. We may write $\theta\setminus x$ if it is clear we are meaning $\theta\setminus\{x\}$.

Given $\with_{i=1}^nA_i$ and a set $\mathcal I$, we define the 
\emph{splitting} and \emph{fusion} terms:
\begin{align}
    \label{Notation:sigma}
        \Split[\with_{i=1}^nA_i]{\mathcal I} &
        \definedas \lambda\nTuple{x_i}_{i=1}^n.\nTuple{\nTuple{x_i}_{i\in \mathcal I},\nTuple{x_i}_{i\notin \mathcal I}}\;,
        &
        \SplitInv[\with_{i=1}^nA_i]{\mathcal I} 
       &\definedas \lambda\nTuple{\nTuple{x_i}_{i\in \mathcal I},\nTuple{x_i}_{i\notin \mathcal I}}.\nTuple{x_i}_{i=1}^n\;. 
  \end{align}
Note that $\Split[\with_{i=1}^nA_i]{\mathcal I} : \with_{i=1}^nA_i\multimap (\with_{i\in\mathcal I}^nA_i)\with(\with_{i\notin\mathcal I}^nA_i)$ and $ \SplitInv[\with_{i=1}^nA_i]{\mathcal I} : (\with_{i\in\mathcal I}^nA_i)\with(\with_{i\notin\mathcal I}^nA_i)\multimap \with_{i=1}^nA_i$.

Linear sum uses prefix notation, but we allow infix notation if $M$ is a pair: $N_1 \dot+ N_2 \approx \sumterm{\lintupleterm{N_1}{N_2}}$. 
We also extend the specialised operators $\zeroterm{}$, $\dot +$ and $\dot *$ to any $\with$-sequence type:
\begin{gather*}
	0_{\with_i H_i}
	\definedas \nTuple{0_{H_i}}_i
	\,,
	\quad
	\dot+_{\with_i H_i}
	\definedas \lambda\nTuple{\nTuple{h_i}_i,\nTuple{h'_i}_i}. \nTuple{\dot+_{H_i}\nTuple{h_i, h'_i}}_i
	\,, 
    \quad
	\dot*_{\with_i H_i}
	\definedas \lambda x.\lambda \nTuple{h_i}_i.{\nTuple{\dot*_{H_i}(x, h_i)}_i}
	\,.
\end{gather*}

Notice that these are closed terms of type: $0_{H}:H$, $\dot +_H:H\with H\multimap H$, $\dot *_H: \typR\multimap H\multimap H$. 

\begin{remark}
\label{rk:linearityLL}
The latter definition highlights the subtle notion of linearity expressed by linear logic typing. Consider $\dot*_{\typR\with\typR} \definedas \lambda x.\lambda \nTuple{h_1,h_2}.{\nTuple{x \dot* h_1, x \dot* h_2}}$,
a term of type $\typR\multimap \typR\with\typR \multimap\typR\with\typR$. 
Here, the parameter $x$ of the outermost abstraction has the linear type $\typR$, even though it occurs twice in the body. This is an instance of additive contraction, where the occurrences of $x$ belong to different components of a $\with$-tuple. The $\beta$-reduction rules of $\lambdaLL$ ensure that these occurrences do not interact during evaluation, so the resulting term eventually depends on $x$ linearly. Categorically, $\typR\with\typR$ corresponds to the categorical product $\typR\times\typR$, with additive contraction given by the diagonal morphism. 

Our paper shows that this notion of linearity --- distinct from the notion of “syntactically occurring exactly once” --- is fully compatible with JAX Autodiff. In particular, Section~\ref{sect:transpose} explains how the definition of the transpose transformation must carefully account for multiple occurrences of a variable arising from additive contractions.
\end{remark}
 
\subsection{$\beta$-reduction}
	\label{subsect:reduction}

The $\beta$-reduction $\rightarrow$ of $\lambdaLL$ is defined by the context closure of the $\beta$-rules given in Figure~\ref{fig:beta_rules}.
We briefly recall standard notions, see Appendix~\ref{app:rwprop} for a more extensive presentation.

The rule $\beta_\lambda$ replaces a pattern $\PatA$ by a term $V$, supposing this latter has a ``structure compatible with $\PatA$''. This is formalised by the notion of \emph{a value $V$ for a pattern $\PatA$}. A value for a variable $x$ of type $A$ is any term of type $A$, a value for $(\,)$ is $(\,)$, a value for $\oc x$ is $\oc M$ for $M$ any term, a value for $(\PatA_1,\PatA_2)$ (resp.~$\nTuple{\PatA_1,\PatA_2}$) is $(V_1,V_2)$ (resp.~$\nTuple{V_1,V_2}$) where $V_i$ is a value for $\PatA_i$.
We then generalise the standard variable substitution $M\{N/x\}$ to the \emph{substitution $M\{V/\PatA\}$ of a pattern $\PatA$ for a value $V$ in a term $M$}, 
by dispatching all components in $V$ to the free occurrences in $M$ of $\FV\PatA$, i.e.:
$M\{(\,)/(\,)\}\definedas M$, 
$M\{\oc N/\oc x\}\definedas M\{N/x\}$, 
$M\{(V_1,V_2)/(\PatA_1,\PatA_2)\}\definedas M\{\nTuple{V_1,V_2}/\nTuple{\PatA_1,\PatA_2}\}\definedas M\{V_1/\PatA_1\} \{V_2/\PatA_2\}$.
 
\begin{figure}
	\begin{align*}
		\beta_\lambda:&
			\appterm{(\absterm{\PatA}{M})}{V}
			\rightarrow
			M\{V/\PatA\}
		&
		\beta_{F}:&
			\underline{f}{(\bangterm{\realterm_1},\bangterm{\realterm_2})}
			\rightarrow
			\bangterm{\underline{f(r_1,r_2)}}
		\\
			\beta_{\sumterm{}}:&
					\dot+ \nTuple{\underline{r_1},  \underline{r_2}}
					\rightarrow
					\underline{(r_1 + r_2)}
			&
					\beta_{\multterm{}}:&
			\dot*\, {\underline{r_1}}\, \underline{r_2}
			\rightarrow
			\underline{(r_1 \cdot r_2)}
	\end{align*}
	\vspace{-.3cm}
	\caption{$\beta$-reduction. $\beta_\lambda$ supposes $V$ to be a value for the pattern $\PatA$.
	Steps $\beta_F$, $\beta_{\sumterm{}}$ and $\beta_{\multterm{}}$ are called \emph{numeric steps} or \emph{flops}.}
	\label{fig:beta_rules}
\end{figure}

We denote by $\rightarrow^*$ and $=_{\beta}$ respectively the reflexive-transitive and the equivalence closure of $\rightarrow$. A \emph{$\beta$-normal form}, $\beta$-nf for short, is a term $M$ s.t.~there is no $N$ s.t.~$M\rightarrow N$.

The $\beta$-reduction is designed on the top of the LL cut-elimination. This yields a well behaving rewriting system, satisfying crucial properties such the following ones. 

\begin{theorem}[Subject Reduction]
	\label{th:subjred}
	Let $\Gamma\vdash M:A$ and $M\rightarrow N$, then $\Gamma\vdash N:A$. 
\end{theorem}

\begin{theorem}[Strong Normalisaton]\label{th:SN}
	Every term in $\lambdaLL$ is strongly normalizing. 
\end{theorem}

\begin{theorem}[Confluence]
	\label{th:confluence}
	If $M'\, { }^*\!\!\leftarrow M \rightarrow^* M''$ then there is $N$ such that $M' \rightarrow^* N\, { }^*\!\!\leftarrow M''$.
\end{theorem} 

\subsection{Logical Equivalence $\sim$}
	\label{subsect:extensional_equivalence}
	
The $\beta$-equivalence is too narrow to compare terms of complex types: our ultimate goal is to compute numeric functions and we are interested whether two terms can be interchanged in a program of ground type without changing the numeric function computed by this latter. A typing system offers a way of extending $\beta$-equivalence by lifting the extensional behaviour over ground types,  using the notion of logical relation.

\begin{definition}[$\sim_A$, $\sim_{\Gamma\vdash A}$]
	\label{def:logical_closed}
Given a type $A$, $\sim_A$ is a binary relation between closed terms of type $A$:
\begin{itemize}
	\item $M \sim_\typR N$ or $M\sim_\typone N$ iff $M=_\beta N$,
	\item $M \sim_{A_1\otimes A_2} N$, iff $M\rightarrow^* (M_1,M_2)$, $N\rightarrow^* (N_1,N_2)$ and $M_i\sim_{A_i} N_i$ for $i\in\{1,2\}$,
	\item $M \sim_{\oc A} N$ iff  $M\rightarrow^* \oc M$, $N\rightarrow^* \oc N$ and $M \sim_{A} N$, 
	\item $M \sim_{\typtop} N$ always,
	\item $M \sim_{A_1\with A_2} N$ iff $(\lambda\nTuple{x_1,x_2}.x_i)M\sim_{A_i}(\lambda\nTuple{x_1,x_2}.x_i)N$, for every $i\in\{1,2\}$,
	\item $M \sim_{A\multimap B} N$ iff for all $M'\sim_A N'$, $MM'\sim_BNN'$.
\end{itemize}
Given typing judgments $\Gamma\vdash M:A$ and $\Gamma\vdash N:A$, with $\Gamma=\PatA_1:A_1,\dots,\PatA_n:A_n$, we set:
    $M\sim_{\Gamma\vdash A} N$ iff $\forall i\leq n,\forall V_i\sim_{A_i}V'_i, M\{V_1/p_1,\dots,V_n/p_n\}\sim_AN\{V'_1/\PatA_1,\dots,V'_n/\PatA_n\}$. 
\end{definition}

Henceforth, we may omit type annotation on $\sim_{A}$ or $\sim_{\Gamma\vdash A}$ whenever clear from the context or irrelevant.  
We refer to Appendix~\ref{app:logicEq} for more details.

\subsection{$\with$-sequence Types as Vector Spaces} 
\label{subsect:examples} 
Let us consider a $\with$-sequence type $H$. Notice that the closed $\beta$-nf of $H$ are nested tuples of real numbers. In fact, the $\beta$-equivalence classes of $H$ define a real vector space of dimension equal to the number of occurrences of $\typR$ in $H$: vector addition is given by $\dot +$ and scalar multiplication by $\dot *$. Normalisation and confluence assure that one can select $\beta$-nf's as canonical representatives of the elements of this vector space, and the rewriting rules lift the algebraic properties of addition and multiplication over $\typR$ to $H$, e.g.~$M_1\dot +(M_2\dot + M_3)=_\beta (M_1\dot +M_2)\dot + M_3$, for closed terms of type $H$. 

Moreover, this vector space $H$ is associated with a canonical base $\mathcal B_H$:
the base cases are $\mathcal B_\typR\definedas\{\underline 1\}$ and
$\mathcal B_{\typtop}\definedas\{\nTuple{\,}\}$, while 
$\mathcal B_{H_1\with H_2}\definedas\{\nTuple{V_1,\underline 0}, \,\text{s.t.}\, V_1\in \mathcal B_{H_1}\}\cup\{\nTuple{\underline 0,V_2}\,\text{s.t.}\, V_2\in \mathcal B_{H_2}\}$.

Similarly, one defines an inner product $\mathcal I_H$ as a closed term of type $H\otimes H\multimap \typR$ by induction on $H$:
$\mathcal I_\typR$ is $\lambda (x,y).\dot* x  y$, 
while $\mathcal I_{\typtop}$ is $\lambda (x,y).\zeroterm{}$, 
and finally $\mathcal I_{H_1\with H_2}
		\definedas\lambda (\nTuple{x_1,x_2},\nTuple{y_1,y_2}).
			\mathcal I_{H_1} (x_1,y_1)
			\dot+
			\mathcal I_{H_2} (x_2,y_2)	
$.

In this way, one can recover syntactically the isomorphism between an euclidean space and its dual, by $\Iso[H]: H\multimap (H\multimap \typR)$ and $\Osi[H]: (H\multimap \typR)\multimap H$:
\begin{align*}
	\Iso[H]&=\lambda h.\lambda h'.\mathcal I_H (h,h')\,,
	&
	\Osi[H]&=\lambda f.\!\!\sum_{V\in\mathcal B_H}\!\! (f(V)) \dot*_H V\,.
\end{align*}
Section~\ref{sect:transpose} compares our transpose transformation with the one obtained by using this isomorphism.

Similar constructions are possible with a generic type $A$, but cannot be defined in general by syntactical terms, in fact  the dimension of a vector space associated with an exponential type $\oc A$ may be infinite. Quantitative semantics (e.g.~\cite{ehrhardfs,Ehrhard02,LairdMM13}) or resource $\lambda$-calculus (e.g.~\cite{difftaylor,Ehrhard11}) provide more suitable frameworks for describing such spaces. We do not explore here these systems, as the transpose transformation is restricted to $\with$-sequence types.

\subsection{Workload}
	\label{subsect:cost_model}

We adapt the notion of workload $\mathcal{W}$ from~\cite[Section 4.3]{radul2023you}, as recalled in Section~\ref{sect:JAX}. 
The goal is to have a reasonable easy static definition of a bound to the number of numeric steps required to evaluate a term $M$, i.e., $\beta_F$, $\beta_{\dot+}$, and $\beta_{\dot*}$ reduction steps. 
While this is complex for full $\beta$-reduction in $\lambdaLL$, we identify a reduction strategy (safe reduction) and conditions on $M$ (Definition~\ref{def:safe_term}) that gives such a definition by a simple induction on the structure of a term. 
 These conditions hold for the terms used in subsequent sections to validate JAX Autodiff transformations, providing quantitative soundness for our JAX Autodiff encoding. 
Proofs are given in Appendix~\ref{app:costmodel}.

The \emph{workload} $\costType{A}$ of a type $A$ is the number of occurrences of $\typR$ not under the scope of a $\oc$, $\Cost{M}$ of a term $M$ is the number of numerical functions not under a $\oc$ as well as the number of possible numerals erased during a reduction, i.e.:
$\Cost{\underline f}
	\definedas\Cost{\dot +}
	\definedas\Cost{\dot *}
	\definedas 1$,
\mbox{ }
$\Cost{x}
	\definedas\Cost{\oc M}
	\definedas\Cost{(\,)}
	\definedas\Cost{\nTuple{\,}}
	\definedas\Cost{\underline r}
	\definedas 0$,
\mbox{}	 
	$\Cost{\lambda \PatA.M}
		\definedas\Cost{M} + \sum_{x:A \in \FV{\PatA}\setminus\FV{M}}\Cost{A}$
and 
\mbox{}	
	$\Cost{MN}\definedas \Cost{\nTuple{M,N}}\definedas\Cost{(M,N)}\definedas\Cost{M}+\Cost{N}$.

A variable of type $A$ is \emph{ground} if $A$ has no arrow.

\begin{definition}[Safe term]
	\label{def:safe_term}
	A term $M$ is \emph{safe} if: 
	(i) for any subterm $\oc M'$ in $M$, $\Cost{M'}=0$; 
	(ii) for any subterm $\nTuple{M_1,M_2}$ in $M$, $\FV{M_1}\cap\FV{M_2}$ has only ground variables.
\end{definition}

Condition (ii) enables the next Proposition~\ref{prop:safe_reduction_is_safe} by restricting additive duplication. 
The condition can be omitted with a more intricate definition of workload that accounts for additive duplication through an appropriate quantitative type system.  
This generalisation is detailed in~\cite[Chapter 7]{thesis}, but we opt for simplicity here, as it suffices to validate Autodiff. 
The workload $\cost{M}$ does not, in general, bound the number of flops in an arbitrary reduction of $M$, not even when $M$ is safe. However, by restricting the set of allowed reduction sequences using the notion of safe reduction, we obtain such a bound, as guaranteed by Proposition~\ref{prop:safe_reduction_is_safe}. This proposition also establishes that safe reduction is complete with respect to full $\beta$-reduction on closed terms of ground type.

The set of \emph{strong values} is defined as (for $\mathtt c\in\{\underline r,\underline f, \dot +, \dot *\}$):
\begin{align*}
	\label{eq:strongval}
	\tag{Strong Values}
	W \bnf&
	x
	\orsintax \mathtt c
	\orsintax \absterm{\PatA}{M} 
	\orsintax\emptytupleterm
	\orsintax \tupleterm{W_1}{W_2}  
	\orsintax \nTuple{\,}
	\orsintax \nTuple{W_1,W_2}
	\orsintax \bangterm{W} 
	\orsintax \dot* W	
\end{align*}

Notice that given a pattern $p$ of type $A$, a strong value $W$ of type $A$ is always a value for $p$, in particular the substitution $M\{W/p\}$ is well-defined. 

The \emph{safe reduction} ($s$-reduction in short) is a \emph{call by closed strong value} reduction: we just replace $\beta_\lambda$ in Fig.~\ref{fig:beta_rules} with
\[
	\redbeta_s:\mbox{ }(\lambda\PatA.M)W \xrightarrow{s} M\{W/\PatA\}\quad\text{for $W$ closed strong value}
\]

\begin{proposition}
	\label{prop:safe_reduction_is_safe}
	\todom{change label 'cor:safe_reduction_is_safe' into 'prop:safe_reduction_is_safe' everywhere}
	A safe closed term $M$ reduces by any maximal safe-reduction sequence to a strong value $W$ in at most $\cost{M}$ numeric steps. 
	If moreover $M$ is of ground type,  then $W$ is a $\beta$-nf.
\end{proposition}

\section{Translation from $\JAX$ to $\lambdaLL$} \label{sect:JAXtoLambdaLL}
We give two translations $\PrimalT$ and $\TangentT$ of Linear A types depending whether these latter refer to primal or tangent data:
\begin{align*}
    \PrimalT(\typrealJAX) & = \typR\,,
    &
    \PrimalT(\typunitJAX) & = \typone\,,
    &
    \PrimalT(\typtensorJAX{\tautypJAX}{\sigmatypJAX}) & = \oc\PrimalT(\tautypJAX)\otimes\oc\PrimalT(\sigmatypJAX)\,,
    \\
    \TangentT(\typrealJAX) & = \typR\,,
    &
    \TangentT(\typunitJAX) & = \typtop\,,
    &
    \TangentT(\typtensorJAX{\tautypJAX}{\sigmatypJAX}) & = \TangentT(\tautypJAX)\with\TangentT(\sigmatypJAX)\,.
\end{align*}

Observe that $\PrimalT(\tautypJAX)$ (resp. $\TangentT(\tautypJAX)$) is in the set of \ref{eq:mult_types} (resp. \ref{eq:mult_add_types}).
\begin{remark} \label{rk:primal_bang_tangent}
    One can prove by induction on a Linear A type $\tautypJAX$ that $\PrimalT(\tautypJAX)$ is a retraction of $\oc\TangentT(\tautypJAX)$, namely~$\PrimalT(\tautypJAX)\dashv\vdash\oc\TangentT(\tautypJAX)$. In fact, in LL we have (see e.g.~\cite{panorama}) the two isomorphisms $\typone\dashv\vdash\oc\typtop$ and $\oc A\otimes \oc B\dashv\vdash\oc(A\with B)$, as well as the retraction pair $\oc A\dashv\vdash\oc \oc A$. This shows that morally one can consider primal types as the exponential promotion of the tangent types, supposing that primal $\typR$ is equivalent to the $\oc$ of tangent  $\typR$.
\end{remark}

We extend $\TangentT$ on \ref{eq:mult_types}: $\TangentT(\typR)=\typR,\TangentT(\typone)=\typtop,\TangentT( \typtensor{\oc D}{\oc E})=\TangentT(D)\with\TangentT(E)$.  
The notion of numeral sequences extend  to sequence types, in the spirit of Section~\ref{sect:JAX}: we will denote $\SeqA$ (resp.~$\oc \SeqA$) for a numeral sequence of a $\with$-sequence (resp.~exponentiated $\otimes$-sequence) type, which is a closed strong value of that type. 

We may silently suppose the immediate correspondence between the numeral sequences of a Linear A type $\tau$, and their $\lambdaLL$ siblings $\TangentT(\tau)$ and $\PrimalT(\tau)$.
Given an environment $\Gamma$ of sequence types, we write $\SeqA\in\Gamma$ for a function mapping every $x:A\in\Gamma$ to a numeral sequence $\SeqA_x$ for the type $A$. 

\subsection{Translation $\SymbTransA$ of \JaxA{} into $\lambdaLL$}

Take a \JaxA{} judgement $x_1:\tau_1,\dots, x_n:\tau_n; \dot y_1:\sigma_1,\dots, \dot y_m:\sigma_m\vdash e: (\tau,\sigma)$. The idea of the translation $\TransA{e}$ is to associate the ``primal operators'' of $e$ with the multiplicative operators of $\lambdaLL$ and the ``tangent operators'' with the additive operators. There are however some subtleties. First, the ``primal part'' is scattered with exponential modalities, enabling the duplication/erasing of primal values, according to the call-by-value translation of $\lambda$-calculus into LL (see e.g.~\cite{girard1987linear,MARAIST1995370}). Namely, a free primal variable $x_i:\tau_i$ of $e$ is associated with a $\oc x_i$ pattern of type $\oc\PrimalT(\tau_i)$ in $\jaxdelta{e}$. Second, the ``tangent part'' of $e$ is represented as a linear map $(\TangentT(\sigma_1)\with\dots\with\TangentT(\sigma_m))\multimap\TangentT(\sigma)$ incorporating the free tangent variables of $e$ as parameters of the map. This map is a kind of matrix representing the tangent computation of $e$. Finally, this map is encapsulated by the affine modality (so getting a final type $\affinebang((\TangentT(\sigma_1)\with\dots\with\TangentT(\sigma_m))\multimap\TangentT(\sigma))$ for the  ``tangent part'' of $e$) allowing for discharging it whenever not necessary.

\begin{figure*}
    \allowdisplaybreaks
    \begin{align*}
        \jaxdelta[\theta]{\retJAX{x}{y}}
        \definedas\mbox{}
            &\trad{\oc x}{\&\linearize{\theta}
            }{y}
         \\
         \jaxdelta[\theta]{\letterm{\retJAX{x}{y}}{e_1}{e_2}}
         \definedas\mbox{}
             &\letterm{\tupleterm{\oc x}{\affinebang f}}{\jaxdelta[\theta\cap\linFV{e_1}]{e_1}}{}
            \letterm{\tupleterm{\oc z}{\affinebang g}}{\jaxdelta[\linvarJAX y,\theta\cap\linFV{e_2}]{e_2}}{}
            \\
            &\tupleterm{\oc z}{\affbangterm{(\absterm{{y^{\&\linearize{\theta}}}}{
                \letterm{\nTuple{y_1,y_2}}{\Split[\&\linearize{\theta}]{\linFV{e_1}} y}
                {
                    \appterm{\affderterm{g}}
                    {(\SplitInv[\&\linearize{\linvarJAX y,\theta\cap\linFV{e_2}}]{\dot{y}}
                    \nTuple{\appterm{\affderterm{f}}{y_1},y_2})}
                }
            })}}
        \\
        \jaxdelta[\theta]{{\emptytupleJAX}}
        \definedas\mbox{}
            & \tupleterm{\oc\emptytupleterm}{\affbangterm{(\lambda y^{\typtop}.\linemptytupleterm)}}
        \\
        \jaxdelta[\theta]{\letterm{{\emptytupleJAX}}{\expvarJAX{z}}{e}}\definedas\mbox{}
            & \letterm{\emptytupleterm}{z}{\jaxdelta[\theta]{e}}
        \\
        \jaxdelta[\theta]{\emptylintupleJAX}\definedas\mbox{}
            & \tradlin{\typtop}{\linemptytupleterm}
        \\
        \jaxdelta[\theta]{\lintupleJAX{x_1}{x_2}}\definedas\mbox{}
            & \tradlin{\&\linearize{\theta}}{y}
        \\
        \jaxdelta[\theta]{\letterm{\emptylintupleJAX}{\linvarJAX{z}}{e}}\definedas\mbox{}
            & \letterm{\tupleterm{\oc x}{\affinebang f}}{\jaxdelta[\theta\setminus\linvarJAX{z}]{e}}{}
            \tupleterm{\oc x}{\affbangterm{(\absterm{y^{\&\linearize{\theta}}}{
                \letterm{\nTuple{z,y'}}{\Split[\&\linearize{\theta}]{\{\theta(\dot z)\}} y}
                {
                    \appterm{\affderterm{f}}{y'}
                }
            })}}
        \\
        \jaxdelta[\theta]{\letterm{\lintupleJAX{x_1}{x_2}}{\linvarJAX{z}}{e}}\definedas\mbox{}
            & \letterm{\tupleterm{\oc x}{\affinebang f}}{\jaxdelta[\linvarJAX{x_1},\linvarJAX{x_2},\theta\setminus\linvarJAX{z}]{e}}{}
            \\
            &
            \tupleterm{\oc x}{\affbangterm{(\absterm{y^{\&\linearize{\theta}}}
            {
                \letterm{\nTuple{\nTuple{x_1,x_2},y'}}{\Split{\{\theta(\dot z)\}} y}{
                    \appterm{\affderterm{f}}
                    {(\SplitInv[\&\linearize{\linvarJAX{x_1},\linvarJAX{x_2},\theta\setminus\linvarJAX{z}}]{\linvarJAX{x_1},\linvarJAX{x_2}} \nTuple{x_1,x_2,y'})}
                }
            })}}
        \\
        \jaxdelta[\theta]{\literalJAX}\definedas\mbox{}
            &\tradexp{\oc\realterm}
        \\
        \jaxdelta[\theta]{\underline f(x_1,x_2)}\definedas\mbox{}
            & \tradexp{\funtermname{f}{(\oc x_1,\oc x_2)}}
        \\
        \jaxdelta[\theta]{\linzeroJAX{\sigmatypJAX}}\definedas \mbox{}
            &\tradlin{\typtop}{\underline{0}_{\linearize{\sigmatypJAX}}}
        \\
        \jaxdelta[\theta]{\linsumJAX{y_1}{y_2}}\definedas \mbox{}
            & \tradlin{\&\linearize{\theta}}{\dot +_{\&\linearize{\theta}} y}
        \\
        \jaxdelta[\theta]{\linmultJAX{x}{y}}\definedas \mbox{}
            & \tradlin{\&\linearize{\theta}}{\dot*_{\&\linearize{\theta}} x y}
        \\
        \jaxdelta[\theta]{\dupJAX{y}}\definedas \mbox{}
            &\tradlin{\&\linearize{\theta}}{\lintupleterm{y}{y}}
        \\
        \jaxdelta[\theta]{\dropJAX{e}}\definedas \mbox{}
            &\letterm{\tupleterm{\oc x}{\affinebang  f}}{\jaxdelta[\theta]{e}}{\trad{\oc \emptytupleterm}{\&\linearize{\theta}}{
            \letterm{z}{\appterm{f}{y}}{\linemptytupleterm}}
            }
    \end{align*}
    \caption{Translation $\jaxdelta[\theta]{e}$ into $\lambdaLL$ of a Linear A expression $\Gamma;\dot\Gamma\vdash e:(\tau,\sigma)$ given en enumeration $\theta$ of $\dot\Gamma$. Note that in some cases $\SplitInv[]{}$ can be the identity or neutrality, in case of identity we can omit it in the following.
    }
    \label{fig:JAXtoLL}
\end{figure*}

Technically, the definition of $\jaxdelta{e}$ depends on a function $\rho$ associating the free primal variables in $e$ to $\lambdaLL$ variables and an enumeration $\theta$ of the set $\linFV{e}$ of the free tangent variables in $e$. Let us ease the notation by adopting the convention of using the same name for the primal variables and their associated $\lambdaLL$ variables, so that we can omit to explicit $\rho$ and simply write $\jaxdelta[\theta]{e}$. Figure~\ref{fig:JAXtoLL} gives the definition of $\jaxdelta[\theta]{e}$ by structural induction on $e$, using the notational conventions of the previous sections.  In particular, given the enumeration $\theta=(\linvarJAX{y_1}:\sigma_1, \ldots,\linvarJAX{y_m}:\sigma_m)$ of $\linFV{e}$, we will denote by 
$\&\linearize\theta$ the type $\&(\linearize{\sigma_1}, \dots,\linearize{\sigma_m})=\linearize{\sigma_1}\& \cdots\&\linearize{\sigma_m}$. All details are in Appendix~\ref{app:translDeltaA}.
 
Tangent computations essentially consist of matrix multiplications, here implemented by the specialised sum $\dot+$ and product $\dot*$ on the elements of the $\with$-sequence types. However, automatic differentiation has an essential feature that makes it different from just implementing matrix multiplication: the matrices considered are structured by blocks determined by the program structure and the multiplications do happen at the level of these blocks, not on the whole matrices. E.g., by taking the notation of the definition of $\jaxdelta[\theta]{\letterm{\retJAX{x}{y}}{e_1}{e_2}}$ in Figure~\ref{fig:JAXtoLL}, we have that the variable $f$ (referring to the tangent computation of $e_1$) applies only on the block of the additive tuple associated with the $\dot y$ input of $e_2$ and not on the whole set of its inputs. 
Notice then that the $\affinebang$ modality wrapping the type of $f$ is essential for well-typing. A type derivation of $\nTuple{\affderterm{f}y_1,y_2}$ must in fact weaken the typing environment of $y_2$ so to introduce $\affinebang f$, possible thanks to the affine modality.

\begin{proposition}[Type $\SymbTransA$]
\label{prop:translation_jax}
    Given 
    $\judgmentJAX{x_1:\tau_1,\dots, x_n:\tau_n}
    {\linvarJAX{y_1}:\sigmatypJAX_1,\dots, \linvarJAX{y_m}:\sigmatypJAX_m}{e}{\tautypJAX}{\sigmatypJAX}$ and an enumeration $\theta=(\linvarJAX{y_1}:\sigmatypJAX_1,\dots, \linvarJAX{y_m}:\sigmatypJAX_m)$ of the set of the free tangent variables in $e$, then
    $\jaxdelta[\theta]{e}$ is a well-typed term in $\lambdaLL$ such that:
    \begin{small}
    \[
    	\oc x_1\!:\!\oc\PrimalT(\tau_1),..., \oc x_n\!:\!\oc\PrimalT(\tau_n)
	\vdash
	\jaxdelta[\theta]{e}\!:\!
	\oc\PrimalT(\tautypJAX)
	\otimes
	\affinebang(\left(\&_{i=1}^m\linearize{\sigmatypJAX_i}\right)\multimap\linearize{ \sigmatypJAX})
    \]
    \end{small}
\end{proposition}

The soundness of $\SymbTransA$ can be formally stated point-wise, by proving that $\TransA{e}$ returns a term computing $\oc \SeqA\mapsto\oc\SemP[\SeqA]{e}$ and $\oc\SeqA,\SeqB\mapsto\SemT[\SeqA;\SeqB]{e}$ for every numeral sequences $\SeqA$ and $\SeqB$ associated with, respectively, the primal and tangent free variables in $e$. 

\begin{proposition}[Soundness $\SymbTransA$]
	\label{prop:sound_transA}
    Given $\Gamma;\dot\Sigma\vdash e:(\tau;\sigma)$, an enumeration $\theta$ of the tangent variables in $\dot\Sigma$, then:
    \begin{itemize}
    \item $\forall\SeqA$ for $\Gamma$:
    $
    	\TransA[\theta]{e}[\oc \SeqA/\PrimalT(\Gamma)] \rightarrow^* (\oc \SemP[\SeqA]{e}, \affbangterm F)
    $,
    \item and $\forall\SeqB$ for the type $\&\theta$: 
    $
    	F\SeqB \rightarrow^* \SemT[\SeqA;\SeqB]{e}
    $.
    \end{itemize} 
\end{proposition}

Finally, we should check that $\TransA{e}$ computes  $\SemP[\SeqA]{e}$ and $\SemT[\SeqA;\SeqB]{e}$ with at most a constant overhead of flops with respect to the original Linear A expression $e$. Notice that $\TransA{e}$ satisfies the conditions of Definition~\ref{def:safe_term}, so by Proposition~\ref{prop:safe_reduction_is_safe} we can use the workload of a term as a bound to the number of numeric steps. The next proposition assures that $\SymbTransA$ preserves the workload.
\begin{proposition}[Workload $\SymbTransA$]
	\label{prop:size_TransA}
Given $\Gamma;\dot\Gamma\vdash e:(\tau;\sigma)$, an enumeration $\theta$ of $\dot\Gamma$, then $\TransA[\theta] e$ is safe and $\cost{\TransA[\theta] e}\leq \cost{e}$. 
\end{proposition}

\paragraph*{The specialisation $\SymbTransB$}
Whenever $\SymbTransA$ is applied to a purely primal expression of $\JaxB$ of type $\judgmentJAX{\Gamma}{}{e^p}{\tautypJAX}{\typone}$, the tangent parts of $\expexprJAX{e}$ are encoded as dummy identities $\top\multimap\top$, so one can discharge all of them and have a translation $\TransB{e^p}$ which returns a $\lambdaLL$ term of type $\oc\PrimalT(\Gamma)\vdash e^p : \oc\PrimalT(\tautypJAX)$, mimicking the shape of $e^p$.
For example,
$
	\TransB{\expvarJAX{x}}\definedas \oc x\,,
	\TransB{\letterm{\expvarJAX{x}}{\expexprJAX{e_1}}{\expexprJAX{e_2}} } \definedas \letterm{\oc x}{\TransB{\expexprJAX{e_1}}}{\TransB{\expexprJAX{e_2}}}\,.
$
We refer to Appendix~\ref{app:translDeltaB} for the complete definition. 

Recall our running example $g(x,y)\definedas(sin(x)*y)+cos(x)$ and the purely primal expression $\expexprJAX{e}\in\ref{linear_B:primal}$ computing $g$ given in Figure~\ref{fig:toyEx_linB}. If one applies the translation $\SymbTransB$ to $\expexprJAX{e}$ and some $\beta_\lambda$-reductions simplifying the stack of let-definitions in $\expexprJAX{e}\in\ref{linear_B:primal}$, one gets the $\lambdaLL$ term $P$ in Figure~\ref{fig:ex_transl}, which is well-typed under the judgement: $\judgment{\oc x: \oc \typR, \oc y: \oc \typR}{P}{\oc \typR}$.

\begin{figure}[htbp]
    \vspace{-0.45cm}
    \centering
    \begin{subfigure}[b]{0.27\textwidth}
        \centering
        \begin{align*}
            &\letterm{\oc v_1}{\underline{sin}\mbox{ }\oc x}{}\\
            &\letterm{\oc v_2}{\oc v_1 \mbox{ }\underline{*}\mbox{ } \oc y}{}\\
            &\letterm{\oc v_3}{\underline{cos}\mbox{ }\oc x}{}\\
            &\letterm{\oc v_4}{\oc v_2 \mbox{ }\underline{+}\mbox{ } \oc v_3}{}\\
            &\oc v_4
        \end{align*}
        \caption{$\lambdaLL$ term $P$ computing $g(x,y)\definedas(sin(x)*y)+cos(x)$.}
        \label{fig:ex_transl}
    \end{subfigure}
    \hfill
    \begin{subfigure}[b]{0.65\textwidth}
        \centering
        \begin{align*} 
            &\letterm{(\oc v_1, \color{blue}\affinebang f_1
            \color{black})}{
                \left(
                    \begin{aligned}
                        &\letterm{\oc w_1}{\underline{cos}\mbox{ }\oc x}{}\\
                        &\tupleterm{\underline{sin}\mbox{ }\oc x}
                        {\color{blue}\affinebang (\lambda u. w_1\mbox{ }\dot{*}\mbox{ }u)\color{black}}
                    \end{aligned}
                \right)
            }{}
            \\[2mm]
            &\letterm{(\oc v_2, \color{blue} \affinebang f_2 \color{black})}{
                \left(
                    \begin{aligned}
                        &\letterm{\oc w_2}{\oc y}{}
                        \letterm{\oc w_3}{\oc v_1}{}\\
                        &\tupleterm{\oc v_1 \mbox{ }\underline{*}\mbox{ } \oc y}
                        {\color{blue}
                        \affinebang (\lambda \nTuple{u_1,u_2}. (w_2\mbox{ }\dot{*}\mbox{ }u_1)
                        \mbox{ }\dot{+}\mbox{ } (w_3\mbox{ }\dot{*}\mbox{ }u_2))
                        \color{black}}
                    \end{aligned}
                \right)
            }{}
            \\[2mm]
            &\letterm{(\oc v_3, \color{blue} \affinebang f_3 \color{black})}{
                \left(
                    \begin{aligned}
                        &\letterm{\oc w_4}{\underline{-sin}\mbox{ }\oc x}{}\\
                        &\tupleterm{\underline{cos}\mbox{ }\oc x}
                        {\color{blue}\affinebang (\lambda u. w_4\mbox{ }\dot{*}\mbox{ }u)\color{black}}
                    \end{aligned}
                \right)
            }{}
            \\[2mm]
            &\letterm{(\oc v_4, \color{blue} \affinebang f_4 \color{black})}{
                \left(
                    \begin{aligned}
                        &\letterm{\oc w_5}{\oc \underline{1}}{}
                        \letterm{\oc w_6}{\oc \underline{1}}{}\\
                        &\tupleterm{\oc v_2 \mbox{ }\underline{+}\mbox{ } \oc v_3}
                        {\color{blue}
                        \affinebang (\lambda \nTuple{u_1,u_2}. (w_5\mbox{ }\dot{*}\mbox{ }u_1)
                        \mbox{ }\dot{+}\mbox{ } (w_6\mbox{ }\dot{*}\mbox{ }u_2))
                        \color{black}}
                    \end{aligned}
                \right)
            }{}
            \\[2mm]
            & 
            \left(
                \oc v_4,
                \color{blue}
                \affinebang \left(\lambda u^{\typR\&\typR}.
                    \letterm{\nTuple{x',y'}}{u}{f_4 \nTuple{f_2 \nTuple{f_1\mbox{ }x', y'}, f_3\mbox{ }x'}}
                \right)
                \color{black}
            \right)
        \end{align*}
        \caption{Application of $\mathcal F$ to $P$, after some $\beta_\lambda$-simplifications.}
        \label{fig:ex_forward}
    \end{subfigure} 
    \hfill 
    \begin{subfigure}{\linewidth}
        \centering
        \begin{align*} 
            &
            \letterm{\oc w_1}{\underline{cos}\mbox{ }\oc x}{}
            \letterm{\oc v_1}{\underline{sin}\mbox{ }\oc x}{}
            \\
            &
            \letterm{\oc w_2}{\oc y}{}
            \letterm{\oc w_3}{\oc v_1}{}            
            \letterm{\oc v_2}{\oc v_1 \mbox{ }\underline{*}\mbox{ } \oc y}{}
            \\
            &
            \letterm{\oc w_4}{\underline{-sin}\mbox{ }\oc x}{}            
            \letterm{\oc v_3}{\underline{cos}\mbox{ }\oc x}{}
            \\
            &
            \letterm{\oc w_5}{\oc\underline{1}}{}
            \letterm{\oc w_6}{\oc\underline{1}}{}            
            \letterm{\oc v_4}{\oc v_2 \mbox{ }\underline{+}\mbox{ } \oc v_3}{}
            \\ 
            &\left(
                \oc v_4,
                \color{blue}
                    \affinebang \left(\quad
                    \begin{aligned}
                        &\letterm{\affinebang f_1^{\mbox{ }\typR \multimap \typR}}
                        {\affinebang (\lambda u. w_1\mbox{ }\dot{*}\mbox{ }u)}{}\\
                        &\letterm{\affinebang f_2^{\mbox{ }(\typR\&\typR) \multimap \typR}}
                        {\affinebang (\lambda \nTuple{u_1,u_2}. (w_2\mbox{ }\dot{*}\mbox{ }u_1)
                        \mbox{ }\dot{+}\mbox{ } (w_3\mbox{ }\dot{*}\mbox{ }u_2))}{}\\
                        &\letterm{\affinebang f_3^{\mbox{ }\typR \multimap \typR}}
                        {\affinebang (\lambda u. w_4\mbox{ }\dot{*}\mbox{ }u)}{}\\
                        &\letterm{\affinebang f_4^{\mbox{ }(\typR\&\typR) \multimap \typR}}
                        {\affinebang (\lambda \nTuple{u_1,u_2}. (w_5\mbox{ }\dot{*}\mbox{ }u_1)
                        \mbox{ }\dot{+}\mbox{ } (w_6\mbox{ }\dot{*}\mbox{ }u_2))}{}\\
                        &\lambda u^{\typR\&\typR}.
                        \begin{aligned}
                            &\letterm{\nTuple{x',y'}}{u}{}
                            f_4 \nTuple{f_2 \nTuple{f_1\mbox{ }x', y'}, f_3\mbox{ }x'}
                        \end{aligned}
                    \end{aligned}
                    \quad\right)
                \color{black}
            \right)
        \end{align*}
        \caption{Application of $\mathcal U$ to the term in Figure~\ref{fig:ex_forward}.}
        \label{fig:ex_unzipping}
    \end{subfigure}
    \hfill  
    \begin{subfigure}{\linewidth}
        \centering
        \begin{align*} 
	   &
            \letterm{\oc w_1}{\underline{cos}\mbox{ }\oc x}{}	   
	   \letterm{\oc v_1}{\underline{sin}\mbox{ }\oc x}{}
            \\
            &
            \letterm{\oc w_2}{\oc y}{}
            \letterm{\oc w_3}{\oc v_1}{}            
            \letterm{\oc v_2}{\oc v_1 \mbox{ }\underline{*}\mbox{ } \oc y}{}
            \\
            &
            \letterm{\oc w_4}{\underline{-sin}\mbox{ }\oc x}{}            
            \letterm{\oc v_3}{\underline{cos}\mbox{ }\oc x}{}
            \\
            &
            \letterm{\oc w_5}{\oc\underline{1}}{}
            \letterm{\oc w_6}{\oc\underline{1}}{}            
            \letterm{\oc v_4}{\oc v_2 \mbox{ }\underline{+}\mbox{ } \oc v_3}{}
            \\
            &\left(
                \oc v_4,
                \color{red}
                    \affinebang \left(\quad
                    \begin{aligned}
                        &\letterm{\affinebang \overleftarrow{f_1}^{\mbox{ }\typR \multimap \typR}}
                        {\affinebang (\lambda {l}. w_1\mbox{ }\dot{*}\mbox{ }{l})}{}\\
                        &\letterm{\affinebang \overleftarrow{f_2}^{\mbox{ } \typR \multimap (\typR\&\typR)}}
                        {\affinebang (\lambda {l}.\nTuple{w_2 \mbox{ }\dot{*}\mbox{ }{l}, w_3 \mbox{ }\dot{*}\mbox{ }{l}})}{}\\
                        &\letterm{\affinebang \overleftarrow{f_3}^{\mbox{ }\typR \multimap \typR}}
                        {\affinebang (\lambda {l}. w_4\mbox{ }\dot{*}\mbox{ }{l})}{}
						\\
                        &\letterm{\affinebang \overleftarrow{f_4}^{\mbox{ } \typR \multimap (\typR\&\typR)}}
                        {\affinebang (\lambda {l}. \nTuple{w_5\mbox{}\dot{*}\mbox{ }{l},w_6\mbox{}\dot{*}\mbox{ }{l}})}{}
						\\
                        &\lambda {z}^{\mbox{ }\typR}.
                            \letterm{\nTuple{z',z''}}{\overleftarrow{f_4}{z}}{}{}
						\\
			&\phantom{\lambda {z}^{\mbox{ }\typR}.}		
			    \letterm{\nTuple{\nTuple{x'_1,y'_1},x'_2}}
			    {
			    	\nTuple{
			    		(\lambda \nTuple{{z_1},{z_2}}.\nTuple{\overleftarrow{f_1}{z_1},{z_2}})(\overleftarrow{f_2}{z'})
					,
					\overleftarrow{f_3}{z''}
				}
			     }{}
			     			\\
			   &\phantom{\lambda {z}^{\mbox{ }\typR}.}\nTuple{x_1'\dot+x_2', y_1'}						     						
                    \end{aligned}
                    \quad\right)
                \color{black}
            \right)
        \end{align*}
        \caption{Application of $\mathcal T$ to the term in Figure~\ref{fig:ex_unzipping}, after some $\beta_\lambda$-simplification.}
        \label{fig:ex_transpose}
    \end{subfigure}
    \vspace{-.6cm}
    \caption{ Application of the AD system to a $\lambdaLL$ term that computes a numerical function $g:\mathbb{R}^2 \to \mathbb{R}$.}
    \label{fig:mainfigure}
\end{figure}
\subsection{Dissecting Linear A into $\lambdaLL$} 
\todom{adding tuples cases for LLA grammar ? cases are commented}
    \begin{figure}
        \begin{small}
        \begin{align}
                \tag{$\PrimalLL$}
                \label{eq:LLPrimalterms}
                P,Q \bnf &    
              \oc x
                \orsintax \oc\realterm 
                \orsintax \oc\emptytupleterm 
            \orsintax \funterm{(\oc x_1,\oc x_2)}                
            \orsintax \appterm{(\absterm{\oc x}{P})}{Q}          
                \\
                \tag{$\linlambdaLL$}
                \label{eq:LLTangentTerms}
                U \bnf&
                u
                \orsintax \underline 0  
                \orsintax \nTuple{\,}
                \orsintax \nTuple{U_1,U_2}
                \orsintax FU               
                \\
                \tag{$\lambdaLL^{\mathtt{f}}$}
                \label{eq:LLAdditiveFuncTerms}
                 F,G \bnf&
                 \affderterm{f}               
                 \orsintax \dot +
                 \orsintax \dot * x                
                 \orsintax \lambda \PatAddA.U
                 \orsintax \letterm{\affinebang f}{\affinebang F}{G}
                 \\
                 \tag{$\LinearBLL$}
                 \label{eq:LLBTerms}
                 R,S \bnf &
                \tupleterm{P}{\affbangterm F}
                \orsintax \letterm{\tupleterm{\oc x}{\affbangterm f}}{S}{R} 
                \orsintax \letterm{\affbangterm f}{\affbangterm F}{R}
                  \orsintax \letterm{\oc x}{P}{R} 
        \end{align}
        \end{small}
        \vspace{-0.3cm}
        \caption{
            The $4$-sorted grammar giving the fragments \ref{eq:LLBTerms}, \ref{eq:LLAdditiveFuncTerms}, \ref{eq:LLTangentTerms} and \ref{eq:LLPrimalterms} of $\lambdaLL$,  together with the metavariables associated with each sort. 
            Variables $x$'s (resp.~$u$, $f$) are supposed of $\otimes$-sequence (resp. $\with$-sequence, tangent function) types. Metavariables $\PatAddA$ vary over patterns of $\with$-sequence types. 
        }
        \label{fig:lambdaLL}
        \end{figure}
        
        Figure~\ref{fig:lambdaLL} defines the fragment $\lambdaLLFM$ of $\lambdaLL$ which strictly contains the image set of $\SymbTransA$. This fragment is build on the top of the fragments $\PrimalLL$ and $\linlambdaLL$ basically giving respectively the purely primal and purely tangent part of the $\SymbTransA$ images. In fact, $\PrimalLL$ corresponds to the restriction to ground types of the call-by-value translation of $\lambda$-calculus into linear logic \cite{MARAIST1995370}, the sort $\PrimalLL$ identifying computations. Notice that the term in Figure~\ref{fig:ex_transl} is in $\explambdaLL$, while \ref{fig:ex_forward}, \ref{fig:ex_unzipping} and~\ref{fig:ex_transpose} are in $\LinearBLL$.
        
        The typing environments of the terms in $\lambdaLLFM$ may contain exponential patterns $\oc x:\oc E$ associated with primal data, or patterns $\affinebang f$ associated with some tangent computation, so of type $\affinebang (L\multimap H)$ for some $\with$-sequence types $L$ and $H$. Henceforth, we denote by $\oc\Sigma$ the environments of exponentiated $\otimes\oc$-sequence patterns, i.e.~$\oc\Sigma$ stands for, e.g.~$\oc x_1:\oc E_1,\dots,\oc x_n:\oc E_n$, and by $\affinebang\Phi$ the environments of the affine patterns of functions between $\with$-sequences, i.e.~$\affinebang\Phi$  stands for, e.g.~$\affinebang f_1:\affinebang(L_1\multimap H_1),\dots,\affinebang  f_m:\affinebang(L_m\multimap H_m)$. We call $\oc\Sigma$ \emph{$\oc$-environment} and $\affinebang\Phi$ \emph{$\affinebang$-environment}.

\todom{adapt appending and the long version to Prop.~\ref{prop:typingLinearBLL}}
\begin{proposition}
	\label{prop:typingLinearBLL}
    \mbox{} 
\begin{enumerate}
\item $\forall P\in\explambdaLL$, $\judgment{\oc\Sigma}{P}{\oc E}$,
\item $\forall U\in\lambdaLL^{\mathtt{t}}$, $\judgment{\oc\Sigma,\affbangterm{\Phi},\PatAddA:L}{U}{H}$,
\item $\forall F\in\lambdaLL^{\mathtt{f}}$, $\judgment{\oc\Sigma,\affbangterm{\Phi}}{F}{\typlollipop{L}{H}}$,
\item $\forall R\in \LinearBLL$, $\judgment{\oc\Sigma,\affbangterm{\Phi}}{R}{\typtensor{\oc E}{\affbangterm{(\typlollipop{L}{H})}}}$,
\end{enumerate}
\noindent for suitable $\oc\Sigma$, $\affbangterm{\Phi}$, $\PatAddA$, $E, L,H$.
\end{proposition}

\todom{adapt appendix to this prop, mixing transA and transB}
\begin{proposition}\label{prop:fragmentTrans}
	Given $e\in\JaxA$ and an enumeration $\theta$ of $\linFV{e}$, $\jaxdelta[\theta]{e}\in\LinearBLL$.
	Given $e^p\in\eqref{linear_B:primal}$, $\TransB{e^p}\in\explambdaLL$.
\end{proposition}

\section{Forward} \label{sect:forward}
Forward AD is defined in Figure~\ref{fig:FMlambda} as a transformation $\mathcal{F}$ mapping a term $P \in \explambdaLL$ and an enumeration $\theta$ of its free variables into a term $\forward{\theta}{P}\in\lambdaLLFM$. The intuition is the same as for the $\mathcal{F}^{\mathtt{Jax}}$ transformation: the only difference is that now the tangent part is represented as a linear map taking in input the tangent siblings of the $P$ free variables. The following theorems state the type, soundness and workload preservation of $\mathcal F$, we refer to Appendix~\ref{app:forward} for proof details. 

\begin{figure}[h]
    \vspace{-0.7cm}
    \begin{align*}
        \forward{(\oc x)}{\oc x}\definedas \mbox{}
            & \FMtransf{\oc x}{u}
        \\
        \forward{(\,)}{\oc\realterm}\definedas \mbox{}
        & \FMtransf{\oc\realterm}{ \zeroterm{}}
        \\
        \forward{(\,)}{\oc\emptytupleterm}\definedas \mbox{}
        & \FMtransf{\oc\emptytupleterm}{\linemptytupleterm}
        \\
        \forward{\contextFM}{\oc\tupleterm{P}{Q}}\definedas 
            & \letterm{\tupleterm{\oc x}{\affinebang f}}{\forward{\contextFM\cap\FV{P}}{P}}{}\\
            & 
            \letterm{\tupleterm{\oc y}{\affinebang g}}{\forward{\contextFM\cap\FV{Q}}{Q}}{}\\
            & \left(\oc\tupleterm{\oc x}{\oc y},
            \affbangterm{\left(\absterm{u^{\with\linearize\theta}}{\quad
                \begin{aligned}
                    & \letterm{\nTuple{u_{PQ},u'}}
                    {\Split{\FV{P}\cap\FV{Q}}u}{}\\
                    &\letterm{\nTuple{u_P, u_Q}}{
                    \Split{\FV{P}\setminus\FV{Q}} u'}{}\\
                    &
                    \nTuple{
                        \affderterm f{\nTuple{u_{PQ},u_P}}, \affderterm g{\nTuple{u_{PQ},u_Q}}
                    }
            \end{aligned} \quad
        }\right)}\right)
        \\
        \forward{\contextFM}{\appterm{(\absterm{\oc x}{P})}{Q}}
            \definedas \mbox{}
            & \letterm{\tupleterm{\oc x}{\affinebang f}}{\forward{\contextFM\cap\FV{Q}}Q}{} \\
            &\letterm{\tupleterm{\oc y}{\affinebang g}}{\forward{x,\contextFM\cap\FV{P}}P}{}\\
            & \left( \oc y,
                \affbangterm{\left(\absterm{u^{\with\linearize\theta}}{\quad
                    \begin{aligned}
                        & \letterm{\nTuple{u_{PQ},u'}}
                        {\Split{\FV{P}\cap\FV{Q}}u}{}\\
                        &\letterm{\nTuple{u_P, u_Q}}{
                        \Split{(\FV{P}\setminus\{\oc x\})\setminus\FV{Q}} u'}{}\\ 
                        &\affderterm g
                        \nTuple{
                            \affderterm f{\nTuple{u_{PQ},u_Q}}, u_{PQ},u_P
                        }
                \end{aligned} \quad
            }\right)}\right)
        \\ 
        \forward{(\oc x_1, \oc x_2)}{\funterm{(\oc x_1,\oc x_2)}}\definedas \mbox{}
                & \letterm{\oc y_1}{\underline{\partial_1 f} (\oc x_1,\oc x_2)}{} 
                \\
                &\letterm{\oc y_2}{\underline{\partial_2 f} (\oc x_1,\oc x_2)}{}
                \\ 
                & \tupleterm{\funterm{(\oc x_1,\oc x_2)}}{{\affbangterm{(\absterm{\nTuple{u_1,u_2}^{\linearize{(\oc x_1, \oc x_2)}}}(y_1 \dot{*} u_1)\dot{+}(y_2 \dot{*}(u_2)))}}}
        \\
        \forward{\contextFM}{\letterm{\PatExpA}{z}{P}}\definedas \mbox{}
            & \letterm{\tupleterm{\oc x}{\affinebang f}}{(z,\affinebang(\lambda u.u))}{}  
            \letterm{\tupleterm{\oc x_1}{\oc x_2}}{x}{}\\
            &\letterm{\tupleterm{\oc y}{\affinebang g}}{\forward{\FV{\PatExpA},\contextFM\cap\FV{P}}{P}}{}\\
            & \left( \oc y,
            \affbangterm{\left(\absterm{u^{\with\linearize\theta}}{
                    \letterm
                    {\nTuple{u_{P},u'}}
                    {\Split{\FV{P}\setminus \{z\}}u}
                    \affderterm g
                    \nTuple{
                        \affderterm f u', u_P
                    } 
            }\right)}\right)
    \end{align*} 
    \vspace{-0.6cm}
    \caption{Forward $\mathcal F$ over $\PrimalLL$ terms. 
        In the $(\lambda \oc x.P)Q$ case, $D_{P,Q,x}$ splits the environment  between the part common to both $P$ and $Q$, and the ones specific to each term, i.e.~
        $D_{P,Q,x}$ is defined as 
        $\lambda u. 
        \letterm
        		{\nTuple{u_{PQ},u'}}
            	{\Split{\FV{P}\cap\FV{Q}}u}
		{\letterm
			{\nTuple{u_P, u_Q}}
			{\Split{(\FV{P}\setminus\{x\})\setminus\FV{Q}} u'}
			{\nTuple{u_{PQ},u_{P},u_{Q}}}
            	}$, where $\Split{}$ is the splitting term given in \eqref{Notation:sigma}.
        }
        \label{fig:FMlambda}
\end{figure}

Consider the $\explambdaLL$ term $P$ in Figure~\ref{fig:ex_transl}, along with the enumeration $\contextFM=(\oc x:\oc \typR,\oc y:\oc \typR)$ of its free variables. When we apply the forward transformation to $P$, we obtain the $\lambdaLLFM$ term depicted in Figure~\ref{fig:ex_forward}, where we applied some $\beta_\lambda$-reductions to keep shallow the stack of let-definitions. This transformed term is well-typed under the judgment:
$\judgmenttransl{\oc x:\oc \typR,\oc y:\oc \typR}{\forward{\theta}{P}}{\oc \typR}{\left(\typR\&\typR\right)}{\typR}$. 
The first component of the output (i.e.~$\oc v_4$) yields the result of the original computation $P$, while the second component expresses the directional derivative of this map at $(x, y)$ as the linear map $(x', y') \mapsto (y*\cos(x) - \sin(x))*x' + \sin(x)*y'$. This result is obtained by collecting the exponential patterns $\oc w_1, \dots, \oc w_6$ corresponding to the partial derivatives of the primitive operations in $P$, transforming them into linear maps via the {\color{blue} blue} part defining the directional derivatives ${\color{blue}\affbangterm{f_1}}, \dots, {\color{blue}\affbangterm{f_4}}$, and composing them according to the let-definition structure of $P$. One may compare Figure~\ref{fig:ex_forward} with the corresponding Linear~A expression in Figure~\ref{fig:ex_forward_JAX}. The two terms are closely related -- indeed, the $\SymbTransA$ translation of the latter is $\sim$-equivalent to the former (Theorem~\ref{th:forward_soundness}). However, while Linear~A propagates tangents through free variables across sub-expressions, our formulation leverages $\lambda$-abstraction to encapsulate all intermediate tangent computations.

\begin{theorem} [Type $\mathcal F$]
\label{th:type_forward}
	Given a judgment $\oc\Sigma\vdash P:\oc E$ of~\ref{eq:LLPrimalterms} and an enumeration $\theta=(\oc x_1:\oc E_1,\dots,\oc x_n:\oc E_n)$ of $\oc\Sigma$, then $\forward{\contextFM}{P}$ is a~\ref{eq:LLBTerms} term of type:
	$\judgmenttransl{\oc\Sigma}{\forward{\theta}{P}}{\oc E}{\left(\with_{i=1}^n\linearize{E_i}\right)}{\linearize{E}}.$
\end{theorem}

The acquainted reader may notice that such typing basically corresponds to the typing of the forward-mode transformation given in CHAD \cite{Vakar2021CHADCH}. What we additionally provide here is the equivalence with JAX transformation (Theorem~\ref{th:forward_soundness}) and the workload estimation (Theorem~\ref{th:cost_forward}).


\begin{theorem}[Soundness $\mathcal F$]
	\label{th:forward_soundness}
    Given a $\JaxB$ expression $\expexprJAX{e}$ in \eqref{linear_B:primal}, an enumeration $\theta=(\expvarJAX{x_1},\dots, \expvarJAX{x_n})$ of the set $\FV{\expexprJAX{e}}$, a renaming $\contextFMJAX = (\expvarJAX{x_1}\mapsto \linvarJAX{y_1},\dots, \expvarJAX{x_n}\mapsto \linvarJAX{y_n})$ of $\FV{\expexprJAX{e}}$ into tangent JAX variables, and let $\theta'=(\linvarJAX{y_1},\dots,\linvarJAX{y_n})$ be the image of $\theta$ under $\contextFMJAX$, we have:
    	$
    		\forward{\contextFM}{\TransB{\expexprJAX{e}}} 
		\sim
		\TransA[\theta']{\forwardJAX{\contextFMJAX}{\expexprJAX{e}}}	
	$.
\end{theorem}

\begin{theorem}[Workload $\mathcal F$]
	\label{th:cost_forward}
	\todom{replace label lemma:cost_forward into th:cost_forward}There is a constant $c$ such that
	$\forall P\in\explambdaLL$ and $\forall \theta$ enumeration of $\FV{P}$, $\cost{\forward{\contextFM}{P}} \leq c \cdot \cost{P}$. If moreover $P$ is safe, then $\forward{\theta}P$ is safe too. 
\end{theorem}  

\section{Unzipping} \label{sect:unzipping}
The unzipping $\mathcal U$ is an endo-transformation of $\lambdaLLFM$, reproducing the splitting between primal and tangent let-definitions in Linear A.
The transformation is defined on the top of a structural decomposition $\unzippingPre S $ of $S$ given in Figure~\ref{fig:unzipping} and producing a triplet $(\ExpContextA{}, P, F)$ of a context $\ExpContextA{}$ of exponential let-definitions, a $\PrimalLL$ term $P$ and a $\lambdaLL^{\mathtt{f}}$ term $F$, so that $\unzipping S \definedas \ExpContextA{(P,\affinebang F)}$. 
All proofs are in Appendix~\ref{app:unzip}.
    \begin{figure*}
        \vspace{-.15cm}
        \begin{small}
            \begin{align*}
            \unzippingPre{\tupleterm{P}{\affinebang F}}
            &\definedas 
                ([],P, F)
            \\
                \unzippingPre{\letterm{\tupleterm{\oc x}{\affinebang f}}{S_1}{S_2}}
                &\definedas 
                (
                    \ExpContextA[1]{\letterm{\oc x}{P_1}{\ExpContextA[2]{}}},
                    P_2,
                    (\letterm{\affinebang f}{\mbox{}\affinebang F_1}{F_2})
                ) 
                \\
                \unzippingPre{\letterm{\oc x}{P}{S_1}}
                &\definedas (
                    \letterm{\oc x}{P}{\ExpContextA[1]{}},
                    P_1,
                    F_1
                )
                \\
                \unzippingPre{\letterm{f}{\mbox{}\affinebang F}{S_1}}
                &\definedas (
                    \ExpContextA[1]{},
                    P_1,
                    (\letterm{f}{\mbox{}\affinebang F}{F_1})
                )
                \\
                \unzippingPre{\letterm{\PatExpA}{z}{S}}
                &\definedas (
                    \letterm{\PatExpA}{z}{\ExpContextA[1]{}},
                    P_1,
                    F_1
                ) 
            \end{align*}
          \end{small}
          \vspace{-0.35cm}
          \caption{
              The unzipping $\unzipping{S}$ is defined as $\unzipping S \definedas \ExpContextA{(P,\affinebang F)}$, where $\ExpContextA{}$, $P$ and $F$ is given by the decomposition $\unzippingPre{S}$ above, with  $\ExpContextA{}$ denoting contexts, i.e.~terms with exactly one hole $[]$. 
          In the inductive cases we suppose $\unzippingPre{S_i}=(\epsilon_i[],P_i, F_i)$.}
          \label{fig:unzipping}
        \end{figure*}
        
Continuing our running example, we apply the unzipping transformation $\mathcal{U}$ to the $\lambdaLLFM$ term $\forward{\theta}{P}$ in Figure~\ref{fig:ex_forward}, obtaining $\unzipping{\forward{\theta}{P}}$ in Figure~\ref{fig:ex_unzipping}, which has the same type as $\forward{\theta}{P}$. The resulting term performs the same computation as $\forward{\theta}{P}$ but separates the primal and tangent components, moving the latter to the end of the term. This structure makes explicit the sequence of exponential patterns (from $\oc w_1$ to $\oc w_6$) that carry the partial derivatives of the primitive operations in $P$ to the tangent computation. This sequence is often named the \emph{tape} in the AD literature.

Notice also that this transformation can be seen basically as a simple let-commutation. 

\begin{proposition} 
\label{prop:unzipping_as_commutation}
Given $S\in\lambdaLLFM$, we have: $S\sim \unzipping{S}$, in particular they have the same type. 
\end{proposition}

\begin{theorem}[Soundness $\mathcal U$] \label{th:unzipping} 
    \label{th:unzipping_soundness}
    Given $\judgmentJAX{\Gamma}{\lincontextJAX{\Gamma}}{e}{\tautypJAX}{\sigmatypJAX}$ and an enumeration $\theta$ of $\lincontextJAX{\Gamma}$, then
    $
        \unzipping{\jaxdelta[\theta]{e}} 
        \;\sim\; 
        \TransA[\theta]{\unzippingJAX{e}}
    $.
\end{theorem}

\begin{theorem}[Workload $\mathcal U$] 
	\label{th:cost_unzip}\todom{replace label lemma:cost_unzip into th:cost_unzip}
    For $S\in\lambdaLLFM$, $\cost{\unzipping{S}}\! \leq\! \cost{S}$. If moreover $S$ is safe, then $\unzipping{S}$ is safe too.
\end{theorem}

\section{Transpose} \label{sect:transpose}
We define the transpose transformation $\mathcal T$  in Figure~\ref{fig:def_transpose}. The definition splits in three subdefinitions, giving, respectively, the action of $\mathcal T$ on the terms of $\lambdaLL^{\mathtt f}$ (Figure~\ref{subfigure:transpose on lambdaLL_f}), on the terms of $\lambdaLL^{\mathtt t}$ (Figure~\ref{subfigure:transpose on lambdaLL_t}) and finally on $\lambdaLL^{\mathtt A}$ (Figure~\ref{subfigure:transpose on lambdaLL_a}). The first two definitions are mutually recursive, while $\mathcal T$ lifts to $\lambdaLL^{\mathtt A}$ by a simple commutation with the exponential constructors. 

The core of the definition is in the case of a term $U\in \lambdaLL^{\mathtt t}$. Let us give some intuitions. 

\begin{figure}
\begin{align*}
	\RenamePat\Rename u
		&\definedas
		 \Rename(u)
		&&\text{if $u\in\Dom\Rename$,}
	\\
	\RenamePat\Rename{\nTuple{p_1,p_2}}
		&\definedas \nTuple{\RenamePat\Rename{p_1}, \RenamePat\Rename{p_2}}
		&&\text{if $\FV{p_1}\cap\Dom\Rename\neq \emptyset$ and $\FV{p_2}\cap\Dom\Rename\neq \emptyset$,} 
	\\
	\RenamePat\Rename{\nTuple{p_1,p_2}}
		&\definedas \RenamePat\Rename{p_i}
		&&\text{if $\FV{p_i}\cap\Dom\Rename\neq \emptyset$ and $\FV{p_{3-i}}\cap\Dom\Rename= \emptyset$} 		
	\\
	\RenamePat\Rename \PatAddA
		&\definedas
		 t
		&&\text{if $\FV \PatAddA\cap\Dom\Rename = \emptyset$,\mbox{ }$t$ fresh variable of type $t:\typtop$}
	\\[-25pt]
\end{align*} 
\caption{Partial renaming of a pattern $\PatAddA$ along a renaming $\Rename$.}
\label{figure:def_pattern_renaming}
\end{figure}
\begin{figure}
\begin{align*}
	\Fusion \PatAddA{\Rename_1}{\Rename_2}
	&\definedas 0_L
	&&\text{if $\FV{\PatAddA}\cap(\Dom{\Rename_1}\cup\Dom{\Rename_2})=\emptyset$,}
	\\
	\Fusion u{\Rename_1}{\Rename_2}
	&\definedas \Rename_1(u) +_L\Rename_2(u)
	&&\text{if $u\in\Dom{\Rename_1}\cap\Dom{\Rename_2}$,}
	\\
	\Fusion u{\Rename_1}{\Rename_2}
	&\definedas \Rename_i(u)
	&&\text{if $u\in\Dom{\Rename_i}\setminus\Dom{\Rename_{3-i}}$,}
	\\
	\Fusion{\nTuple{p_1,p_2}}{\Rename_1}{\Rename_2}
	&\definedas \nTuple{\Fusion{p_1}{\Rename_1}{\Rename_2}, \Fusion{p_2}{\Rename_1}{\Rename_2}}
	&&\text{otherwise.}
	\\[-25pt]
\end{align*} 
\caption{Definition of a ``zero-parsimonious'' sum of  two renamings $\Rename_1$ and $\Rename_2$ of a pattern $\PatAddA$.}
\label{subfigure:def_fusion_renaming}
\end{figure}

By Proposition~\ref{prop:typingLinearBLL}, $U$ has a $\with$-sequence type $H$ and have at most one free $\with$-sequence pattern $\PIn:L$. Recall Subsection~\ref{subsect:examples}, the term $U$ can be seen as a linear map from the vector space associated with $L$ to the vector space associated with $H$.
So that we can transpose it by:

\begin{equation}
	\label{eq:transpose_function}
	\overleftarrow{U} = 
		\Osi[L]\left(
			\lambda {\PIn}.\Iso[H](\POut)U
			\right)		
\end{equation}
which is a term of type $\POut:H\vdash \overleftarrow{U}: L$, reversing $U$. 
This is the syntactic counterpart of the fact that a Cartesian differential category equipped with a dagger structure is a Cartesian reverse derivative category \cite{cockett_et_al:LIPIcs.CSL.2020.18}. 
So, why not simply define $\mathcal{T}(U)$ as $\overleftarrow{U}$? Because the term $\overleftarrow{U}$ is highly inefficient in terms of the number of flops required to compute a reverse derivative.
More precisely, $\Osi[L]\definedas\lambda f^{L\multimap \typR}.\sum_{V\in\mathcal B_L} (f(V)) \dot*_L V$ is not a safe term as it replicates $f$ as many times as the dimension of the space associated with the input type $L$, which can be exceedingly large. If evaluating $f(V)$ will require a number of flops linear in $L$, then evaluating the full sum in $\Osi[L]$ will be at least quadratic. 

The essence of JAX Autodiff is to exploit the syntactic structure of $U$ to construct a term $\mathcal{T}(U)$ extensionally equivalent to $\overleftarrow{U}$, yet with a comparable workload as $U$. This section shows how such a transformation can be expressed within $\lambdaLL$ via a careful handling of variable renaming.

The definition of $\lintranspose{}U$ (Figure~\ref{subfigure:transpose on lambdaLL_t}) should take into account two crucial features of the $\with$-pattern $\PIn:L$ in the typing environment of $U$: variables in $\PIn$ may occur several times in $U$ (because of additive contraction) or do not occur at all (because of $\with$ elimination). Different occurrences should be renamed into different variables as the transpose of additive contraction is addition. For example, if $\PIn=\nTuple{u,u'}$ and $U=\nTuple{\nTuple{u,u},u'}$, then $\lintranspose{}U$ is a term  $\beta_\lambda$-equivalent to $\nTuple{u_1+u_2,u'}$\todom{add below an example}. This is implemented by a \emph{variable renaming} $\Rename$ (i.e.~a bijection between two sets of variables $\Dom\Rename$ and $\Codom\Rename$ preserving types) and of an action $\RenameTerm\Rename M$ of the renaming on a term $M$, replacing any $u\in\Dom{\Rename}\cap\FV M$ with $\Rename(u)$. Specifically, the definition of $\lintranspose{}{\nTuple{U_1,U_2}}$ in Figure~\ref{subfigure:transpose on lambdaLL_t} recursively applies $\mathcal T$ to $\RenameTerm{\Rename_1}{U_1}$ and $\RenameTerm{\Rename_2}{U_2}$, separating occurrences of the same variable in $U_1$ and $U_2$. The results are then combined in the term $\Fusion{\PIn}{\Rename_1}{\Rename_2}$, as explained below.

On the other side, the variables in $\PIn$ not occurring in $U$ will be associated with $\underline 0$ terms, as the transpose of weakening is the empty sum, i.e.~zero. However, we must be  parsimonious in adding such $\underline 0$, as if they were summed with other terms, they would cost some useless numerical additions. Our notion of renaming is then partial, in the sense that $\Dom{\Rename}$ can be strictly smaller than the set $\FV{\PIn}$, in fact it will be $\FV{\PIn}\cap\FV U$. 
Formally, such as ``zero-parsimonious'' sum is implemented by defining in Figure~\ref{subfigure:def_fusion_renaming} a $\lambdaLL^{\mathtt{t}}$ term $\Fusion{\PIn}{\Rename_1}{\Rename_2}$, given a pattern $\PIn$ and two partial renamings $\Rename_1$ and $\Rename_2$ of disjoint codomain. For instance, take $\PIn=\nTuple{\nTuple{x,y},u}$ and $\Rename_1\definedas\{x\mapsto x_1, y\mapsto y_1\}$ and $\Rename_2\definedas\{x\mapsto x_2\}$, we have:  
$\Fusion{\PIn}{\Rename_1}{\Rename_2}=\nTuple{\nTuple{x_1\dot+x_2,y_1},\underline 0}$.

Finally, notice that the type of $\lintranspose{}U$ may be a type $L'$ that is “smaller” than the type $L$ of the input pattern $\PIn$ in $U$, for instance in the cases of $U=\underline 0$ or $U=\nTuple\,$ of Figure~\ref{subfigure:transpose on lambdaLL_t}. In general, $L'$ is obtained from $L$ by removing the types associated with the variables in $\PIn$ that do not occur free in $U$. Such a type $L'$ is formally given by the \emph{partial} renaming operation $\RenamePat{\Rename}{\PIn}$, defined in Figure~\ref{figure:def_pattern_renaming}. 
Note that $\RenamePat{\Rename}{\PIn}\neq\RenameTerm{\Rename}{\PIn}$. For example, let $\PIn=\nTuple{\nTuple{x,y},u}$ and $\Rename\definedas\{x\mapsto x_1, y\mapsto y_1\}$. Then:
$\RenamePat{\Rename}{\PIn}=\nTuple{x_1,y_1}$, while $\RenameTerm{\Rename}{\PIn}=\nTuple{\nTuple{x_1,y_1},u}$.

The original type $L$ of $\PIn$ is recovered from $\lintranspose{}U$ of type $L'$ in the definition of $\lintranspose{\affcontext{\overleftarrow{\Phi}}}{\lambda \PIn.U}$ in Figure~\ref{subfigure:transpose on lambdaLL_f}. In this case, the recursive call $\lintranspose{\affcontext{\overleftarrow{\Phi}}, \PIn}{U}$ is assigned to $\RenamePat{\Rename}{\PIn}$, where $\Rename$ is the identity renaming restricted to $\FV{U}\cap\FV{\PIn}$, i.e. $\Dom\Rename=\FV{\PatAddA}\cap\FV{U}$ and $\Rename(u)=u$. This result is then injected into $\Fusion{\PIn}{\Rename}{\emptyset}$, with $\emptyset$ denoting the empty renaming, i.e.~$\Dom\emptyset=\emptyset$. The term $\Fusion{\PIn}{\Rename}{\emptyset}$ reconstructs the original type $L$ by inserting zero terms for the components of $L$ erased in $L'$.
For example, if $\PIn=\nTuple{\nTuple{x,y},u}$ and $U=\nTuple{x,y}$, with typing judgment $\PIn:(L_1\with L_2)\with L_3\vdash U:L_1\with L_2$, then $\Rename=\{x\mapsto x,, y\mapsto y\}$, and:
$
	\lintranspose{}{\lambda\PIn.U}=\lambda \nTuple{x,y}.\letterm{\nTuple{x,y}}{\lintranspose{\PIn}{U}}{\nTuple{\nTuple{x,y},0_{L_3}}}
$,	
where $\lintranspose{\PIn}{U}=\nTuple{x,y}$. Notice in particular that $\nTuple{x,y}:L_2\with L_2\vdash\lintranspose{\PIn}{U}:L_2\with L_2$ while $\vdash\lintranspose{}{\lambda\PIn.U}:(L_1\with L_2)\multimap (L_1\with L_2)\with L_3$, as expected. We refer to Appendix~\ref{app:transpose} for more details.

\begin{figure*}
	\begin{subfigure}{\textwidth}
	\begin{align*}
		\lintranspose{\affcontext{\overleftarrow{\Phi}}}{\lambda \PatAddA.U}
	&
	\definedas 
	\lambda\POut.
	\letterm{\RenamePat\Rename\PatAddA}{\lintranspose{\affcontext{\overleftarrow{\Phi}},\PatAddA}{U}}{\Fusion \PatAddA{\Rename}{\emptyset}}
	\\
	\lintranspose{\affcontext{\overleftarrow{\Phi}},\affinebang\overleftarrow f}{\affderterm f}
	&
	\definedas 
	\affderterm{\overleftarrow f}
	\\		
	\lintranspose{\affcontext{\overleftarrow{\Phi}}}{\letterm{\affinebang f
	}{\affinebang F}{G}}
	&
	\definedas 
	\begin{cases}
	\letterm{\affinebang \overleftarrow{f}
	}{\affinebang \lintranspose{\affcontext{\overleftarrow{\Phi}}}{F}}{\lintranspose{\affcontext{\overleftarrow{\Phi}},\affinebang\overleftarrow f}{G}}
	&\text{if $f\in\FV{G}$,}\\
	\lintranspose{\affcontext{\overleftarrow{\Phi}}}{G}	
	&\text{otherwise.}
	\end{cases}
	\\
		\lintranspose{\affcontext{\overleftarrow{\Phi}}}{\dot +}
	&
	\definedas 
	\lambda u.\nTuple{u,u}
	\\
	\lintranspose{\affcontext{\overleftarrow{\Phi}}}{\dot * x}
	&
	\definedas 
	\dot * x
	\end{align*}
	\caption{Definition of $\mathcal T$ on $\lambdaLL^{\mathtt f}$. 
	If $\oc\Sigma, \affcontext{\Phi} \vdash F:L\multimap H$, then $\oc\Sigma, \affcontext{\overleftarrow{\Phi}}\vdash \lintranspose{\affcontext{\overleftarrow{\Phi}}}{F} : H\multimap L$.
	In the case of $\lambda \PatAddA.U$, the pattern
	$\POut$ is the one free in $\lintranspose{\affcontext{\overleftarrow{\Phi}},\PatAddA}{U} $, the renaming
	$\Rename$ is the identity restricted to $\FV{\PatAddA}\cap\FV{U}$, i.e. $\Dom\Rename=\FV{\PatAddA}\cap\FV{U}$ and $\Rename(u)=u$, and $\emptyset$ denotes the empty renaming, i.e.~$\Dom\emptyset=\emptyset$.
	In the case of $\letterm{\affinebang f}{\affinebang F}{G}$, $\mathcal T$ simply eliminate $\affinebang F$ if $f\notin\FV{G}$, to avoid useless computation. 
	}	
	\label{subfigure:transpose on lambdaLL_f}
	\end{subfigure}
	
	\begin{subfigure}{\textwidth}
	\begin{align*}
		\lintranspose{\affcontext{\overleftarrow{\Phi}},\PIn}{u}
	&
		\definedas u
	\\
	 \lintranspose{\affcontext{\overleftarrow{\Phi}},\PIn}{FU'}
	 &
		\definedas 
	(\lambda{\POut}'.\lintranspose{\affcontext{\overleftarrow{\Phi}},\PIn}{U'})
	(\lintranspose{\affcontext{\overleftarrow{\Phi}}}{F}\POut)
	\\
		\lintranspose{\affcontext{\overleftarrow{\Phi}},\PIn}{\underline 0}
	&
		\definedas \lintranspose{\affcontext{\overleftarrow{\Phi}},\PIn}{\nTuple{\,}}
	\definedas \nTuple{\,}
	\\
		\lintranspose{\affcontext{\overleftarrow{\Phi}},\PIn}{\nTuple{U_1,U_2}}
	&
		\definedas 
		\letterm{
		\nTuple{
			\RenamePat{\Rename_1}\PatAddA, 
			\RenamePat{\Rename_2}\PatAddA
		}
		}
		{
			\nTuple{
				\lintranspose{\affcontext{\overleftarrow{\Phi}},\RenameTerm{\Rename_1}{\PIn}}
				{
					\RenameTerm{\Rename_1}{U_1}
				},
				\lintranspose{\affcontext{\overleftarrow{\Phi}},\RenameTerm{\Rename_2}\PIn}{
					\RenameTerm{\Rename_2}{U_2}
				}
			}
		}
		{
			\Fusion{\PIn}{\Rename_1}{\Rename_2}		
		}
\end{align*}
	\caption{
		Definition of $\mathcal T$ on $\lambdaLL^{\mathtt t}$. 
		Given $\oc\Sigma, \affcontext{\Phi}, \PIn:L\vdash U:H$, 
		we have 
		$\oc\Sigma, \affcontext{\overleftarrow{\Phi}}, \POut : H\vdash \lintranspose{\affcontext{\overleftarrow{\Phi}},\PIn}{U} : L'$, for a suitable pattern $\POut : H$
		and a type $L'$ erasing from $L$ the components of $\PIn$ not in $\FV{U}$. 
		In the $FU'$ case, ${\POut}'$ is the pattern associated with $\lintranspose{\affcontext{\overleftarrow{\Phi}},\PIn}{U'}$.
		In the $\nTuple{U_1,U_2}$ case,  $\Rename_1$, $\Rename_2$ are two renamings of disjoint codomains s.t.~$\Dom{\Rename_i}=\FV{U_i}\cap\FV{\PIn}$.
	}
	\label{subfigure:transpose on lambdaLL_t}		
	\end{subfigure}
 
	\begin{subfigure}{\textwidth}
	\begin{align*}
	 \lintranspose{\affcontext{\overleftarrow{\Phi}}}{\tupleterm{P}{\affinebang F}}
		 &
		 \definedas 
		 \tupleterm{P}{\affinebang\lintranspose{\affcontext{\overleftarrow{\Phi}}}{F}}
	\\
		\lintranspose{\affcontext{\overleftarrow{\Phi}}}{\letterm{(\oc x,\affinebang f)}{R}{S}}
		&
		\definedas 
		\begin{cases}
			\letterm{\oc x}{\ExpContextA{P}}{\lintranspose{\affcontext{\overleftarrow{\Phi}}}{S}}, \text{ for } \unzippingPre{R}=(\ExpContextA{},P, F) & \text{if } f\notin \FV{S}\\
			\letterm{(\oc x,\affinebang\overleftarrow{f})}
			{\lintranspose{\affcontext{\overleftarrow{\Phi}}}{R}}
			{\lintranspose{\affcontext{\overleftarrow{\Phi}}, \affinebang\overleftarrow{f}}{S}} & \text{otherwise}
		\end{cases}
		\\
		\lintranspose{\affcontext{\overleftarrow{\Phi}}}{\letterm{\PatExpA}{z}{S}} 
		&
		\definedas 
		\letterm{\PatExpA}{z}{\lintranspose{\affcontext{\overleftarrow{\Phi}}}{S}}
		\\
		\lintranspose{\affcontext{\overleftarrow{\Phi}}}{\letterm{\oc x}{P}{S}}
		&
		\definedas 
		\letterm{\oc x}{P}{\lintranspose{\affcontext{\overleftarrow{\Phi}}}{S}}
		\\
		\lintranspose{\affcontext{\overleftarrow{\Phi}}}{\letterm{\affinebang f}{\mbox{}\affinebang F}{S}}
		&
		\definedas 
		\begin{cases}
			\letterm{\affinebang\overleftarrow{f}}
			{\affinebang \lintranspose{\affcontext{\overleftarrow{\Phi}}}{F}}
			{\lintranspose{\affcontext{\overleftarrow{\Phi}},\affinebang\overleftarrow{f}}{S}}    
			& \text{if } f\notin \FV{S}\\
			\lintranspose{\affcontext{\overleftarrow{\Phi}},\affinebang\overleftarrow{f}}{S}
			& \text{otherwise}
		\end{cases}
	\end{align*}
	\caption{Definition of $\mathcal T$ on $\lambdaLL^{\mathtt A}$. If $\oc\Sigma, \affcontext{\Phi}\vdash S: \oc E\otimes\affinebang(L\multimap H)$, then $\oc\Sigma, \affcontext{\overleftarrow\Phi}\vdash \lintranspose{\affcontext{\overleftarrow{\Phi}}}{S}: \oc E\otimes\affinebang(H\multimap L)$.
	As before, $\lintranspose{\affcontext{\overleftarrow{\Phi}}}{\letterm{\affinebang f}{\affinebang F}{S}}$ supposes $f\in\FV{S}$, otherwise 
	$\mathcal T$ simply eliminate $\affinebang F$.}
	\label{subfigure:transpose on lambdaLL_a}		
	\end{subfigure}
\caption{
	Definition of the transpose transformation $\mathcal T$. 
}
\label{fig:def_transpose}
\end{figure*}

Let us come back to our running example. Consider the $\lambdaLLFM$ term $\unzipping{\forward{\theta}{P}}$ in Figure~\ref{fig:ex_unzipping}: by applying $\mathcal{T}$ and after some $\beta_\lambda$-simplifications for readability\footnote{For instance, take the application $f_3\ x'$ from the last line of Figure~\ref{fig:ex_unzipping}. According to the $FU'$ case in Figure~\ref{subfigure:transpose on lambdaLL_t}, the transposed form of this term is $(\lambda u'. u')\ (\overleftarrow{f_3}\ x')$. In Figure~\ref{fig:ex_transpose}, we have simplified this expression using a $\beta_\lambda$-reduction to eliminate the identity function.
}, we obtain the $\lambdaLLFM$ term in Figure~\ref{fig:ex_transpose}. The full description can be found in Appendix~\ref{subsect:exTransp}.
According to Theorem~\ref{th:transpose}, this term has type $\oc\typR\otimes\affinebang(\typR\multimap\typR\&\typR)$ and free patterns $\oc x:\oc \typR,\oc y:\oc \typR$. 
The term begins with the primal computation which is equal to that in Figure~\ref{fig:ex_unzipping}: $\mathcal T$ keeps untouched all exponential subexpressions.  
The computation in Figure~\ref{fig:ex_transpose} proceeds then with the reverse-mode differentiation pass, which is highlighted in {\color{red} red}. The goal is to aggregate the derivatives of the primitive operations in order to compute the gradient. This part splits between the transposition of the directional derivatives specific to the four let-definitions of the original term $P$ (Figure~\ref{fig:ex_transl}), here implemented by the definitions of ${\color{red}\overleftarrow{f_1}}$ to ${\color{red}\overleftarrow{f_4}}$ and a final ``aggregation'' recovering the dependence graph of the different let-definitions in $P$ backwardly. 
Let us focus on the body of ${\color{red}\lambda z^{\typR}}$. 
The variable ${\color{red}z}$ represents the cotangent associated with the output of $P$. It is applied to ${\color{red}\overleftarrow{f_4}}$, which is defined in the $\mathtt{let}$ above as the transpose of the addition (in fact the term ${\color{red}\lambda {l}. \nTuple{w_5\mbox{}\dot{*}\mbox{ }{l},w_6\mbox{}\dot{*}\mbox{ }{l}}}$ will reduce to the diagonal $\lambda {l}. \nTuple{{l},{l}}$ as $w_5$ and $w_6$ value $1$). This application takes the role of a fanout operation, duplicating the value carried by ${\color{red}z}$ and propagating it to the components of the gradient of $P$, through the variables ${\color{red}z'}$ and ${\color{red}z''}$. The latter traces the effect of the derivative of $\underline{cos}\mbox{ }\oc x$ back to the $x$ component of the gradient, while the former flows through the $\underline{sin}\mbox{ }\oc x \mbox{ }\underline{*}\mbox{ } y$ term, contributing to both partial derivatives of $g$ with respect to $x$ and $y$. 
These ``aggregations'' or ``flows'' are implemented by $\mathcal T$ by using the terms $\RenamePat\Rename\PatAddA$ and $\Fusion \PatAddA{\Rename_1}{\Rename_2}$ described above. For readability, we have simplified by $\beta_\lambda$-reduction many of these terms in Figure~\ref{fig:ex_transpose}, however we have kept untouched the ones generated by the pattern ${\color{blue}\nTuple{x',y'}}$ during the action of $\mathcal T$ over the subexpression ${\color{blue}f_4\nTuple{f_2\nTuple{f_1 x', y'}, f_3x'}}$ in Figure~\ref{fig:ex_unzipping}.
In fact, setting $\PIn=\nTuple{x',y'}$ and $\Rename_1=\{x'\to x'_1,y'\to y'_1\}$ and $\Rename_2=\{x'\to x'_2\}$, we have $\nTuple{\RenamePat{\Rename_1}{\PIn},\RenamePat{\Rename_2}{\PIn}}=\nTuple{\nTuple{x_1',y_1'},x_2'}$ and  $\Fusion{\PIn}{\Rename_1}{\Rename_2}=\nTuple{x_1'\dot+ x_2', y'_1}$.

One can compare Figure~\ref{fig:ex_transpose} with its counterpart, Figure~\ref{fig:ex_transpose_JAX}, in Linear A. The definition of ${\color{red}\overleftarrow{f_4}}$ morally corresponds to the $\delta$ encoding of the first two red lines in Figure~\ref{fig:ex_transpose_JAX}, ${\color{red}\overleftarrow{f_3}}$ to the third, ${\color{red}\overleftarrow{f_2}}$ to the fourth and fifth, and ${\color{red}\overleftarrow{f_1}}$ to the sixth. These components are then composed backwardly within the ${\color{red}\lambda z \dots}$ term in $\lambdaLL$, which encompasses the last two red lines in Figure~\ref{fig:ex_transpose_JAX}.


\begin{theorem}[Type $\mathcal T$]\label{th:transpose}
	Let $\judgmenttransl{\oc\Sigma,\affcontext{\Phi}}{R}{\oc E}{L}{H}$ in $\lambdaLLFM$, then $\judgmenttransl{\oc\Sigma,\affcontext{\transptyp{\Phi}}}{\lintranspose{\affcontext{\transptyp{\Phi}}}{R}}{\oc E}{H}{L}$.
\end{theorem}

\begin{theorem}[Soundness $\mathcal T$]
	\label{th:transpose_JAX_LL}
	Given $\judgmentJAX{\Gamma}{\lincontextJAX{\Gamma}}{d}{\sigmatypJAX}{\tautypJAX}$ and an enumeration $\theta$ of $\lincontextJAX{\Gamma}$, then
	$
	   \transpose{\TransA[\theta]{d}} 
		\; \sim\;
		\TransA[\linvarJAX{u}:\tautypJAX]{\transpJAX{\theta}{\linvarJAX{u}:\tautypJAX}{d}}
	$.
\end{theorem}

The following is the analogous of claim 2 of~\cite[Th.~7.2.1]{radul2023you}. 
\begin{theorem}[Workload $\mathcal T$]
	\label{th:cost_transp}
	Given $\oc\Sigma \vdash R: \oc E\otimes\affinebang(L\multimap H)$, we have  and $\cost{\transpose{R}}+\costType{L} \leq \cost{R}+\costType{H}$. If moreover $R$ is safe, then $\transpose{R}$ is safe too.
\end{theorem}

\begin{remark}
	Theorem~\ref{th:cost_transp} refines the intuition that transposition preserves workload by introducing an amortised cost analysis, following~\cite[Sect. 4.3]{radul2023you}. While $\mathcal{W}(\mathcal{T}(R)) \leq \mathcal{W}(R)$ suggests that transposition does not increase cost, it overlooks how erasures in $R$ are transformed into zero terms in $\mathcal{T}(R)$. 
	By definition, workload already accounts for the cost of erasing inputs, but what is missing is the cost of erasing outputs.
	To account for this, the amortised analysis introduces a refined inequality: 
	$\cost{\transpose{R}}+\costType{L} \leq \cost{R}+\costType{H}$ where $\costType{L}$ in the LHS (resp. $\costType{H}$ in the RHS) accounts for erasing tangent outputs of $\mathcal{T}(R)$ (resp. $R$).
\end{remark}

Moreover, we show that our transpose transformation on $U\in\lambdaLL^{\mathtt t}$ produces a term which is extensionally equivalent to $\small \overleftarrow{U}$ of Equation~\ref{eq:transpose_function} but satisfying the condition of Proposition~\ref{prop:safe_reduction_is_safe}. 
\begin{lemma}
	\label{lemma:transpose_vs_function_on_U}
	Given a term $U \in \TangentLL$ such that $\judgment{\PIn:L}{U}{H}$ and let $\POut:H$ be the free additive pattern in $\lintranspose{\PatAddA}{U}$, then we have that:
	$\small \overleftarrow{U}
	\sim_{\POut:H\vdash L} 
	\letterm{\RenamePat\Rename\PatAddA}{\lintranspose{\affcontext{\overleftarrow{\Phi}},\PatAddA}{U}}{\Fusion \PatAddA{\Rename}{\emptyset}}$
	where $\Rename$ is the identity renaming restricted to $\FV{\PIn}\cap \FV{U}$ and $\emptyset$ is the empty renaming. 
\end{lemma}
Note that the term $\Fusion \PatAddA{\Rename}{\emptyset}$ in the statement above enables the type $L'$ of $\lintranspose{\PatAddA}{U}$ to be lifted to the type $L$ by inserting $\underline{0}$ in the null components of the gradient.


\paragraph*{Skipping Unzipping} Recall from Section~\ref{sect:JAX} that the JAX Autodiff transpose is defined only on the fragment Linear B of Linear A, so that unzipping is a necessary step before the transpose. On the contrast, our $\lambdaLL$ transpose is defined on \ref{eq:LLBTerms}, containing the whole image set of Linear A along the $\SymbTransA$ encoding. 
The following proposition states that applying or not the $\lambdaLL$ unzipping yields equivalent terms, so that our formalisation allows for avoiding unzipping. 

\begin{proposition}
	\label{cor:skipping_unzipping}
	\label{prop:unzipping_sim}
	\label{prop:transpose_sim}
	Given $R, R' \in \LinearBLL$, if $R\sim R'$, then: $\unzipping{R} \sim\unzipping{R'}$, $\transpose{R} \sim\transpose{R'}$, and $\transpose{R} \sim\transpose{\unzipping{R}}$.
\end{proposition}

The unzipping transformation can obscure the parallel structure of a program as it is not modular. 
By skipping it as described above, we can preserve the program’s inherent parallel structure. 
Let's illustrate this with an example. 
Consider $P = \underline{f}(Q_1, Q_2)$ where $Q_1$ and $Q_2$ are two complex, independent subprograms of $P$, and $\underline{f}$ representing a binary numeric function, for example multiplication. The program $\mathcal{T}(\mathcal{U}(\mathcal{F}(P)))$ expresses a computation of the gradient of $\Sem 
P$ which first computes all primal values of the intermediate computations of $Q_1$ and $Q_2$, and \emph{then} all tangents of  $Q_1$ and $Q_2$ backwardly. Our system allows however to skip unzipping: $\mathcal{T}(\mathcal{F}(P))$ expresses another computation of the gradient of $\Sem P$, which basically computes $\mathcal{T}(\mathcal{F}(Q_1))$ and $\mathcal{T}(\mathcal{F}(Q_2))$ \emph{independently} and then gather the two results together in order to get the gradient associated with $\Sem P$ (see Appendix~\ref{subsect:exMod}).

\section{Comparison with reverse AD based on back-propagators}\label{sect:comparison}
  

As discussed in the introduction, recent literature offers a wealth of alternative formalisations of reverse-mode automatic differentiation. 
As an example of the benefits of using a general language like $\lambdaLL$, we compare JAX Autodiff with the approach presented in~\cite{POPL2020}, which introduces a linear type system that accounts for the dual number approach to AD.

The original paper~\cite{POPL2020} focuses on the simply typed $\lambda$-calculus, which can be represented in
$\lambdaLL$ via the call-by-value translation, mapping $A \to B$ to $\oc A \multimap \oc B$ \cite{girard1987linear,MARAIST1995370}. The backward transformation in~\cite{POPL2020}  is written by $\Backprop_d$ and it is based on the notion of \emph{back-propagator} (a terminology dating back to \cite{Pearlmutter:2008}), which is a term of a special type $\typR^{\bot_d}$. The subscript $d$ is a natural number and refers to the dimension of the global gradient, which corresponds to the number of inputs of the global program under consideration, and the type $\typR^{\bot_d}$ stands for:
\[
  \typR\multimap\underbrace{\typR\&\dots\&\typR}_{d \text{ times}}.
\]
The $\Backprop_d$ transformation applied to a program $P$ taking $d$ inputs pairs every variable $\oc x : \oc\typR$ occurring in $P$ with a sibling variable\footnote{In fact, $\Backprop_d$ assumes a correspondence between each free variable $x$ of ground type  and its counterpart $x^* : \typR^{\bot_d}$.}  $x^* : \typR^{\bot_d}$. Intuitively, $x^*$ will be replaced with a subroutine giving how much a perturbation in the $d$ inputs of $P$ is required to produce a given perturbation at $x$. The gradient is then read back by applying the back-propagator associated with the output of $P$ to $\underline{1}$.

\begin{figure}
\begin{align*}
	\Backprop_d(\underline{f}(\oc x_1, \oc x_2))\,
	\definedas
	&\mbox{ }
	\letterm{\oc w_1}{\underline{\partial_1 f} (\oc x_1,\oc x_2)}{} 
    \letterm{\oc w_2}{\underline{\partial_2 f} (\oc x_1,\oc x_2)}{}
    \\ 
    & \mbox{ } 
    \tupleterm
    {\funterm{(\oc x_1,\oc x_2)}}
    {\oc(\lambda u^\typR.x_1^*(w_1 \dot{*} u)\mbox{ } \dot{+}_{\typR^d}\mbox{ } x_2^*(w_2 \dot{*}u))}
\end{align*}

\caption{Backprop transformation over binary numeric functions, as defined in~\cite{POPL2020}.}
\label{fig:funct_brunel_mazza_pagani} 
\end{figure}

Figure~\ref{fig:funct_brunel_mazza_pagani} defines the action of $\Backprop_d$ on the basic binary functions $\underline f$. We hope the reader will allow some adaptation of the definitions from~\cite{POPL2020} to fit $\lambdaLL$ notation, in particular by assuming the already mentioned call-by-value translation of simply typed terms into linear logic. The judgment $\oc x_1:\oc\typR ,\oc x_2 :\oc\typR\vdash \funterm{(\oc x_1,\oc x_2)}: \oc\typR$ is then transformed along  $\Backprop_d$ into:
\[
	\oc x_1:\oc\typR,\oc x_1^*:\oc(\typR^{\bot_d}), \oc x_2 :\oc\typR,\oc x_2^*:\oc(\typR^{\bot_d})\vdash \Backprop_d(\underline{f}(\oc x_1, \oc x_2)): \oc\typR\otimes\oc(\typR^{\bot_d})
\]
Let us compare $\Backprop_d$ with the JAX Autodiff transformation described so far.
As already mentioned in the introduction, $\Backprop_d$ is defined as a stand-alone transformation, whereas reverse JAX Autodiff is expressed as the composition of forward mode $\mathcal{F}$, unzipping $\mathcal{U}$, and transposition $\mathcal{T}$. Moreover, $\Backprop_d$ is parameterised by $d$, the dimension of the domain of the global function. By contrast, JAX Autodiff transformations are defined solely with respect to the free variables of the expressions to which the transformation is applied, which guarantees a greater modularity.

Let us now focus on the cotangent part of the transformations, we have:
\begin{align}
	\label{align:backprop}
	&\lambda u^\typR.x_1^*(w_1 \dot{*} u)\mbox{ } \dot{+}_{\typR^d}\mbox{ } x_2^*(w_2 \dot{*}u)
	&&\text{from $\Backprop_d(\underline f(\oc x_1, \oc x_2))$}\\
	\label{align:autodiff}
	&\lambda u^\typR.\nTuple{w_1 \dot{*} u, w_2 \dot{*}u}
	&&\text{from $\mathcal{T}(\mathcal{U}(\mathcal{F}(\underline f(\oc x_1, \oc x_2))))$}	
\end{align}
The JAX Autodiff transformation \eqref{align:autodiff} gives, as expected, the transpose of the jacobian of $f$, computing the gradient of $f$ whenever fed by $1$. In the contrast, the $\Backprop_d$ transformation \eqref{align:backprop} tags the components of such a jacobian with the two variables $x_1^*$ and $x_2^*$. One can easily read back the gradient of $f$ by supposing $d=2$ and replacing $x_1^*$ with the injection $\lambda u. \nTuple{u,0}$ and $x_2^*$ with $\lambda u. \nTuple{0,u}$, but these variables may be used to inject the values of the partial derivatives of $f$ to different components of larger vectors (if $d>2$), e.g. the gradient of a global function enveloping $\underline f$. 

\begin{figure}
    \begin{subfigure}{\linewidth}
        \centering
        \begin{align*} 
            &\letterm{(\oc v_1, \color{orange}\oc v_1^*
            \color{black})}{
                \left(
                        \letterm{\oc w_1}{\underline{cos}\mbox{ }\oc x}
                        {
                        	\tupleterm{\underline{sin}\mbox{ }\oc x}
                        	{\color{orange}\oc (\lambda u. x^*(w_1\mbox{ }\dot{*}\mbox{ }u))\color{black}}
                        }
                \right)
            }{}
            \\[1mm]
            &\letterm{(\oc v_2, \color{orange} \oc v_2^* \color{black})}{
                \left(
                        \letterm{\oc w_2}{\oc y}{}
                        \letterm{\oc w_3}{\oc v_1}{}
                        \tupleterm{\oc v_1 \underline{*} \oc y}
                        {\color{orange}
                        \oc (\lambda u. v_1^*(w_2\dot{*}\mbox{ }u)
                        \mbox{ }\dot{+}\mbox{ } y^*(w_3\mbox{ }\dot{*}\mbox{ }u))
                        \color{black}}
                \right)
            }{}
            \\[1mm]
            &\letterm{(\oc v_3, \color{orange} \oc v_3^* \color{black})}{
                \left(
                        \letterm{\oc w_4}{\underline{-sin}\mbox{ }\oc x}{}
                        \tupleterm{\underline{cos}\mbox{ }\oc x}
                        {\color{orange}\oc (\lambda u. x^*(w_4\mbox{ }\dot{*}\mbox{ }u))\color{black}}
                \right)
            }{}
            \\[1mm]
            &\letterm{(\oc v_4, \color{orange} \oc v_4^* \color{black})}{
                \left(
                        \letterm{\oc w_5}{\oc \underline{1}}{}
                        \letterm{\oc w_6}{\oc \underline{1}}{}
                        \tupleterm{\oc v_2 \mbox{ }\underline{+}\mbox{ } \oc v_3}
                        {\color{orange}
                        \oc (\lambda u. v_2^*(w_5\mbox{ }\dot{*}\mbox{ }u)
                        \mbox{ }\dot{+}\mbox{ } v_3^*(w_6\mbox{ }\dot{*}\mbox{ }u))
                        \color{black}}
                \right)
            }{
            }
            \\[1mm]
            & 
            \left(
                \oc v_4,
                \color{orange} \oc v_4^*
            \right)
        \end{align*}
        \vspace{-0.7cm}
        \caption{Application of $\Backprop_d$ to the $\lambdaLL$ term defined in Figure~\ref{fig:ex_transl}.}
        \label{subfig:ex_backprop_litteral}
    \end{subfigure}              
   \medskip
    \begin{subfigure}{\linewidth}
        \centering
        \begin{align*} 
        &
        \letterm{\oc w_1}{\underline{cos}\mbox{ }\oc x}{}        
        \letterm{\oc v_1}{\underline{sin}\mbox{ }\oc x}{}
        \letterm{\oc w_2}{\oc y}{}
        \letterm{\oc w_3}{\oc v_1}{}
        \letterm{\oc v_2}{\oc v_1 \mbox{ }\underline{*}\mbox{ } \oc y}{}
        \\
        &       
        \letterm{\oc w_4}{\underline{-sin}\mbox{ }\oc x}{}
	\letterm{\oc v_3}{\underline{cos}\mbox{ }\oc x}{}        
	\letterm{\oc w_5}{\oc \underline{1}}{}
        \letterm{\oc w_6}{\oc \underline{1}}{}      	
        \letterm{\oc v_4}{\oc v_2 \mbox{ }\underline{+}\mbox{ } \oc v_3}{}
        \\
         &
         \left(
                \oc v_4,
                \color{orange}
                    \oc \left(\quad
                    \begin{aligned}
                        &
                        \letterm{\color{orange}\oc v_1^*\color{black}}{
                        	{\color{orange}\oc (\lambda u. x^*(w_1\mbox{ }\dot{*}\mbox{ }u))\color{black}}
         		}{}
			\letterm{\color{orange} \oc v_2^* \color{black}}{
                        {\color{orange}
                        \oc (\lambda u. v_1^*(w_2\dot{*}\mbox{ }u)
                        \mbox{ }\dot{+}\mbox{ } y^*(w_3\mbox{ }\dot{*}\mbox{ }u))
                        \color{black}}}{}                      
                        \\                     
                        &
			\letterm{\color{orange} \oc v_3^* \color{black}}{
                        {\color{orange}\oc (\lambda u. x^*(w_4\mbox{ }\dot{*}\mbox{ }u))\color{black}}}{}                        
			\letterm{\color{orange} \oc v_4^* \color{black}}{
                        {\color{orange}
                        \oc (\lambda u. (w_5\mbox{ }\dot{*}\mbox{ }u)
                        \mbox{ }\dot{+}\mbox{ } (w_6\mbox{ }\dot{*}\mbox{ }u)
                        )
                        \color{black}}}{\color{orange}\oc v_4^*\color{black}}
                    \end{aligned}
                    \quad\right)
                \color{black}
            \right)
        \end{align*}
       \vspace{-0.1cm}
        \caption{Application of the unzipping and then some $\beta_\lambda$-simplification to the term in Figure~\ref{subfig:ex_backprop_litteral}.}
        \label{subfig:ex_backprop_simply}
    \end{subfigure}       
            \vspace{-0.3cm}      
        \caption{
        Application of the backprop transformation as defined in \cite[Figure 3.(b)]{POPL2020} to the $\lambdaLL$ term given in Figure~\ref{fig:ex_transl}, computing the numerical function $g(x,y)=(sin(x)*y)+cos(x)$.
        }
        \label{fig:ex_brunel_mazza_pagani}
  \end{figure} 
 
To better illustrate this point, let us compare $\Backprop_d$ and JAX Autodiff using our running example from Figure~\ref{fig:mainfigure}. 
Applying $\Backprop_d$ to the term in Figure~\ref{fig:ex_transl} yields the term shown in Figure~\ref{subfig:ex_backprop_litteral}, with the cotangent part highlighted in orange. 
By subsequently applying the unzipping transformation, we obtain the term in Figure~\ref{subfig:ex_backprop_simply}, which can then be directly compared to the term in Figure~\ref{fig:ex_transpose}, obtained via the JAX Autodiff transformation.
 
By evaluating in Figure~\ref{subfig:ex_backprop_simply} the intermediate back-propagators $v_i^*$'s by $\beta_\lambda$-reduction we get:
\begin{equation}
\label{eq:backprop_running_cotangent}
	\lambda u.
		x^*(w_1 \dot* (w_2 \dot* (w_5  \dot * u)))
		\,\dot+_{\typR^2}\, y^*(w_3 \dot* (w_5  \dot * u)) 
		\,\dot+_{\typR^2}\, x^*(w_4  \dot * (w_6  \dot * u))
\end{equation}
Notice that the above $\beta_\lambda$-reduction duplicates the variable $x^*$, which justifies the design choice in $\Backprop_d$ of encapsulating back-propagators within the exponential modality $\oc$. 
This marks a first point of divergence from Figure~\ref{fig:ex_transpose}, where tangents can be instead wrapped by using only the affine modality $\affinebang{}$.
On the side of Figure~\ref{fig:ex_transpose}, by $\beta_\lambda$-replacing the intermediate $\overleftarrow{f_i}$'s and $\beta_\lambda$-simplifying the red subexpressions, we get\todom{add $w_5$ and $w_6$ to Figure~\ref{fig:ex_transpose}}:
\begin{equation}
\label{eq:autodiff_running_cotangent}
	\lambda u.
		\nTuple{w_1 \dot* (w_2 \dot* (w_5  \dot * u)) 
		\,\dot+_{\typR}\, w_4 \dot* (w_6  \dot * u), w_3\dot* (w_5  \dot * u)}
\end{equation}

Clearly, both terms compute the gradient of the overall program in reverse order, accumulating partial derivatives from output to input. For example, they first compute\footnote{
The $\beta_\lambda$-reductions leading to \eqref{eq:backprop_running_cotangent} and \eqref{eq:autodiff_running_cotangent} are not sage. \emph{Safe reductions} do not substitute expressions like $w_5 \dot * u$ directly but first reduce them to numerals—closed strong values of type $\typR$ (Section~\ref{subsect:cost_model}). This distinction matters when $w_5 \dot * u$ is shared across gradient components. Nonetheless, for this discussion, we keep variables $w_i$ explicit to ease comparison with $\Backprop_d$.

}
 $w_5 \dot * u$, then multiply the result by $w_2$, and finally by $w_1$.

However, the two terms differ in how they handle the various components of the gradient. 
The term \eqref{eq:autodiff_running_cotangent} already displays the tuple representing the global gradient in the body of the $\lambda u$ abstraction. 
The intermediate results have been routed to the appropriate components via the terms 
 in Figure~\ref{figure:def_pattern_renaming} and~\ref{subfigure:def_fusion_renaming} generated by the $\mathcal T$ transformation. 

By contrast, the term \eqref{eq:backprop_running_cotangent} does not make the tuple explicit. 
Instead, it encodes the gradient as a formal sum labeled by the back-propagators $x^*$ and $y^*$. 
The gradient can then be reconstructed by substituting these variables with their associated injections.
However, evaluating \eqref{eq:backprop_running_cotangent} into \eqref{eq:autodiff_running_cotangent} by substituting $x^*$ with $\lambda u.\nTuple{u, 0}$ and $y^*$ with $\lambda u.\nTuple{0, u}$ can be costly in terms of numerical operations.

Here is another difference with respect to the JAX Autodiff approach. 
The correct workload of backpropagation is ensured in \cite{POPL2020} through a custom operational semantics that carefully manages the evaluation of expressions of type $\typR^{\bot_d}$.
In particular, this semantics adopts a rewriting rule called \emph{linear factoring}, which transforms expressions of the form $x^*(e_1) \dot +_{\typR^d} x^*(e_2)$ into $x^*(e_1 \dot +_{\typR} e_2)$ --- thereby converting additions over gradient vectors into additions over real numbers.

The papers \cite{10.1145/3498710, Vakar_linear} also explore a form of reverse-mode automatic differentiation similar to $\Backprop_d$. However, instead of relying on custom symbolic rewriting for efficiency, they achieve this latter through a specialised implementation of the data type ${\typR^d}$ and the operation $\dot +_{\typR^d}$. In particular, \cite{Vakar_linear} presents a range of optimizations for ${\typR^d}$, progressing from a straightforward algebraic definition to a much lower-level implementation based on mutable arrays, thereby illustrating a trade-off between abstraction and performance.

%


\section{Conclusion} \label{sect:conclusion}

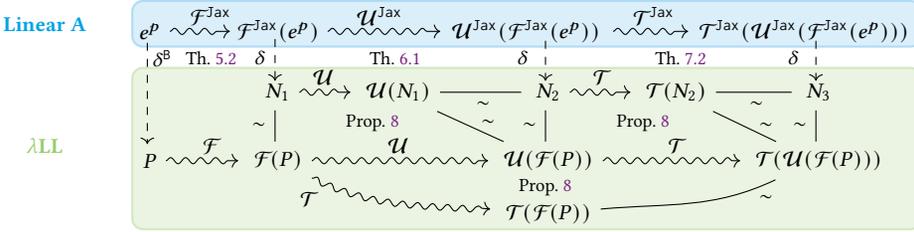
\begin{figure}[h!]
    \centering
    \begin{tikzpicture}[black]
      \draw[fill=YellowGreen!40,YellowGreen!40,thick,rounded corners,
        fill opacity=0.45](-5,4.32) rectangle (5.5,2.2);
      
      \draw[fill=Cerulean!30,Cerulean!30,thick,rounded corners,
        fill opacity=0.45](-5,5.2) rectangle (5.5,4.6);

       \node[align=flush right, Cerulean] at (-6.2,4.9) {\mysizehead{Linear A}};
 
      \node at (-4.8,4.8) {\mysizebody{$e^p$}};

      \node at (-3.1,4.8) {\mysizebody{$\forwardJAX{}{e^p}$}};

      \node at (0.2,4.8) {\mysizebody{$\unzippingJAX{\forwardJAX{}{e^p}}$}};

      \node (UJ) at (0.5,4.8) { };
      
      \node at (3.9,4.8) {\mysizebody{${\mathcal{T}^{\mathtt{Jax}}(\unzippingJAX{\forwardJAX{}{e^p}}})$}};

      \node (TJ) at (4.1,4.8){};

      \draw [->,decorate,decoration={snake,amplitude=.4mm,
      segment length=2mm,post length=1mm}] (-4.5,4.8) -- (-3.7,4.8);
      \node  at (-4,5) {\mysizebody{$\mathcal{F}^{\mathtt{Jax}}$}};

      \draw [->,decorate,decoration={snake,amplitude=.4mm,
      segment length=2mm,post length=1mm}] (-2.4,4.8) -- (-0.9,4.8);
      \node  at (-1.75,5) {\mysizebody{$\mathcal{U}^{\mathtt{Jax}}$}};

      \draw [->,decorate,decoration={snake,amplitude=.4mm,
      segment length=2mm,post length=1mm}] (1.4,4.8) -- (2.4,4.8);
      \node  at (1.9,5) {\mysizebody{$\mathcal{T}^{\mathtt{Jax}}$}};

      \node[align=flush right, YellowGreen]  at (-6.2,3.3) {\mysizehead{$\mathbf{\lambda}$LL}};
      
      \node (P) at (-4.8,3.1) {\mysizebody{$P$}};

      \node  (FM) at (-3.1,3.1) {\mysizebody{$\mathcal{F}(P)$}};

      \node  (U) at (0.5,3.1) {\mysizebody{$\mathcal{U}(\mathcal{F}(P))$}};

      \node  (T) at (4.1,3.1) {\mysizebody{$\mathcal{T}(\mathcal{U}(\mathcal{F}(P)))$}};

      \draw [->,decorate,decoration={snake,amplitude=.4mm,
      segment length=2mm,post length=1mm}]  (P) -- (FM);
      \node  at (-4,3.3) {\mysizebody{$\mathcal{F}$}};

      \draw [->,decorate,decoration={snake,amplitude=.4mm,
      segment length=2mm,post length=1mm}]  (FM) -- (U);
      \node  at (-1.5,3.3) {\mysizebody{$\mathcal{U}$}};

      \draw [->,decorate,decoration={snake,amplitude=.4mm,
      segment length=2mm,post length=1mm}] (U) -- (T);
      \node  at (2.2,3.3) {\mysizebody{$\mathcal{T}$}};

      \draw [->,dashed] (-4.8,4.65) -- (P);
      \node  at (-4.6,4.45) {\mysizetiny{$\SymbTransB$}};

      \draw [->,dashed] (-3.1,4.7) -- (-3.1,4.15);
      \node  at (-3.3,4.45) {\mysizetiny{$\SymbTransA$}};
      
      \node  (N1) at (-3.1,4) {\mysizebody{$N_1$}};

      \node  at (-3.95,4.45) {\mysizetiny{Th.~\ref{th:forward_soundness}}};

      \draw (N1) -- (FM) node [midway,left] {\mysizebody{$\sim$}}; 

      \node  (N2) at (0.5,4) {\mysizebody{$N_2$}};

      \draw [->,dashed] (UJ) -- (0.5,4.15);
      \node  at (0.2,4.45) {\mysizetiny{$\SymbTransA$}};

      \node  (UN1) at (-1.5,4) {\mysizebody{$\mathcal{U}(N_1)$}};

      \draw [->,decorate,decoration={snake,amplitude=.4mm,
      segment length=2mm,post length=1mm}]  (N1) -- (UN1);
      \node  at (-2.5,4.2) {\mysizebody{$\mathcal{U}$}};

      \draw  (UN1) -- (N2) node [midway,below] {\mysizebody{$\sim$}}; 

      \draw  (N2)-- (U) node [midway,left] {\mysizebody{$\sim$}};

      \draw (U) -- (UN1);
      \node at (-0.3,3.6) {\mysizebody{$\sim$}};

      \node  at (-1.5,4.45) {\mysizetiny{Th.~\ref{th:unzipping_soundness}}};

      \node  at (-1.8,3.6) {\mysizetiny{Prop.~\ref{prop:unzipping_sim}}};
      

      \node  (N3) at (4.1,4) {\mysizebody{$N_3$}};

      \node  (TN2) at (2.2,4) {\mysizebody{$\mathcal{T}(N_2)$}};

      \draw [->,decorate,decoration={snake,amplitude=.4mm,
      segment length=2mm,post length=1mm}]  (N2) -- (TN2);
      \node  at (1.2,4.2) {\mysizebody{$\mathcal{T}$}};

      \draw [->,dashed] (TJ) -- (4.1,4.15);
      \node  at (3.8,4.45) {\mysizetiny{$\SymbTransA$}};

      \draw  (TN2) -- (T)node [midway,right] {\mysizebody{$\sim$}};
      
      \draw (N3) -- (T) node [midway,left] {\mysizebody{$\sim$}};

      \draw (TN2) -- (N3)  node [midway,below] {\mysizebody{$\sim$}};

      \node  at (2.3,4.45) {\mysizetiny{Th.~\ref{th:transpose_JAX_LL}}};

      \node  at (1.8,3.6) {\mysizetiny{Prop.~\ref{prop:transpose_sim}}};
      \node (TFM) at (0.5,2.4) {\mysizebody{$\mathcal{T}(\mathcal{F}(P))$}};

      \draw [->,decorate,decoration={snake,amplitude=.4mm,
      segment length=2mm,post length=1mm}] (FM) to [out=-25, in=-185] (TFM);
      \node  at (-2.7,2.6) {\mysizebody{$\mathcal{T}$}}; 

      \draw (TFM) to [out=5, in=-155] (T);
      \node at (3.4,2.6) {\mysizebody{$\sim$}};
      \node  at (0.5,2.75) {\mysizetiny{Prop.~\ref{prop:transpose_sim}}};
    \end{tikzpicture}
    \vspace{-.35cm}
    \caption{Comparing Linear A and $\lambdaLL$}
    \label{fig:final_picture}  
  \end{figure}

Figure~\ref{fig:final_picture} summarises our main contributions. 
We have a linear $\lambda$-calculus $\lambdaLL$ with well-behaved $\beta$-reduction and a logical relation $\sim$ that compares programs with respect to their extensional behaviour on the ground types.
We have defined a $\lambdaLL$ encoding $\SymbTransA$ of the Linear A system and the three transformations (forward $\mathcal{F}^{\mathtt{Jax}}$, unzipping $\mathcal{U}^{\mathtt{Jax}}$, transpose $\mathcal{T}^{\mathtt{Jax}}$) formalising AD implementation in libraries like JAX \cite{radul2023you}.

We have proven the soundness both qualitatively and quantitatively: qualitatively, because all transformations commute with $\SymbTransA$ modulo $\sim$ (Theorems~\ref{th:forward_soundness},~\ref{th:unzipping_soundness},~\ref{th:transpose_JAX_LL}); quantitatively, because they all preserve the numerical workload of the original ones (Theorems~\ref{th:cost_forward}, \ref{th:cost_unzip}, \ref{th:cost_transp}). We have also proven that unzipping can be skipped in $\lambdaLL$, giving more modularity (Proposition~\ref{prop:transpose_sim}).

The transformations $\mathcal{F}$, $\mathcal{U}$, and $\mathcal{T}$ were defined on the fragment \ref{eq:LLBTerms} of $\lambdaLL$, which includes the $\SymbTransA$ image of Linear A. Our goal was to compare this encoding with the original Linear A transformations. We plan to consider how these transformations might extend to general $\lambdaLL$ terms—a challenging task if one wishes to preserve a reasonable computational workload. One could go further and explore extending this setting to formalising Autodiff in Dex, a library built on mutable arrays and effect systems.

A natural question is: what practical benefits could such a theoretical framework bring to libraries like JAX? 
Beyond providing a proof-theoretical foundation for formal certification, one may wonder whether unzipping can be skipped in JAX too. 
Our definition of $\mathcal T$ relies on the presence of first-order variables representing linear maps. 
However, such variables are not available in the Linear A grammar, which essentially corresponds to JAX’s expression module, \texttt{Jaxpr} \cite{jax2018github}. 
To make unzipping optional in JAX, one would likely need either to extend \texttt{Jaxpr} so that it encompasses the entire \ref{eq:LLBTerms} grammar, or to devise a workaround that restores the encapsulation provided by these first-order variables within Linear A --- for instance, by using variables that encode matrices.


As mentioned in Section~\ref{sect:introduction}, many formalisations of AD have recently been introduced, raising questions about their interrelations. 
In particular, some of these systems are based on the linear $\lambda$-calculus.
Section~\ref{sect:comparison} compares our approach with those based on the notion of backpropagator, such as in \cite{POPL2020, MazzaP21, 10.1145/3498710, Vakar_linear}. 
Section~\ref{sect:comparison} compares our approach with those based on the notion of back-propagator, such as in \cite{POPL2020, MazzaP21, 10.1145/3498710, Vakar_linear}. 
This serves as an illustration of the kinds of comparisons that $\lambdaLL$ enables. 
Another group of works presents AD as program transformations that share similar types as ours~\cite{Vakar2021CHADCH,kerjeanPedrotDelta}.
Namely, a recent paper~\cite{Vakar2024CHAD} describes an efficient implementation of such a system, CHAD. We plan to use $\lambdaLL$ to provide a precise comparison between this system and JAX Autodiff.

The diagrams in Figure~\ref{fig:final_picture} are closed under the logical relation $\sim$, which captures extensional equivalence. However, a closer look at our running examples (Figure~\ref{fig:exampleJAX} and Figure~\ref{fig:mainfigure}) suggests that a stronger closure may hold. We conjecture that $\sim$ could be replaced by the equivalence generated solely by the non-exponential cut-elimination steps of linear logic and the $\sigma$-commutation rules, which commute let-definitions and induce the equational classes of $\lambda$-terms through their linear logic proof-net representations \cite{Regnier_sigma,CarraroG14}.

\newpage

\begin{acks}
Partially supported by Fondation CFM, Bourse "Jean-Pierre Aguilar", and ENS de Lyon.
\end{acks}

\bibliographystyle{ACM-Reference-Format}
\bibliography{sample}

\clearpage
\appendix
\section*{Appendices}
\DoToC
\clearpage

\section{JAX} \label{app:JAX}
The grammar of Linear A expressions  can be conveniently described by the following grammar.
\allowdisplaybreaks
    \begin{equation}
        \tag{Linear A}
        \begin{split}
            e \bnf& 
        \retJAX{x}{y} 
            \orsintax \letterm{\retJAX{x}{y}}{e_1}{e_2} 
             \\&
            \orsintax \emptytupleJAX 
            \orsintax \tupleJAX{x_1}{x_2}       	
        \orsintax \letterm{{\emptytupleJAX}}{\expvarJAX{z}}{e} 
        \orsintax \letterm{\tupleJAX{x_1}{x_2}}{\expvarJAX{z}}{e} 
        \\&
        \orsintax \emptylintupleJAX
            \orsintax \lintupleJAX{x_1}{x_2}
        \orsintax \letterm{\emptylintupleJAX}{\linvarJAX{z}}{e}
        \orsintax \letterm{\lintupleJAX{x_1}{x_2}}{\linvarJAX{z}}{e} 
        \\&  
            \orsintax \literalJAX 
            \orsintax \underline{f}(x_1,\ldots,x_n)
               \orsintax \linzeroJAX{\tautypJAX} 
        \orsintax \linsumJAX{x}{y}
             \orsintax \linmultJAX{x}{y} 
             \orsintax \dupJAX{x}
             \orsintax \dropJAX{e} 
        \end{split}
        \label{eq:linearAsyntax}
    \end{equation}
    
In addition to the typing rules of Figure~\ref{fig:jaxtyprules}, we give in Figure~\ref{fig:jaxtyprules_binary_primal} the rules for the binary primal tuples, which has been omitted in Figure~\ref{fig:jaxtyprules} for lack of space. 
\begin{figure*}[h]
        \begin{center} 
	    \AxiomC{\phantom{$\dot A_1$}}
            \UnaryInfC{$\judgmentJAX{\expvarJAX{x}_1:\tautypJAX_1,\expvarJAX{x}_2:\tautypJAX_2}{}
            {\tupleJAX{x_1}{x_2}}{\typtensorJAX{\tautypJAX_1}{\tautypJAX_2}}{\typunitJAX}$}
            \DisplayProof            
            \qquad
            \AxiomC{$\judgmentJAX{\Gamma,\expvarJAX{x}_1:\tautypJAX_1,
            \expvarJAX{x}_2:\tautypJAX_2}{\lincontextJAX{\Gamma}}
            {e}{\tautypJAX}{\sigmatypJAX}$}
            \UnaryInfC{$\judgmentJAX{\Gamma,\expvarJAX{z}:\typtensorJAX{\tautypJAX_1}{\tautypJAX_2}}
            {\lincontextJAX{\Gamma}}
            {\letterm{\tupleJAX{x_1}{x_2}}{\expvarJAX{z}}{e}}
            {\tautypJAX}{\sigmatypJAX}$}
            \DisplayProof
        \end{center} 
        \caption{Linear A Typing Rules for Primal Binary Tuples}
        \label{fig:jaxtyprules_binary_primal}
\end{figure*}   

The typing rules for the syntactic sugar of JAX are derived from the rules of Figure~\ref{fig:jaxtyprules} and are given in Figure~\ref{fig:jaxDerTypRules}. In addition to the syntactic sugar for pairs of primal/tangent and tangent/tangent expressions, we add here that of primal/primal expressions: $\otimes(e_1,e_2)\approx \letJAX{\expvarJAX{x}}{e_1} {\letJAX{\expvarJAX{y}}{e_2}{\tupleJAX{\expvarJAX{x}}{\expvarJAX{y}}}}$.

\begin{figure*}
    \begin{center} 
        \AxiomC{}
        \UnaryInfC{$\judgmentJAX{\expvarJAX{x}:\tautypJAX}{}{\expvarJAX{x}}
        {\tautypJAX}{\typunitJAX}$}
                \DisplayProof
    \end{center} 

	\medskip
	
    \begin{center} 
        \AxiomC{$\judgmentJAX{\Gamma_1}{\lincontextJAX{\Gamma}_1}{e_1}
        {\tautypJAX_1}{\typunitJAX}$}
        \AxiomC{$\judgmentJAX{\Gamma_2,\expvarJAX{x}:\tautypJAX_1}
        {\lincontextJAX{\Gamma}_2}{e_2}{\tautypJAX}
        {\sigmatypJAX}$}
        \BinaryInfC{$\judgmentJAX{\Gamma_1 \cup \Gamma_2}
        {\lincontextJAX{\Gamma}_1, \lincontextJAX{\Gamma}_2}
        {\letterm{\expvarJAX{x}}{e_1}{e_2}}{\tautypJAX}{\sigmatypJAX}$}
        \DisplayProof
    \end{center}

	\medskip

    \begin{center} 
        \AxiomC{}
        \UnaryInfC{$\judgmentJAX{}{\linvarJAX{x}:\tautypJAX}{\linvarJAX{x}}
        {\typunitJAX}{\tautypJAX}$}
        \DisplayProof       
    \end{center} 
    
	\medskip

    \begin{center} 
        \AxiomC{$\judgmentJAX{\Gamma_1}{\lincontextJAX{\Gamma}_1}{e_1}
        {\typunitJAX}{\sigmatypJAX_1}$}
        \AxiomC{$\judgmentJAX{\Gamma_2}
        {\lincontextJAX{\Gamma}_2,\linvarJAX{y}:\sigmatypJAX_1}{e_2}{\tautypJAX}
        {\sigmatypJAX}$}
        \BinaryInfC{$\judgmentJAX{\Gamma_1 \cup \Gamma_2}
        {\lincontextJAX{\Gamma}_1, \lincontextJAX{\Gamma}_2}
        {\letterm{\linvarJAX{y}}{e_1}{e_2}}{\tautypJAX}{\sigmatypJAX}$}
        \DisplayProof        
    \end{center}

	\medskip

    \begin{center}
        \AxiomC{$\forall i\in\{1,2\},\quad\judgmentJAX{\Gamma_i}
        {\dot\Gamma_i}{e_i}{\sigmatypJAX_i}{\typunitJAX}$}
        \UnaryInfC{$\judgmentJAX{\Gamma_1\cup\Gamma_2}
        {\dot\Gamma_1, \dot\Gamma_2}{\otimes(e_1,e_2)}{\sigmatypJAX_1\otimes\sigmatypJAX_2}{\typunitJAX}$}
        \DisplayProof
    \end{center}
            
	\medskip
	
    \begin{center}
        \AxiomC{$\forall i\in\{1,2\},\quad\judgmentJAX{\Gamma_i}
        {\dot\Gamma_i}{e_i}{\typunitJAX}{\tautypJAX_i}$}
        \UnaryInfC{$\judgmentJAX{\Gamma_1\cup\Gamma_2}
        {\dot\Gamma_1, \dot\Gamma_2}{\dot\otimes(e_1,e_2)}{\typunitJAX}{\tautypJAX_1\otimes\tautypJAX_2}$}
        \DisplayProof
    \end{center}
    \caption{Derived JAX Typing Rules for Syntactic Sugar.}
    \label{fig:jaxDerTypRules}
\end{figure*}

The grammar of Linear B expressions, completed with the primal binary tuples is the following:
    \begin{align*}
        \tag{Linear B}
            d \bnf& 
            \retexpJAX{\expexprJAX{e}}{\linexprJAX{e}}
            \orsintax \letterm{\expvarJAX{x}}{\expexprJAX{e}}{d} 
            \orsintax \letterm{{\emptytupleJAX}}{\expvarJAX{z}}{d}
            \orsintax \letterm{\tupleJAX{\expvarJAX{x}_1}{\expvarJAX{x}_2}}{\expvarJAX{z}}{d} 
            \\
            \tag{Primal}
            \expexprJAX{e} \bnf& 
            \expvarJAX{x} 
            \orsintax \letterm{\expvarJAX{x}}{\expexprJAX{e_1}}{\expexprJAX{e_2}} 
            \orsintax\literalJAX 
             \orsintax \underline{f}(\expvarJAX{x}_1,\ldots, \expvarJAX{x}_n) 
             \orsintax \dropJAX{\expexprJAX{e}}\\&
        \orsintax \emptytupleJAX 
            \orsintax \tupleJAX{\expexprJAX{e_1}}{\expexprJAX{e_2}} 
            \orsintax \letterm{{\emptytupleJAX}}{\expvarJAX{z}}{\expexprJAX{e}}
            \orsintax \letterm{\tupleJAX{\expvarJAX{x}_1}{\expvarJAX{x}_2}}{\expvarJAX{z}}{\expexprJAX{e}}
        \\
            \tag{Tangent}
            \linexprJAX{e} \bnf& 
            \linvarJAX{x} 
            \orsintax \letterm{\linvarJAX{x}}{\linexprJAX{e_1}}{\linexprJAX{e_2}} 
            \orsintax \dupJAX{x}
            \orsintax \linzeroJAX{\tautypJAX} 
            \orsintax \linsumJAX{x}{y} 
            \orsintax  \linmultJAX{x}{y} 
            \orsintax \dropJAX{\linexprJAX{e}}
            \\&
        \orsintax \emptylintupleJAX 
            \orsintax \lintupleJAX{e_1}{e_2} 
            \orsintax \letterm{\emptylintupleJAX}{\linvarJAX{z}}{\linexprJAX{e}}
            \orsintax \letterm{\tupleJAX{\linvarJAX{x}_1}{\linvarJAX{x}_2}}{\linvarJAX{z}}{\linexprJAX{e}}
    \end{align*}

    We denote $n$-fold tangent tuples $\dot{\otimes}(e_1, \dot{\otimes}(e_2,\dots, e_n),\dots)$ as an $n$-ary tangent tuple $\dot{\otimes}(e_1,\dots,e_n)$. We can use shortcut like $\dot{\otimes}(e_i)^n_{i=1}$, or even $\dot{\otimes}(e_i)_i$ if 1 and $n$ are clear from the context or irrelevant.
    We adopt similar writings for types:  $\dot{\otimes}(\tau_i)^n_{i=1}$ or $\dot{\otimes}(\tau_i)_i$.

    Given $\theta=(\dot{x}_1,\dots,\dot{x}_n)$ where $\dot{x}_i:\tau_i$, we define the syntactic sugar $\dot{\otimes}\theta$ by induction on $\theta$ as follows
    \begin{align*}
        \dot{\otimes}\theta \approx
        \begin{cases}
            \dot{\otimes}()  &\text{ if }\theta=(\,)\\
            \dot{x}          &\text{ if }\theta=(\dot{x})\\
            \dot{\otimes}(\dot{x},\dot{\otimes}\theta')      &\text{ otherwise we can suppose }\theta=\dot{x},\theta'\\
        \end{cases}
    \end{align*}
    Similarly, we can define the type $\otimes \theta$.
    Moreover, we define the syntactic sugar $\letterm{\dot{\otimes}\theta}{\dot{z}}{e}$ by induction on $\theta$ as follows
    \begin{align*}
        \letterm{\dot{\otimes}\theta}{\dot{z}}{e} \approx
        \begin{cases}
            \letterm{\dot{\otimes}()}{\dot{z}}{e}  &\text{ if }\theta=(\,)\\
            \letterm{\dot{x}}{\dot{z}}{e}          &\text{ if }\theta=(\dot{x})\\
            \letterm{\dot{\otimes}(\dot{x},\dot{y})}{\dot{z}}{} 
            \letterm{\dot{\otimes}\theta'}{\dot{y}}{e}     &\text{ otherwise we can suppose }\theta=\dot{x},\theta'\\
        \end{cases}
    \end{align*}

    Given $\theta=(\dot{x}_1,\dots,\dot{x}_n)$ and $\theta_i \subseteq \theta$ with $i\in\{1,2\}$, let $\dot{y}_1:\otimes\theta_1$ and $\dot{y}_2:\otimes\theta_2$, we define the fusion expression as 
    \begin{equation}
        \label{eq:fusion_expr_JAX}
        \SplitInvJAX[\theta]{\dot{y}_1,\dot{y}_2}=
        \begin{aligned}
            &\letterm{\dot{\otimes}\theta_1}{\dot{y}_1}{}\\
            &\letterm{\dot{\otimes}\theta_2}{\dot{y}_2}{\dot{\otimes}\theta}\\
        \end{aligned}
    \end{equation}
    Observe that $ \SplitInvJAX[\theta]{\dot{y}_1,\dot{y}_2}$ is well-typed as 
    $\judgmentJAX{}{\dot{y}_1:\otimes\theta_1,\dot{y}_2:\otimes\theta_2}
    {\SplitInvJAX{\dot{y}_1,\dot{y}_2}{\theta} }{\typone}{\otimes\theta}$.

The rules for the forward mode transformation of JAX can be found in Figure~\ref{fig:FMJAX}.

The unzipping transformation of JAX is defined with the rules in Figure~\ref{fig:UnzipJAX}.

Finally, the transpose transformation is described in Figure~\ref{fig:TlinJAX} and in Figure~\ref{fig:TJAX}.

\begin{figure*}
    \centering
        \begin{align*}
            \forwardJAX{\expvarJAX{x} \to \linvarJAX{y}}{\expvarJAX{x}}\definedas \mbox{}&
            \retJAX{x}{y}
            \\
            \forwardJAX{\contextFMJAX_1,\contextFMJAX_2,\{\expvarJAX{z}_i \to 
            \linvarJAX{u}_i\}^k_{i=1}}{\letterm{\expvarJAX{x}}{\expexprJAX{e}_1}{\expexprJAX{e_2}}}\definedas \mbox{}& \letterm{\linvarJAX{a}_1}{\dupJAX{u_1}}{}\\
            & \letterm{\lintupleJAX{w_1}{v_1}}{\linvarJAX{a}_1}{}\\
            &\ldots\\
            &\letterm{\linvarJAX{a}_k}{\dupJAX{u_k}}{}\\
            &\letterm{\lintupleJAX{w_k}{v_k}}{\linvarJAX{a}_k}{}\\
            &\letterm{\retJAX{x}{y}}{\forwardJAX{\contextFMJAX_1,\{\expvarJAX{z}_i \to \linvarJAX{w}_i\}^k_{i=1}}{e_1}}{}\\
            &\forwardJAX{\contextFMJAX_2,\{\expvarJAX{z}_i \to \linvarJAX{v}_i\}^k_{i=1}}{e_2}\\
            \text{where } dom(\contextFMJAX_i)= \FV{\expexprJAX{e}_i} \setminus (\FV{\expexprJAX{e}_1} &\cap \FV{\expexprJAX{e}_2}) \text{ and }
            \{\expvarJAX{z}_i\}^k_{i=1} = \FV{\expexprJAX{e}_1} \cap \FV{\expexprJAX{e}_2}
            \\
            \forwardJAX{\emptyset}{{\emptytupleJAX}}\definedas \mbox{}&\retexpJAX{{\emptytupleJAX}}{\emptylintupleJAX}
            \\
            \forwardJAX{\{\expvarJAX{x}_i \to \linvarJAX{y}_i\}^2_{i= 1}}{\tupleJAX{x_1}{x_2}}\definedas \mbox{}&\retexpJAX{\tupleJAX{x_1}{x_2}}{\lintupleJAX{y_1}{y_2}}
            \\
            \forwardJAX{\contextFMJAX, \expvarJAX{z}\to\linvarJAX{w}}{\letterm{{\emptytupleJAX}}{\expvarJAX{z}}{\expexprJAX{e}}}\definedas \mbox{}& \forwardJAX{\contextFMJAX}{\expexprJAX{e}}
            \\
            \forwardJAX{\contextFMJAX, \expvarJAX{z}\to\linvarJAX{w}}{\letterm{\tupleJAX{x_1}{x_2}}{\expvarJAX{z}}{\expexprJAX{e}}}\definedas \mbox{}&\letterm{\tupleJAX{x_1}{x_2}}{\expvarJAX{z}}{}
            \\
            &\letterm{\lintupleJAX{y_1}{y_2}}{\linvarJAX{w}}{}\\
            &\forwardJAX{\contextFMJAX,\{\expvarJAX{x}_i \to \linvarJAX{y}_i\}^2_{i=1}}{\expexprJAX{e}}\\
            \forwardJAX{\emptyset}{\literalJAX}\definedas \mbox{}& \retexpJAX{\literalJAX}{\linzeroJAX{\typrealJAX}}\\
            \forwardJAX{\contextFMJAX,\{\expvarJAX{x}_i \to \linvarJAX{y}_i\}^n_{i=1}}{\underline{f}(\expvarJAX{x}_1, \dots, \expvarJAX{x}_n)}\definedas \mbox{}& \letterm{\expvarJAX{w_1}}{\underline{\partial_1 f}(\expvarJAX{x}_1, \dots, \expvarJAX{x}_n)}{}\\
            & \ldots\\
            & \letterm{\expvarJAX{w_n}}{\underline{\partial_n f}(\expvarJAX{x}_1, \dots, \expvarJAX{x}_n)}{}\\
            & \letterm{\linvarJAX{z_1}}{\linmultJAX{w_1}{y_1}}{}\\
            &\ldots\\
            & \letterm{\linvarJAX{z_n}}{\linmultJAX{w_n}{y_n}}{}\\
            &\retexpJAX{\underline{f}(\expvarJAX{x}_1, \dots, \expvarJAX{x}_n)}{\linvarJAX{z_1} \dot{+}\ldots\dot{+}\linvarJAX{z_n}}\\
            \forwardJAX{\contextFMJAX}{\dropJAX{\expexprJAX{e}}}\definedas \mbox{}&\dropJAX{\forwardJAX{\contextFMJAX}{\expexprJAX{e}}}
        \end{align*}
    
        \caption{Forward Tranformation in $\JAX$}
        \label{fig:FMJAX} 
\end{figure*}

\begin{figure*}
    \begin{prooftree}
        \AxiomC{}
        \UnaryInfC{$\unzippingJAX{\retJAX{x}{y}}\definedas \retJAX{x}{y}$}
    \end{prooftree}

    \begin{prooftree}
        \AxiomC{$\unzippingJAX{e_1}\definedas E_1 \texttt{ in } \retexpJAX{\expexprJAX{e_1}}{\linexprJAX{e_1}}$}
        \AxiomC{$\unzippingJAX{e_2}\definedas E_2 \texttt{ in } \retexpJAX{\expexprJAX{e_2}}{\linexprJAX{e_2}}$}
        \BinaryInfC{$\unzippingJAX{\letterm{\retJAX{x}{y}}{e_1}{e_2}}\definedas E_1 \texttt{ in } \letterm{x}{\expexprJAX{e_1}} E_2 \texttt{ in } \retexpJAX{\expexprJAX{e_2}}{\letterm{\linvarJAX{y}}{\linexprJAX{e_1}}{\linexprJAX{e_2}}}$}
    \end{prooftree}
    \begin{center}
        \AxiomC{}
        \UnaryInfC{$\unzippingJAX{\emptylintupleJAX} \definedas \expunzipJAX{{\emptytupleJAX}}$}
        \DisplayProof
        \quad
        \AxiomC{}
        \UnaryInfC{$\expunzipJAX{\tupleJAX{x_1}{x_2}}$}
        \DisplayProof
    \end{center}

    \begin{prooftree}
        \AxiomC{$\unzippingJAX{e_1}\definedas E  \texttt{ in } \retexpJAX{\expexprJAX{e_1}}{\linexprJAX{e_1}}$}
        \UnaryInfC{$\unzippingJAX{\letterm{{\emptytupleJAX}}{\expvarJAX{z}}{e_1}}\definedas \letterm{{\emptytupleJAX}}{\expvarJAX{z}} E  \texttt{ in } \retexpJAX{\expexprJAX{e_1}}{\linexprJAX{e_1}}$}
    \end{prooftree}

    \begin{prooftree}
        \AxiomC{$\unzippingJAX{e_1}\definedas E  \texttt{ in } \retexpJAX{\expexprJAX{e_1}}{\linexprJAX{e_1}}$}
        \UnaryInfC{$\unzippingJAX{\letterm{\tupleJAX{x_1}{x_2}}{\expvarJAX{z}}{e_1}}\definedas \letterm{\tupleJAX{x_1}{x_2}}{\expvarJAX{z}}{E  \texttt{ in } \retexpJAX{\expexprJAX{e_1}}{\linexprJAX{e_1}}$}}
    \end{prooftree}

    \begin{prooftree}
        \AxiomC{}
        \UnaryInfC{$\linunzipJAX{\lintupleJAX{x_1}{x_2}}$}
    \end{prooftree}

    \begin{prooftree}
        \AxiomC{$\unzippingJAX{e_1}\definedas E  \texttt{ in } \retexpJAX{\expexprJAX{e_1}}{\linexprJAX{e_1}}$}
        \UnaryInfC{$\unzippingJAX{\letterm{\emptylintupleJAX}{\linvarJAX{z}}{e_1}}\definedas E  \texttt{ in } \retexpJAX{\expexprJAX{e_1}}{\letterm{\emptylintupleJAX}{\linvarJAX{z}}{\linexprJAX{e_1}}}$}
    \end{prooftree}

    \begin{prooftree}
        \AxiomC{$\unzippingJAX{e_1}\definedas E  \texttt{ in } \retexpJAX{\expexprJAX{e_1}}{\linexprJAX{e_1}}$}
        \UnaryInfC{$\unzippingJAX{\letterm{\lintupleJAX{x_1}{x_2}}{\linvarJAX{z}}{e_1}}\definedas E  \texttt{ in } \retexpJAX{\expexprJAX{e_1}}{\letterm{\lintupleJAX{x_1}{x_2}}{\linvarJAX{z}}{\linexprJAX{e_1}}}$} 
    \end{prooftree}

    \begin{center}
        \AxiomC{}
        \UnaryInfC{$\expunzipJAX{\literalJAX}$}
        \DisplayProof
        \quad
        \AxiomC{}
        \UnaryInfC{$\expunzipJAX{\expfunJAX}$}
        \DisplayProof
    \end{center}

    \begin{center}
        \AxiomC{}
        \UnaryInfC{$\linunzipJAX{\linzeroJAX{\tautypJAX}}$}
        \DisplayProof
        \quad
        \AxiomC{}
        \UnaryInfC{$\linunzipJAX{\linsumJAX{x}{y}}$}
        \DisplayProof
    \end{center}

    \begin{center}
        \AxiomC{}
        \UnaryInfC{$\linunzipJAX{\linmultJAX{x}{y}}$}
        \DisplayProof
        \quad
        \AxiomC{}
        \UnaryInfC{$\linunzipJAX{\dupJAX{x}}$}
        \DisplayProof
    \end{center}

    \begin{prooftree}
        \AxiomC{$\unzippingJAX{e_1}\definedas E  \texttt{ in } \retexpJAX{\expexprJAX{e_1}}{\linexprJAX{e_1}}$}
        \UnaryInfC{$\unzippingJAX{\dropJAX{e_1}}\definedas E  \texttt{ in } \retexpJAX{ \dropJAX{\expexprJAX{e_1}}}{\dropJAX{\linexprJAX{e_1}}}$} 
    \end{prooftree}

    \caption{Unzipping Tranformation in $\JAX$}
    \label{fig:UnzipJAX}
\end{figure*}

\begin{figure*}
    \centering
    \begin{align*}
        \transpJAX{\linvarJAX{x}:\tautypJAX}{\linvarJAX{u}:\tautypJAX}{\linvarJAX{x}}\definedas \mbox{}&
            \linvarJAX{u}\\
        \transpJAX{\theta}{\linvarJAX{u}:\tautypJAX}{\letterm{\linvarJAX{x}}{\linexprJAX{e_1}}{\linexprJAX{e_2}}}\definedas \mbox{}& 
            \letterm{\lintupleJAX{x}{u_2}}
            {\transpJAX{\linvarJAX{x}:\sigmatypJAX,\theta\cap \linFV{\linexprJAX{e_2}}}{\linvarJAX{u}:\tautypJAX}{\linexprJAX{e_2}}}
            \\
            &\mbox{ }
            {\letterm{\linvarJAX{u}_1}
            {\transpJAX{\theta\cap \linFV{\linexprJAX{e_1}}}{\linvarJAX{x}:\sigmatypJAX}{\linexprJAX{e_1}}}
            {\SplitInvJAX[\theta]{\linvarJAX{u}_1,\linvarJAX{u}_2}}} \\
        \transpJAX{\emptyset}{\linvarJAX{u}:\typunitJAX}{\emptylintupleJAX}\definedas \mbox{}&
            \letterm{\emptylintupleJAX}{\linvarJAX{u}}{\emptylintupleJAX}\\
        \transpJAX{\theta}{\linvarJAX{u}:\typtensorJAX{\tautypJAX}{\sigmatypJAX}}{\dot{\otimes}(\linexprJAX{e_1},\linexprJAX{e_2})}\definedas \mbox{}&
            \letterm{\lintupleJAX{u_1}{u_2}}{\linvarJAX{u}}{\dot{\otimes}(\transpJAX{\theta\cap \linFV{\linexprJAX{e_1}}}{\linvarJAX{u_1}:\tautypJAX}{\linexprJAX{e_1}},\transpJAX{\theta\cap \linFV{\linexprJAX{e_2}}}{\linvarJAX{u_2}:\sigmatypJAX}{\linexprJAX{e_2}})}\\
        \transpJAX{\theta}{\linvarJAX{u}:\tautypJAX}{\letterm{\emptylintupleJAX}{\linvarJAX{z}}{\linexprJAX{e}}}\definedas \mbox{}&
            \dot{\otimes}(\emptylintupleJAX,\transpJAX{\theta\cap \linFV{\linexprJAX{e}}}{\linvarJAX{u}:\tautypJAX}{\linexprJAX{e}})\\
        \transpJAX{\theta}{\linvarJAX{u}:\tautypJAX}{\letterm{\tupleJAX{\linvarJAX{x}_1}{\linvarJAX{x}_2}}{\linvarJAX{z}}{\linexprJAX{e}}}\definedas \mbox{}&
            \letterm{\linvarJAX{w}}{\transpJAX{\linvarJAX{x_1}:\tautypJAX_1,\linvarJAX{x_2}:\tautypJAX_2,\theta\cap \linFV{\linexprJAX{e}}}{\linvarJAX{u}:\tautypJAX}{\linexprJAX{e}}}{}\\
            &\letterm{\lintupleJAX{y_1}{z}}{\linvarJAX{w}}{}\\
            &\letterm{\lintupleJAX{y_2}{y_3}}{\linvarJAX{z}}{{\dot{\otimes}(\lintupleJAX{y_1}{y_2},\linvarJAX{y_3})}}\\
        \transpJAX{\linvarJAX{x}:\tautypJAX}{\linvarJAX{u}:\typtensorJAX{\tautypJAX}{\tautypJAX}}{\dupJAX{x}}\definedas \mbox{}&
            \letterm{\lintupleJAX{z}{w}}{\linvarJAX{u}}{\linvarJAX{z}\dot{+}\linvarJAX{w}}\\
        \transpJAX{\{\linvarJAX{x}:\tautypJAX,\linvarJAX{y}:\tautypJAX\}}{\linvarJAX{u}:\tautypJAX}{\linvarJAX{x}\dot{+}\linvarJAX{y}}\definedas \mbox{}&
            \dupJAX{u}\\
        \transpJAX{\emptyset}{\linvarJAX{u}:\tautypJAX}{\linzeroJAX{\tautypJAX}}\definedas \mbox{}&
            \dropJAX{\linvarJAX{u}}\\
        \transpJAX{\theta}{\linvarJAX{u}:\typunitJAX}{\dropJAX{\linexprJAX{e}}}\definedas \mbox{}&
           \linzeroJAX{\otimes \theta}\\
        \transpJAX{\linvarJAX{y}:\tautypJAX}{\linvarJAX{u}:\tautypJAX}{x\dot{*}\linvarJAX{y}}\definedas \mbox{}&
           x\dot{*}\linvarJAX{u}
    \end{align*} 
    \caption{Transpose on purely tangent expressions of $\JAX$}
    \label{fig:TlinJAX}
\end{figure*}

\begin{figure*}
    \centering
    \begin{align*}
        \transpJAX{\theta}{\linvarJAX{u}:\tautypJAX}{\retexpJAX{\expexprJAX{e}}{\linexprJAX{e}}}\definedas \mbox{}&
            \retexpJAX{\expexprJAX{e}}{\transpJAX{\theta}{\linvarJAX{u}:\tautypJAX}{\linexprJAX{e}}}\\
        \transpJAX{\theta}{\linvarJAX{u}:\tautypJAX}{\letterm{\expvarJAX{x}}{\expexprJAX{e}}{d}}\definedas \mbox{}&
            \letterm{\expvarJAX{x}}{\expexprJAX{e}}{\transpJAX{\theta}{\linvarJAX{u}:\tautypJAX}{d}}\\
        \transpJAX{\theta}{\linvarJAX{u}:\tautypJAX}{\letterm{\tupleJAX{\expvarJAX{x}_1}{\expvarJAX{x}_2}}{\expvarJAX{z}}{d}}\definedas \mbox{}&
            \letterm{\tupleJAX{\expvarJAX{x}_1}{\expvarJAX{x}_2}}{\expvarJAX{z}}{\transpJAX{\theta}{\linvarJAX{u}:\tautypJAX}{d}}
    \end{align*} 

    \caption{Transpose Transformation in $\JAX$}
    \label{fig:TJAX}
\end{figure*}

\clearpage
\section{$\lambdaLL$}
We recall that we say that a pattern is \emph{exponential} whenever it is of the form $\oc x$ and $\oc \Gamma$ denotes a set of exponential pattern. We use meta-variables $\PatExpA, \PatExpB$ (resp.~$\PatAddA, \PatAddB$) for denoting \emph{patterns of $\otimes$-sequence types} (resp.~\emph{$\with$-sequence types}). 

\subsection{$\beta$-reduction and Rewriting Properties}\label{app:rwprop}
\begin{figure*} [h!]
    \begin{align*} 
      \ContextA \bnf\!\!& 
          [\,] 
            \orsintax \absterm{p^A}{  \ContextA} 
            \orsintax \appterm{  \ContextA}{N}
            \orsintax \appterm{M}{  \ContextA} 
            \orsintax \tupleterm{  \ContextA}{N} 
            \orsintax \tupleterm{M}{  \ContextA} 
            \orsintax \bangterm{\ContextA} 
            \orsintax \nTuple{\ContextA, M}
            \orsintax \nTuple{M, \ContextA} 
    \end{align*} 
    \caption{Grammar of one-hole contexts. Moreover, we call \emph{exponential safe context} a context generated by the sub-grammar not having $\bangterm{\ContextA}$.}
    \label{fig:grammar_contexts}
\end{figure*} 
    
All reductions are closed by one-hole contexts.  A \emph{one-hole context} $\ContextA$ is a term with a sole occurrence of an hole $[\,]$, this latter being a place holder which will be replaced for a term $M$ (with possible capture of free variables) generating a new term denoted by $\ContextA[M]$. The grammar of the one-hole contexts of $\lambdaLL$ is given in Figure~\ref{fig:grammar_contexts}. 

\medskip \subsubsection{Subject Reduction.}
We prove subject reduction by means of a pattern substitution lemma (Lemma~\ref{lemma:substitution}).
Given a type derivation $\Pi$, we define the size of such derivation $s(\Pi)$ as the number of derivation rules of $\Pi$. 

\begin{lemma}
	\label{lemma:fv_in_environment}
	If $\Gamma \vdash M:A$, then $\FV{M}\subseteq\FV{\Gamma}$.
\end{lemma}
\begin{proof}
	By induction on a derivation of $\Gamma \vdash M:A$.
\end{proof}

\begin{lemma}
	\label{lemma:weakening_subst}
	If $\FV{p}\cap\FV{M}=\emptyset$, then $M\{V/p\}=M$.
\end{lemma}
\begin{proof}
	By induction on $M$.
\end{proof}

\begin{lemma}[Pattern Substitution]
	\label{lemma:substitution}
	Given two derivable judgments $\oc\Gamma_1,\Delta_1,\PatA:A\vdash M:B$ and $\oc\Gamma_2,\Delta_2\vdash V:A$ such that 
	\begin{enumerate}
	\item $V$ is a value for $\PatA$,
	\item $\FV{\oc\Gamma_i,\Delta_i}\cap\FV{\Delta_{3-i}}=\emptyset$ for $i \in\{1,2\}$,
	\end{enumerate}
	then we have that the judgment $\oc\Gamma_1\cup\oc\Gamma_2, \Delta_1,\Delta_2\vdash M\{V/\PatA\}:B$ is derivable. 
\end{lemma}

\begin{sproof}
	Taking the judgments as in the hypotheses of the lemma, for any derivation $\Pi_1$ of $\oc\Gamma_1,\Delta_1,\PatA:A\vdash M:B$ and $\Pi_2$ of $\oc\Gamma_2,\Delta_2\vdash V:A$, we give a derivation of $\oc\Gamma_1\cup\oc\Gamma_2, \Delta_1,\Delta_2\vdash M\{V/p\}:B$ by induction on the lexicographically ordered pair $(s(\Pi_2), s(\Pi_1))$.
	Notice that the condition $1$ in the lemma hypothesis is necessary to assure that the substitution $M\{V/p\}$ is well-defined, while condition $2$ assure that the typing environment $\oc\Gamma_1\cup\oc\Gamma_2, \Delta_1,\Delta_2$ is a set of patterns of pairwise distinct variables. 
	
	We split depending on the last derivation rule in $\Pi_1$ or $\Pi_2$. 
    We apply Lemma~\ref{lemma:fv_in_environment} and Lemma~\ref{lemma:weakening_subst} when the last rule $r$ of $\Pi_1$ is $\typbangW$ or $\typrwithL{i}$ and it is acting on the pattern $\PatA:A$. In both cases, Lemma~\ref{lemma:fv_in_environment} is used to show that certain variables (e.g., x in the $\typbangW$ case or components of the unused pattern in the $\typrwithL{i}$ case) are not free in the term $M$ because they do not appear in its typing environment. This allows us to apply Lemma~\ref{lemma:weakening_subst}, which ensures that substituting values for these variables has no effect on $M$. Thus, unnecessary substitutions can be safely ignored, preserving the correctness of the derivation.
\end{sproof}

\begin{remark}
	Let us remark that the proof above of the substitution lemma (Lemma~\ref{lemma:substitution}) uses the hypothesis that an exponential pattern (i.e.~a pattern belonging to $\oc\Gamma$) must have the basic form $\oc x$ for a variable $x$.
	If we have relaxed our definition of patterns, allowing for e.g. $\oc(\PatA_1, \PatA_2)$, then substitution lemma would have failed (and hence subject reduction). 
	In fact, in this case we would have for the terms $V= \oc ((x,x'),(x,x'))$  and $M= ((x,z),(x',z))$ and the type for $B=(A\otimes A')\otimes (A\otimes A')$ the possible judgements:
	\begin{enumerate}
		\item $\oc(x,x'):\oc (A\otimes A')\vdash V:\oc B$
		\item $\oc(x,x'):\oc (A\otimes A'), \oc z:\oc B\vdash M: (A\otimes B)\otimes  (A'\otimes B)$
	\end{enumerate}
	while the term $M\{V/\oc z\} = ((x,((x,x'),(x,x'))),(x',((x,x'),(x,x'))))$ could not be typed under the environment $\oc(x,x'):\oc (A\otimes A')$ because there is no possible between the $x$'s and $(x,x')$. 
\end{remark}

Finally, we prove SR by using the following lemma and the substitution lemma.

\begin{lemma}\label{lemma:focusW}
	If $\oc\Gamma,\Delta \vdash \lambda p.M:A\multimap B$ is derivable, then $\oc\Gamma,\Delta,p:A \vdash M:B$.
	\begin{proof}
		By induction on the size of the derivation $\oc\Gamma,\Delta,p:A \vdash M:B$.
	\end{proof}
\end{lemma}

More precisely, the proof of Subject Reduction is the following

\begin{sproof}[Proof Theorem~\ref{th:subjred}]
    Let $\Pi$ be the derivation for $\oc\Gamma,\Delta \vdash M:A$, we proceed by induction on $s(\Pi)$. We split depending on the last derivation rule in $\Pi$. 
    We apply Lemma~\ref{lemma:substitution} when the last rule $r$ of $\Pi$ acting on the term $M$ is of type $\typrlollipopR$ then $M=M_1M_2$ and we have that: $\oc\Gamma=\oc\Gamma_1\cup\oc\Gamma_2$ and $\Delta=\Delta_1,\Delta_2$. More precisely, when the redex is $M_1 M_2$, then we proceed by induction on the reduction step, so we analyze the cases of the reduction rules in Figure~\ref{fig:beta_rules} as follows:
    \begin{itemize}
		\item If the reduction rule is $\beta_{\lambda}$, then $M_1=\lambda p.M'_1$ and $M_2$ is a value $V$ for the pattern $p$. Moreover, $N$ is in the form $M'_1\{V/p\}$.
					
		By hypothesis we have that $\Pi$ is the derivation for $\oc\Gamma_1\cup\oc\Gamma_2,\Delta_1,\Delta_2 \vdash (\lambda p.M'_1) V :A$. Therefore we have two sub-derivations $\Pi_1$ and $\Pi_2$ above $r$ for $\oc\Gamma_1,\Delta_1 \vdash  \lambda p.M'_1 :B \multimap A$ and $\oc\Gamma_2,\Delta_2 \vdash  V :B$, respectively.
	
		By Lemma~\ref{lemma:focusW} on  $\oc\Gamma_1,\Delta_1 \vdash  \lambda p.M'_1 :B \multimap A$ we have a derivation for  $\oc\Gamma_1,\Delta_1,p:B \vdash M'_1:A$.
	
		We can conclude by applying Lemma~\ref{lemma:substitution}, getting a derivation for the judgement $\oc\Gamma_1\cup\oc\Gamma_2,\Delta_1,\Delta_2 \vdash M'_1\{V/p\} :A$.
	
		\item If the reduction rule is $\beta_{F}$, $\beta_{\dot{*}}$ or $\beta_{\dot{+}}$, then the proof is simple and direct.
	\end{itemize}
\end{sproof}

\medskip \subsubsection{Progress Property.}
In addition to the properties outlined in Subsection~\ref{subsect:reduction}, the $\lambdaLL$ calculus also satisfies the progress property.
The progress property identifies a grammar to the $\beta$-normal forms and corresponds in proof-theory to the sub-formula property. We express here this grammar only for the closed terms as this is what we need and the generalisation to open terms is more involved.
\begin{proposition}[Progress]
	\label{prop:progress_property}
	The set of closed $\beta$-nf is given by:
		\begin{align*}
			W\bnf&
			\absterm{p^A}{M} 
			\orsintax\emptytupleterm
			\orsintax \tupleterm{W_1}{W_2} 
			\orsintax \nTuple{\,}
			\orsintax \nTuple{W_1,W_2}
			\orsintax \bangterm{W} 
			\orsintax \realterm
			\orsintax \underline f
			\orsintax \dot + 
			\orsintax \dot * (\oc\realterm) 
		\end{align*}
	where $M$, $W$ or $W_i$ are  $\beta$-nf ($W$ and $W_i$ are moreover closed).
\end{proposition}
\begin{proof}
	By induction on $W$.
\end{proof}

\medskip \subsubsection{Strong Normalisation.}
In a rewriting system, strong normalization (SN) ensures that no term appears in an infinite reduction sequence. We prove strong normalization for $\lambdaLL$ by using the notion of reducibility~\cite{girard1989proofs}.

More precisely, strong normalisation (Theorem~\ref{th:SN}) is obtained by using the notion of reducibility (see e.g.~\cite{girard1989proofs}, Definition~\ref{def:reducibility} and Corollary~\ref{cor:reducibility}), which should be adapted in order to deal with the multiplicative connectives ($\otimes$, $\typone$) without having the involutive negation $A^\perp$ of classical linear logic.

Let $\mathscr{T}_A$ be the set of terms of type $A$, for some typing environment. A typical way to define reducibility for linear logic formulas is by using orthogonality \cite{accattoli2013linear} ---
a map $\mathcal X\mapsto\mathcal X^\bot$ over sets of terms which formalises the notion of ``passing a test''. In this scenario, the reducibility for the type tensor $\RED{\typtensor{A}{B}}$ is defined as $\{(M,N) \subseteq \mathscr{T}_{\typtensor{A}{B}}\mbox{ }|\mbox{ } M \subseteq \RED{A} \land N \subseteq \RED{B}\}^{\bot\bot}$. 
Unfortunately, we cannot apply this method immediately in $\lambdaLL$, as we have not an involutive negation (a type operator $(\,)^\perp$ such that $A^{\perp\perp}=A$). However, we can overcome the difficulty by using let-expressions and the ground type $\typR$. 

\begin{definition}[Reducibility]
	\label{def:reducibility}
	We define the sets $\testSN{}$ and $\RED{}$ by mutual recursion as follows
	\begin{align*}
		\testSN{x:A} &\definedas 
		\{ \judgment{\Delta, x:A}{N}{\typR} \orsintax \forall M \in \RED{A}.\mbox{ }N\{\sfrac{M}{x}\}\in \RED{\typR}\}\\ 
		\testSN{!x:!A} &\definedas 
		\{ \judgment{\Delta,!x:!A}{N}{\typR} \orsintax  \forall M\in \RED{A}.\mbox{ }\letterm{!x}{!M}{N} \in \RED{\typR}\}\\
		\testSN{():\typone} &\definedas 
		\{ \judgment{\Delta}{N}{\typR} \orsintax N\in \RED{\typR}\}\\
		\testSN{(x_1,x_2):A_1 \otimes A_2} &\definedas 
		\{
			\judgment{\Delta,(x_1,x_2):A_1 \otimes A_2}{N}{\typR}.\mbox{ }
			\letterm{(x_1,x_2)}{(V_1,V_2)}{N} \in \RED{\typR}
		\}
		\\  
		\testSN{\nTuple{x_1,x_2}:A_1\&A_2} &\definedas 
		\{
			\judgment{\Delta,\nTuple{x_1,x_2}:A_1 \& A_2}{N}{\typR} \text{ s.t. } \nTuple{V_1,V_2} \text{ typable}.\mbox{ }
			\letterm{\nTuple{x_1,x_2}}{\nTuple{V_1,V_2}}{N} \in \RED{\typR}
		\}
		\\[0.5cm]
		\RED{\typR}&\definedas 
		\{M\in \mathscr{T}_\typR \orsintax M \text{ is SN}\}\\
		\RED{\typone}&\definedas
		\{M \in \mathscr{T}_\typone \orsintax \forall N\in \testSN{():\typone}.\mbox{ }\letterm{\emptytupleterm}{M}{N}\in \RED{\typR}\}\\
		\RED{\typtensor{A}{B}}&\definedas
		\{M \in \mathscr{T}_{\typtensor{A}{B}} \orsintax \forall N\in \testSN{(x_1,x_2):\typtensor{A}{B}}.\mbox{ }\letterm{(x_1,x_2)}{M}{N} \in \RED{\typR} \}\\
		\RED{\typtop}&\definedas
		\{M\in \mathscr{T}_\typtop \orsintax M \text{ is SN}\}\\
		\RED{\typwith{A}{B}}&\definedas
		\{M \in \mathscr{T}_{\typwith{A}{B}} \orsintax \forall N\in \testSN{\nTuple{x_1,x_2}:\typwith{A}{B}}.\mbox{ }\letterm{\nTuple{x_1,x_2}}{M}{N} \in \RED{\typR} \}\\
		\RED{!A}&\definedas
		\{M \in \mathscr{T}_{!A} \orsintax \forall N\in \testSN{!x:!A}.\mbox{ }\letterm{!x}{M}{N} \in \RED{\typR} \}\\
		\RED{\typlollipop{A}{B}}&\definedas
		\{M \in \mathscr{T}_{\typlollipop{A}{B}} \orsintax \forall N\in \RED{A}.\mbox{ }MN \in \RED{B} \}
	\end{align*}
\end{definition}

We call \emph{neutral} a term generated by the following grammar:
\begin{align}
	\label{neutral_terms}
	\mathcal{N} \bnf x \orsintax \appterm{M}{N} \tag{Neutral}
\end{align}
Let $\nu(M)$ be the number which bounds the length of every normalisation sequence beginning from $M$.

\begin{lemma}[Properties of Reducibility]
	\label{lemma:propRED}
	Given a type $A$, $\RED{A}$ enjoys the following properties:
	\begin{enumerate}[start=0,label={(PR\arabic*)}]
		\item $\testSN{x:A}$ is not empty.
		\item If $M\in \RED{A}$ then $M$ is SN.
		\item If $M\in \RED{A}$ and $M \rightarrow N$ then $N\in \RED{A}$.
		\item If $M$ is \ref{neutral_terms} and $\forall N \in \RED{A}.M \rightarrow N $ then $M \in \RED{A}$.
	\end{enumerate}
	\begin{sproof}
		By induction on $A$. 
	\end{sproof}
\end{lemma}

Now we are ready to conclude that all terms are SN by proving that all terms are reducible, to do so we proceed by proving the following auxiliary lemmas
\begin{lemma}
	\label{lemma:SN1}
	Let $\judgment{\Delta}{M}{A}$, if $\forall N \in \testSN{x:A}.N\{\sfrac{M}{x}\}\in \RED{\typR}$, then $M\in\RED{A}$.
	\begin{sproof}
		By induction on $A$.
	\end{sproof}
\end{lemma}
\begin{lemma}
	\label{lemma:SN_tensor}
	If $(M_1,M_2)\in\RED{\typtensor{A_1}{A_2}}$, then $M_i\in\RED{A_i}$ for $1 \leq i\leq 2$.
\end{lemma}
\begin{lemma}
	\label{lemma:SN_bang}
	If $!M \in \RED{!A}$ then $M\in \RED{A}$.
\end{lemma}
\begin{lemma}
	\label{lemma:SN_with}
	If $\nTuple{M_1,M_2}\in\RED{\typwith{A_1}{A_2}}$, then $M_i\in\RED{A_i}$ for $1 \leq i\leq 2$.
\end{lemma}

Finally, we are able to prove the following lemma and Theorem~\ref{th:SN} follows as a corollary
\begin{lemma}\label{lemma:key_red_candidates}
	Given $\judgment{p_1:A_1,\ldots,p_n:A_n}{M}{B}$, $\forall V_i \in \RED{A_i}.\mbox{ } \overline{M}=M\{V_1/p_1, \ldots, V_n/p_n\}\in\RED{B} $, where $V_i$ value for the pattern $p_i$.
\end{lemma}
\begin{sproof}
	 Let $\Pi$ be the derivation for the judgement $\judgment{p_1:A_1,\ldots,p_n:A_n}{M}{B}$, we proceed by induction on $s(\Pi)$. We split depending on the last derivation rule in $\Pi$ and we use the auxiliary lemmas above in the corresponding cases.
\end{sproof}
\begin{corollary} \label{cor:reducibility}
	If $M$ is a term in $\lambdaLL$, then $M$ is reducible.
\end{corollary} 

\medskip \subsubsection{Confluence.}
The confluence property (Theorem~\ref{th:confluence}) is achieved by a case study of weak confluence (Lemma~\ref{lemma:WC}) and Newman's lemma (e.g.~\cite{terese}).  Let us then check the weak confluence of $\rightarrow$. 
\begin{lemma}\label{lemma:WC1}
	If $M \rightarrow M'$ then $M\{V/p\}\rightarrow M'\{V/p\}$.
\end{lemma}
\begin{lemma}\label{lemma:WC2}
	If $V \rightarrow^* V'$ then $M\{V/p\}\rightarrow^* M\{V'/p\}$.
\end{lemma}
\begin{lemma}[Weak Confluence] \label{lemma:WC}
	If $M' \leftarrow M \rightarrow M''$ then there exists $N$ such that $M' \rightarrow^* N$ and $M''\rightarrow^* N$.
\end{lemma}
\begin{sproof}
	By induction on $M$. The case for $M=\appterm{(\absterm{p}{M_1})}{V}$ is proved by using Lemma \ref{lemma:WC1} when $M'=M\{V/p\}\leftarrow M=\appterm{(\absterm{p}{M_1})}{V} \rightarrow \appterm{(\absterm{p}{M'_1})}{V}=M''$ and  Lemma \ref{lemma:WC2} when $M'=M\{V/p\}\leftarrow M=\appterm{(\absterm{p}{M_1})}{V} \rightarrow \appterm{(\absterm{p}{M_1})}{V'}=M''$.
\end{sproof}

\subsection{Logical Equivalence $\sim$} \label{app:logicEq}
Thanks to  the strong normalisation and progress properties (Theorem~\ref{th:SN} and Proposition~\ref{prop:progress_property}), the Definition~\ref{def:logical_closed} of $\sim$ for the $\typtop$ and $\with$ connectives is analogous to ones for the multiplicative connectives. 

Recall that $\sim_{\Gamma\vdash A}$ is the extension of $\sim$ to open terms.
We achieve the standard properties of the  logical relation $\sim$ defined in Definition~\ref{def:logical_closed}.
\begin{lemma}
	\label{lemma:beta-sim}
	Given $M =_\beta N$, then $M\sim N$. 
\end{lemma}
\begin{proof}
	First one prove the statement for closed term of a type $A$ by induction on $A$. 
	The extension to open terms follows because $=_\beta$-equivalence is contextual. 
\end{proof}

\begin{proposition}
	\label{prop:sim_congruence}
	The relation $\sim$ is an equivalence relation extending $=_\beta$ and context closed, i.e.~$M\sim N$ implies $\ContextA[M]\sim\ContextA[N]$ for every $\ContextA$. 
\end{proposition}
\begin{proof}
	Lemma~\ref{lemma:beta-sim} implies $=_\beta\subseteq\sim$ and hence reflexivity of $\sim$. The symmetry and transitive properties of $=_\beta$ are lifted to $\sim$ by induction on the definition of $\sim$. Context closure is proven by induction on $\ContextA$. 
\end{proof}

\begin{lemma}
	\label{lemma:sim_on_tuples}
	Given an exponential sequence type $E$ (resp.~additive sequence type $H$) we have that $\sim_E$ (resp.~$\sim_H$) coincides with $\beta$-equivalence of closed terms of that type $E$ (resp.~$H$). 
\end{lemma}
\begin{proof}
	By Proposition~\ref{prop:sim_congruence}, we need to prove only that $\sim_E$ (or~$\sim_H$) is included in $=_\beta$. This follows easily by induction on the definition of $\sim$.
\end{proof} 

The following lemma gives examples of $\sim$-equivalent terms which are not in general $\beta$-equivalent. 
\begin{lemma}[let-commutaton]
	\label{lemma:let_commutation}
		Given terms $M$ and $N$ and an exponential safe context $\ContextA$ (i.e. a context generated by the grammar in Figure~\ref{fig:grammar_contexts} without $\bangterm{\ContextA}$) s.t.~no free variable of $M$ can be captured by binders in $\ContextA$, as well as free variables of $p,q$ which are not free in $\ContextA$, we have: 
		$(\lambda p.\ContextA[N])M\sim\ContextA[(\lambda p.N)M]$.
\end{lemma}
The above let-commutation gives exactly the extension to $\beta$-reduction we need to achieve our results, summarised in the diagram of Figure~\ref{fig:final_picture}. In fact, we can replace $\sim$ by  extending $\beta$-equivalence with an adaptation of the $\sigma$-equivalences given in \cite{REGNIER1994} and achieve the same results. 

\begin{remark}
	Notice that $\letterm{(p,q)}{M}{N}$ (resp. $\letterm{(\,)}{M}{N}$) is a special case of $MN$, so we have also that $\letterm{(p,q)}{M}{\ContextA[N]}\sim\ContextA[\letterm{(p,q)}{M}{N}]$ (resp. $\letterm{()}{M}{\ContextA[N]}\sim\ContextA[\letterm{()}{M}{N}]$).
\end{remark}

\subsection{Workload} \label{app:costmodel}
In this appendix our goal is to show that any safe closed term $M$ reaches its $\beta$-normal form in at most $\Cost{M}$ numeric $\beta$-steps by using safe reduction, to do this we need to prove some auxiliary lemmas.
First, we show that strong values of ground type have a null workload
\begin{lemma}
	\label{lemma:cost_strong_value}
	Let $W$ be a strong value of ground type. We have $\Cost{W}=0$ and all free variables of $W$ are ground.
\end{lemma}
\begin{sproof}
	By induction on $W$. 
	Notice in particular that the strong value $W=\dot *W'$ has type $\typR\multimap\typR$, so it is not ground. 
\end{sproof}
Notice that the hypothesis $W$ be of ground type is important: for example $\dot +$ is a (closed) strong value with a non null workload. 

The following lemmas help to formalize the connection between the different reduction strategies involved in our system, $\beta$-reduction and $s$-reduction, and the concept of strong values, offering insight into when a term reaches its final, irreducible state under these reduction strategies.

\begin{lemma}
	\label{lemma:s-normal_strong_value}
	If $M$ is a closed normal form for safe reduction, then $M$ is a closed strong value. 
\end{lemma} 
\begin{proof}
	By induction on $M$.
	\begin{itemize}
		\item Case $M=M_1M_2$:\\
		By hypothesis  $M$ is a closed normal form for safe reduction, so $M_1$ and $M_2$ are closed normal forms for safe reduction. By induction hypotheses they can be supposed closed strong values. 
		We split into sub-cases, depending on $M_1$.
		Let us consider the case in which $M_1$ is of arrow type. By typing $M_1$ cannot be a tuple (additive or multiplicative), neither an exponential $\oc$, nor a numeral. Moreover, since $M$ is closed by hypothesis we have that $M_1$ cannot be a free variable.
		Therefore, the remaining cases are abstraction, numeric function ($\underline f$, $\dot +$, $\dot *$) or $\dot *W$. We details these cases as follows:
		\begin{itemize}
			\item Subcase $M_1=\lambda p.M_1'$:\\ 
			By inductive hypothesis $M_2$ is a closed strong value, so $M=(\lambda p.M_1')M_2$ is a $\beta_s$ redex, which is contrary to the hypothesis of $M$ be a $s$-normal form. 

			\item Subcase $M_1=\underline f$ or $M_1=\dot+$:\\
			By typing $M_2$ is a closed term of type $\oc\typR$ (for unary $\underline f$) or $\oc\typR\otimes\oc\typR$ (for binary $\underline f$) or $\typR\with\typR$ (for $\dot +$). 
			By induction hypothesis, $M_2$ is a closed strong value of type $\oc\typR$ or $\oc\typR\otimes\oc\typR$ or $\typR\with\typR$.
			One can check that the only closed strong values of these types are tuples (multiplicative or additive) of numerals. 
			Therefore, $M_1M_2$ is a $\beta_s$ redex, which is contrary to the hypothesis $M$ is a normal form for the safe reduction. 

			\item Subcase $M_1= \dot*$:\\
			By inductive hypothesis $M_2$ is a closed strong value, so $\dot *M_2$ is a closed strong value and we can conclude.

			\item Subcase $M_1= \dot *W$:\\
			By inductive hypothesis $M_2$ is a closed strong value and by typing it is of type $\typR$, so it is a numeral. Moreover, $W$ also is a closed value of type $\typR$, so $M_1M_2$ is a numerical, hence safe, redex. 	
		\end{itemize}
		\item All the other cases are similar or immediate. 
	\end{itemize}
\end{proof}

\begin{lemma}
	\label{lemma:s-normal_is_beta_normal}
	Let $M$ be a ground closed term. The following are equivalent:
	\begin{enumerate}
	\item $M$ is a closed strong value,
	\item $M$ is a closed normal form for the whole reduction $\rightarrow$,
	\item $M$ is a closed normal form for the safe reduction.
	\end{enumerate}
\end{lemma}
\begin{sproof}
	The implication $(1)\Rightarrow(2)$ is by induction on the grammar of~\ref{eq:strongval}, remarking that the hypothesis of $M$ ground implies that $M$ is not an abstraction. The implication $(2)\Rightarrow (3)$ is immediate, as safe redexes are also $\beta_\lambda$-redexes. The implication $(3)\Rightarrow (1)$ is by Lemma~\ref{lemma:s-normal_strong_value}.
\end{sproof}

The main auxiliary lemma is related to the properties retained by the substitution in the context of safe reduction. More precisely, the following statement is both qualitative, as it guarantees that safeness and typing are preserved during substitution, and quantitative, ensuring  that the workload does not increase with respect to the sum of the cost related to the analysed term and the cost of the substituted value.

\begin{remark}\label{rmk:cond_closed_safe_sub}
	It is worth noting that in the statement of Safe Substitution we must require the strong value 
	$W$ to be closed, since this is the only way to ensure that the substitution  
	$M\{W/p\}$ is a safe term. Specifically, when $M=\nTuple{M_1,M_2}$, this requirement prevents the substitution from introducing free variables that could compromise the safeness of $M\{W/p\}$. Without assuming that $W$ is closed, the intersection $\FV{M_1\{W/\PatA \}}\cap\FV{M_2\{W/\PatA \}}$ could contain non-ground variables, violating the safeness condition in item (ii) of Definition~\ref{def:safe_term}. 
	The closure of $W$ is required solely to preserve the safeness of the substitution.
\end{remark}

\begin{lemma}[Safe Substitution]
	\label{lemma:safe_sub}
	Given a safe term $M$ such that $\oc\Gamma,\Delta,\PatA :A \vdash M:B$ and a safe closed strong value $W$ for the pattern $\PatA $ such that $\vdash W:A$, we have:
	\begin{enumerate}
		\item  $M\{W/\PatA \}$ is a safe term;
		\item $\oc\Gamma,\Delta  \vdash M\{W/\PatA \}:B$
		\item $\Cost{M\{W/\PatA \}}\leq\Cost{W}+\Cost{M}$.
	\end{enumerate}
\end{lemma}

\begin{sproof}
    Claim 1 is proved by induction on $M$, using the properties of safe terms listed in Definition~\ref{def:safe_term} and the definition of $\Cost{M}$. The two delicate cases are $M=\oc M'$ and $M=\nTuple{M_1,M_2}$, namely those related to the conditions in Definition~\ref{def:safe_term}. Let us details these two cases:
	\begin{itemize}
		\item Case $M=\oc M'$:\\
		By definition of substitution $M\{W/\PatA \}=(\oc M')\{W/\PatA \}=\oc (M'\{W/\PatA \})$.
		Moreover, Definition~\ref{def:safe_term} of safe term item $i$ we have to prove that $\Cost{M'\{W/\PatA \}}=0$ in order to conclude that $M\{W/\PatA \}$ is safe.

		Recall that, by hypothesis $M=\oc M'$ is a safe term, so by item $i$ of Definition~\ref{def:safe_term} we know that $\Cost{M'}=0$.
		Moreover, by typing the pattern $\PatA $ is of exponential type this means that $\PatA =x$ and $W=!W'$ for some strong value $W'$. Hence by definition of workload $\Cost{W}=\Cost{!W'}=0$.

		By item 3 of this lemma we have that $\Cost{M'\{W/\PatA \}}\leq\Cost{W}+\Cost{M'}$ which is equal to zero and so we can conclude.

		\item Case $M=\nTuple{M_1,M_2}$:\\
		By definition of substitution $M\{W/\PatA \}=(\nTuple{M_1,M_2})\{W/\PatA \}=\nTuple{M_1\{W/\PatA \},M_2\{W/\PatA \}}$. Moreover, Definition~\ref{def:safe_term} of safe term item $ii$ we have to prove that $\FV{M_1\{W/\PatA \}}\cap\FV{M_2\{W/\PatA \}}$ has only ground variables in order to conclude that $M\{W/\PatA \}$ is safe.

		Recall that, by hypothesis $M=\nTuple{M_1,M_2}$ is a safe term, so by item $ii$ of Definition~\ref{def:safe_term} we know that $\FV{M_1}\cap\FV{M_2}$ has only ground variables.
		Moreover, by hypothesis $W$ is closed and substituting a closed value cannot introduce any new free variables, so we can conclude that $\FV{M_1\{W/\PatA \}}\cap\FV{M_2\{W/\PatA \}}$ has only ground variables. 
	\end{itemize} 
	
    Furthermore, we prove Claim 2 and Claim 3 by a similar approach we used for the Pattern Substitution Lemma (Lemma~\ref{lemma:substitution}). More precisely, for any derivation $\Pi_1$ of $\oc\Gamma,\Delta, \PatA :A \vdash M:B$ and $\Pi_2$ of $\vdash W:A$, we give a derivation of $\oc\Gamma,\Delta\vdash M\{W/\PatA \}:B$ by induction on the lexicographically ordered pair $(s(\Pi_2), s(\Pi_1))$, where $s(\Pi_i)$ is the number of derivation rules of $\Pi_i$. 
	We split depending on the last derivation rule in $\Pi_1$ or $\Pi_2$. 
	\begin{itemize}
		\item If the last rule of $\Pi_1$ is a rule $r$ among $\{\typbangW, \typrwithL{i}, \typrtensorL, \typroneL\}$ acting on a pattern in $\oc\Gamma,\Delta$, then the immediate subderivation of $\Pi_1$ is $\Pi_1'$ of $\oc\Gamma',\Delta',\PatA :A\vdash M:B$.
		We can conclude by induction hypothesis on $(s(\Pi_2), s(\Pi'_1))$ getting
		a) type derivation for $\oc\Gamma',\Delta'\vdash M\{W/\PatA \}:B$; b) $\Cost{M\{W/\PatA \}}\leq \Cost{M}+\Cost{W}$.

		\item For the other cases, we can then suppose that the last rules of $\Pi_1$ is not acting on $\oc\Gamma,\Delta$.
		We then split in further sub-cases depending if the last rules of $\Pi_1$ acts on the pattern $\PatA:A$ or acts on the term $M$. 

		Let us consider first the cases of a last rule $r$ of $\Pi_1$ acting on the pattern $\PatA:A$.
		\begin{itemize}
		\item If $r$ is of type $\typbangW$, then $\PatA=\oc x$ and $W = \oc W'$ for some safe closed strong value $W'$. 
		
		Notice that the subderivation $\Pi'_1$ above $r$ in $\Pi_1$ has conclusion $\oc\Gamma,\Delta\vdash M:B$.

		By Lemma~\ref{lemma:fv_in_environment} $x\notin \FV{M}$ and by applying Lemma~\ref{lemma:weakening_subst} we have: $M\{\oc W'/\oc x\}=M$, so item 2 of the lemma holds by taking $\Pi'_1$.

		By definition of workload we have $\Cost{W}=0$ and so item 3 of the lemma holds as
		$\Cost{M\{W/p\}}=\Cost{M}\leq\Cost{M}+\Cost{W}=\Cost{M}$.

		\item If $r$ is of type $\typrwithL{i}$, then $\PatA=\nTuple{\PatA_1,\PatA_2}:A_1\with A_2$ and $W = \nTuple{W_1,W_2}$ for some safe strong values $W_i$ for $\PatA_i$.
		
		Notice that the subderivation $\Pi'_1$ above $r$ in $\Pi_1$ has conclusion $\oc\Gamma,\Delta,p_i:A_i\vdash M:B$.

		By Lemma~\ref{lemma:fv_in_environment} $\FV{\PatA_{3-i}}\notin \FV{M}$ so by Lemma~\ref{lemma:weakening_subst}, $M\{W_i/\PatA_i\}\{W_{3-i}/\PatA_{3-i}\}=M\{W_i/\PatA_i\}$.

		By cases inspection, one can infer that the last rule of $\Pi_2$ is a $\typrwithR$.
		Therefore we have a subderivation $\Pi_2'$ above such rule for the judgement $\vdash W_i:A_i$. 

		Let us suppose $i=1$ (the other case being similar), so we have: $M\{W_1/\PatA_1\}\{W_2/\PatA_2\} =M\{W_1/\PatA_1\}$.

		By induction hypothesis on $(s(\Pi'_2), s(\Pi'_1))$ we have: a) a derivation for $\oc\Gamma,\Delta\vdash M\{W_1/p_1\}:B$; b) $\Cost{M\{W_1/p_1\}}\leq \Cost{M}+\Cost{W_1}$.

		The item 2 of the lemma holds because $M\{W/p\} =M\{W_1/\PatA_1\}$ and by point a of the induction hypothesis.

		We show that item 3 of the lemma holds as follows
		\begin{align*}
			\Cost{M\{W/p\}}&=\Cost{M\sfrac{\nTuple{W_1,W_2}}{\nTuple{p_1,p_2}}}\\
			&=\Cost{M\{W_1 /p_1\}\{W_2 /p_2\}}\\
			&=\Cost{M\{W_1 /p_1\}}\\
			&\overset{\tiny \text{IH}}{\leq} \Cost{M}+\Cost{W_1}\\
			&\leq \Cost{M}+\Cost{W}\\
			&=\Cost{M}+\Cost{\nTuple{W_1,W_2}}\\
			&=\Cost{M}+\Cost{W_1}+\Cost{W_2}
		\end{align*}
		
		\item If $r$ is of type $\typrtensorL$, then $\PatA=(\PatA_1,\PatA_2):A_1\otimes A_2$ and $W = (W_1,W_2)$ for some values $W_i$ of $\PatA_i$. 
		
		Notice that the subderivation $\Pi'_1$ above $r$ in $\Pi_1$ has conclusion $\oc\Gamma,\Delta,p_1:A_1,p_2:A_2\vdash M:B$.

		By cases inspection, one can infer that the last rule of $\Pi_2$ is a $\typrtensorR$.
		Therefore we have two subderivations $\Pi_{2,1}$ and $\Pi_{2,2}$ above such rule for $\vdash W_1:A_1$ and $\vdash W_2:A_2$, respectively. 

		We have to prove that: 1) $M\{W/p\}=\Cost{M\sfrac{(W_1,W_2)}{(p_1,p_2)}}=M\{W_1/p_1\}\{W_2/p_2\}$ is well-typed as $\oc\Gamma,\Delta\vdash M\{W_1/p_1\}\{W_2/p_2\}:B$;
		2) $\Cost{M\{W_1 /p_1\}\{W_2 /p_2\}}\leq \Cost{M}+\Cost{W_1}+\Cost{W_2}$
		
		By induction hypothesis on $(s(\Pi_{2,1}), s(\Pi'_1))$ we have: a) a derivation for $\oc\Gamma,\Delta,p_2:A_2\vdash M\{W_1/p_1\}:B$; b) $\Cost{M\{W_1/p_1\}}\leq \Cost{M}+\Cost{W_1}$.

		We proceed by applying the induction hypothesis on $(s(\Pi_{2,2}), s(\Pi'_1))$ obtaining: a) a derivation for $\oc\Gamma,\Delta\vdash M\{W_1/p_1\}\{W_2/p_2\}:B$; b) 
		$\Cost{(M\{W_1 /p_1\})\{W_2 /p_2\}}\leq\Cost{M\{W_1 /p_1\}}+\Cost{W_2}$.

		We can conclude as item 2 holds directly from item a of IH on  $(s(\Pi_{2,2}), s(\Pi'_1))$ and item 3 holds as 
		$$\Cost{M\{W_1 /p_1\}}+\Cost{W_2} \overset{\tiny \text{IH on $(s(\Pi_{2,1}), s(\Pi'_1))$}}{\leq} \Cost{M}+\Cost{W_1}+\Cost{W_2}$$

		\item If $r$ is of type $\typroneL$, then $\PatA=(\,):\typone$ and $W = (\,)$. 
		
		Notice that the subderivation $\Pi'_1$ above $r$ in $\Pi_1$ has conclusion $\oc\Gamma,\Delta\vdash M:B$.
		
		By cases inspection, one can infer that the last rule of $\Pi_2$ is a $\typroneR$. 
		
		By definition of substitution we have $M\{(\,)/(\,)\}=M$ and by definition of workload we have $\Cost{(\,)}=0$, so we can conclude.
		\end{itemize}

		\item Let us consider now the cases in which the last rule $r$ in $\Pi_1$ acts on the subject $M$.
		
		\begin{itemize}
			\item If $r$ is of type $\typrvar$, then $M=\PatA=x:A$. Moreover, $A=B$ and $\oc \Gamma,\Delta$ is empty.
			
			By definition of substitution $M\{W/p\}=W$.

			By definition of workload we have $\Cost{x}=0$, so item 3 of the lemma holds.

			Moreover, item 2 of the lemma holds because by hypothesis we have a derivation $\Pi_2$ for $\vdash W:A$.

			\item  If $r$ is of type $\typrbangL$, then $M=x$ and $\PatA=\oc x$. Moreover, $B=!A$ and $\oc \Gamma,\Delta$ is empty.
			
			By definition of value for a pattern $W=\oc W'$ for some safe closed strong value $W'$.

			By hypothesis $W$ is safe and by item $i$ in Definition~\ref{def:safe_term} of safeness we have $\Cost{W'}=0$.

			By cases inspection, one can infer that the last rule of $\Pi_2$ is a $\typrbangr$. 
			Therefore we have an immediate subderivation $\Pi_2'$ above such rule for the judgement $\vdash W':A$.

			By definition of substitution $M\{W/p\}=W'$, so item 2 of the lemma holds by taking the derivation $\Pi_2'$.

			By definition od workload we have $\Cost{\oc W'}=0$. We can conclude that item 3 of the lemma holds as follows
			\begin{align*}
				\Cost{M\{W/p\}}
				&=\Cost{x\{\oc W'/\oc x\}}\\
				&=\Cost{x\{W'/x\}}\\
				&=\Cost{W'}\\
				&=0\\
				&\leq \Cost{M}+\Cost{W}\\
				&=\Cost{x}+\Cost{\oc W'}\\
				&=0
			\end{align*}

			\item If $r$ is of type $\typrwithR$, then $M=\nTuple{M_1,M_2}$ and $B=B_1 \& B_2$.
			
			The immediate subderivations of $\Pi_1$ are $\Pi_{1,1}$ and $\Pi_{1,2}$ above $r$ for $\oc\Gamma, \Delta, p:A \vdash M_1: B_1$ and $\oc\Gamma, \Delta,p:A\vdash M_2: B_2$, respectively. 

			By induction hypothesis on  $(s(\Pi_2), s(\Pi'_{1,i}))$ with $i\in\{1,2\}$ we have: a) a type derivation for $\oc\Gamma, \Delta \vdash M_i\{W/p\}: B_i$; b) $\Cost{M_i\{W/p\}}=\Cost{M_i}+\Cost{W}$.

			By definition of substitution we have that $M\{W/p\}=\nTuple{M_1,M_2}\{W/p\}=\\\nTuple{M_1\{W/p\},M_2\{W/p\}}$.

			We can conclude that item 2 of the lemma holds by using $\typrwithR$ and the inductive hypotheses.

			By item 1 of the lemma $M\{W/p\}$ is safe, so in this case $\nTuple{M_1,M_2}\{W/p\}$ is safe and by item $ii$ in Definition~\ref{def:safe_term} of safe term we have that $\FV{M_1}\cap\FV{M_2}$ contains only ground variables.
			Moreover, $W$ is a strong value of ground type and by Lemma~\ref{lemma:cost_strong_value} we have $\Cost{W}=0$.

			We show that item 3 of the lemma holds as follows
			\begin{align*}
				\Cost{M\{W/p\}}&=\Cost{(\nTuple{M_1,M_2})\{W/p\}}\\
				&=\Cost{\nTuple{M_1\{W/p\},M_2\{W/p\}}}\\
				&=\Cost{M_1\{W/p\}}+\Cost{M_2\{W/p\}}\\ 
				&\overset{\tiny \text{IHs}}{\leq} 
				\Cost{W}+\Cost{M_1}+\Cost{W}+\Cost{M_2}\\
				&=\Cost{M_1}+\Cost{M_2}\\
				&\leq \Cost{M}+\Cost{W}\\
				&= \Cost{\nTuple{M_1,M_2}}+\Cost{W}\\
				&=\Cost{M_1}+\Cost{M_2}+\Cost{W}\\
				&=\Cost{M_1}+\Cost{M_2}
			\end{align*}
			\item All other cases are similar or immediate.
		\end{itemize}
	\end{itemize}
\end{sproof}

Moreover, we proceed by showing that the workload decreases along safe reduction (Proposition~\ref{prop:safe-reduction_decreases_measures}) and the safeness of a term is preserved along safe reduction (Lemma~\ref{lemma:safe_invariance}).

\begin{lemma}
	\label{lemma:safe_implies_no_numerical}
	If $M$ is safe, then there is no numerical operation (i.e.~$\dot +$, $\dot *$ or $\underline f$) under a $\oc$.
\end{lemma}
\begin{sproof}
	We proceed by strengthening the statement, proving by induction on $M$ the following two claims:
	\begin{enumerate}
	    \item If $M$ is safe, then there is not numerical operation under a $\oc$;
	    \item If moreover $\Cost M=0$, then there is no numerical operation at all in $M$.
	\end{enumerate} 
\end{sproof}

\begin{proposition}
	\label{prop:safe-reduction_decreases_measures}
    Let $M$ be a safe term. If $M\rightarrow N$ is a safe step, then $\Cost{N}\leq \Cost{M}$. If moreover the step is numerical, then $\Cost{N}< \Cost{M}$.
\end{proposition}
\begin{proof}
	By induction on the evaluation context $\ContextA$. The induction step splits according to the cases of Figure~\ref{fig:grammar_contexts}, while the base case splits according to Figure~\ref{fig:beta_rules}.

    In the base case of induction, if the $\beta$-step is $\beta_s$ then we use the Safe Substitution Lemma (Lemma~\ref{lemma:safe_sub}) and we conclude. 

	In the induction step, if $\ContextA=\oc\gamma'[\,]$, so that $M = \oc\gamma'[M_0]$ and $N= \oc\gamma'[N_0]$, 
	we then have $\Cost{N}=0$ by definition and the two inequalities $\leq$ hold trivially. As for the strict inequality in case of numerical steps: since $M$ is safe, by Lemma~\ref{lemma:safe_implies_no_numerical} there is no numerical operator in $\gamma'[M_0]$, so the step $M\rightarrow N$ cannot be numerical.
\end{proof}

\begin{lemma}[Safeness Invariance]
	\label{lemma:safe_invariance}
	If $M$ is safe and $M\xrightarrow{s} N$, then $N$ is safe too.
\end{lemma}
\begin{sproof}
	By induction on the evaluation context $\ContextA$ of the reduction step $M\xrightarrow{s} N$.

    In the base case of induction, if the $\beta$-step is $\beta_s$ then we conclude by using item $1$ of Lemma~\ref{lemma:safe_sub}.
	
	One case of the induction step is subtle: if $\ContextA=\oc\gamma'[\,]$, so $M=\oc\gamma'[M_0]$, $N=\oc\gamma'[N_0]$ and $\gamma'[M_0]\xrightarrow{s}\gamma'[N_0]$. 
	By induction hypothesis $\gamma'[N_0]$ is safe. 
	In order to prove that $\oc\gamma'[N_0]$ is safe too, we must prove that $\Cost{\gamma'[N_0]}=0$. 
	By Proposition~\ref{prop:safe-reduction_decreases_measures}, we have $\Cost{\gamma'[N_0]}\leq\Cost{\gamma'[M_0]}$. 
	By safeness of $M$ (Definition~\ref{def:safe_term}), we have $\Cost{\gamma'[M_0]}=0$ and we can conclude. 
\end{sproof}

We are finally able to prove that a safe closed term $M$ normalizes by using safe reduction in at most $\Cost{M}$ steps. Formally, this is stated as follows 
 
\begin{proof}[Proof of Proposition~\ref{prop:safe_reduction_is_safe}]
    Consider a maximal safe-reduction sequence  $(M_i)_{i=0}^n$ starting from $M$, i.e. $M_0=M$ and $M_n$ is a safe-normal form. 

	By Subject Reduction, all $M_i$'s are closed.
	In particular, Lemma~\ref{lemma:s-normal_strong_value} gives that $M_n$ is a closed strong value. 

	By Lemma~\ref{lemma:safe_invariance}, all $M_i$'s are safe too. So we can apply Proposition~\ref{prop:safe-reduction_decreases_measures} to each reduction step and getting that the sequence $\Cost{M_0},\Cost{M_1},\dots$ is decreasing, moreover it strictly decreases if the step is numeric. We conclude that $\Cost{M}$ bounds the number of numeric steps of this sequence. 

	Since $M_n$ is a closed strong value of ground type, Lemma~\ref{lemma:s-normal_is_beta_normal} assures that $M_n$ is also a $\beta$-normal form. 
\end{proof}

\bigskip
The following lemma will be useful to show that our transformations are work preserving.

\begin{lemma}\label{lemma:cost_affine}
	$\cost{\affinebang M}=\cost{M}$.
\end{lemma}
\begin{proof}
	By notational convention defined in Section~\ref{sect:lambdaLL} we have $\affinebang M = \nTuple{(\,),M}$, so we can conclude as follows
	$$\cost{\affinebang M}=\cost{\nTuple{(\,),M}}=\cost{(\,)}+\cost{M}=\cost{M}$$
	because by definition of workload $\cost{(\,)}=0$.
\end{proof}

\clearpage
\section{Translation}
\subsection{Translation $\SymbTransA$}\label{app:translDeltaA}
Translation $\SymbTransA$ is defined in Figure~\ref{fig:JAXtoLL}. We add in Figure~\ref{fig:JAXtoLL_app} the cases of the primal binary tuples. 
\begin{figure*}[h!]
            \begin{align*}
                \jaxdelta[\theta]{\tupleJAX{x_1}{x_2}}
                \definedas\mbox{}
                    & 
                    \tupleterm{\bangterm{\tupleterm{\oc x_1}{\oc x_2}}}{\affbangterm{(\lambda y^{\typtop}.\linemptytupleterm)}}
                \\
                \jaxdelta[\theta]{\letterm{\tupleJAX{x_1}{x_2}}{\expvarJAX{z}}{e}}\definedas\mbox{}
                    & \letterm{\tupleterm{\oc x_1}{\oc x_2}}{z}{\jaxdelta[\theta]{e}}
            \end{align*}
            \caption{Translation $\jaxdelta[\theta]{e}$ for $e$ primal binary tuples.}
            \label{fig:JAXtoLL_app}
\end{figure*}

We prove the Soundness of $\SymbTransA$ as follows
\begin{proof}[Proof Proposition~\ref{prop:sound_transA}]
    Let us recall the statement we want to prove:\\
    \textit{
    Given $\Gamma;\dot\Sigma\vdash e:(\tau;\sigma)$, an enumeration $\theta$ of the tangent variables in $\dot\Sigma$, then:
    }

    \begin{itemize}
    \item \textit{$\forall\SeqA$ for $\Gamma$:
    $
    	\TransA[\theta]{e}[\oc \SeqA/\PrimalT(\Gamma)] \rightarrow^* (\oc \SemP[\SeqA]{e}, \affbangterm F)
    $,}
    \item \textit{$\forall\SeqB$ for the type $\&\theta$: 
    $
    	F\SeqB \rightarrow^* \SemT[\SeqA;\SeqB]{e}
    $.}
    \end{itemize}  
    \smallskip
    We proceed by induction on $e$. An interesting case is $e=\letterm{(x;\dot y)}{e_1}{e_2}$ which is well-typed by the judgement $\Gamma_1\cup\Gamma_2;\dot\Gamma_1, \dot{\Gamma_2}\vdash^{\text{\tiny{Jax}}} \letterm{(x;\dot y)}{e_1}{e_2}:(\tau;\sigma)$. We fix a numeral sequence $\oc\SeqA$ for $\PrimalT(\Gamma_1),\PrimalT(\Gamma_2)$ and we can observe that $\oc\SeqA\vert\FV{e_i}$ is a numeral sequence for $\PrimalT(\Gamma_i)$.
We recall from Section~\ref{sect:JAX} that 
    \begin{align}
        \label{eq:soundAsemPLet}
        \SemP[\vec{\underline{r}}]{\letterm{(x;\dot y)}{e_1}{e_2}}
        &= \SemP[\vec{\underline{r}}\vert\FV{e_2}, x\mapsto {\SemP[\vec{\underline{r}}\vert\FV{e_1}]{e_1}}]{e_2}
        \\
        \label{eq:soundAsemTLet}
        \SemT[
            \vec{\underline{r}};\vec{\underline{s}}
        ]{\letterm{(x;\dot y)}{e_1}{e_2}}
        &= 
        \SemT[
            \vec{\underline{r}}\vert\FV{e_2}, 
            x\mapsto {\SemP[\vec{\underline{r}}\vert\FV{e_1}]{e_1}}
            ; 
            \vec{\underline{s}}\vert\linFV{e_2}, 
            \dot y\mapsto{\SemT[\vec{\underline{s}}\vert\linFV{e_1}]{e_1}}
            ]{e_2}
    \end{align}
    and we proceed as follows to prove the first claim
    \begin{align}
        \TransA[\theta]{e}[\oc\SeqA/\PrimalT(\Gamma)]
            &\definedas \left(
                \begin{aligned}
                    &\letterm{\tupleterm{\oc x}{\affinebang f}}{\jaxdelta[\theta\cap\linFV{e_1}]{e_1}}{}\\
                     &\letterm{\tupleterm{\oc z}{\affinebang g}}{\jaxdelta[\linvarJAX y,\theta\cap\linFV{e_2}]{e_2}}{}\\
                    &\tupleterm{\oc z}{\affbangterm{F}}
                \end{aligned}
            \right)[\oc\SeqA/\PrimalT(\Gamma)]\\[3mm]
            &=\label{eq:soundAbeforeIH} 
            \begin{aligned}
                    &\letterm{\tupleterm{\oc x}{\affinebang f}}{\jaxdelta[\theta\cap\linFV{e_1}]{e_1}[\oc\SeqA\vert\FV{e_1}/\PrimalT(\Gamma_1)]}{}\\
                     &\letterm{\tupleterm{\oc z}{\affinebang g}}{\jaxdelta[\linvarJAX y,\theta\cap\linFV{e_2}]{e_2}[\oc\SeqA\vert\FV{e_2}/\PrimalT(\Gamma_2)]}{}\\
                    &\tupleterm{\oc z}{\affbangterm{F}}
                \end{aligned}\\[3mm]
            &\hspace*{-2mm}\rightarrow^* \label{eq:soundAafterIH} 
                \begin{aligned}
                    &\letterm{\tupleterm{\oc x}{\affinebang f}}{(\oc\SemP[\SeqA\vert\FV{e_1}]{e_1}, \affbangterm F_1)}{}\\
                    &\letterm{\tupleterm{\oc z}{\affinebang g}}{(\oc\SemP[\SeqA\vert\FV{e_2}]{e_2}, \affbangterm F_2)}{}\\
                    &\tupleterm{\oc z}{\affbangterm{F}}
                \end{aligned}\\[3mm]
            &\hspace*{-2mm}\rightarrow^* \label{eq:soundAafterbeta1}
                \begin{aligned}
                    &\letterm{\tupleterm{\oc x}{\affinebang f}}{(\oc\SemP[\SeqA\vert\FV{e_1}]{e_1}, \affbangterm F_1)}{}\\
                    &\tupleterm{\oc\SemP[\SeqA\vert\FV{e_2}]{e_2}}{\affbangterm{F}\{\affbangterm{F_2}/g\}}
                \end{aligned}\\[3mm]
            &\hspace*{-2mm}\rightarrow^* \label{eq:soundAafterbeta2}
            \begin{aligned}
                &\tupleterm{\oc\SemP[\vec{\underline{r}}\vert\FV{e_2}, x\mapsto {\SemP[\vec{\underline{r}}\vert\FV{e_1}]{e_1}}]{e_2}}{\affbangterm{F}\{\affbangterm{F_2}/g,\affbangterm{F_1}/f\}}
            \end{aligned}
    \end{align}
    where $\affbangterm{F}\{\affbangterm{F_2}/g,\affbangterm{F_1}/f\}=
    \affbangterm{(\absterm{{y^{\&\linearize{\theta}}}}{
        \letterm{\nTuple{y_1,y_2}}{\Split[\&\linearize{\theta}]{\linFV{e_1}} y}
        {
            \appterm{F_2}{\nTuple{\appterm{F_1}{y_1},y_2}}
        }
    })}$.
    The passage from line~\eqref{eq:soundAbeforeIH} to~\eqref{eq:soundAafterIH} is by induction hypothesis and the passage from line~\eqref{eq:soundAafterIH} (resp.~\eqref{eq:soundAafterbeta1}) to~\eqref{eq:soundAafterbeta1} (resp.~\eqref{eq:soundAafterbeta2}) is obtained by applying $\redbeta_\lambda$. We can conclude by observing that the first term of the tuple in~\eqref{eq:soundAafterbeta2} is equal to~\eqref{eq:soundAsemPLet}. 
    
    In order to prove the second claim, we fix $\SeqB$ for the type $\&\theta$ and and we can observe that $\SeqB\vert\linFV{e_i}$ is a numeral sequence for $\TangentT(\dot\Gamma_i)$. Formally, we proceed as follows
    \begin{align}
        \label{eq:soundA_F_1}
        (
            \lambda y^{\&\linearize{\theta}}.
            &\letterm{\nTuple{y_1,y_2}}{\Split[\&\linearize{\theta}]{\linFV{e_1}} y}
            {\appterm{F_2}{\nTuple{\appterm{F_1}{y_1},y_2}}}
        )\SeqB\\
        &\label{eq:soundA_F_2} 
        \rightarrow  
        \letterm{\nTuple{y_1,y_2}}
        {\Split[\&\linearize{\theta}]{\linFV{e_1}} \SeqB}
        {\appterm{F_2}{\nTuple{\appterm{F_1}{y_1},y_2}}}\\
        &\label{eq:soundA_F_3}
        =\letterm{\nTuple{y_1,y_2}}
        {(\lambda\nTuple{x_i}_{i=1}^n.\nTuple{\nTuple{x_i}_{i\in \linFV{e_1}},\nTuple{x_i}_{i\notin \linFV{e_1}}}) \SeqB}
        {\appterm{F_2}{\nTuple{\appterm{F_1}{y_1},y_2}}}\\
        &\label{eq:soundA_F_4}
        =\letterm{\nTuple{y_1,y_2}}
        {(\lambda\nTuple{x_i}_{i=1}^n.\nTuple{\nTuple{x_i}_{i\in \linFV{e_1}},\nTuple{x_i}_{i\in \linFV{e_2}}}) \SeqB}
        {\appterm{F_2}{\nTuple{\appterm{F_1}{y_1},y_2}}}\\
        &\label{eq:soundA_F_5}
        \rightarrow  
        \letterm{\nTuple{y_1,y_2}}
        {\nTuple{\SeqB\vert\linFV{e_1},\SeqB\vert\linFV{e_2}}}
        {\appterm{F_2}{\nTuple{\appterm{F_1}{y_1},y_2}}}\\
        &\label{eq:soundA_F_6}
        \rightarrow^*   
        {\appterm{F_2}{\nTuple{\appterm{F_1}{\SeqB\vert\linFV{e_1}},\SeqB\vert\linFV{e_2}}}}\\
        &\label{eq:soundA_F_7}
        \rightarrow^*   
        {\appterm{F_2}{\nTuple{\SemT[\SeqA;\SeqB\vert\linFV{e_1}]{e_1},\SeqB\vert\linFV{e_2}}}}\\
        &\label{eq:soundA_F_8}
        \rightarrow^*   
        \SemT[
            \vec{\underline{r}}\vert\FV{e_2}, 
            x\mapsto {\SemP[\vec{\underline{r}}\vert\FV{e_1}]{e_1}}
            ; 
            \vec{\underline{s}}\vert\linFV{e_2}, 
            \dot y\mapsto{\SemT[\vec{\underline{s}}\vert\linFV{e_1}]{e_1}}
            ]{e_2}
    \end{align}
    The passage from line~\eqref{eq:soundA_F_1} to~\eqref{eq:soundA_F_2} is obtained by applying $\redbeta_\lambda$.\\
    The passage from line~\eqref{eq:soundA_F_2} to~\eqref{eq:soundA_F_3} is obtained by definition of splitting term $\sigma$.\\
    The passage from line~\eqref{eq:soundA_F_3} to~\eqref{eq:soundA_F_4} is obtained by observing that a variable $x$ such that $x\in\linearize{\theta}$ and $x\notin\linFV{e_1}$ is a free tangent variable in $e_2$.\\
    The passage from line~\eqref{eq:soundA_F_4} (resp.~\eqref{eq:soundA_F_5}) to~\eqref{eq:soundA_F_5} (resp.~\eqref{eq:soundA_F_6}) is obtained by applying $\redbeta_\lambda$.\\
    The passage from line~\eqref{eq:soundA_F_6} (resp.~\eqref{eq:soundA_F_7}) to~\eqref{eq:soundA_F_7} (resp.~\eqref{eq:soundA_F_8}) is obtained by inductive hypothesis on $e_1$  (resp. $e_2$).\\
    We can conclude by observing that~\eqref{eq:soundA_F_8} is equal to~\eqref{eq:soundAsemTLet}.
\end{proof}

Let us define the workload of a JAX type $\tau$, denoted by $\CostJAX{\tau}$, as the number of occurrences of the base type $\mathbb{R}$ within $\tau$. The following lemma establishes a correspondence between the type translation $\linearize{\cdot}$ and the workload.
\begin{lemma}\label{lemma:cost_transl_type}
    Given $\tau$ in the grammar of \ref{eq:JAXtypes}, then $\CostJAX{\tau}=\Cost{\linearize{\tau}}$.    
\end{lemma}
\begin{proof}
    By induction on $\tau$.
\end{proof}

We must also ensure that the translation $\TransA{e}$ computes $\SemP[\SeqA]{e}$ and $\SemT[\SeqA;\SeqB]{e}$ efficiently --- specifically, without performing more flops than those required by the original Linear A expression $e$. This property is essential to guarantee that the translation preserves the computational efficiency. Notice that $\TransA{e}$ satisfies the conditions of Definition~\ref{def:safe_term}, so by Proposition~\ref{prop:safe_reduction_is_safe} we can use the workload of a term as a bound to the number of numeric steps. 
\begin{proof}[Proof Proposition~\ref{prop:size_TransA}]
    The safeness of $\jaxdelta[\theta]{e}$ is easy to prove by induction on $e$ simply checking the items in Definition~\ref{def:safe_term}. Let us focus on the proof related to work preservation of $\TransA[\theta]{e}$, we proceed by induction on $e$.
    \begin{itemize}
        \item Case $e=\letterm{\retJAX{x}{y}}{e_1}{e_2}$:
        \begin{align*}
            \CostJAX{e}&=\CostJAX{\letterm{\retJAX{x}{y}}{e_1}{e_2}}=\CostJAX{e_1}+\CostJAX{e_2}\\
            \Cost{\jaxdelta[\theta]{e}}&=
            \mathcal{W}\left(
                \begin{aligned}
                    &\letterm{\tupleterm{\oc x}{\affinebang f}}{\jaxdelta[\theta\cap\linFV{e_1}]{e_1}}{}\\
	 	            &\letterm{\tupleterm{\oc z}{\affinebang g}}{\jaxdelta[\linvarJAX y,\theta\cap\linFV{e_2}]{e_2}}{}\\
		            &\tupleterm{\oc z}{\affbangterm{(\absterm{{y^{\&\linearize{\theta}}}}{
			            \letterm{\nTuple{y_1,y_2}}{\Split[\&\linearize{\theta}]{\linFV{e_1}} y}
			        {\appterm{\affderterm{g}}{\nTuple{\appterm{\affderterm{f}}{y_1},y_2}}}
		            })}} 
                \end{aligned}
            \right)
        \end{align*}
        Observe that our workload essentially counts the number of numerical operations not under a ! and the number of possible numerals erased during a reduction. In this case nothing is erased, so the sums related to the workload for the two let-constructs and for the $\lambda$-abstraction are equal to zero. This means that $\Cost{\jaxdelta[\theta]{e}}$ is equal to $\Cost{\jaxdelta[\theta\cap\linFV{e_1}]{e_1}}+\Cost{\jaxdelta[\linvarJAX y,\theta\cap\linFV{e_2}]{e_2}}$ and we can conclude by inductive hypotheses.

        \item Case $e=\dropJAX{e_1}$:\\
        By hypothesis $\judgmentJAX{\Gamma}{\dot\Gamma}{\dropJAX{e_1}}{\typunitJAX}{\typunitJAX}$ and by typing $\judgmentJAX{\Gamma}{\dot\Gamma}{e_1}{\tau_1}{\sigma_1}$.\\
        By definition we have:
            \begin{align*}
                \CostJAX{e}&=\CostJAX{\dropJAX{e_1}}=\CostJAX{e_1}+\CostJAX{\tau_1}+\CostJAX{\sigma_1}\\[2mm]
                \Cost{\jaxdelta[\theta]{e}}&=
                \Cost{\letterm{\tupleterm{\oc x}{\affinebang  f}}{\jaxdelta[\theta]{e}}{\trad{\emptytupleterm}{\&\linearize{\theta}}{
                    \letterm{z}{\appterm{f}{y}}{\linemptytupleterm}}
                }}
                \\ &=\Cost{\jaxdelta[\theta\cap\linFV{e_1}]{e_1}}+\Cost{\linearize{\sigma_1}}
            \end{align*}
        
        Observe that we do not count the workload associated to the primal output of $e_1$.\\
        By Lemma~\ref{lemma:cost_transl_type} we have that $\CostJAX{\sigma_1}=\Cost{\linearize{\sigma_1}}$ and we can conclude.

        \item Case $e=\literalJAX$:
        \begin{align*}
            \CostJAX{e}&=\CostJAX{\literalJAX}=1\\
            \Cost{\jaxdelta[\theta]{e}}&=
            \Cost{\tradexp{\oc\realterm}}=0
        \end{align*} 

        \item Case $e=\linzeroJAX{\sigmatypJAX}$:
        \begin{align*}
            \CostJAX{e}&=\CostJAX{\linzeroJAX{\sigmatypJAX}}=1+\CostJAX{\sigmatypJAX}\\
            \Cost{\jaxdelta[\theta]{e}}&=
            \Cost{\tradlin{\typtop}{\underline{0}_{\linearize{\sigmatypJAX}}}}=0
        \end{align*}

        \item All the other cases are simple and direct.
    \end{itemize}
\end{proof}

It is worth noting that the cost of evaluating an expression is preserved by AD transformations in Autodiff.

\subsection{Translation $\SymbTransB$}\label{app:translDeltaB}
We can simplify the definition of $\SymbTransA$ on the fragment \JaxB, by taking advantage of its three-sorted grammar.
The three-sorted grammar of \JaxB{} identifies a class of ``purely primal expressions'' $\expexprJAX{e}$ in the subgrammar \eqref{linear_B:primal} and a class of ``purely tangent expressions'' $\linexprJAX{e}$ in the subgrammar \eqref{linear_B:tanget} and then the mixing  $d$  of the two in the grammar \eqref{linear_B}. One can take advantage of this structure in order to define a translation $\SymbTransB$ of  \JaxB{} into $\lambdaLL$  that is more lightweight than the one provided by $\SymbTransA$. Specifically, $\SymbTransB$ may be obtained by extracting a single component from the output pair produced by $\SymbTransA$.

Figure~\ref{fig:B_toLL} defines the translation $\TransB[\theta]{d}$ of a \JaxB{} expression $d$, on the top of the definitions of $\TransB{\expexprJAX{e}}$, given a primal expression $\expexprJAX{e}$ (Figure~\ref{fig:JAXPrimal_to_LL}) and of $\TransB[\theta]{\linexprJAX{e}}$, given a tangent expression $\linexprJAX{e}$ (Figure~\ref{fig:JAXTangent_to_LL}). Notice that $\TransB{\expexprJAX{e}}$ omits the index $\theta$ on $\expexprJAX{e}$ as primal expressions have no free tangent variables. 
The definition of $\SymbTransB$ on purely primal expressions $\expexprJAX{e}$ is simple and just commutes with all operators  except on tuples where one has to manage exponentials (see Figure~\ref{fig:JAXPrimal_to_LL}).
The cases for the purely tangent $\dot e$ expression are more involved and depends on an enumeration $\theta$ of the free tangent variables, exactly as $\SymbTransA$ (see Figure~\ref{fig:JAXTangent_to_LL}) 

    \begin{figure}[p]
        \centering
        \small
        \begin{subfigure}{\textwidth}
            \vspace{-0.9cm}
            \begin{align*}
                \TransB{\expvarJAX{x}}\definedas \mbox{}
                    &  \oc x
                \\
                \TransB{\letterm{\expvarJAX{x}}{\expexprJAX{e_1}}{\expexprJAX{e_2}} } \definedas \mbox{}
                    & \letterm{\oc x}{\TransB{\expexprJAX{e_1}}}{\TransB{\expexprJAX{e_2}}}
                \\
                \TransB{{\emptytupleJAX}}\definedas \mbox{}
                    &\oc \emptytupleterm
                \\
                \TransB{\tupleJAX{\expexprJAX{e_1}}{\expexprJAX{e_2}}}\definedas \mbox{}
                    & \oc\tupleterm{\TransB{\expexprJAX{e_1}}}{\TransB{\expexprJAX{e_2}}}
                \\
                \TransB{\letterm{{\emptytupleJAX}}{\expvarJAX{z}}{\expexprJAX{e}}}\definedas \mbox{}
                    &\letterm{\emptytupleterm}{\expvarJAX{z}}{\TransB{\expexprJAX{e}}}
                \\
                \TransB{\letterm{\tupleJAX{\expvarJAX{x}_1}{\expvarJAX{x}_2}}{\expvarJAX{z}}{\expexprJAX{e}}}\definedas \mbox{}
                    & \letterm{\tupleterm{\oc x_1}{\oc x_2}}{z}{\TransB{\expexprJAX{e}}}
                \\
                \TransB{\literalJAX}\definedas \mbox{}
                    & \oc\realterm
                \\
                \TransB{\mathrm{f}(\expvarJAX{x}_1, \dots, \expvarJAX{x}_n)}\definedas \mbox{}
                    & \underline{f} (\oc x_1,\ldots,\oc x_n)
                \\
                \TransB{\dropJAX{\expexprJAX{e}}}\definedas \mbox{}
                    & \letterm{\oc x}{\TransB{\expexprJAX{e}}}{\oc \emptytupleterm}
            \end{align*} 
            \vspace{-6mm}
            \caption{
                Definition of $\TransB{\expexprJAX{e}}$, given an expression $\expexprJAX{e}$ in \eqref{linear_B:primal}. The enumeration $\theta$ of $\linFV{\expexprJAX{e}}$ is empty. 
            }
            \label{fig:JAXPrimal_to_LL}
        \end{subfigure}
        
        \begin{subfigure}{\textwidth}
            \begin{align*}
                \TransB[\theta]{\linvarJAX{x}}\definedas \mbox{}
                    &  \lambda y^{\with\linearize{\theta}}.{y}
                \\
                \TransB[\theta]{\letterm{\linvarJAX{y} }{\linexprJAX{e_1}}{\linexprJAX{e_2}}}\definedas \mbox{}
                    & \letterm{\affinebang f}{\affinebang\TransB[\theta\cap\linFV{e_1}] {\linexprJAX{e_1}}}{}\\
                    & \letterm{\affinebang g}{\affinebang\TransB[\linvarJAX{y},\theta\cap\linFV{e_2}] {\linexprJAX{e_2}}}{}\\
                    & \absterm{y^{\&\linearize{\theta}}}{
                        \letterm{\nTuple{y_1,y_2}}
                                {\Split{\linFV{e_1}}y}{
                                \appterm{\affderterm{g}}
                                {(\SplitInv[\&\linearize{\linvarJAX y,\theta\cap\linFV{e_2}}]{\dot{y}}
                                \nTuple{\appterm{\affderterm{f}}{y_1},y_2})}
                               }
                    } 
                \\
                \TransB[\theta]{\emptylintupleJAX}\definedas \mbox{}
                    & \lambda y^{\typtop}.{\linemptytupleterm}
                \\
                \TransB[\theta]{\lintupleJAX{e_1}{e_2}}\definedas \mbox{}
                    & \letterm{\affinebang f}{\affinebang\TransB[\theta\cap\linFV{e_1}] {\linexprJAX{e_1}}}{}\\
                    & \letterm{\affinebang g}{\affinebang\TransB[\theta\cap\linFV{e_2}] {\linexprJAX{e_2}}}{}\\ 
                    & \absterm{y^{\&\linearize{\theta}}}{
                        \letterm{\nTuple{y_1,y_2}}
                                {\Split{\linFV{e_1}}y}{
                                \nTuple{\affderterm{f} y_1,\affderterm{g} y_2}
                                }
                    }
                \\
                \TransB[\theta]{\letterm{\emptylintupleJAX}{\linvarJAX{z}}{e}}\definedas \mbox{}
                    & \letterm{\affinebang f}{\affinebang \TransB[\theta\setminus\linvarJAX{z}]{e}}{}\\
                    & \absterm{y^{\&\linearize{\theta}}}{
                        \letterm{\nTuple{z,y'}}
                                {\Split{\{\theta(\dot z)\}}y}{
                                    \affderterm{f} y'
                                }
                    }
                \\
                \TransB[\theta]{\letterm{\lintupleJAX{x_1}{x_2}}{\linvarJAX{z}}{e}}\definedas \mbox{}
                    & \letterm{\affinebang f}{\affinebang \TransB[\linvarJAX{x_1},\linvarJAX{x_2},\theta\setminus\linvarJAX{z}]{e}}{}\\
                    & \absterm{y^{\&\linearize{\theta}}}{
                        \letterm{\nTuple{\nTuple{x_1,x_2},y'}}{\Split{\{\theta(\dot z)\}}y}{
                        \appterm{\affderterm{f}}
                            {(\SplitInv[\&\linearize{\linvarJAX{x_1},\linvarJAX{x_2},\theta\setminus\linvarJAX{z}}]{\linvarJAX{x_1},\linvarJAX{x_2}} \nTuple{x_1,x_2,y'})}
                        }
                    }
                \\
                \TransB[\theta]{\dupJAX{x}}\definedas \mbox{}
                    &\lambda y^{\with\linearize{\theta}}.{\lintupleterm{y}{y}}
                \\
                \TransB[\theta]{\linzeroJAX{\sigmatypJAX}}\definedas \mbox{}
                    &\lambda y^{\typtop}.{0_{\linearize{\sigmatypJAX}}}
                \\
                \TransB[\theta]{\linsumJAX{x_1}{x_2}}\definedas \mbox{}
                    & \lambda y^{\with\linearize{\theta}}.{\sumterm
                    {y}}
                \\
                \TransB[\theta]{\linmultJAX{x}{y}}\definedas \mbox{}
                    & \lambda y^{\with\linearize{\theta}}.{\multterm 
                    {\tupleterm{\derterm x}{y}}}
                \\
                \TransB[\theta]{\dropJAX{\linexprJAX{e}}}\definedas \mbox{}
                    &\letterm{\affinebang f}{\affinebang\TransB[\theta]{ \linexprJAX{e}}}{}\\
                    &\lambda y^{\with\linearize{\theta}}.\letterm{z}{\appterm{f}{y}}{\linemptytupleterm}
            \end{align*}
            \vspace{-6mm}
            \caption{Definition of $\TransB[\theta]{\linexprJAX{e}}$, given an expression $\linexprJAX{e}$ in \ref{linear_B:tanget} and en enumeration $\theta$ of $\linFV{\linexprJAX{e}}$.}
            \label{fig:JAXTangent_to_LL}
        \end{subfigure}
        
        \begin{subfigure}{\textwidth}
            \begin{align*}
                \TransB[\theta]{\retexpJAX{\expexprJAX{e}}{\linexprJAX{e}}} \definedas  \mbox{}
                    &(\TransB{\expexprJAX{e}},\affinebang\TransB[\theta]{\linexprJAX{e}})
                \\
                        \TransB[\theta]{\letterm{\expvarJAX{x}}{\expexprJAX{e}}{d}} \definedas \mbox{}
                    & \letterm{\oc x}{\TransB{\expexprJAX{e}}}{\TransB[\theta]{d}}
                \\
                        \TransB[\theta]{\letterm{{\emptytupleJAX}}{\expvarJAX{z}}{d}}\definedas \mbox{}
                    &\letterm{\emptytupleterm}{z}{\TransB[\theta]{d}}
                \\
                \TransB[\theta]{\letterm{\tupleJAX{\expvarJAX{x}_1}{\expvarJAX{x}_2}}{\expvarJAX{z}}{d}}\definedas \mbox{}
                    & \letterm{\tupleterm{\oc x_1}{\oc x_2}}{z}{\TransB[\theta]{d}}
            \end{align*}
            \vspace{-5mm}
            \caption{Definition of $\TransB[\theta]{d}$, given an expression $d$ in \ref{linear_B} and en enumeration $\theta$ of $\linFV{d}$.}
        \end{subfigure} 
        \vspace{-7mm}
        \caption{Translation $\SymbTransB$ of \JaxB{} expressions.
        Note that in this case $\SplitInv[]{}$ can be the identity or neutrality, in case of identity we can omit it.}
        \label{fig:B_toLL}
    \end{figure}

\medskip    
The following proposition states the type of the translation $\SymbTransB$ and relates translation $\SymbTransA$ with translation $\SymbTransB$  

\begin{proposition}[Type $\SymbTransB$]
	\label{prop:transB}
    Given a $\JaxB$ expression of type 
    $\judgmentJAX{x_1:\tau_1,\dots, x_n:\tau_n}
    {\linvarJAX{y_1}:\sigmatypJAX_1,\dots, \linvarJAX{y_m}:\sigmatypJAX_m}
    {d}{\tautypJAX}{\sigmatypJAX}$ and an enumeration $\theta=(\linvarJAX{y_1}:\sigmatypJAX_1,\dots, \linvarJAX{y_m}:\sigmatypJAX_m)$ of the free tangent variables of $d$, then $\TransB[\theta]{d}\in\lambdaLL$ such that:
    \begin{small}
    \[
    	\judgment{\oc x_1:\oc \PrimalT(\tau_1),..., \oc x_n:\oc \PrimalT(\tau_n)}{\TransB[\theta]{d}}{\oc \PrimalT(\tautypJAX)\otimes\affbangterm(\typlollipop{\left(\&_{i=1}^m\linearize{\sigmatypJAX_i}\right)}
    {\linearize{ \sigmatypJAX}})}\,.
    \]
    \end{small}
   Moreover, $\TransA[\theta]{d}\sim\TransB[\theta]{d}$.
\end{proposition}

In order to prove the proposition above we need the following two auxiliary lemmas on the translation $\SymbTransB$ applied to~\eqref{linear_B:primal} and to~\eqref{linear_B:tanget}.

\begin{lemma}[Type Primal $\SymbTransB$] 
    \label{lemma:primal_transB}
    Given a Primal $\JaxB$ expression of type $\judgmentJAX{x_1:\tau_1, \dots, x_n:\tau_n}{}{\expexprJAX{e}}{\tautypJAX}{\typunitJAX}$, then $\TransB{\expexprJAX{e}}$
    is a well-typed $\lambdaLL$ term such that:
    $
        \judgment{\oc x_1:\oc \PrimalT(\tau_1), \dots, \oc x_n:\oc \PrimalT(\tau_n)}{\TransB{\expexprJAX{e}}}{\oc \PrimalT(\tautypJAX)}\,.
     $
    Moreover, 
	$\TransA[(\,)]{\expexprJAX{e}} \sim\tradexp{\TransB{\expexprJAX{e}}}$.
\end{lemma}
\begin{sproof}
    Notice that the definition of $\TransB{\expexprJAX{e}}$ on a primal expression $\expexprJAX{e}$ is basically the identity on almost all operators, but the proof of this lemma is not immediate as the left hand-side of the definition in Figure~\ref{fig:JAXPrimal_to_LL} uses JAX syntactical sugar, while on the right-hand side we have true $\lambdaLL$ terms. 
\end{sproof}

\begin{lemma}[Type Tangent $\SymbTransB$] 
    \label{lemma:tangent_transB}
    Given a Tangent \JaxB{} expression of type 
    $\judgmentJAX{x_1:\tau_1,\dots, x_n:\tau_n}
    {\linvarJAX{y_1}:\sigmatypJAX_1,\dots, \linvarJAX{y_m}:\sigmatypJAX_m}
    {\linexprJAX{e}}{\typunitJAX}{\sigmatypJAX}$ and an enumeration $\theta=(\linvarJAX{y_1}:\sigmatypJAX_1,\dots, \linvarJAX{y_m}:\sigmatypJAX_m)$ of the free tangent variables of $\linexprJAX{e}$, then $\TransB[\theta]{\linexprJAX{e}}$
    is a well-typed expression in $\lambdaLL$ such that:
    $
    	\judgment{\oc x_1:\oc \PrimalT(\tau_1),\dots, \oc x_n:\oc \PrimalT(\tau_n)}{\TransB[\theta]{\linexprJAX{e}}}{\affbangterm(\typlollipop{\left(\&_{i=1}^m\linearize{\sigmatypJAX_i}\right)}
    {\linearize{ \sigmatypJAX}})}\,.
    $
   Moreover, $\TransA[\theta]{\linexprJAX{e}}\sim\tupleterm{\oc \emptytupleterm}{\affinebang\TransB[\theta]{\linexprJAX{e}}}$.
\end{lemma} 
\begin{sproof}
    By induction on $\linexprJAX{e}$.
\end{sproof}

\medskip
Finally, the following lemma establishes a connection between $\SymbTransB$ and the stack $E$  of primal let-definitions employed in the definition of the unzipping transformation for Autodiff. This correspondence is instrumental in facilitating the proof of soundness for our unzipping transformation, which will be demonstrated in Section~\ref{sect:unzipping}.
\begin{lemma}
    \label{lemma:deltaBcontext}
    We have that $\TransB{E[(e^p;\dot e)]}=\TransB{E}[(\TransB{e^p},\TransB{\dot e})]$.
\end{lemma}
\begin{sproof}
    By immediate induction on $E[]$.  
\end{sproof}

The next two lemmas play a key role in the proof of the soundness theorem for the transpose transformation in our setting. The first lemma establishes a correspondence between the fusion expression in JAX and the fusion term in $\lambdaLL$, mediated by the translation $\SymbTransB$.

\begin{lemma}
    \label{lemma:TransB_fusion}
    Given $\theta=(\dot{x}_1,\dots,\dot{x}_n)$ and two partitions $\theta_1$ and $\theta_2$ of $\theta$ such that $\dot{y}_1:\otimes\theta_1$ and $\dot{y}_2:\otimes\theta_2$, then 
    $\TransB[\dot{y}_1,\dot{y}_2]{\SplitInvJAX[\theta]{\dot{y}_1,\dot{y}_2}}\sim \SplitInv[\theta]{\theta_1}$
\end{lemma}

Furthermore, since the transpose transformation in Autodiff is defined using the syntactic sugar introduced for Linear B, while $\SymbTransB$ operates directly on the core grammar of Linear B (excluding syntactic sugar), it is important to establish a correspondence between these two formulations. This relationship is formalized in the following lemma.

\begin{lemma}
    \label{lemma:TransB_synt_sugar}
    We have the following:
    \begin{enumerate}
        \item Given two tangent JAX expressions $\judgmentJAX{\Gamma_1}{\dot{\Gamma}_1}{\dot{e}_1}{\typone}{\tau_1 \dot{\otimes}\sigma_1}$ and $\judgmentJAX{\Gamma_2}{\dot{\Gamma}_2,\dot{x}_1:\tau_1,\dot{x}_2:\sigma_1}{\dot{e}_2}{\typone}{\tau_2 \dot{\otimes}\sigma_2}$, and $\theta$ is an enumeration of $\dot{\Gamma}_1,\dot{\Gamma}_2$, then 
        \begin{align*}
            \TransB[\theta]{\letterm{\lintupleJAX{x_1}{x_2}}{\dot{e}_1}{\dot{e}_2}}
            \sim\mbox{ }
            \begin{aligned}
                &\letterm{\affinebang f_1}
                {\affinebang \TransB[\theta\cap\linFV{\dot{e}_1}]{\dot{e}_1}}{}\\
                &\letterm{\affinebang f_2}
                {\affinebang \TransB[\dot{x}_1,\dot{x}_2,\theta\cap\linFV{\dot{e}_2}]{\dot{e}_2}}{}\\
                &\lambda u^{\&\linearize{\theta}}. 
                \letterm{\nTuple{u_1,u_2}}{\Split[\theta]{\linFV{\dot{e}_1}} u}{f_2 
                (\SplitInv[\theta]{(\dot{x}_1,\dot{x}_2)} \nTuple{f_1 u_1, u_2})}
            \end{aligned}
        \end{align*}
        where $\affinebang f_1$ and $\affinebang f_2$ have the following types:
        \begin{align*}
            \affinebang f_1 &: \affinebang(\&\linearize{\theta\cap\linFV{\dot{e}_1}}\multimap\linearize{\tau_1}\&\linearize{\sigma_1})\\
            \affinebang f_2 &: \affinebang(\&\linearize{\dot{x}_1,\dot{x}_2,\theta\cap\linFV{\dot{e}_2}}\multimap\linearize{\tau_2}\&\linearize{\sigma_2})
        \end{align*}

        \item  Given two tangent JAX expressions $\judgmentJAX{\Gamma_1}{\dot{\Gamma}_1}{\dot{e}_1}{\typone}{\tau_1 \dot{\otimes}\sigma_1}$ and $\judgmentJAX{\Gamma_2}{\dot{\Gamma}_2}{\dot{e}_2}{\typone}{\tau_2 \dot{\otimes}\sigma_2}$, and $\theta$ is an enumeration of $\dot{\Gamma}_1,\dot{\Gamma}_2$, then 
        \begin{align*}
            \TransB[\theta]{\lintupleJAX{e_1}{e_2}}
            \sim\mbox{ }
            \begin{aligned}
                &\letterm{\affinebang f_1}
                {\affinebang \TransB[\theta\cap\linFV{\dot{e}_1}]{\dot{e}_1}}{}\\
                &\letterm{\affinebang f_2}
                {\affinebang \TransB[\theta\cap\linFV{\dot{e}_2}]{\dot{e}_2}}{}\\
                &\lambda u^{\&\linearize{\theta}}. 
                \letterm{\nTuple{u_1,u_2}}{\Split[\theta]{\linFV{\dot{e}_1}} u}
                {\nTuple{f_1 u_1, f_2 u_2}}
            \end{aligned}
        \end{align*}
        where $\affinebang f_1$ and $\affinebang f_2$ have the following types:
        \begin{align*}
            \affinebang f_1 &: \affinebang(\&\linearize{\theta\cap\linFV{\dot{e}_1}}\multimap\linearize{\tau_1}\&\linearize{\sigma_1})\\
            \affinebang f_2 &: \affinebang(\&\linearize{\theta\cap\linFV{\dot{e}_2}}\multimap\linearize{\tau_2}\&\linearize{\sigma_2})
        \end{align*}
    \end{enumerate}
\end{lemma} 

\medbreak
Similarly to $\SymbTransA$, soundness can be proved for $\SymbTransB$. More precisely, soundness of $\SymbTransB$ follows as a corollary from Proposition~\ref{prop:transB} and the soundness of $\SymbTransA$ (Proposition~\ref{prop:sound_transA}).

\begin{corollary}[Soundness of $\SymbTransB$]
	\label{prop:sound_transB}
    Given a \JaxB{} expression of type $\Gamma;\dot\Sigma\vdash^{\text{\tiny{Jax}}} d:(\tau;\sigma)$, an enumeration $\theta$ of the tangent variables in $\dot\Sigma$, then:
    \begin{itemize}
    \item for every numeral sequence $\SeqA \text{ for }\Gamma$:
    $
    	\TransB[\theta]{d}[\oc \SeqA/\PrimalT(\Gamma)] \rightarrow^* (\oc \SemP[\SeqA]{d}, \affbangterm F)
    $,
    \item and moreover, for every numeral sequence $\SeqB$ for the type $\&\theta$: 
    $
    	F\SeqB \rightarrow^* \SemT[\SeqA;\SeqB]{e}
    $.
    \end{itemize} 
\end{corollary}

\medbreak
Similarly to Proposition~\ref{prop:size_TransA}, one can check the workload preservation property for $\SymbTransB$ as in Proposition~\ref{prop:size_TransB} by first proving the workload preservation of $\SymbTransB$ on Primal and Tangent.

\begin{lemma}[Workload Primal $\SymbTransB$] 
    \label{lemma:cost_primal_transB}
    Given a Primal $\JaxB$ expression of type $\judgmentJAX{x_1:\tau_1, \dots, x_n:\tau_n}{}{\expexprJAX{e}}{\tautypJAX}{\typunitJAX}$, then $\TransB{\expexprJAX{e}}$  is safe and $\TransB{\expexprJAX{e}}\leq \CostJAX{\expexprJAX{e}} $.
\end{lemma}
\begin{sproof}
    By induction on $\linexprJAX{e}$.
\end{sproof}

\begin{lemma}[Workload Tangent $\SymbTransB$] 
    \label{lemma:cost_tangent_transB}
    Given a Tangent \JaxB{} expression of type 
    $\judgmentJAX{x_1:\tau_1,\dots, x_n:\tau_n}
    {\linvarJAX{y_1}:\sigmatypJAX_1,\dots, \linvarJAX{y_m}:\sigmatypJAX_m}
    {\linexprJAX{e}}{\typunitJAX}{\sigmatypJAX}$ and an enumeration $\theta=(\linvarJAX{y_1}:\sigmatypJAX_1,\dots, \linvarJAX{y_m}:\sigmatypJAX_m)$ of the free tangent variables of $\linexprJAX{e}$, then $\TransB[\theta]{\linexprJAX{e}}$ is safe and $\TransB{\expexprJAX{e}}\leq \CostJAX{\expexprJAX{e}}$.
\end{lemma}
\begin{sproof}
    By induction on $\expexprJAX{e}$.
\end{sproof}

\begin{proposition}[Workload $\SymbTransB$]
	\label{prop:size_TransB}
    Given a \JaxB{} expression of type $\Gamma;\dot\Gamma\vdash^{\text{\tiny{Jax}}} d:(\tau;\sigma)$, an enumeration $\theta$ of the tangent variables in $\dot\Gamma$, then $\TransB[\theta] d$ is safe and $\Cost{\TransB[\theta] d}\leq \CostJAX{d} $.
\end{proposition}
\begin{sproof}
    The safeness of $\TransB[\theta]{d}$ is easy to prove by induction on $d$ simply checking the items in Definition~\ref{def:safe_term}. Let us focus on the proof related to workload preservation of $\TransB[\theta]{d}$, we proceed by induction on $d$ and we use Lemma~\ref{lemma:cost_primal_transB} and Lemma~\ref{lemma:cost_tangent_transB} in the case $d=\retexpJAX{\expexprJAX{e}}{\linexprJAX{e}}$.
\end{sproof}

\section{Forward}\label{app:forward}
\subsection{Soundness Forward}
The forward transformation in $\lambdaLL$ is proved to be sound (Theorem~\ref{th:forward_soundness}), by means of the following auxiliary lemmas

\begin{lemma}\label{lemma:addlemFW}
    Given $\judgment{\Gamma,u:L}{M}{H}$ for some $\&$-sequence types $L$ and $H$, then $\absterm{u}{M}\sim_{\Gamma,\typlollipop{L}{H}} \absterm{p}{M\{p/u\}}$ where $\FV{p}\cap\FV{M}=\emptyset$.
\end{lemma}
\begin{sproof}
     The proof shows that replacing a variable $u$ with a fresh pattern  $p$ in a function $\absterm{u}.M$ yields a logically equivalent function $\absterm{p}{M\{p/u\}}$, assuming $p$ doesn't capture variables in $M$.
    By the definition of the logical relation on open terms, we reduce to comparing applications of these functions to logically related arguments. After $\beta$-reduction and substitution, both sides evaluate to versions of $M$ with logically related values substituted for $u$. If $p$ is a tuple pattern, we handle it inductively by decomposing the values and applying the relation component-wise.
    Hence, both abstractions behave identically under logical equivalence.
\end{sproof}

More precisely, the following lemma $\sim$-relates the $\SymbTransA$ translation of $\dupJAX{u}$ with the additive contraction of $\lambdaLL$ terms, it will be useful to prove the soundness of our forward mode.

\begin{lemma} \label{lemma:dup_delta_a}
    Let $\Gamma;\dot{u_1}:\tau, \dot{u_2}:\tau,\dot\Gamma\vdash^{\text{\tiny{Jax}}} e:(\sigma;\sigma')$, and let $\theta$ be an enumeration of the tangent variables in $\dot u:\tau, \dot\Gamma$, then:
    \begin{align*}
        \SymbTransA_{\theta}
        \left(
            \begin{aligned}
                &\letterm{\dot a}{\dupJAX{u}}{}\\
                &\letterm{\lintupleJAX{u_1}{u_2}}{\dot a}{e}	
            \end{aligned}
        \right) 
        \sim
        \begin{aligned}
            &\letterm{(\oc x,\affinebang f)}
            {\TransA[\dot{u_1},\dot{u_2},\theta\setminus \dot u]{e}}{}\\
            &
            \left(
                \oc x,
                \affbangterm{\left(\absterm{{y^{\&\linearize{\theta}}}}{\quad
                    \begin{aligned}
                        &\letterm{\nTuple{\nTuple{u_1,u_2},y'}}{\Split{\{\theta(\dot u)\}} y}{}\\
                        &\appterm{\affderterm{f}}
                        {(\SplitInv[\&\linearize{\linvarJAX{u_1},\linvarJAX{u_2},\theta\setminus\linvarJAX{u}}]{\linvarJAX{u_1},\linvarJAX{u_2}} \nTuple{x_1,x_2,y'})}
                    \end{aligned}
                }\quad\right)}
            \right)
            \\
        \end{aligned} 
    \end{align*} 
\end{lemma}
\begin{sproof}
	By applying the definition of $\SymbTransA$ and by using Lemma~\ref{lemma:let_commutation} and Proposition~\ref{prop:sim_congruence}.
\end{sproof}

Finally, we prove the soundness property for forward mode as follows 

\begin{proof}[Proof of Theorem~\ref{th:forward_soundness}]
	 We proceed by induction on $\expexprJAX{e}$. The only delicate case is $\expexprJAX{e}=\letterm{\expvarJAX{x}}{\expexprJAX{e}_1}{\expexprJAX{e_2}}$. For the sake of simplicity we assume that $\expexprJAX{e}_1$ and $\expexprJAX{e}_2$ only share one primal variable, denoted by $z$.
        \begin{align*}
            \forward{\contextFM}{\TransB{\expexprJAX{e}}}
            =&\mbox{ } \forward{\contextFM}{\TransB{\letterm{\expvarJAX{x}}{\expexprJAX{e}_1}{\expexprJAX{e_2}}}}\\[3mm]
            =&\mbox{ }\forward{\contextFM}{\letterm{\oc x}{\TransB{\expexprJAX{e}_1}}{\TransB{\expexprJAX{e_2}}}} 
            \approx \forward{\contextFM}{(\absterm{\oc x}{\TransB{\expexprJAX{e_2}}})\TransB{\expexprJAX{e_1}}}\\[3mm]
            =&\mbox{ } 
            \begin{aligned}
                &\letterm{\tupleterm{\oc x}{\affinebang f}}{\forward{\contextFM\cap\FV{\TransB{\expexprJAX{e_1}}}}{\TransB{\expexprJAX{e_1}}}}{}\\
                & \letterm{\tupleterm{\oc y}{\affinebang g}}{\forward{x,\contextFM\cap\FV{\TransB{\expexprJAX{e_2}}}}{\TransB{\expexprJAX{e_2}}}}{}\\
                & \left( \oc y,
                    \affbangterm{\left(\absterm{u^{\with\linearize\theta}}{\quad
                        \begin{aligned}
                            & \letterm{\nTuple{u_{PQ},u'}}
                            {\Split{\FV{\TransB{\expexprJAX{e_1}}}\cap\FV{\TransB{\expexprJAX{e_2}}}}u}{}\\
                            &\letterm{\nTuple{u_P, u_Q}}{
                            \Split{(\FV{\TransB{\expexprJAX{e_1}}}\setminus\{x\})\setminus\FV{\TransB{\expexprJAX{e_2}}}} u'}{}\\ 
                            &\affderterm g
                            \nTuple{
                                \affderterm f{\nTuple{u_{PQ},u_Q}}, u_{PQ},u_P
                            }
                        \end{aligned} 
                    \quad
                }\right)}\right)
            \end{aligned}\\[3mm]
            =&\mbox{ } 
            \begin{aligned}
                &\letterm{\tupleterm{\oc x}{\affinebang f}}{\forward{\contextFM\cap\FV{\TransB{\expexprJAX{e_1}}}}{\TransB{\expexprJAX{e_1}}}}{}\\
                & \letterm{\tupleterm{\oc y}{\affinebang g}}{\forward{x,\contextFM\cap\FV{\TransB{\expexprJAX{e_2}}}}{\TransB{\expexprJAX{e_2}}}}{}\\
                & \left(\oc y,
                    \affbangterm{\left(\absterm{u^{\with\linearize\theta}}{\quad
                    \myhighlight{SpringGreen}{
                        \begin{aligned}
                            & \letterm{\nTuple{u_{PQ},u'}}
                            {\Split{\FV{\expexprJAX{e_1}}\cap\FV{\expexprJAX{e_2}}}u}{}\\
                            &\letterm{\nTuple{u_P, u_Q}}{
                            \Split{(\FV{\expexprJAX{e_1}}\setminus\{x\})\setminus\FV{\expexprJAX{e_2}}} u'}{}\\ 
                            &\affderterm g
                            \nTuple{
                                \affderterm f{\nTuple{u_{PQ},u_Q}}, u_{PQ},u_P
                            }
                        \end{aligned} 
                    } \quad
                }\right)}\right)\\
                & \hspace{6cm}\textcolor{SpringGreen}{H}
            \end{aligned}
        \end{align*}
        where the last line is obtained by observing that $\FV{\TransB{\expexprJAX{e}}}=\FV{ \expexprJAX{e}}$ which is an immediate consequence of Lemma~\ref{lemma:primal_transB}.

        \begin{align*} 
            \TransA[\theta']{\forwardJAX{\contextFMJAX}{\expexprJAX{e}}}
            =&\mbox{ } \TransA[\theta']{\forwardJAX{\contextFMJAX}{\letterm{\expvarJAX{x}}{\expexprJAX{e}_1}{\expexprJAX{e_2}}}}\\[3mm]
            \sim&\quad
            \begin{aligned}
                &\letterm{\tupleterm{\oc x}{\affinebang f}}
                {\TransA[\theta'\cap\linFV{\forwardJAX{\contextFMJAX_1,\{z\mapsto \dot w_1\}}{\expexprJAX{e}_1}}]{\forwardJAX{\contextFMJAX_1,\{z\mapsto \dot w_1\}}{\expexprJAX{e}_1}}}{}\\
                &\letterm{\tupleterm{\oc y}{\affinebang g}}
                {\TransA[\theta'\cap\linFV{\forwardJAX{\contextFMJAX_2,\{z\mapsto \dot w_2\}}{\expexprJAX{e}_2}}]{\forwardJAX{\contextFMJAX_2,\{z\mapsto \dot w_2\}}{\expexprJAX{e}_2}}}{}\\
                &\tupleterm{\oc y}{
                    \affbangterm{(\absterm{{y^{\&\linearize{\theta'}}}}{
                    \myhighlight{Apricot}{
                        \letterm{\nTuple{y_1,y_2}}{\Split[\&\linearize{\theta}]{\theta''\cap\linFV{\forwardJAX{\contextFMJAX_1,\{z\mapsto \dot w_1\}}{\expexprJAX{e}_1}}} y}
                        {
                            \appterm{\affderterm{g}}{\nTuple{\appterm{\affderterm{f}}{y_1},y_2}}
                        }
                    }
                })}}\\
                & \hspace{6cm}\textcolor{Apricot}{H'}
            \end{aligned}
        \end{align*} 
        where the last line is obtained by applying Lemma~\ref{lemma:dup_delta_a} and some $\beta$-steps.

        We can observe that $\contextFMJAX(\FV{e})=\linFV{e}$ and then by inductive hypotheses we have that 
        $$\forward{\contextFM\cap\FV{\TransB{\expexprJAX{e_i}}}}{\TransB{\expexprJAX{e_i}}} 
        \sim
        \TransA[\theta'\cap\linFV{\forwardJAX{\contextFMJAX_i,\{z\mapsto \dot w_i\}}{\expexprJAX{e}_i}}]{\forwardJAX{\contextFMJAX_i,\{z\mapsto \dot w_i\}}{\expexprJAX{e}_i}} 
        \qquad  
        \text{with $i\in\{1,2\}$.}$$

        In order to conclude the proof we have to show that
        $ \absterm{u^{\with\linearize\theta}}{\textcolor{SpringGreen}{H}}\sim \absterm{{y^{\&\linearize{\theta'}}}}{\textcolor{Apricot}{H'}}$.
        Let $p$ be a complete pattern of type $\&\linearize{\theta}=\&\linearize{\theta'}$, then by Lemma~\ref{lemma:addlemFW} we have:
        \begin{align*}
            \absterm{u^{\with\linearize\theta}}{\textcolor{SpringGreen}{H}} 
            &\sim \lambda p.\textcolor{SpringGreen}{H}\{p/u\}\\
            \absterm{u^{\with\linearize\theta'}}{\textcolor{Apricot}{H'}} 
            &\sim \lambda p.\textcolor{Apricot}{H'}\{p/u\}
        \end{align*}
       so it is easy to see that 
       $\lambda p.\textcolor{SpringGreen}{H}\{p/u\} 
       \Rightarrow_{\beta}
       \lambda p.\textcolor{Apricot}{H'}\{p/u\}$
       and we can conclude.

\end{proof}

\subsection{Work Preservation Forward}
Theorem~\ref{th:cost_forward} show that our forward transformation is work preserving up to a constant factor. In fact, $\forward{\contextFM}{P}$ introduces a constant number of numerical operations in case of $P$ numeric function.  

\begin{proof}[Proof Theorem~\ref{th:cost_forward}]
    We proceed by induction on $P$.
The part of the statement related to the safeness of $\forward{\theta}{P}$ is easy to prove by induction on $P$ simply checking the items in Definition~\ref{def:safe_term}. In contrast, the work preservation aspect of the statement requires a more careful analysis. 
Let us consider the two most interesting cases:

    \begin{itemize}
        \item Case $P=(\absterm{\oc x}{Q_1}){Q_2}$:
        \begin{align*}
            \Cost{P}&=\Cost{(\absterm{\oc x}{Q_1}){Q_2}}\\
            &=\Cost{\absterm{\oc x}{Q_1}}+\Cost{Q_2}\\
            &=\Cost{Q_1}+\Cost{Q_2}
        \end{align*}
        where the last line follows from the observation that $\oc x$ is an exponential pattern, meaning that all occurrences of $\typR$ within its type appear under the scope of a bang modality ($!$) and are therefore excluded from the workload calculation.
        \begin{align*}
            \Cost{\forward{\theta}{P}}&=\Cost{\forward{\theta}{(\absterm{\oc x}{Q_1}){Q_2}}} \\[2mm]
            &=\mathcal{W} 
            \left(
                \begin{aligned}
                    &\letterm{(\oc x, \affinebang f)}{\forward{\theta\cap\FV{Q_2}}{Q_2}}{}\\
                    &\letterm{(\oc y, \affinebang g)}{\forward{\FV{\oc x},\theta\cap\FV{Q_1}}{Q_1}}{}\\
                    &( \oc y,
                    \affbangterm{(\absterm{u^{\with\linearize\theta}}{ 
                            \letterm
                            {\nTuple{u_{1,2},u_1 ,u_2}}
                            {D_{Q_1,Q_2,\oc x} \mbox{ }u}
                            \affderterm g
                            \nTuple{
                                \affderterm f \nTuple{u_{1,2},u_1}, 
                                u_{1,2},
                                u_2
                            } 
                    })})
                \end{aligned}
            \right)
        \end{align*}
        Observe that our workload accounts for the number of numerical operations not occurring under a $!$ modality, as well as the numerals potentially erased during reduction. In this case, no erasure occurs; consequently, the workload contributions from the two let-constructs and the $\lambda$-abstraction are equal to zero. Therefore, it follows that $\Cost{\forward{\theta}{P}}$ is equal to $\Cost{\forward{\theta\cap\FV{Q_2}}{Q_2}}+\Cost{\forward{\FV{\oc x},\theta\cap\FV{Q_1}}{Q_1}}$.
        
        We conclude by inductive hypotheses and by taking $c=1$.

        \item Case $P=\funterm{(\oc x_1,\oc x_2)}$:
        \begin{align*}
            \Cost{\funterm{(\oc x_1,\oc x_2)}}&=1\\[3mm]
            \Cost{\Cost{\forward{\theta}{P}}}
            &=\Cost{\forward{(\oc x_1, \oc x_2)}{\funterm{(\oc x_1,\oc x_2)}}}\\[2mm]
            &=\mathcal{W} 
            \left(
                \begin{aligned}
                    & \letterm{\oc y_1}{\underline{\partial_1 f} (\oc x_1,\oc x_2)}{}\\
                    &\letterm{\oc y_2}{\underline{\partial_2 f} (\oc x_1,\oc x_2)}{}\\
                    &\tupleterm{\funterm{(\oc x_1,\oc x_2)}}{{\affbangterm{(\absterm{\nTuple{u_1,u_2} 
                    }(y_1 \dot{*} u_1)\dot{+}(y_2 \dot{*}(u_2)))}}}
                \end{aligned}
            \right)\\[2mm]
            &=\sum_{i=1}^{2} \Cost{\underline{\partial_i f} (\oc x_1,\oc x_2)} 
            + \Cost{\funterm{(\oc x_1,\oc x_2)}}
            + \Cost{(y_1 \dot{*} u_1)\dot{+}(y_2 \dot{*}(u_2))}\\
            &=2+1+3 = 6
        \end{align*}
        We can conclude by taking $c=6$.

        Observe that this is the only case in which we use $c>1$.

        Moreover, it is interesting to details also the case of $n$-ary function $\funterm{(\oc x_1,\ldots,\oc x_n)}$ for which we have to suppose that the maximal arity of numeric function primitive of $\lambdaLL$ is bounded by a constant $b$.
        We have to fix this constant because in that case we have:
        \begin{align*}
            \Cost{\forward{(\oc x_1, \ldots , \oc x_n)}{&\funterm{(\oc x_1, \ldots , \oc x_n)}}}\\
            &=\sum_{i=1}^{n} \Cost{\underline{\partial_i f} (\oc x_1,\oc x_2)} 
            + \Cost{\funterm{(\oc x_1, \ldots , \oc x_n)}}
            +n-1
            + \sum_{i=1}^{n} \Cost{y_1 \dot{*} u_1 }\\
            &=n+1+n-1+n = 3n
        \end{align*}
        where $n$ is the arity of the numerical function $\funterm$ and the cost $n-1$ is for the binary sums performed by the forward transformation. 
        We take $c>3b$ and we can conclude as $b$, unlike $n$, does not depend on the term but it is fixed once for the language.
    \end{itemize}
    \vspace{-0.4cm}
\end{proof}

\clearpage
\section{Unzipping} \label{app:unzip}
\subsection{Soundness Unzipping} 
The unzipping transformation in $\lambdaLL$ is shown to be sound by proving that it commutes with the $\delta$ translation, modulo the equivalence relation $\sim$.

\begin{proof}[Proof of Theorem~\ref{th:unzipping_soundness}]
    The equivalence 
    $\TransB[\theta]{\unzippingJAX{e}}\sim\TransA[\theta]{\unzippingJAX{e}}$ is a consequence of Proposition~\ref{prop:transB}. The first equivalence is proven by induction on $e$. Let us consider the two most delicate cases.

    First, let $e=\letterm{\retJAX{x}{y}}{e_1}{e_2}$, so that 
        \[
            \jaxdelta[\theta]{e} \definedas 
            \letterm{\tupleterm{\oc x}{\affinebang f}}{\jaxdelta[\theta\cap\linFV{e_1}]{e_1}}{
            \letterm{\tupleterm{\oc z}{\affinebang g}}{\jaxdelta[\linvarJAX y,\theta\cap\linFV{e_2}]{e_2}}{
            \tupleterm{\oc z}{F_{gf}}}}
    \]
    where $F_{gf}=\affbangterm{(\absterm{{y^{\&\linearize{\theta}}}}{
        \letterm{\nTuple{y_1,y_2}}{\Split[\&\linearize{\theta}]{\linFV{e_1}} y}
        {
            \appterm{\affderterm{g}}{\nTuple{\appterm{\affderterm{f}}{y_1},y_2}}
        }
    })}$.
    By induction hypothesis, we have that:
    $\unzipping{\jaxdelta[\theta\cap\linFV{e_i}]{e_i}} \sim
    \TransB[\theta\cap\linFV{e_i}]{\unzippingJAX{e_i}}$, for $i\in\{1,2\}$.
    
    Let us write:
    $\unzippingJAXPre{e_i} \definedas  (E_i[\,], e_i^p, \dot e_i)$
    and
    $\unzippingPre{\jaxdelta[\theta\cap\linFV{e_i}]{e_i}} \definedas  (\ExpContextA[i]{}, P_i, F_i)$.
    We have:
    \begin{align}
        &
        \unzipping{\TransA{e}}
        \\
        \label{eq:unzipping_before_eta} 
        &\definedas \ExpContextA[1]{
            \letterm{\oc x}{P_1}{
                \ExpContextA[2]{
                    (P_2,
                        \letterm{\affinebang f}{F_1}{
                            \letterm{\affinebang g}{F_2}{
                                F_{gf}
                            }
                        }
                    )
                    }
                } 
            }
        \\
        &
        \label{eq:unzip_comm_1}
        \sim
        \ExpContextA[1]{
            \letterm{\oc x}{P_1}{
	            \letterm{\affinebang f}{F_1}{
        		        \ExpContextA[2]{
		        		\letterm{\affinebang g}{F_2}{
                    			(P_2,F_{gf})
				}
                    	}
                	} 
            }
         }
        \\
        &
        \label{eq:unzip_after_beta}
        =_\beta
        \ExpContextA[1]{
            \letterm{(\oc x,\affinebang f)}{(P_1,F_1)}{
        		        \ExpContextA[2]{
		        		\letterm{(\oc z,\affinebang g)}{(P_2,F_2)}{
                    			(\oc z,F_{gf})
                    		}
                		} 
            }
        }
        \\
        &
        \label{eq:unzip_comm_2}
        \sim
            \letterm{(\oc x,\affinebang f)}{\ExpContextA[1]{(P_1,\affinebang F_1)}}{        		    
		        		\letterm{(\oc z,\affinebang g)}{\ExpContextA[2]{(P_2,\affinebang F_2)}}{
                    			(\oc z,F_{gf})
                    		}
                		} 
        \\
        \label{eq:unzip_before_IH}
        &
        =
        	\letterm{(\oc x,\affinebang f)}{
            \unzipping{\jaxdelta{e_1}}
        }{
            \letterm{(\oc z,\affinebang g)}{
                \unzipping{\jaxdelta{e_2}}
            }{
                (\oc z,F_{gf})
            }
        }
        \\
        \label{eq:unzip_after_IH}
        &\sim
        \letterm{(\oc x,\affinebang f)}{
            \TransB{\unzippingJAX{e_1}}
        }{
            \letterm{(\oc z,\affinebang g)}{
                \TransB{\unzippingJAX{e_2}}
            }{
                (\oc z,F_{gf})
            }
        }
        \\
        \label{eq:unzipping_lineTop_deltaBonContextE_i}
        &=
        \letterm{(\oc x,\affinebang f)}{
            \TransB{E_1[(e_1^p,\dot e_1)]}
        }{
            \letterm{(\oc z,\affinebang g)}{
                \TransB{E_2[(e_2^p,\dot e_2)]}
            }{
                (\oc z,F_{gf})
            }
        }
        \\
        \label{eq:unzipping_lineBottom_deltaBonContextE_i}
        &\sim
        \letterm{(\oc x,\affinebang f)}{
            \TransB{E_1}[(\TransB{e_1^p},\TransB{\dot e_1})]
        }{
            \letterm{(\oc z,\affinebang g)}{
                \TransB{E_1}[(\TransB{e_2^p},\TransB{\dot e_2})]
            }{
                (\oc z,F_{gf})
            }
        }
        \\
        &\sim
        \TransB{E_1}[
            \letterm{\oc x}{\TransB{e_1^p}}{
                \TransB{E_2}[
                    (
                        \TransB{e_2^p},
                        \letterm
                        {\affinebang f}
                        {\TransB{\dot e_1}}
                        {\letterm{\affinebang g}{\TransB{\dot e_2}}{F_{gf}}}
                    )
                ]
            }
        ]
        \\
        \label{eq:unzipping_lineTop_deltaBonContextE}
        &=
        \TransB{E_1}[
            \letterm{\oc x}{\TransB{e_1^p}}{
                \TransB{E_2}[
                    (
                        \TransB{e_2^p},
                        \TransB{
                            \letterm
                            {\dot y}
                            {\dot e_1}
                            {\dot e_2}
                        }
                    )
                ]
            }
        ]
        \\
        \label{eq:unzipping_lineBottom_deltaBonContextE}
        &=
        \TransB{E_1[
            \letterm{\oc x}{e_1^p}{
                E_2[
                    (
                        e_2^p,
                        \letterm{\dot y}{\dot e_1}{\dot e_2}
                    )
                ]
            }
        ]
        }
        \\
        &=
        \TransB[\theta]{\unzippingJAX{e}}
    \end{align}
    The passage from line~\eqref{eq:unzipping_before_eta} (resp.~\eqref{eq:unzip_after_beta}) to~\eqref{eq:unzip_comm_1}  (resp.\eqref{eq:unzip_comm_2}) uses Proposition~\ref{prop:typingLinearBLL} and Lemma~\ref{lemma:let_commutation}, and the line~\eqref{eq:unzip_before_IH} to~\eqref{eq:unzip_after_IH} is the induction hypothesis.
    At the end, the passage from line~\eqref{eq:unzipping_lineTop_deltaBonContextE_i}
    (resp.~\eqref{eq:unzipping_lineTop_deltaBonContextE}) 
    to~\eqref{eq:unzipping_lineBottom_deltaBonContextE_i}
    (resp.~\eqref{eq:unzipping_lineBottom_deltaBonContextE}) uses Lemma~\ref{lemma:deltaBcontext}.

    We detail also the case is $e=\dropJAX{e_1}$ too. 
    Suppose $\unzippingJAX{e_1}=E_1[(e_1^p,\dot e_1)]$ as well as $\unzipping{\TransA{e_1}}=\ExpContextA[1]{(P_1,\affinebang F_1)}$. We have:
    \begin{align}
        \unzipping{\TransA{e}} 
        & \definedas
        \unzipping{\letterm{(\oc x,\affinebang f)}{\TransA{e_1}}{(\oc \emptytupleterm, \affbangterm{\lambda y.\lambda y.\letterm{z}{fy}{\linemptytupleterm}})}}
        \\
        \label{eq:unzipping:drop_before_IH}
        &\rightarrow^* \unzipping{\letterm{(\oc x,\affinebang f)}{\TransA{e_1}}{(\oc \emptytupleterm, \affbangterm{\linemptytupleterm})}}
        \\
        &=
        \label{eq:unzipping:drop_before_let_comm}
        \ExpContextA[1]{\letterm{\oc x}{P_1}{(\oc\emptytupleterm,\letterm{\affinebang f}{F_1}{\affbangterm{\lambda y.\linemptytupleterm}})}}
        \\
        &\sim
        \label{eq:unzipping:drop_after_let_comm}
        \letterm
            {(\oc x,\affinebang f)}
            {\ExpContextA[1]{(P_1,F_1)}}
            {(\oc (\,),\affbangterm{\lambda y.\linemptytupleterm})}
        \\
        \label{eq:unzipping:drop_before_IH}
        &=
        \letterm
            {(\oc x,\affinebang f)}
            {\unzipping{\TransA{e_1}}}
            {(\oc (\,),\affbangterm{\lambda y.\linemptytupleterm})}
        \\
        \label{eq:unzipping:drop_after_IH}
        &\sim
        \letterm
            {(\oc x,\affinebang f)}
            {\TransB{\unzippingJAX{e_1}}}
            {(\oc (\,),\affbangterm{\lambda y.\linemptytupleterm})}
        \\
        \label{eq:unzipping:drop_before_trans_local1}
        &=
        \letterm
            {(\oc x,\affinebang f)}
            {\TransB{E_1[(e_1^p,\dot e_1)]}}
            {(\oc (\,),\affbangterm{\lambda y.\linemptytupleterm})}
        \\
        \label{eq:unzipping:drop_after_trans_local1}
        &=
        \letterm
            {(\oc x,\affinebang f)}
            {\TransB{E_1}[(\TransB{e_1^p},\TransB{\dot e_1})]}
            {(\oc (\,),\affbangterm{\lambda y.\linemptytupleterm})}
        \\
        &\sim
        \TransB{E_1}[(\letterm{\oc x}{\TransB{e_1^p}}{\oc (\,)};\letterm{\affinebang f}{\TransB{\dot e_1}}{\affbangterm{\lambda y.\linemptytupleterm}})]
        \\
        \label{eq:unzipping:drop_before_trans_local2}
        &=
        \TransB{E_1}[(\TransB{\dropJAX{e_1^p}};\TransB{\dropJAX{\dot e_1}})]
        \\
        \label{eq:unzipping:drop_after_trans_local2}
        &=
        \TransB{E_1[(\dropJAX{e_1^p};\dropJAX{\dot e_1})]}=\TransB{\unzippingJAX{e}}
    \end{align}
    where the passage from line~\eqref{eq:unzipping:drop_before_let_comm} to~\eqref{eq:unzipping:drop_after_let_comm} uses Proposition~\ref{prop:typingLinearBLL} and Lemma~\ref{lemma:let_commutation}, the passage from line~\eqref{eq:unzipping:drop_before_trans_local1}
    (resp.~\eqref{eq:unzipping:drop_before_trans_local2})
    to line 
    \eqref{eq:unzipping:drop_after_trans_local1}
    (resp.~\eqref{eq:unzipping:drop_after_trans_local2})
    is given  by Lemma~\ref{lemma:deltaBcontext}, and line~\eqref{eq:unzipping:drop_before_IH} to \eqref{eq:unzipping:drop_after_IH} is the induction hypothesis.
\end{proof}

\subsection{Work Preservation Unzipping} 
We also establish that the unzipping transformation in $\lambdaLL$ preserves the workload.

\begin{proof}[Proof of Theorem~\ref{th:cost_unzip}]
    The safeness of $\mathcal{U}$ is easy to prove by induction on $S$ simply checking the items in Definition~\ref{def:safe_term}.
Let us focus on the proof related to work preservation of $\mathcal{U}$, we proceed by induction on $S$. The only two delicate cases:

\begin{itemize}
        \item Case $S=(P,\affinebang F)$:\\
        Let $\unzippingPre{S}=([],P,F)$ and by definition $\unzipping{S}=[(P,\affinebang F)]=(P,\affinebang F)$, so in this case we have that $S=\unzipping{S}$ and we can conclude.
        
        \item Case $S=\letterm{\tupleterm{\oc x}{\affinebang f}}{S_1}{S_2}$:\\
        Let $\unzippingPre{S_i}=(\ExpContextA[i]{},P_i,F_i)$ and by definition $\unzipping{S_i}=\ExpContextA[i]{(P_i,\affinebang F_i)}$.

        By induction hypothesis on $S_i$ we have that
        $$\Cost{\unzipping{S_i}}=\Cost{\ExpContextA[i]{(P_i,\affinebang F_i)}}\leq \Cost{S_i}.$$

        Moreover, $\unzippingPre{S}=(\ExpContextA[1]{\letterm{\oc x}{P_1}{\ExpContextA[2]}},P_2,(\letterm{\affinebang f}{\affinebang F_1}{F_2}))$ so we have:
        \begin{equation}\label{eq:WUnzip_let}
            \begin{aligned}
                \unzipping{S}
                &\definedas \ExpContextA[1]{\letterm{\oc x}{P_1}{\ExpContextA[2]{(P_2,\letterm{\affinebang f}{\affinebang F_1}{\affinebang F_2})}}} \\
                &\hspace{-13mm}\overset{\tiny \text{Prop.~\ref{prop:typingLinearBLL} + Lemma~\ref{lemma:let_commutation}}}{\sim}
                \ExpContextA[1]{\letterm{\oc x}{P_1}{\letterm{\affinebang f}{\affinebang F_1}{{\ExpContextA[2]{(P_2,\affinebang F_2})}}}}\\
                &\hspace{-1mm}=_{\beta} 
                \ExpContextA[1]{\letterm{\tupleterm{\oc x}{\affinebang f}}{\tupleterm{P_1}{\affinebang F_1}}{{\ExpContextA[2]{(P_2,\affinebang F_2})}}}\\
                &\hspace{-13mm}\overset{\tiny \text{Prop.~\ref{prop:typingLinearBLL} + Lemma~\ref{lemma:let_commutation}}}{\sim}
                \letterm{\tupleterm{\oc x}{\affinebang f}}{\ExpContextA[1]{\tupleterm{P_1}{\affinebang F_1}}}{{\ExpContextA[2]{(P_2,\affinebang F_2})}}
            \end{aligned}
        \end{equation}
        and we can conclude as follows:
        \begin{align*}
            \Cost{\unzipping{S}}
            & \overset{\tiny \text{Eq.~\ref{eq:WUnzip_let}}}{=}
            \Cost{\letterm{\tupleterm{\oc x}{\affinebang f}}{\ExpContextA[1]{\tupleterm{P_1}{\affinebang F_1}}}{{\ExpContextA[2]{(P_2,\affinebang F_2})}}}\\
            &\hspace{3mm} \overset{\tiny \text{IHs}}{\leq}
            \Cost{\letterm{\tupleterm{\oc x}{\affinebang f}}{S_1}{S_2}} 
            =\Cost{S}
        \end{align*}
    \end{itemize}  
\end{proof}

Moreover, the lemma below follows directly form the work-preservation of the unzipping transformation (Theorem~\ref{th:cost_unzip}) and will be useful in the following section to prove that the transpose transformation is work preserving.
\begin{lemma}\label{lemma:cost_transp_UnzipPre}
	Given $R\in\lambdaLL^{\mathtt{A}}$ and $\unzippingPre{R}=(\ExpContextA[]{\,}, P, F)$, we have:
	$\Cost{\ExpContextA[]{P}}\leq\Cost{R}$.
\end{lemma}

\clearpage
\section{Transpose} \label{app:transpose} 
The following lemmas, concerning the properties of the renamings defined in Figure~\ref{figure:def_pattern_renaming} and the ``zero-parsimonious'' sum defined in Figure~\ref{subfigure:def_fusion_renaming}, are useful for gaining a clearer understanding of the transpose transformation and for demonstrating that it is work preserving.
\begin{lemma}
	\label{lemma:typing_weak}
	If $\Gamma, \PatAddA:L\vdash M:A$, then for every renaming $\Rename$ such that $\FV{M}\cap\FV{\PatAddA}\subseteq\Dom\Rename\subseteq\FV{\PatAddA}$, we have that: 
	\begin{enumerate}
		\item $\Gamma,\RenameTerm\Rename{\PatAddA}:L\vdash \RenameTerm\Rename M:A$, 
		\item $\Gamma,\RenamePat\Rename{\PatAddA}:L'\vdash \RenameTerm\Rename M:A$, where $L'$ is the type of $\RenamePat\Rename{\PatAddA}$,
		\item $\Cost{\RenameTerm\Rename M}=\Cost{M}$.
	\end{enumerate}
\end{lemma}
\begin{sproof}
    By induction on a derivation of $\Gamma, \PatAddA:L\vdash M:A$. 
    The condition $\Dom\Rename\subseteq\FV{\PatAddA}$ is necessary to avoid the renaming in $\RenameTerm\Rename M$ of variables in $\Gamma$. 
\end{sproof}

\begin{lemma}
	\label{lemma:cost_fusion_renaming}
	Let $\PatAddA:L$ be a pattern, $\Rename_1$ and $\Rename_2$ be two renamings with disjoint codomains and let $L_i$ be the type of $\RenamePat{\Rename_i}\PatAddA$.
	We have that:
	\[
		\lambda\nTuple{\RenamePat{\Rename_1}\PatAddA,\RenamePat{\Rename_2}\PatAddA}.\Fusion \PatAddA{\Rename_1}{\Rename_2}
	\]
	is a closed term of type $(L_1\with L_2)\multimap L$ and of workload $\Cost{\Dom{\Rename_1}\cap\Dom{\Rename_2}\cap\FV{\PatAddA}}$.
\end{lemma}

\begin{sproof}
	Notice that we are supposing that the fresh variables of type $\typtop$ introduced by $\RenamePat{\Rename_1}\PatAddA$ and $\RenamePat{\Rename_2}\PatAddA$ are pairwise different, so that the hypothesis of $\Codom{\Rename_1}\cap\Codom{\Rename_2}=\emptyset$ guarantees that $\nTuple{\RenamePat{\Rename_1}\PatAddA,\RenamePat{\Rename_2}\PatAddA}$ is a well-defined pattern, i.e. there are no different occurrences of the same variable. 
	We then prove by induction on $\PatAddA$ that:
    
	\begin{itemize}
	 \item $\nTuple{\RenamePat{\Rename_1}\PatAddA,\RenamePat{\Rename_2}\PatAddA}:L_1\with L_2 \vdash \Fusion \PatAddA{\Rename_1}{\Rename_2}:L$ is derivable,
	 \item any variable in $\RenamePat{\Rename_1}\PatAddA$, $\RenamePat{\Rename_2}\PatAddA$ of type different from $\typtop$ occurs free in $\Fusion \PatAddA{\Rename_1}{\Rename_2}$,
	 \item $\Cost{\Fusion \PatAddA{\Rename_1}{\Rename_2}}=\Cost{\Dom{\Rename_1}\cap\Dom{\Rename_2}\cap\FV{\PatAddA}}$
	\end{itemize}
\end{sproof}

\subsection{Example Transpose} \label{subsect:exTransp}
In this appendix we want to detail how we obtain the term in Figure~\ref{fig:ex_transpose} by applying the transpose transformation defined in Figure~\ref{fig:def_transpose} and some $\beta_\lambda$-simplifications for readability.
Moreover, we hope to make the definition of transpose more accessible by grounding it in a concrete step-by-step example.

According to the definition of transpose on $\lambdaLL^{\mathtt A}$ (Figure~\ref{subfigure:transpose on lambdaLL_a}), specifically the rule related to exponential let-definitions, the forward and tape computations (first four lines of Figure~\ref{fig:ex_forward}) remains unchanged in Figure~\ref{fig:ex_transpose}.

The core of the transformation resides in the application of the transpose to the {\color{blue}blue} part in Figure~\ref{fig:ex_forward}:

\begin{align*}
    \mathcal{T}_{\affcontext{\overleftarrow{\Phi_0}}}
    \quad \left(
    \begin{aligned}
        &\letterm{\affinebang f_1^{\mbox{ }\typR \multimap \typR}}
        {\affinebang (\lambda u. w_1\mbox{ }\dot{*}\mbox{ }u)}{}\\
        &\letterm{\affinebang f_2^{\mbox{ }(\typR\&\typR) \multimap \typR}}
        {\affinebang (\lambda \nTuple{u_1,u_2}. (w_2\mbox{ }\dot{*}\mbox{ }u_1)
        \mbox{ }\dot{+}\mbox{ } (w_3\mbox{ }\dot{*}\mbox{ }u_2))}{}\\
        &\letterm{\affinebang f_3^{\mbox{ }\typR \multimap \typR}}
        {\affinebang (\lambda u. w_4\mbox{ }\dot{*}\mbox{ }u)}{}\\
        &\letterm{\affinebang f_4^{\mbox{ }(\typR\&\typR) \multimap \typR}}
        {\affinebang (\lambda \nTuple{u_1,u_2}. (w_5\mbox{ }\dot{*}\mbox{ }u_1)
        \mbox{ }\dot{+}\mbox{ } (w_6\mbox{ }\dot{*}\mbox{ }u_2))}{}\\
        &\lambda u^{\typR\&\typR}.
        \begin{aligned}
            &\letterm{\nTuple{x',y'}}{u}{}
            f_4 \nTuple{f_2 \nTuple{f_1\mbox{ }x', y'}, f_3\mbox{ }x'}
        \end{aligned}
    \end{aligned}
    \quad \right)
    \qquad\quad
    \text{where $\affcontext{\overleftarrow{\Phi_0}}$ is empty.}
\end{align*}

First, let us focus on the transposition of the affine let-bindings, to which we apply the last rule of Figure~\ref{subfigure:transpose on lambdaLL_a}. More precisely, we examine the transposition of two of them in detail: one concerning a unary affine function, and the other a binary one.

We detail the transposition of the let-binding of $\affinebang f_1$ as follows:
\begin{align*}
    \mathcal{T}_{\affcontext{\overleftarrow{\Phi_0}}}(&
        \letterm{\affinebang f_1^{\mbox{ }\typR \multimap \typR}}
        {\affinebang (\lambda u. w_1\mbox{ }\dot{*}\mbox{ }u)}{S_1}
    ) 
    \quad \text{where $\affcontext{\overleftarrow{\Phi_0}}$ is empty}
    \\
    &\overset{\tiny \mathcal{T}(\letterm{\affinebang f}{\affinebang F}{S})}{=}
    \letterm{\affinebang \overleftarrow{f_1}}
    {\affinebang \mathcal{T}_{\affcontext{\overleftarrow{\Phi_0}}}(\lambda u. w_1\mbox{ }\dot{*}\mbox{ }u)}{\mathcal{T}_{\affcontext{\overleftarrow{\Phi_1}}}(S_1)} \quad \text{where $\affcontext{\overleftarrow{\Phi_1}}=\affinebang \overleftarrow{f_1}$}
    \\
    &\hspace{5mm}\overset{\tiny \mathcal{T}(\lambda \PatAddA.U)}{=}
    \letterm{\affinebang \overleftarrow{f_1}}
    {\affinebang \left(\lambda l.
    \letterm{\RenamePat{\alpha'}{u}}{\lintranspose{\affcontext{\overleftarrow{\Phi}},u}{w_1 \dot{*} u}}{\Fusion{u}{\alpha'}{\emptyset}}\right)}
    {\mathcal{T}_{\affcontext{\overleftarrow{\Phi_1}}}(S_1)}
    \\
    &\hspace{15mm} \text{where $\alpha'=(u\mapsto l')$ and we have that $\RenamePat{\alpha'}{u}=\alpha'(u)=l'$ and}\\
    &\hspace{15mm} \text{$\Fusion{u}{\alpha'}{\emptyset}=\alpha'(u)=l'$}
    \\
    &\hspace{10mm} =
    \letterm{\affinebang \overleftarrow{f_1}}
    {\affinebang \left(\lambda l. \letterm{l'}{\lintranspose{\affcontext{\overleftarrow{\Phi}},u}{w_1 \dot{*} u}}{l'} \right)}
    {\mathcal{T}_{\affcontext{\overleftarrow{\Phi_1}}}(S_1)}
    \\
    &\hspace{10mm}\approx
    \letterm{\affinebang \overleftarrow{f_1}}
    {\affinebang \left(\lambda l.\mbox{ }(\lambda l'.l')\mbox{ }\lintranspose{\affcontext{\overleftarrow{\Phi}},u}{w_1 \dot{*} u} \right)}
    {\mathcal{T}_{\affcontext{\overleftarrow{\Phi_1}}}(S_1)}
    \\
    &\hspace{8mm} \xrightarrow{\beta_\lambda}
    \letterm{\affinebang \overleftarrow{f_1}}
    {\affinebang  \left(\lambda l.  \lintranspose{\affcontext{\overleftarrow{\Phi}},u}{w_1 \dot{*} u}\right)}
    {\mathcal{T}_{\affcontext{\overleftarrow{\Phi_1}}}(S_1)}
    \\
    &\hspace{10mm}\approx
    \letterm{\affinebang \overleftarrow{f_1}}
    {\affinebang  \left(\lambda l.  \lintranspose{\affcontext{\overleftarrow{\Phi}},u}{(\dot{*}w_1) u} \right)}
    {\mathcal{T}_{\affcontext{\overleftarrow{\Phi_1}}}(S_1)}
    \\
    &\hspace{7mm}\overset{\tiny \mathcal{T}(FU')}{=}
    \letterm{\affinebang \overleftarrow{f_1}}
    {\affinebang  
    \left(
    \lambda l. 
    \left(\lambda u. \lintranspose{\affcontext{\overleftarrow{\Phi}},u}{u}\right)
    \left(\lintranspose{\affcontext{\overleftarrow{\Phi}}}{(\dot{*}w_1) l}\right)
    \right)}
    {\mathcal{T}_{\affcontext{\overleftarrow{\Phi_1}}}(S_1)}
    \\
    &\hspace{8mm}\overset{\tiny \mathcal{T}(u)}{=}
    \letterm{\affinebang \overleftarrow{f_1}}
    {\affinebang  
    \left(
    \lambda l. 
    \left(\lambda u.u \right)
    \left(\lintranspose{\affcontext{\overleftarrow{\Phi}}}{(\dot{*}w_1) l}\right)
    \right)}
    {\mathcal{T}_{\affcontext{\overleftarrow{\Phi_1}}}(S_1)}
    \\
    &\hspace{8mm}\overset{\tiny \mathcal{T}(\dot{*} x)}{=}
    \letterm{\affinebang \overleftarrow{f_1}}
    {\affinebang  
    \left(
    \lambda l. 
    \left( \lambda u.u \right)
    \left( (\dot{*}w_1) l \right)
    \right)}
    {\mathcal{T}_{\affcontext{\overleftarrow{\Phi_1}}}(S_1)}
    \\
    &\hspace{8mm} \xrightarrow{\beta_\lambda}
    \letterm{\affinebang \overleftarrow{f_1}}
    {\affinebang (\lambda l. (\dot{*}w_1) l)}
    {\mathcal{T}_{\affcontext{\overleftarrow{\Phi_1}}}(S_1)}
\end{align*}
Therefore, we have that 
\begin{equation}
    \label{eq:example_transpose_1}
    \mathcal{T}_{\affcontext{\overleftarrow{\Phi_0}}}(
        \letterm{\affinebang f_1^{\mbox{ }\typR \multimap \typR}}
        {\affinebang (\lambda u. w_1\mbox{ }\dot{*}\mbox{ }u)}{S_1}
    ) 
    \rightarrow^*
    \letterm{\affinebang \overleftarrow{f_1}}
    {\affinebang (\lambda l. (\dot{*}w_1) l)}
    {\mathcal{T}_{\affcontext{\overleftarrow{\Phi_1}}}(S_1)}
\end{equation}

We detail the transposition of the let-binding of $\affinebang f_2$ as follows:
\begin{align*}
    \mathcal{T}_{\affcontext{\overleftarrow{\Phi_1}}}(&
        \letterm{\affinebang f_2^{\mbox{ }(\typR\&\typR) \multimap \typR}}
        {\affinebang (\lambda \nTuple{u_1,u_2}. (w_2\mbox{ }\dot{*}\mbox{ }u_1)
        \mbox{ }\dot{+}\mbox{ } (w_3\mbox{ }\dot{*}\mbox{ }u_2))}{S_2}
    ) 
    \quad \text{where $\affcontext{\overleftarrow{\Phi_1}}=\affinebang \overleftarrow{f_1}$}
    \\ 
    &\overset{\tiny \mathcal{T}(\letterm{\affinebang f}{\affinebang F}{S})}{=}
    \letterm{\affinebang f_2}
    {\affinebang \mathcal{T}_{\affcontext{\overleftarrow{\Phi_1}}}(\lambda \nTuple{u_1,u_2}. (w_2\mbox{ }\dot{*}\mbox{ }u_1)\mbox{ }\dot{+}\mbox{ } (w_3\mbox{ }\dot{*}\mbox{ }u_2))}
    {\mathcal{T}_{\affcontext{\overleftarrow{\Phi_2}}}(S_2)}
    \\
    &\hspace{20mm}
    \text{where $\affcontext{\overleftarrow{\Phi_2}}=\affinebang \overleftarrow{f_1},\affinebang \overleftarrow{f_2}$}
    \\
    &\hspace{5mm}\overset{\tiny \mathcal{T}(\lambda \PatAddA.U)}{=}
    \letterm{\affinebang \overleftarrow{f_2}}
    {\affinebang \left(\lambda l.\left(\hspace{2mm}
    \begin{aligned}
        &\letterm{\RenamePat{\alpha''}{\nTuple{u_1,u_2}}}
        {\lintranspose{\affcontext{\overleftarrow{\Phi_1}},\nTuple{u_1,u_2}}{(w_2\mbox{ }\dot{*}\mbox{ }u_1)\mbox{ }\dot{+}\mbox{ } (w_3\mbox{ }\dot{*}\mbox{ }u_2)}}{}\\
        &\Fusion{\nTuple{u_1,u_2}}{\alpha''}{\emptyset}
    \end{aligned} 
    \hspace{2mm}\right)\right)}{}\\
    &\hspace{20mm}
    {\mathcal{T}_{\affcontext{\overleftarrow{\Phi_2}}}(S_2)}
    \\
    &\hspace{15mm} \text{where $\alpha''$ is the identity renaming on $\{u_1, u_2\}$ and we have that}
    \\
    &\hspace{15mm} \text{$\RenamePat{\alpha''}{\nTuple{u_1,u_2}}=\nTuple{\RenamePat{\alpha''}{u_1},\RenamePat{\alpha''}{u_2}}
    =\nTuple{\alpha''(u_1),\alpha''(u_2)}=\nTuple{u_1,u_2}$ and}
    \\
    &\hspace{15mm} \text{$\Fusion{\nTuple{u_1,u_2}}{\alpha''}{\emptyset}=\nTuple{\Fusion{u_1}{\alpha''}{\emptyset},\Fusion{u_2}{\alpha''}{\emptyset}}=\nTuple{\alpha''(u_1),\alpha''(u_2)}=\nTuple{u_1,u_2}$}
    \\
    &\hspace{10mm} =
    \letterm{\affinebang \overleftarrow{f_2}}
    {\affinebang \left(\lambda l.\left(\hspace{2mm}
    \begin{aligned}
        &\letterm{\nTuple{u_1,u_2}}
        {\lintranspose{\affcontext{\overleftarrow{\Phi_1}},\nTuple{u_1,u_2}}{(w_2\mbox{ }\dot{*}\mbox{ }u_1)\mbox{ }\dot{+}\mbox{ } (w_3\mbox{ }\dot{*}\mbox{ }u_2)}}{}\\
        &\nTuple{u_1,u_2}
    \end{aligned} 
    \hspace{2mm}\right)\right)}
    {\mathcal{T}_{\affcontext{\overleftarrow{\Phi_2}}}(S_2)}
    \\
    &\hspace{10mm} \approx
    \letterm{\affinebang \overleftarrow{f_2}}
    {\affinebang \left(\lambda l.  
    \left(
    \begin{aligned}
        (\lambda \nTuple{u_1,u_2}.&\nTuple{u_1,u_2})
        \\
        &\lintranspose{\affcontext{\overleftarrow{\Phi_1}},\nTuple{u_1,u_2}}{(w_2\mbox{ }\dot{*}\mbox{ }u_1)\mbox{ }\dot{+}\mbox{ } (w_3\mbox{ }\dot{*}\mbox{ }u_2)}
    \end{aligned}
    \right)\right)}
    {\mathcal{T}_{\affcontext{\overleftarrow{\Phi_2}}}(S_2)}
    \\
    &\hspace{8mm} \xrightarrow{\beta_\lambda}
    \letterm{\affinebang \overleftarrow{f_2}}
    {\affinebang \left(\lambda l.  
    \lintranspose{\affcontext{\overleftarrow{\Phi_1}},\nTuple{u_1,u_2}}{(w_2\mbox{ }\dot{*}\mbox{ }u_1)\mbox{ }\dot{+}\mbox{ } (w_3\mbox{ }\dot{*}\mbox{ }u_2)}\right)}
    {\mathcal{T}_{\affcontext{\overleftarrow{\Phi_2}}}(S_2)}
    \\
    &\hspace{10mm}\approx
    \letterm{\affinebang \overleftarrow{f_2}}
    {\affinebang \left(\lambda l.  
    \lintranspose{\affcontext{\overleftarrow{\Phi_1}},\nTuple{u_1,u_2}}{\dot{+}\nTuple{w_2\mbox{ }\dot{*}\mbox{ }u_1, w_3\mbox{ }\dot{*}\mbox{ }u_2}}\right)}
    {\mathcal{T}_{\affcontext{\overleftarrow{\Phi_2}}}(S_2)}
    \\
    &\hspace{7mm}\overset{\tiny \mathcal{T}(FU')}{=}
    \letterm{\affinebang \overleftarrow{f_2}}
    {\affinebang \left(\lambda l.  
    \left(
        \lambda \nTuple{l_1,l_2}.
        \lintranspose{\affcontext{\overleftarrow{\Phi_1}},\nTuple{u_1,u_2}}{\nTuple{w_2\mbox{ }\dot{*}\mbox{ }u_1, w_3\mbox{ }\dot{*}\mbox{ }u_2}}
    \right)
    \left(
        \lintranspose{\affcontext{\overleftarrow{\Phi_1}}}{\dot{+}} l
    \right)
    \right)}
    {\mathcal{T}_{\affcontext{\overleftarrow{\Phi_2}}}(S_2)}
    \\
    &\hspace{7mm}\overset{\tiny \mathcal{T}(\dot{+})}{=}
    \letterm{\affinebang \overleftarrow{f_2}}
    {\affinebang \left(\lambda l.  
    \left(
        \lambda \nTuple{l_1,l_2}.
        \lintranspose{\affcontext{\overleftarrow{\Phi_1}},\nTuple{u_1,u_2}}{\nTuple{w_2\mbox{ }\dot{*}\mbox{ }u_1, w_3\mbox{ }\dot{*}\mbox{ }u_2}}
    \right)
    \left(
        (\lambda u.\nTuple{u,u})l
    \right)
    \right)}
    {\mathcal{T}_{\affcontext{\overleftarrow{\Phi_2}}}(S_2)}
    \\
    &\hspace{8mm} \xrightarrow{\beta_\lambda}
    \letterm{\affinebang \overleftarrow{f_2}}
    {\affinebang \left(\lambda l.  
    \left(
        \lambda \nTuple{l_1,l_2}.
        \lintranspose{\affcontext{\overleftarrow{\Phi_1}},\nTuple{u_1,u_2}}{\nTuple{w_2\mbox{ }\dot{*}\mbox{ }u_1, w_3\mbox{ }\dot{*}\mbox{ }u_2}}
    \right)\nTuple{l,l}
    \right)}
    {\mathcal{T}_{\affcontext{\overleftarrow{\Phi_2}}}(S_2)}
    \\
    &\hspace{-3mm}\overset{\tiny \mathcal{T}(\nTuple{U_1,U_2})}{=}
    \letterm{\affinebang \overleftarrow{f_2}}
    {\affinebang \left(\lambda l.  
    \left(
    \begin{aligned}
        \lambda &\nTuple{l_1,l_2}.
        \mathtt{let}\mbox{ }\nTuple{\RenamePat{\overline{\alpha}'}{\nTuple{u_1,u_2}},\RenamePat{\overline{\alpha}''}{\nTuple{u_1,u_2}}} = 
        \\
        &\nTuple{
            \lintranspose{\affcontext{\overleftarrow{\Phi_1}},\overline{\alpha}'\nTuple{\nTuple{u_1,u_2}}}
            {\overline{\alpha}'[w_2\mbox{ }\dot{*}\mbox{ }u_1]}, 
            \lintranspose{\affcontext{\overleftarrow{\Phi_1}},\overline{\alpha}''\nTuple{\nTuple{u_1,u_2}}}
            {\overline{\alpha}''[w_3\mbox{ }\dot{*}\mbox{ }u_2]}
        }
        \\
        &\mathtt{in}\mbox{ }
        \Fusion{\nTuple{u_1,u_2}}{\overline{\alpha}'}{\overline{\alpha}''}
    \end{aligned}
    \right)
    \nTuple{l,l}
    \right)}
    {}
    \\
    &\hspace{10mm}  \mathcal{T}_{\affcontext{\overleftarrow{\Phi_2}}}(S_2)
    \\
    &\hspace{10mm} \text{where $\overline{\alpha}'=(u_1 \mapsto  l_1)$ and $\overline{\alpha}''=(u_2 \mapsto l_2)$.}
    \\
    &\hspace{10mm} \text{We can observe that $\Dom{\overline{\alpha}'}=\{u_1\}\cap\Dom{\overline{\alpha}''}=\{u_2\}=\emptyset$, so we have that}
    \\
    &\hspace{10mm} \text{$\RenamePat{\overline{\alpha}'}{\nTuple{u_1,u_2}}=\RenamePat{\overline{\alpha}'}{u_1} 
    =\overline{\alpha}'(u_1)=l_1$ and $\RenamePat{\overline{\alpha}''}{\nTuple{u_1,u_2}}=\RenamePat{\overline{\alpha}''}{u_2} =\overline{\alpha}''(u_2)=l_2$ and}
    \\
    &\hspace{10mm}
    \Fusion{\nTuple{u_1,u_2}}{\overline{\alpha}'}{\overline{\alpha}''}=\nTuple{\Fusion{u_1}{\overline{\alpha}'}{\overline{\alpha}''},\Fusion{u_2}{\overline{\alpha}'}{\overline{\alpha}''}}
    =\nTuple{\overline{\alpha}'(u_1), \overline{\alpha}''(u_2)}
    =\nTuple{l_1,l_2}
    \\
    &\hspace{5mm} =
   \letterm{\affinebang \overleftarrow{f_2}}
    {\affinebang \left(\lambda l.  
    \left(
    \begin{aligned}
        \lambda &\nTuple{l_1,l_2}.
        \mathtt{let}\mbox{ }\nTuple{l_1,l_2} = 
        \\
        &\nTuple{
            \lintranspose{\affcontext{\overleftarrow{\Phi_1}},\overline{\alpha}'\nTuple{\nTuple{u_1,u_2}}}
            {\overline{\alpha}'[w_2\mbox{ }\dot{*}\mbox{ }u_1]}, 
            \lintranspose{\affcontext{\overleftarrow{\Phi_1}},\overline{\alpha}''\nTuple{\nTuple{u_1,u_2}}}
            {\overline{\alpha}''[w_3\mbox{ }\dot{*}\mbox{ }u_2]}
        }
        \\
        &\mathtt{in}\mbox{ }
        \nTuple{l_1,l_2}
    \end{aligned}
    \right)
    \nTuple{l,l}
    \right)}
    {}
    \\
    &\hspace{10mm}  \mathcal{T}_{\affcontext{\overleftarrow{\Phi_2}}}(S_2)
    \\
    &\hspace{5mm} \approx
    \letterm{\affinebang \overleftarrow{f_2}}
    {\affinebang \left(\lambda l.  
    \left(
    \begin{aligned}
        \lambda &\nTuple{l_1,l_2}.
        (\lambda \nTuple{l_1,l_2}.\nTuple{l_1,l_2})
        \\
        &\nTuple{
            \lintranspose{\affcontext{\overleftarrow{\Phi_1}},\overline{\alpha}'\nTuple{\nTuple{u_1,u_2}}}
            {\overline{\alpha}'[w_2\mbox{ }\dot{*}\mbox{ }u_1]}, 
            \lintranspose{\affcontext{\overleftarrow{\Phi_1}},\overline{\alpha}''\nTuple{\nTuple{u_1,u_2}}}
            {\overline{\alpha}''[w_3\mbox{ }\dot{*}\mbox{ }u_2]}
        }
    \end{aligned}
    \right)
    \nTuple{l,l}
    \right)}
    {}
    \\
    &\hspace{10mm}  \mathcal{T}_{\affcontext{\overleftarrow{\Phi_2}}}(S_2)
    \\
    &\hspace{5mm} =_\alpha
    \letterm{\affinebang \overleftarrow{f_2}}
    {\affinebang \left(\lambda l.  
    \left(
    \begin{aligned}
        \lambda &\nTuple{l_1,l_2}.
        (\lambda \nTuple{x_1,x_2}.\nTuple{x_1,x_2})
        \\
        &\nTuple{
            \lintranspose{\affcontext{\overleftarrow{\Phi_1}},\overline{\alpha}'\nTuple{\nTuple{u_1,u_2}}}
            {\overline{\alpha}'[w_2\mbox{ }\dot{*}\mbox{ }u_1]}, 
            \lintranspose{\affcontext{\overleftarrow{\Phi_1}},\overline{\alpha}''\nTuple{\nTuple{u_1,u_2}}}
            {\overline{\alpha}''[w_3\mbox{ }\dot{*}\mbox{ }u_2]}
        }
    \end{aligned}
    \right)
    \nTuple{l,l}
    \right)}
    {}
    \\
    &\hspace{10mm}  \mathcal{T}_{\affcontext{\overleftarrow{\Phi_2}}}(S_2)
    \\
    &\hspace{5mm} =
    \letterm{\affinebang \overleftarrow{f_2}}
    {\affinebang \left(\lambda l.  
    \left(\begin{aligned}
        \lambda \nTuple{l_1,l_2}.&
        (\lambda \nTuple{x_1,x_2}.\nTuple{x_1,x_2})\\
        &\nTuple{
            \lintranspose{\affcontext{\overleftarrow{\Phi_1}},l_1}
            { w_2\mbox{ }\dot{*}\mbox{ }l_1}, 
            \lintranspose{\affcontext{\overleftarrow{\Phi_1}},l_2}
            {w_3\mbox{ }\dot{*}\mbox{ }l_2}
        }
    \end{aligned}\right)
    \nTuple{l,l}
    \right)}
    {\mathcal{T}_{\affcontext{\overleftarrow{\Phi_2}}}(S_2)}
    \\ 
    &\hspace{5mm} \rightarrow^*
    \letterm{\affinebang \overleftarrow{f_2}}
    {\affinebang \left(\lambda l.  
    \left(\begin{aligned}
        \lambda \nTuple{l_1,l_2}.
        (\lambda \nTuple{x_1,x_2}.\nTuple{x_1,x_2})
        \nTuple{ 
            {w_2\mbox{ }\dot{*}\mbox{ }l_1},  
            {w_3\mbox{ }\dot{*}\mbox{ }l_2}
        }
    \end{aligned}\right)
    \nTuple{l,l}
    \right)}
    {\mathcal{T}_{\affcontext{\overleftarrow{\Phi_2}}}(S_2)}
    \\
    &\hspace{5mm} \xrightarrow{\beta_\lambda}
    \letterm{\affinebang \overleftarrow{f_2}}
    {\affinebang \left(\lambda l.  
    \left(\begin{aligned}
        \lambda \nTuple{l_1,l_2}. 
        \nTuple{ 
            {w_2\mbox{ }\dot{*}\mbox{ }l_1},  
            {w_3\mbox{ }\dot{*}\mbox{ }l_2}
        }
    \end{aligned}\right)
    \nTuple{l,l}
    \right)}
    {\mathcal{T}_{\affcontext{\overleftarrow{\Phi_2}}}(S_2)}
    \\
    &\hspace{5mm} \xrightarrow{\beta_\lambda}
    \letterm{\affinebang \overleftarrow{f_2}}
    {\affinebang \left(\lambda l.   
        \nTuple{ 
            {w_2\mbox{ }\dot{*}\mbox{ }l},  
            {w_3\mbox{ }\dot{*}\mbox{ }l}
        } 
    \right)}
    {\mathcal{T}_{\affcontext{\overleftarrow{\Phi_2}}}(S_2)}
\end{align*}

Therefore, we have that 
\begin{equation}
    \label{eq:example_transpose_2}
    \begin{aligned}
    \mathcal{T}_{\affcontext{\overleftarrow{\Phi_1}}}(
        \letterm{\affinebang f_2^{\mbox{ }(\typR\&\typR) \multimap \typR}}
        {\affinebang (\lambda \nTuple{u_1,u_2}. (w_2\mbox{ }\dot{*}\mbox{ }u_1)
        \mbox{ }&\dot{+}\mbox{ } (w_3\mbox{ }\dot{*}\mbox{ }u_2))}{S_2}
    ) 
    \rightarrow^* \\
    &\letterm{\affinebang \overleftarrow{f_2}}
    {\affinebang \left(\lambda l.   
        \nTuple{ 
            {w_2\mbox{ }\dot{*}\mbox{ }l},  
            {w_3\mbox{ }\dot{*}\mbox{ }l}
        } 
    \right)}
    {\mathcal{T}_{\affcontext{\overleftarrow{\Phi_2}}}(S_2)}
    \end{aligned}
\end{equation}

We can proceed in a similar way to transpose the let-bindings related to $f_3$ and $f_4$, obtaining:
\begin{equation}
    \label{eq:example_transpose_3}
    \begin{aligned}
    \mathcal{T}_{\affcontext{\overleftarrow{\Phi_2}}}(
        \letterm{\affinebang f_3^{\mbox{ }\typR \multimap \typR}}
        {\affinebang (\lambda u. w_4\mbox{ }\dot{*}\mbox{ }u)}{S_3}
    ) 
    \rightarrow^*  
    \letterm{\affinebang \overleftarrow{f_3}}
    {\affinebang (\lambda l. w_4 \mbox{ }\dot{*}\mbox{ } l)}
    {\mathcal{T}_{\affcontext{\overleftarrow{\Phi_3}}}(S_3)}
    \end{aligned}
\end{equation}

\begin{equation}
    \label{eq:example_transpose_4}
    \begin{aligned}
    \mathcal{T}_{\affcontext{\overleftarrow{\Phi_3}}}(
        \letterm{\affinebang f_4^{\mbox{ }(\typR\&\typR) \multimap \typR}}
        {\affinebang (\lambda \nTuple{u_1,u_2}. (w_5\mbox{ }\dot{*}\mbox{ }u_1)
        \mbox{ }&\dot{+}\mbox{ } (w_6\mbox{ }\dot{*}\mbox{ }u_2))}{S_4}
    ) 
    \rightarrow^* \\
    &\letterm{\affinebang \overleftarrow{f_4}}
    {\affinebang \left(\lambda l.   
        \nTuple{ 
            {w_5\mbox{ }\dot{*}\mbox{ }l},  
            {w_6\mbox{ }\dot{*}\mbox{ }l}
        } 
    \right)}
    {\mathcal{T}_{\affcontext{\overleftarrow{\Phi}}}(S_4)}
    \end{aligned}
\end{equation}
where $\affcontext{\overleftarrow{\Phi_2}}=\affinebang \overleftarrow{f_1},\overleftarrow{f_2}$, $\affcontext{\overleftarrow{\Phi_3}}=\affinebang \overleftarrow{f_1},\overleftarrow{f_2}, \overleftarrow{f_3}$ and $\affcontext{\overleftarrow{\Phi}}=\affinebang \overleftarrow{f_1},\overleftarrow{f_2}, \overleftarrow{f_3}, \overleftarrow{f_4}$.

By combining Equations~\ref{eq:example_transpose_1}-\ref{eq:example_transpose_2}-\ref{eq:example_transpose_3}-\ref{eq:example_transpose_4} we obtain that 

\begin{equation}
    \label{eq:example_transpose_initial}
    \begin{aligned}
    &\mathcal{T}_{\affcontext{\overleftarrow{\Phi_0}}}
    \quad \left(
    \begin{aligned}
        &\letterm{\affinebang f_1^{\mbox{ }\typR \multimap \typR}}
        {\affinebang (\lambda u. w_1\mbox{ }\dot{*}\mbox{ }u)}{}\\
        &\letterm{\affinebang f_2^{\mbox{ }(\typR\&\typR) \multimap \typR}}
        {\affinebang (\lambda \nTuple{u_1,u_2}. (w_2\mbox{ }\dot{*}\mbox{ }u_1)
        \mbox{ }\dot{+}\mbox{ } (w_3\mbox{ }\dot{*}\mbox{ }u_2))}{}\\
        &\letterm{\affinebang f_3^{\mbox{ }\typR \multimap \typR}}
        {\affinebang (\lambda u. w_4\mbox{ }\dot{*}\mbox{ }u)}{}\\
        &\letterm{\affinebang f_4^{\mbox{ }(\typR\&\typR) \multimap \typR}}
        {\affinebang (\lambda \nTuple{u_1,u_2}. (w_5\mbox{ }\dot{*}\mbox{ }u_1)
        \mbox{ }\dot{+}\mbox{ } (w_6\mbox{ }\dot{*}\mbox{ }u_2))}{}\\
        &\lambda u^{\typR\&\typR}.
        \begin{aligned}
            &\letterm{\nTuple{x',y'}}{u}{}
            f_4 \nTuple{f_2 \nTuple{f_1\mbox{ }x', y'}, f_3\mbox{ }x'}
        \end{aligned}
    \end{aligned}
    \quad \right)
    \\[4mm]
    &\hspace{3.5cm}\rightarrow^* 
    \begin{aligned}
        &\letterm{\affinebang \overleftarrow{f_1}^{\mbox{ }\typR \multimap \typR}}
        {\affinebang (\lambda {l}. w_1\mbox{ }\dot{*}\mbox{ }{l})}{}\\
        &\letterm{\affinebang \overleftarrow{f_2}^{\mbox{ } \typR \multimap (\typR\&\typR)}}
        {\affinebang (\lambda {l}.\nTuple{w_2 \mbox{ }\dot{*}\mbox{ }{l}, w_3 \mbox{ }\dot{*}\mbox{ }{l}})}{}\\
        &\letterm{\affinebang \overleftarrow{f_3}^{\mbox{ }\typR \multimap \typR}}
        {\affinebang (\lambda {l}. w_4\mbox{ }\dot{*}\mbox{ }{l})}{}
		\\
        &\letterm{\affinebang \overleftarrow{f_4}^{\mbox{ } \typR \multimap (\typR\&\typR)}}
        {\affinebang (\lambda {l}. \nTuple{w_5\mbox{}\dot{*}\mbox{ }{l},w_6\mbox{}\dot{*}\mbox{ }{l}})}{}\\
        &\mathcal{T}_{\affcontext{\overleftarrow{\Phi}}}
        \left(
            \lambda u^{\typR\&\typR}. 
            \letterm{\nTuple{x',y'}}{u}
            {f_4 \nTuple{f_2 \nTuple{f_1\mbox{ }x', y'}, f_3\mbox{ }x'}}
        \right)
    \end{aligned} 
\end{aligned}
\end{equation}
where $\affcontext{\overleftarrow{\Phi}}=\affinebang \overleftarrow{f_1},\overleftarrow{f_2}, \overleftarrow{f_3}, \overleftarrow{f_4}$.

Let us focus now on the transposition of the last line
\begin{align*}
    \mathcal{T}_{\affcontext{\overleftarrow{\Phi}}}
    &\left(
        \lambda u^{\typR\&\typR}. 
        \letterm{\nTuple{x',y'}}{u}
        {f_4 \nTuple{f_2 \nTuple{f_1\mbox{ }x', y'}, f_3\mbox{ }x'}}
    \right)
    \\
    &\hspace{7mm}\approx
    \mathcal{T}_{\affcontext{\overleftarrow{\Phi}}}
    \left(
        \lambda u^{\typR\&\typR}.
        \left(
            \lambda \nTuple{x',y'}.{f_4 \nTuple{f_2 \nTuple{f_1\mbox{ }x', y'}, f_3\mbox{ }x'}}
        \right)u
    \right)
    \\
    &\hspace{3mm}\overset{\tiny \mathcal{T}(\lambda \PatAddA.U)}{=}
    \lambda z.  
    \letterm{\RenamePat{\overline{\alpha}}{u}}
    {
        \mathcal{T}_{\affcontext{\overleftarrow{\Phi}}}\left(
            \left(
                \lambda \nTuple{x',y'}.{f_4 \nTuple{f_2 \nTuple{f_1\mbox{ }x', y'}, f_3\mbox{ }x'}}
            \right)u \right) 
    }
    {\Fusion{u}{\overline{\alpha}}{\emptyset}}
    \\
    &\hspace{15mm}\text{where $\overline{\alpha}$ is the identity renaming on $u$, so we have that}
    \\
    &\hspace{15mm}
    \text{$\RenamePat{\overline{\alpha}}{u}=\overline{\alpha}(u)=u$ and $\Fusion{u}{\overline{\alpha}}{\emptyset}=\overline{\alpha}(u)=u$}
    \\
    &\hspace{7mm}=
    \lambda z.  
    \letterm{u}
    {
        \mathcal{T}_{\affcontext{\overleftarrow{\Phi}}}\left(
            \left(
                \lambda \nTuple{x',y'}.{f_4 \nTuple{f_2 \nTuple{f_1\mbox{ }x', y'}, f_3\mbox{ }x'}}
            \right)u \right) 
    }
    {u}
    \\
    &\hspace{7mm}\approx
    \lambda z. (\lambda u.u) \mbox{ }
        \mathcal{T}_{\affcontext{\overleftarrow{\Phi}}}\left(
            \left(
                \lambda \nTuple{x',y'}.{f_4 \nTuple{f_2 \nTuple{f_1\mbox{ }x', y'}, f_3\mbox{ }x'}}
            \right)u\right)  
    \\
    &\hspace{7mm}\xrightarrow{\beta_\lambda}
    \lambda z. 
    \mathcal{T}_{\affcontext{\overleftarrow{\Phi}}}\left(
        \left(
            \lambda \nTuple{x',y'}.{f_4 \nTuple{f_2 \nTuple{f_1\mbox{ }x', y'}, f_3\mbox{ }x'}}
        \right)u \right)
    \\
    &\hspace{5mm}\overset{\tiny \mathcal{T}(FU')}{=}
    \lambda z. 
    \left(\lambda u.\lintranspose{\affcontext{\overleftarrow{\Phi}},u}{u}\right)
    \left(
        \mathcal{T}_{\affcontext{\overleftarrow{\Phi}}}\
        \left(
            \lambda \nTuple{x',y'}.{f_4 \nTuple{f_2 \nTuple{f_1\mbox{ }x', y'}, f_3\mbox{ }x'}}
        \right) \mbox{ } z
    \right)
    \\
    &\hspace{5mm}\overset{\tiny \mathcal{T}(u)}{=}
    \lambda z. 
    \left(\lambda u.u\right)
    \left(
        \mathcal{T}_{\affcontext{\overleftarrow{\Phi}}}\
        \left(
            \lambda \nTuple{x',y'}.{f_4 \nTuple{f_2 \nTuple{f_1\mbox{ }x', y'}, f_3\mbox{ }x'}}
        \right) \mbox{ } z
    \right)
    \\
    &\hspace{3mm}\overset{\tiny \mathcal{T}(\lambda \PatAddA.U)}{=}
    \lambda z. 
    \left(\lambda u.u\right)
    \left( 
        \left(
            \lambda j. \left(
            \begin{aligned}
                &\letterm{\RenamePat{\overline{\alpha_1}}{\nTuple{x',y'}}}
                { 
                    \lintranspose{\affcontext{\overleftarrow{\Phi}},\nTuple{x',y'}} {{f_4 \nTuple{f_2 \nTuple{f_1\mbox{ }x', y'}, f_3\mbox{ }x'}}}
                }{}\\
                &\Fusion{\nTuple{x',y'}}{\overline{\alpha_1}}{\emptyset}
            \end{aligned}
            \right)
        \right)
         z
    \right)
    \\
    &\hspace{15mm}\text{where $\overline{\alpha_1}$ in the identity renaming on $\nTuple{x',y'}$ and so we have that}
    \\
    &\hspace{15mm}
    \text{$\RenamePat{\overline{\alpha_1}}{\nTuple{x',y'}}=\nTuple{\RenamePat{\overline{\alpha_1}}{x'},\RenamePat{\overline{\alpha_1}}{y'}}=\nTuple{\overline{\alpha_1}(x'),\overline{\alpha_1}(y')}=\nTuple{x',y'}$ and }
    \\
    &\hspace{15mm}
    \text{$\Fusion{\nTuple{x',y'}}{\overline{\alpha_1}}{\emptyset}=\nTuple{\Fusion{x'}{\overline{\alpha_1}}{\emptyset},\Fusion{y'}{\overline{\alpha_1}}{\emptyset}}=\nTuple{\overline{\alpha_1}(x'),\overline{\alpha_1}(y')}=\nTuple{x',y'}$.}
    \\
    &\hspace{7mm}=
    \lambda z. 
    \left(\lambda u.u\right)
    \left( 
        \left(
            \lambda j. 
            \letterm{\nTuple{x',y'}}
            { 
                \lintranspose{\affcontext{\overleftarrow{\Phi}},\nTuple{x',y'}} {{f_4 \nTuple{f_2 \nTuple{f_1\mbox{ }x', y'}, f_3\mbox{ }x'}}}
            }
            {\nTuple{x',y'}}
        \right)
         z
    \right)
    \\
    &\hspace{7mm}\approx
    \lambda z. 
    \left(\lambda u.u\right)
    \left( 
        \left(
            \lambda j. 
            (\lambda \nTuple{\nTuple{x',y'}}.\nTuple{x',y'})
            \mbox{ }
            \lintranspose{\affcontext{\overleftarrow{\Phi}},\nTuple{x',y'}} {{f_4 \nTuple{f_2 \nTuple{f_1\mbox{ }x', y'}, f_3\mbox{ }x'}}}
        \right)
         z
    \right)
\end{align*}

Therefore, we have that 
\begin{equation}
    \label{eq:example_transpose_initial_-1}
    \begin{aligned}
        &\begin{aligned}
            \mathcal{T}_{\affcontext{\overleftarrow{\Phi}}}
            \left(
                \lambda u^{\typR\&\typR}. 
                \letterm{\nTuple{x',y'}}{u}
                {f_4 \nTuple{f_2 \nTuple{f_1\mbox{ }x', y'},        f_3\mbox{ }x'}}
            \right)
        \end{aligned}
        \rightarrow
        \\
        &\hspace{0.7cm}
        \begin{aligned}
            \lambda z. 
            \left(\lambda u.u\right)
            \left( 
                \left(
                    \lambda j. 
                    (\lambda \nTuple{\nTuple{x',y'}}.\nTuple{x',y'})
                    \mbox{ }
                    \lintranspose{\affcontext{\overleftarrow{\Phi}},\nTuple{x',y'}} {{f_4 \nTuple{f_2 \nTuple{f_1\mbox{ }x', y'}, f_3\mbox{ }x'}}}
                \right)
                z
            \right)
        \end{aligned}
    \end{aligned}
\end{equation}

We proceed by focusing on the transposition of $\judgment{\oc\Sigma, \affcontext{\overleftarrow{\Phi}}, j:\typR}{\lintranspose{\affcontext{\overleftarrow{\Phi}},\nTuple{x',y'}} {{f_4 \nTuple{f_2 \nTuple{f_1\mbox{ }x', y'}, f_3\mbox{ }x'}}}}{\typR\&\typR}$ as follows
\begin{align*}
    &\lintranspose{\affcontext{\overleftarrow{\Phi}},\nTuple{x',y'}} 
    {{f_4 \nTuple{f_2 \nTuple{f_1\mbox{ }x', y'}, f_3\mbox{ }x'}}}
    \\
    &\hspace{7mm}\overset{\tiny \mathcal{T}(FU')}{=}
    \left(
        \lambda \nTuple{z',z''}.
        \lintranspose{\affcontext{\overleftarrow{\Phi}},\nTuple{x',y'}} 
        {\nTuple{f_2 \nTuple{f_1\mbox{ }x', y'}, f_3\mbox{ }x'}}
    \right)
    \left(
         \left( \lintranspose{\affcontext{\overleftarrow{\Phi}}}{f_4} \right) j
    \right)
    \\
    &\hspace{9mm}\overset{\tiny \mathcal{T}(f)}{=}
    \left(
        \lambda \nTuple{z',z''}.
        \lintranspose{\affcontext{\overleftarrow{\Phi}},\nTuple{x',y'}} 
        {\nTuple{f_2 \nTuple{f_1\mbox{ }x', y'}, f_3\mbox{ }x'}}
    \right)
    \left(
        \overleftarrow{f_4} \mbox{ }j
    \right)
    \\
    &\hspace{7mm}\overset{\tiny \mathcal{T}(\nTuple{U_1,U_2})}{=}
    \left(\lambda \nTuple{z',z''}.\left(
        \begin{aligned}
            & \mathtt{let}\mbox{ }
            \nTuple{\RenamePat{\alpha_1}{\nTuple{x',y'}},
            \RenamePat{\alpha_2}{\nTuple{x',y'}}}
            =
            \\
            &\hspace{3mm}
            \nTuple{
                \lintranspose{\affcontext{\overleftarrow{\Phi}},\alpha_1\nTuple{\nTuple{x',y'}}}
                {\alpha_1 [f_2 \nTuple{f_1\mbox{ }x', y'}]},
                \lintranspose{\affcontext{\overleftarrow{\Phi}},\alpha_2\nTuple{\nTuple{x',y'}}}
                {\alpha_2 [f_3\mbox{ }x']}
            }
            \\
            &\mathtt{in}\mbox{ } \Fusion{\nTuple{x',y'}}{\alpha_1}{\alpha_2}
        \end{aligned}
     \right)\right)
    \left(
        \overleftarrow{f_4} \mbox{ }j
    \right)
    \\
    &\hspace{15mm}\text{where $\Rename_1\definedas(x'\mapsto x'_1, y'\mapsto y'_1)$ and $\Rename_2\definedas(x'\mapsto x'_2)$}.
    \\
    &\hspace{12mm}=
    \left(\lambda \nTuple{z',z''}.\left(
        \begin{aligned}
            & \mathtt{let}\mbox{ }
            \nTuple{\RenamePat{\alpha_1}{\nTuple{x',y'}},
            \RenamePat{\alpha_2}{\nTuple{x',y'}}}
            =
            \\
            &\hspace{3mm}
            \nTuple{
                \lintranspose{\affcontext{\overleftarrow{\Phi}},\nTuple{x'_1,y'_1}}
                { f_2 \nTuple{f_1\mbox{ }x'_1, y'_1}},
                \lintranspose{\affcontext{\overleftarrow{\Phi}},x'_2}
                {f_3\mbox{ }x'_2}
            }
            \\
            &\mathtt{in}\mbox{ } \Fusion{\nTuple{x',y'}}{\alpha_1}{\alpha_2}
        \end{aligned}
     \right)\right)
    \left(
        \overleftarrow{f_4} \mbox{ }j
    \right) 
\end{align*}

Therefore, we have that 
\begin{equation}
    \label{eq:example_transpose_initial_0}
    \begin{aligned}
        &\begin{aligned}
            \lintranspose{\affcontext{\overleftarrow{\Phi}},\nTuple{x',y'}} 
            {{f_4 \nTuple{f_2 \nTuple{f_1\mbox{ }x', y'}, f_3\mbox{ }x'}}}
        \end{aligned}
        =
        \\
        &\hspace{0.7cm}
        \begin{aligned}
            \left(\lambda \nTuple{z',z''}.\left(
                \begin{aligned}
                    & \mathtt{let}\mbox{ }
                    \nTuple{\RenamePat{\alpha_1}{\nTuple{x',y'}},
                    \RenamePat{\alpha_2}{\nTuple{x',y'}}}
                    =
                    \\
                    &\hspace{3mm}
                    \nTuple{
                        \lintranspose{\affcontext{\overleftarrow{\Phi}},\nTuple{x'_1,y'_1}}
                        { f_2 \nTuple{f_1\mbox{ }x'_1, y'_1}},
                        \lintranspose{\affcontext{\overleftarrow{\Phi}},x'_2}
                        {f_3\mbox{ }x'_2}
                    }
                    \\
                    &\mathtt{in}\mbox{ } \Fusion{\nTuple{x',y'}}{\alpha_1}{\alpha_2}
                \end{aligned}
            \right)\right)
            \left(
                \overleftarrow{f_4} \mbox{ }j
            \right) 
        \end{aligned}
    \end{aligned}
\end{equation}

We proceed by focusing on the transposition of $\judgment{\oc\Sigma, \affcontext{\overleftarrow{\Phi}}, z':\typR}{\lintranspose{\affcontext{\overleftarrow{\Phi}},\nTuple{x'_1,y'_1}}
{f_2 \nTuple{f_1\mbox{ }x'_1, y'_1}}}{\typR\&\typR}$ as follows
\begin{align*}
    &\lintranspose{\affcontext{\overleftarrow{\Phi}},\nTuple{x'_1,y'_1}}
    {f_2 \nTuple{f_1\mbox{ }x'_1, y'_1}}
    \\
    &\hspace{7mm}\overset{\tiny \mathcal{T}(FU')}{=}
    \left(
        \lambda \nTuple{z_1,z_2}.\lintranspose{\affcontext{\overleftarrow{\Phi}},\nTuple{x'_1,y'_1}}{\nTuple{f_1\mbox{ }x'_1, y'_1}}
    \right)
    \left(
        \lintranspose{\affcontext{\overleftarrow{\Phi}}}{f_2} \mbox{ } z'
    \right)
    \\
    &\hspace{15mm} \text{where $\nTuple{z_1,z_2}$ is the pattern associated with $\lintranspose{\affcontext{\overleftarrow{\Phi}},\nTuple{x'_1,y'_1}}{\nTuple{f_1\mbox{ }x'_1, y'_1}}$.}
    \\
    &\hspace{7mm}\overset{\tiny \mathcal{T}(f)}{=}
    \left(
        \lambda \nTuple{z_1,z_2}.\lintranspose{\affcontext{\overleftarrow{\Phi}},\nTuple{x'_1,y'_1}}{\nTuple{f_1\mbox{ }x'_1, y'_1}}
    \right)
    \left(
        \overleftarrow{f_2} \mbox{ } z'
    \right)
    \\
    &\hspace{7mm}\overset{\tiny \mathcal{T}(\nTuple{U_1,U_2})}{=}
    \left(\lambda \nTuple{z_1,z_2}. \left(
        \begin{aligned}
            &\mathtt{let}\mbox{ }
            \nTuple{\RenamePat{\alpha_3}{\nTuple{x'_1, y'_1}},\RenamePat{\alpha_4}{\nTuple{x'_1, y'_1}}}
            =\\
            &\hspace{2mm} \nTuple{
                \lintranspose{\affcontext{\overleftarrow{\Phi}},\alpha_3\nTuple{\nTuple{x'_1,y'_1}}}{\alpha_3[f_1\mbox{ }x'_1]},
                \lintranspose{\affcontext{\overleftarrow{\Phi}},\alpha_4\nTuple{\nTuple{x'_1,y'_1}}}{\alpha_4[y'_1]}
            }
            \\
            &\mathtt{in}\mbox{ }\Fusion{\nTuple{x'_1, y'_1}}{\alpha_3}{\alpha_4}
        \end{aligned}
    \right)\right)
    \left(
        \overleftarrow{f_2} \mbox{ } z'
    \right)
    \\
    &\hspace{15mm} \text{Since $\nTuple{z_1,z_2}$ is the pattern associated to $\lintranspose{\affcontext{\overleftarrow{\Phi}},\nTuple{x'_1,y'_1}}{\nTuple{f_1\mbox{ }x'_1, y'_1}}$,
    we take the two }
    \\
    &\hspace{15mm} \text{renamings as $\alpha_3=(x'_1\mapsto z_1)$  and $\alpha_4=(y'_1\mapsto z_2)$. We can observe that }
    \\
    &\hspace{15mm} \text{$\Dom{\alpha_3}\cap \Dom{\alpha_4}=\emptyset$, so we have that $\RenamePat{\alpha_3}{\nTuple{x'_1, y'_1}}=z_1$, }
    \\
    &\hspace{15mm} \text{$\RenamePat{\alpha_4}{\nTuple{x'_1, y'_1}}=z_2$ and $\Fusion{\nTuple{x'_1, y'_1}}{\alpha_3}{\alpha_4}=\nTuple{z_1,z_2}$}
    \\
    &\hspace{10mm}=
    \left(\lambda \nTuple{z_1,z_2}. \left(
        \begin{aligned}
            &\mathtt{let}\mbox{ }
            \nTuple{z_1,z_2}
            =\\
            &\hspace{2mm} \nTuple{
                \lintranspose{\affcontext{\overleftarrow{\Phi}},\alpha_3\nTuple{\nTuple{x'_1,y'_1}}}{\alpha_3[f_1\mbox{ }x'_1]},
                \lintranspose{\affcontext{\overleftarrow{\Phi}},\alpha_4\nTuple{\nTuple{x'_1,y'_1}}}{\alpha_4[y'_1]}
            }
            \\
            &\mathtt{in}\mbox{ }\nTuple{z_1,z_2}
        \end{aligned}
    \right)\right)
    \left(
        \overleftarrow{f_2} \mbox{ } z'
    \right)
    \\
    &\hspace{10mm}=
    \left(\lambda \nTuple{z_1,z_2}. 
            \letterm{\nTuple{z_1,z_2}}
            {\nTuple{
                \lintranspose{\affcontext{\overleftarrow{\Phi}},z_1}{f_1\mbox{ }z_1},
                \lintranspose{\affcontext{\overleftarrow{\Phi}},z_2}{z_2}
            }}
            {\nTuple{z_1,z_2}}
    \right)
    \left(
        \overleftarrow{f_2} \mbox{ } z'
    \right)
    \\
    &\hspace{10mm} \approx_\alpha
    \left(
        \begin{aligned}
            \lambda \nTuple{z_1,z_2}. & (\lambda \nTuple{s_1,s_2}.\nTuple{s_1,s_2})\\
            & \nTuple{
                \lintranspose{\affcontext{\overleftarrow{\Phi}},z_1}{f_1\mbox{ }z_1},
                \lintranspose{\affcontext{\overleftarrow{\Phi}},z_2}{z_2}
            }
        \end{aligned}
    \right)
    \left(
        \overleftarrow{f_2} \mbox{ } z'
    \right)
    \\
    &\hspace{7mm}\overset{\tiny \mathcal{T}(FU')}{=}
    \left(
        \begin{aligned}
            \lambda \nTuple{z_1,z_2}. & (\lambda \nTuple{s_1,s_2}.\nTuple{s_1,s_2})\\
            & \nTuple{
                \left(\lambda z_1. \lintranspose{\affcontext{\overleftarrow{\Phi}},z_1}{z_1} \right)\left(\lintranspose{\affcontext{\overleftarrow{\Phi}}}{f_1}\mbox{ }z_1\right),
                \lintranspose{\affcontext{\overleftarrow{\Phi}},z_2}{z_2}
            }
        \end{aligned}
    \right)
    \left(
        \overleftarrow{f_2} \mbox{ } z'
    \right)
    \\
    &\hspace{9mm}\overset{\tiny \mathcal{T}(f)}{=}
    \left(
        \begin{aligned}
            \lambda \nTuple{z_1,z_2}. & (\lambda \nTuple{s_1,s_2}.\nTuple{s_1,s_2})\\
            & \nTuple{
                \left(\lambda z_1. \lintranspose{\affcontext{\overleftarrow{\Phi}},z_1}{z_1} \right)\left(\overleftarrow{f_1}\mbox{ }z_1\right),
                \lintranspose{\affcontext{\overleftarrow{\Phi}},z_2}{z_2}
            }
        \end{aligned}
    \right)
    \left(
        \overleftarrow{f_2} \mbox{ } z'
    \right)
    \\
    &\hspace{9mm}\overset{\tiny \mathcal{T}(u)}{=}
    \left(
        \begin{aligned}
            \lambda \nTuple{z_1,z_2}. & (\lambda \nTuple{s_1,s_2}.\nTuple{s_1,s_2})\\
            & \nTuple{
                \left(\lambda z_1.z_1\right)\left(\overleftarrow{f_1}\mbox{ }z_1\right),
                \lintranspose{\affcontext{\overleftarrow{\Phi}},z_2}{z_2}
            }
        \end{aligned}
    \right)
    \left(
        \overleftarrow{f_2} \mbox{ } z'
    \right)
    \\
    &\hspace{12mm}{\xrightarrow{\beta_\lambda}}
    \left(
            \lambda \nTuple{z_1,z_2}. (\lambda \nTuple{s_1,s_2}.\nTuple{s_1,s_2})\mbox{ }
            \nTuple{
                \overleftarrow{f_1}\mbox{ }z_1,
                \lintranspose{\affcontext{\overleftarrow{\Phi}},z_2}{z_2}
            }
    \right)
    \left(
        \overleftarrow{f_2} \mbox{ } z'
    \right)
    \\
    &\hspace{9mm}\overset{\tiny \mathcal{T}(u)}{=}
   \left(
            \lambda \nTuple{z_1,z_2}. (\lambda \nTuple{s_1,s_2}.\nTuple{s_1,s_2})\mbox{ }
            \nTuple{
                \overleftarrow{f_1}\mbox{ }z_1,
                z_2
            }
    \right)
    \left(
        \overleftarrow{f_2} \mbox{ } z'
    \right)
    \\
    &\hspace{12mm}{\xrightarrow{\beta_\lambda}}
    \left(
        \lambda \nTuple{z_1,z_2}.\nTuple{\overleftarrow{f_1}\mbox{ }z_1, z_2}
    \right)
    \left(
        \overleftarrow{f_2} \mbox{ } z'
    \right)
\end{align*}

Therefore, we have that 
\begin{equation}
    \label{eq:example_transpose_initial_1}
    \begin{aligned}
        &\begin{aligned}
            \lintranspose{\affcontext{\overleftarrow{\Phi}},\nTuple{x'_1,y'_1}}
            {f_2 \nTuple{f_1\mbox{ }x'_1, y'_1}}
        \end{aligned}
        \rightarrow^*
        \begin{aligned}
            \left(
                \lambda \nTuple{z_1,z_2}.\nTuple{\overleftarrow{f_1}\mbox{ }z_1, z_2}
            \right)
            \left(
                \overleftarrow{f_2} \mbox{ } z'
            \right)
        \end{aligned}
    \end{aligned}
\end{equation}

Now, we focus on the transposition of $\judgment{\oc\Sigma, \affcontext{\overleftarrow{\Phi}}, z'':\typR}{\lintranspose{\affcontext{\overleftarrow{\Phi}},x'_2}
{f_3 \mbox{ } x'_2}}{\typR}$ as follows
\begin{align*}
    &\lintranspose{\affcontext{\overleftarrow{\Phi}},x'_2}
    {f_3 \mbox{ } x'_2}
    \\
    &\hspace{7mm}\overset{\tiny \mathcal{T}(FU')}{=}
    \left(
        \lambda x'_2.\lintranspose{\affcontext{\overleftarrow{\Phi}},x'_2}
        {x'_2}
    \right)
    \left(
       \lintranspose{\affcontext{\overleftarrow{\Phi}}}
        {f_3} \mbox{ } z''
    \right)
    \\
    &\hspace{9mm}\overset{\tiny \mathcal{T}(f)}{=}
    \left(
        \lambda x'_2.\lintranspose{\affcontext{\overleftarrow{\Phi}},x'_2}
        {x'_2}
    \right)
    \left(
       \overleftarrow{f_3} \mbox{ } z''
    \right)
    \\
    &\hspace{9mm}\overset{\tiny \mathcal{T}(u)}{=}
    \left(
        \lambda x'_2.x'_2
    \right)
    \left(
       \overleftarrow{f_3} \mbox{ } z''
    \right)
    \\
    &\hspace{12mm}{\xrightarrow{\beta_\lambda}} \quad
    \overleftarrow{f_3} \mbox{ } z''
\end{align*}

Therefore, we have that 
\begin{equation}
    \label{eq:example_transpose_initial_2}
    \begin{aligned}
        &\begin{aligned}
            \lintranspose{\affcontext{\overleftarrow{\Phi}},x'_2}
            {f_3 \mbox{ } x'_2}
        \end{aligned}
        \rightarrow
        \begin{aligned}
            \overleftarrow{f_3} \mbox{ } z''
        \end{aligned}
    \end{aligned}
\end{equation}

Therefore, by combining the equations above we have the following 
\begin{align*}
    &\lintranspose{\affcontext{\overleftarrow{\Phi}},\nTuple{x',y'}} 
    {{f_4 \nTuple{f_2 \nTuple{f_1\mbox{ }x', y'}, f_3\mbox{ }x'}}}   
    \\
    &\hspace{7mm} \overset{\text{Eq.}~\ref{eq:example_transpose_initial_0}}{=}    
    \left(\lambda \nTuple{z',z''}.\left(
       \begin{aligned}
            & \mathtt{let}\mbox{ }
            \nTuple{\RenamePat{\alpha_1}{\nTuple{x',y'}},
            \RenamePat{\alpha_2}{\nTuple{x',y'}}}
            =
            \\
            &\hspace{3mm}
            \nTuple{
                \lintranspose{\affcontext{\overleftarrow{\Phi}},\nTuple{x'_1,y'_1}}
                { f_2 \nTuple{f_1\mbox{ }x'_1, y'_1}},
                \lintranspose{\affcontext{\overleftarrow{\Phi}},x'_2}
                {f_3\mbox{ }x'_2}
            }
            \\
            &\mathtt{in}\mbox{ } \Fusion{\nTuple{x',y'}}{\alpha_1}{\alpha_2}
        \end{aligned}
    \right)\right)
    \left(
        \overleftarrow{f_4} \mbox{ }j
    \right)                      
    \\
    &\hspace{7mm} \overset{\text{Eq.}~\ref{eq:example_transpose_initial_1}}{\rightarrow^*}  
    \left(\lambda \nTuple{z',z''}.\left(
       \begin{aligned}
            & \mathtt{let}\mbox{ }
            \nTuple{\RenamePat{\alpha_1}{\nTuple{x',y'}},
            \RenamePat{\alpha_2}{\nTuple{x',y'}}}
            =
            \\
            &\hspace{3mm}
            \nTuple{
                \left(
                \lambda \nTuple{z_1,z_2}.\nTuple{\overleftarrow{f_1}\mbox{ }z_1, z_2}
                \right)
                \left(
                    \overleftarrow{f_2} \mbox{ } z'
                \right),
                \lintranspose{\affcontext{\overleftarrow{\Phi}},x'_2}
                {f_3\mbox{ }x'_2}
            }
            \\
            &\mathtt{in}\mbox{ } \Fusion{\nTuple{x',y'}}{\alpha_1}{\alpha_2}
        \end{aligned}
    \right)\right)
    \left(
        \overleftarrow{f_4} \mbox{ }j
    \right) 
    \\
    &\hspace{7mm} \overset{\text{Eq.}~\ref{eq:example_transpose_initial_2}}{\rightarrow}  
    \left(\lambda \nTuple{z',z''}.\left(
       \begin{aligned}
            & \mathtt{let}\mbox{ }
            \nTuple{\RenamePat{\alpha_1}{\nTuple{x',y'}},
            \RenamePat{\alpha_2}{\nTuple{x',y'}}}
            =
            \\
            &\hspace{3mm}
            \nTuple{
                \left(
                \lambda \nTuple{z_1,z_2}.\nTuple{\overleftarrow{f_1}\mbox{ }z_1, z_2}
                \right)
                \left(
                    \overleftarrow{f_2} \mbox{ } z'
                \right),
                \overleftarrow{f_3} \mbox{ } z''
            }
            \\
            &\mathtt{in}\mbox{ } \Fusion{\nTuple{x',y'}}{\alpha_1}{\alpha_2}
        \end{aligned}
    \right)\right)
    \left(
        \overleftarrow{f_4} \mbox{ }j
    \right)
    \\
    &\hspace{10mm} \approx 
    \letterm{\nTuple{z',z''}}{\overleftarrow{f_4} \mbox{ }j}
    {\left(
    \begin{aligned}
        (\lambda \nTuple{\RenamePat{\alpha_1}{\nTuple{x',y'}},&
          \RenamePat{\alpha_2}{\nTuple{x',y'}}}.
          \Fusion{\nTuple{x',y'}}{\alpha_1}{\alpha_2}
        )
        \\
        &\nTuple{
                    \left(
                        \lambda \nTuple{z_1,z_2}.\nTuple{\overleftarrow{f_1}\mbox{ }z_1, z_2}
                    \right)
                    \left(
                        \overleftarrow{f_2} \mbox{ } z'
                    \right),
                    \overleftarrow{f_3} \mbox{ } z''
            }    
    \end{aligned}
    \right)
    }
    \\
    &\hspace{12mm}
    \text{Recall that we have defined $\Rename_1\definedas(x'\mapsto x'_1, y'\mapsto y'_1)$ and $\Rename_2\definedas(x'\mapsto x'_2)$ so we have that}
    \\
    &\hspace{12mm}
    \begin{aligned} 
        \lambda\nTuple{\alpha_1\nTuple{\nTuple{x',y'}},\alpha_2\nTuple{\nTuple{x',y'}}}.&
        \Fusion{\nTuple{x',y'}}{\alpha_1}{\alpha_2}
        \\
        &\hspace{-1mm}\overset{\text{$\alpha_1$ and $\alpha_2$}}{=}
        \lambda\nTuple{\nTuple{x'_1,y'_2},x'_2}.
        \Fusion{\nTuple{x',y'}}{\alpha_1}{\alpha_2}
        \\
        &\hspace{1mm}\overset{\text{Def. }\nu}{=}
        \lambda\nTuple{\nTuple{x'_1,y'_2},x'_2}.
        \nTuple{\Fusion{x'}{\alpha_1}{\alpha_2},\Fusion{y'}{\alpha_1}{\alpha_2}} 
        \\
        &\hspace{1mm}\overset{\text{Def. }\nu}{=}
        \lambda\nTuple{\nTuple{x'_1,y'_2},x'_2}.
        \nTuple{\alpha_1(x')\mbox{ }\dot{+}\mbox{ }\alpha_2(x'),\alpha_1(y')} 
        \\
        &\hspace{-1mm}\overset{\text{$\alpha_1$ and $\alpha_2$}}{=}
        \lambda\nTuple{\nTuple{x'_1,y'_2},x'_2}.
        \nTuple{x'_1\mbox{ }\dot{+}\mbox{ }x'_2,y'_1} 
    \end{aligned}
    \\
    &\hspace{10mm} =
    \letterm{\nTuple{z',z''}}{\overleftarrow{f_4} \mbox{ }j}
    {\left(
    \begin{aligned}
        (\lambda\nTuple{\nTuple{x'_1,y'_2},x'_2}.&
        \nTuple{x'_1\mbox{ }\dot{+}\mbox{ }x'_2,y'_1} 
        )
        \\
        &\nTuple{
                    \left(
                        \lambda \nTuple{z_1,z_2}.\nTuple{\overleftarrow{f_1}\mbox{ }z_1, z_2}
                    \right)
                    \left(
                        \overleftarrow{f_2} \mbox{ } z'
                    \right),
                    \overleftarrow{f_3} \mbox{ } z''
            }    
    \end{aligned}
    \right)
    }
    \\
    &\hspace{10mm} \approx
    \left(\lambda \nTuple{z',z''}.\left(
       \begin{aligned}
            & \letterm{\nTuple{\nTuple{x'_1,y'_2},x'_2}}  
            {
                \nTuple{
                    \left(
                    \lambda \nTuple{z_1,z_2}.\nTuple{\overleftarrow{f_1}\mbox{ }z_1, z_2}
                    \right)
                    \left(
                        \overleftarrow{f_2} \mbox{ } z'
                    \right),
                    \overleftarrow{f_3} \mbox{ } z''
                }
            }
            {}
            \\
            &\nTuple{x'_1\mbox{ }\dot{+}\mbox{ }x'_2,y'_1} 
        \end{aligned}
    \right)\right)
    \left(
        \overleftarrow{f_4} \mbox{ }j
    \right)
    \\[3mm]
    &\hspace{10mm} \approx
    \begin{aligned}
        &\letterm{\nTuple{z',z''}}{\overleftarrow{f_4} \mbox{ }j}{}\\
        &\letterm{\nTuple{\nTuple{x'_1,y'_2},x'_2}}  
        {
            \nTuple{
                \left(
                        \lambda \nTuple{z_1,z_2}.\nTuple{\overleftarrow{f_1}\mbox{ }z_1, z_2}
                \right)
                \left(
                    \overleftarrow{f_2} \mbox{ } z'
                \right),
                \overleftarrow{f_3} \mbox{ } z''
            }
        }{}
        \\
        &\nTuple{x'_1\mbox{ }\dot{+}\mbox{ }x'_2,y'_1} 
    \end{aligned}
\end{align*}

Therefore, we have that 
\begin{equation}\label{eq:example_transpose_initial_3}
    \begin{aligned}
        &\begin{aligned}
            \lintranspose{\affcontext{\overleftarrow{\Phi}},\nTuple{x',y'}} 
            {{f_4 \nTuple{f_2 \nTuple{f_1\mbox{ }x', y'}, f_3\mbox{ }x'}}}   
        \end{aligned}
        \rightarrow^*
        \\
        &
        \hspace{3cm}\begin{aligned}
        &\letterm{\nTuple{z',z''}}{\overleftarrow{f_4} \mbox{ }j}{}\\
        &\letterm{\nTuple{\nTuple{x'_1,y'_2},x'_2}}  
        {
            \nTuple{
                \left(
                        \lambda \nTuple{z_1,z_2}.\nTuple{\overleftarrow{f_1}\mbox{ }z_1, z_2}
                \right)
                \left(
                    \overleftarrow{f_2} \mbox{ } z'
                \right),
                \overleftarrow{f_3} \mbox{ } z''
            }
        }{}
        \\
        &\nTuple{x'_1\mbox{ }\dot{+}\mbox{ }x'_2,y'_1} 
        \end{aligned}
    \end{aligned}
\end{equation}

Finally, we can conclude that the transposition of the last line can be obtained as follows
\begin{align*}
    &\mathcal{T}_{\affcontext{\overleftarrow{\Phi}}}
        \left(
            \lambda u^{\typR\&\typR}. 
            \letterm{\nTuple{x',y'}}{u}
            {f_4 \nTuple{f_2 \nTuple{f_1\mbox{ }x', y'}, f_3\mbox{ }x'}}
        \right)
    \\
    &\hspace{7mm} \overset{\text{Eq.}~\ref{eq:example_transpose_initial_-1}}{\rightarrow} 
    \lambda z. 
    \left(\lambda u.u\right)
    \left( 
        \left(
            \lambda j. 
            (\lambda \nTuple{\nTuple{x',y'}}.\nTuple{x',y'})
            \mbox{ }
            \lintranspose{\affcontext{\overleftarrow{\Phi}},\nTuple{x',y'}} {{f_4 \nTuple{f_2 \nTuple{f_1\mbox{ }x', y'}, f_3\mbox{ }x'}}}
        \right)
        z
    \right)       
    \\
    &\hspace{7mm} \xrightarrow{\beta_\lambda}
    \lambda z. 
    \left(\lambda u.u\right)
    \left( 
        \left(
            \lambda j.  
            \lintranspose{\affcontext{\overleftarrow{\Phi}},\nTuple{x',y'}} {{f_4 \nTuple{f_2 \nTuple{f_1\mbox{ }x', y'}, f_3\mbox{ }x'}}}
        \right)
        z
    \right)       
    \\
    &\hspace{7mm} \overset{\text{Eq.}~\ref{eq:example_transpose_initial_3}}{\rightarrow^*}
    \begin{aligned}
        \lambda z. 
    \left(\lambda u.u\right)
    \left(  
            \lambda j.  
            \left(
            \begin{aligned}
                &\letterm{\nTuple{z',z''}}{\overleftarrow{f_4} \mbox{ }j}{}\\
                &\letterm{\nTuple{\nTuple{x'_1,y'_2},x'_2}}  
                {
                    \nTuple{
                        \left(
                                \lambda \nTuple{z_1,z_2}.\nTuple{\overleftarrow{f_1}\mbox{ }z_1, z_2}
                        \right)
                        \left(
                            \overleftarrow{f_2} \mbox{ } z'
                        \right),
                        \overleftarrow{f_3} \mbox{ } z''
                    }
                }{}
                \\
                &\nTuple{x'_1\mbox{ }\dot{+}\mbox{ }x'_2,y'_1} 
            \end{aligned}
            \right) 
        z
    \right)
    \end{aligned}
    \\ 
    &\hspace{7mm} \xrightarrow{\beta_\lambda}
    \begin{aligned}
    \lambda z. 
    \left(\lambda u.u\right) 
        \left(
            \begin{aligned}
                &\letterm{\nTuple{z',z''}}{\overleftarrow{f_4} \mbox{ }z}{}\\
                &\letterm{\nTuple{\nTuple{x'_1,y'_2},x'_2}}  
                {
                    \nTuple{
                        \left(
                                \lambda \nTuple{z_1,z_2}.\nTuple{\overleftarrow{f_1}\mbox{ }z_1, z_2}
                        \right)
                        \left(
                            \overleftarrow{f_2} \mbox{ } z'
                        \right),
                        \overleftarrow{f_3} \mbox{ } z''
                    }
                }{}
                \\
                &\nTuple{x'_1\mbox{ }\dot{+}\mbox{ }x'_2,y'_1} 
            \end{aligned}
        \right)  
    \end{aligned}
    \\
    &\hspace{7mm} \xrightarrow{\beta_\lambda}
    \begin{aligned}
    \lambda z.  
        \left(
            \begin{aligned}
                &\letterm{\nTuple{z',z''}}{\overleftarrow{f_4} \mbox{ }z}{}\\
                &\letterm{\nTuple{\nTuple{x'_1,y'_2},x'_2}}  
                {
                    \nTuple{
                        \left(
                                \lambda \nTuple{z_1,z_2}.\nTuple{\overleftarrow{f_1}\mbox{ }z_1, z_2}
                        \right)
                        \left(
                            \overleftarrow{f_2} \mbox{ } z'
                        \right),
                        \overleftarrow{f_3} \mbox{ } z''
                    }
                }{}
                \\
                &\nTuple{x'_1\mbox{ }\dot{+}\mbox{ }x'_2,y'_1} 
            \end{aligned}
        \right)  
    \end{aligned}
\end{align*}

Therefore, we have that the transposition of the last line is
\begin{equation}
    \label{eq:example_transpose_final}
    \begin{aligned}
        &\begin{aligned}
        \mathcal{T}_{\affcontext{\overleftarrow{\Phi}}}
        \left(
            \lambda u^{\typR\&\typR}. 
            \letterm{\nTuple{x',y'}}{u}
            {f_4 \nTuple{f_2 \nTuple{f_1\mbox{ }x', y'}, f_3\mbox{ }x'}}
        \right)
        \end{aligned}
        \rightarrow^*
        \\
        &
        \hspace{1cm}\begin{aligned}
        \lambda z.  
        \left(
            \begin{aligned}
                &\letterm{\nTuple{z',z''}}{\overleftarrow{f_4} \mbox{ }z}{}\\
                &\letterm{\nTuple{\nTuple{x'_1,y'_2},x'_2}}  
                {
                    \nTuple{
                        \left(
                                \lambda \nTuple{z_1,z_2}.\nTuple{\overleftarrow{f_1}\mbox{ }z_1, z_2}
                        \right)
                        \left(
                            \overleftarrow{f_2} \mbox{ } z'
                        \right),
                        \overleftarrow{f_3} \mbox{ } z''
                    }
                }{}
                \\
                &\nTuple{x'_1\mbox{ }\dot{+}\mbox{ }x'_2,y'_1} 
            \end{aligned}
        \right)  
        \end{aligned}
    \end{aligned}
\end{equation}

Summing up, we can conclude that 
\begin{align*}
    &\mathcal{T}_{\affcontext{\overleftarrow{\Phi_0}}}
    \quad \left(
    \begin{aligned}
        &\letterm{\affinebang f_1^{\mbox{ }\typR \multimap \typR}}
        {\affinebang (\lambda u. w_1\mbox{ }\dot{*}\mbox{ }u)}{}\\
        &\letterm{\affinebang f_2^{\mbox{ }(\typR\&\typR) \multimap \typR}}
        {\affinebang (\lambda \nTuple{u_1,u_2}. (w_2\mbox{ }\dot{*}\mbox{ }u_1)
        \mbox{ }\dot{+}\mbox{ } (w_3\mbox{ }\dot{*}\mbox{ }u_2))}{}\\
        &\letterm{\affinebang f_3^{\mbox{ }\typR \multimap \typR}}
        {\affinebang (\lambda u. w_4\mbox{ }\dot{*}\mbox{ }u)}{}\\
        &\letterm{\affinebang f_4^{\mbox{ }(\typR\&\typR) \multimap \typR}}
        {\affinebang (\lambda \nTuple{u_1,u_2}. (w_5\mbox{ }\dot{*}\mbox{ }u_1)
        \mbox{ }\dot{+}\mbox{ } (w_6\mbox{ }\dot{*}\mbox{ }u_2))}{}\\
        &\lambda u^{\typR\&\typR}.
        \begin{aligned}
            &\letterm{\nTuple{x',y'}}{u}{}
            f_4 \nTuple{f_2 \nTuple{f_1\mbox{ }x', y'}, f_3\mbox{ }x'}
        \end{aligned}
    \end{aligned}
    \quad \right) 
    \\[3mm]
    &\hspace{7mm} \overset{\text{Eq.}~\ref{eq:example_transpose_initial}}{\rightarrow^*}
    \quad
    \begin{aligned}
        &\letterm{\affinebang \overleftarrow{f_1}^{\mbox{ }\typR \multimap \typR}}
        {\affinebang (\lambda {l}. w_1\mbox{ }\dot{*}\mbox{ }{l})}{}\\
        &\letterm{\affinebang \overleftarrow{f_2}^{\mbox{ } \typR \multimap (\typR\&\typR)}}
        {\affinebang (\lambda {l}.\nTuple{w_2 \mbox{ }\dot{*}\mbox{ }{l}, w_3 \mbox{ }\dot{*}\mbox{ }{l}})}{}\\
        &\letterm{\affinebang \overleftarrow{f_3}^{\mbox{ }\typR \multimap \typR}}
        {\affinebang (\lambda {l}. w_4\mbox{ }\dot{*}\mbox{ }{l})}{}
		\\
        &\letterm{\affinebang \overleftarrow{f_4}^{\mbox{ } \typR \multimap (\typR\&\typR)}}
        {\affinebang (\lambda {l}. \nTuple{w_5\mbox{}\dot{*}\mbox{ }{l},w_6\mbox{}\dot{*}\mbox{ }{l}})}{}\\
        &\mathcal{T}_{\affcontext{\overleftarrow{\Phi}}}
        \left(
            \lambda u^{\typR\&\typR}. 
            \letterm{\nTuple{x',y'}}{u}
            {f_4 \nTuple{f_2 \nTuple{f_1\mbox{ }x', y'}, f_3\mbox{ }x'}}
        \right)
    \end{aligned}
    \\[3mm]
    &\hspace{7mm} \overset{\text{Eq.}~\ref{eq:example_transpose_final}}{\rightarrow^*}
    \quad
    \begin{aligned}
        &\letterm{\affinebang \overleftarrow{f_1}^{\mbox{ }\typR \multimap \typR}}
        {\affinebang (\lambda {l}. w_1\mbox{ }\dot{*}\mbox{ }{l})}{}\\
        &\letterm{\affinebang \overleftarrow{f_2}^{\mbox{ } \typR \multimap (\typR\&\typR)}}
        {\affinebang (\lambda {l}.\nTuple{w_2 \mbox{ }\dot{*}\mbox{ }{l}, w_3 \mbox{ }\dot{*}\mbox{ }{l}})}{}\\
        &\letterm{\affinebang \overleftarrow{f_3}^{\mbox{ }\typR \multimap \typR}}
        {\affinebang (\lambda {l}. w_4\mbox{ }\dot{*}\mbox{ }{l})}{}
		\\
        &\letterm{\affinebang \overleftarrow{f_4}^{\mbox{ } \typR \multimap (\typR\&\typR)}}
        {\affinebang (\lambda {l}. \nTuple{w_5\mbox{}\dot{*}\mbox{ }{l},w_6\mbox{}\dot{*}\mbox{ }{l}})}{}\\
        &  
        \begin{aligned}
        \lambda z. 
        &\letterm{\nTuple{z',z''}}{\overleftarrow{f_4} \mbox{ }z}{}\\
        &\letterm{\nTuple{\nTuple{x'_1,y'_2},x'_2}}  
                {
                    \nTuple{
                        \left(
                                \lambda \nTuple{z_1,z_2}.\nTuple{\overleftarrow{f_1}\mbox{ }z_1, z_2}
                        \right)
                        \left(
                            \overleftarrow{f_2} \mbox{ } z'
                        \right),
                        \overleftarrow{f_3} \mbox{ } z''
                    }
                }{}
        \\ 
        &\nTuple{x'_1\mbox{ }\dot{+}\mbox{ }x'_2,y'_1} 
        \end{aligned}
    \end{aligned}
\end{align*}
where the last equation is equal to the {\color{red}red} part of Figure~\ref{fig:ex_transpose}.

\subsection{Soundness Transpose}
In order to show that the soundness property for the transpose transformation holds we need to prove the following auxiliary lemma about the soundness of the transpose transformation on tangent expressions of Linear B.
\begin{lemma}[Soundness Transpose on~\ref{linear_B:tanget}] 
	\label{lemma:transp_sound_tangent}
	Given a well-typed \ref{linear_B:tanget} expression in Linear B $\judgmentJAX{\Gamma}{\lincontextJAX{\Gamma}}{\linexprJAX{e}}{\typone}{\tautypJAX}$ and an enumeration $\theta$ for $\lincontextJAX{\Gamma}$, then 
	$\transpose{\TransB[\theta]{\linexprJAX{e}}}
	\; \sim\;
	\TransB[\linvarJAX{u}:\tautypJAX]{\transpJAX{\theta}{\linvarJAX{u}:\tautypJAX}{\linexprJAX{e}}}$.
\end{lemma}
\begin{sproof}
	By induction on $\linexprJAX{e}$. 
\end{sproof}
 
\begin{proof}[Proof of Theorem~\ref{th:transpose_JAX_LL}]
    The first equivalence is a consequence of Proposition~\ref{prop:transpose_sim} and the last equivalence is by Proposition~\ref{prop:transB}. We should then prove $\transpose{\TransB[\theta]{d}}
\; \sim\;
\TransB[\linvarJAX{u}:\tautypJAX]{\transpJAX{\theta}{\linvarJAX{u}:\tautypJAX}{d}}$
and we proceed by induction on $d$ using Lemma~\ref{lemma:transp_sound_tangent}.
\end{proof}

\subsection{Work Preservation Transpose}
Given a finite set of variables $\mathcal V$, we will write $\Cost{\mathcal V}$  for the sum $\sum_{x:A\in\mathcal V}\Cost{A}$.  In the case of a set $\affinebang\Phi$ of variables of type $f':\affinebang(L'\multimap H')$ and a term $M$, we also use the notation:
\begin{align*}
	\Cost{\affinebang\Phi^{\mathrm{in}}_M} &\definedas\sum_{f:\affinebang(L'\multimap H')\in\affinebang\Phi\cap\FV{M}}\CostType{L'}
	&
	\Cost{\affinebang\Phi^{\mathrm{out}}_M} &\definedas\sum_{f:\affinebang(L'\multimap H')\in\affinebang\Phi\cap\FV{M}}\CostType{H'}
\end{align*}
\smallskip
Work preservation of the transpose transformation in $\lambdaLL$ follows directly as a corollary of the following lemma 

\begin{lemma}
\label{lemma:cost_transp}
    We have the following:
    \begin{enumerate}
    \item if $\oc\Sigma, \affinebang\Phi, \PIn:L \vdash U: H$ and $\Rename$ is the identity renaming restricted to $\FV{\PIn}\cap\FV{U}$, then:
    \begin{multline*}
        \Cost{
            \lambda \POut.
			\letterm{\RenamePat{\alpha}{\PIn}}{\lintranspose{\affcontext{\overleftarrow{\Phi}},\PIn}{U}}
			{\Fusion{\PIn}{\alpha}{\emptyset}}
        }
        +\CostType{L}
        +\Cost{\affinebang\overleftarrow{\Phi}^{\mathrm{in}}_{\lintranspose{\affcontext{\overleftarrow{\Phi}},\PIn}{U}}} \\
		\leq
        \Cost{\lambda {\PIn}.U}
        +\CostType{H}
        +\Cost{\affinebang\overleftarrow{\Phi}^{\mathrm{out}}_{\lintranspose{\affcontext{\overleftarrow{\Phi}},\PIn}{U}}}
    \end{multline*}
    \item if $\oc\Sigma, \affinebang\Phi \vdash F: L\multimap H$, then:\\
    $$
        \Cost{\lintranspose{\affcontext{\overleftarrow{\Phi}}}{F}}+\CostType{L}+\Cost{\affinebang\overleftarrow{\Phi}^{\mathrm{in}}_{\lintranspose{\affcontext{\overleftarrow{\Phi}}}{F}}}
    \leq\Cost{F}+\CostType{H}+\Cost{\affinebang\overleftarrow{\Phi}^{\mathrm{out}}_{\lintranspose{\affcontext{\overleftarrow{\Phi}}}{F}}}
    $$
    \item if $\oc\Sigma, \affinebang\Phi\vdash R:\oc E\otimes\affinebang(L\multimap H)$, then:\\
    $$
        \Cost{\lintranspose{\affcontext{\overleftarrow{\Phi}}}{R}}+\CostType{L}+\Cost{\affinebang\overleftarrow{\Phi}^{\mathrm{in}}_{\lintranspose{\affcontext{\overleftarrow{\Phi}}}{R}}}
    \leq\Cost{R}+\CostType{H}+\Cost{\affinebang\overleftarrow{\Phi}^{\mathrm{out}}_{\lintranspose{\affcontext{\overleftarrow{\Phi}}}{R}}}
    $$
    \end{enumerate}
\end{lemma}

\begin{sproof}[Proof Claim 1: Cases of $\mathcal{T}$ on $\lambdaLL^{\mathtt t}$]
	By typing of $\lambdaLL^{\mathtt t}$ we have that a term $U\in \lambdaLL^{\mathtt t}$ is well-typed as: $\oc\Sigma, \affinebang\Phi, \PIn:L \vdash U: H$, so we are in the first case of the lemma.

	Observe that $\FV{\POut}\cap\FV{\Fusion \PIn {\Rename}{\emptyset}}=\emptyset$ so we have that 
	\begin{align*}
		\lambda \POut.&
		\letterm{\RenamePat{\alpha}{\PIn}}{\lintranspose{\affcontext{\overleftarrow{\Phi}},\PIn}{U}}
		{\Fusion{\PIn}{\alpha}{\emptyset}}
		\\
		&=
		\letterm{\RenamePat{\alpha}{\PIn}}
		{\lambda \POut.\lintranspose{\affcontext{\overleftarrow{\Phi}},\PIn}{U}}
		{\Fusion{\PIn}{\alpha}{\emptyset}}
		\\
		&\approx
		(\lambda \RenamePat{\alpha}{\PIn}.\Fusion{\PIn}{\alpha}{\emptyset})
		(\lambda \POut.\lintranspose{\affcontext{\overleftarrow{\Phi}},\PIn}{U})
	\end{align*}

	Therefore, in terms of workload, we have that
	\begin{align*}
		\Cost{
			\lambda \POut.&
			\letterm{\RenamePat{\alpha}{\PIn}}{\lintranspose{\affcontext{\overleftarrow{\Phi}},\PIn}{U}}
			{\Fusion{\PIn}{\alpha}{\emptyset}}
		}
		\\
		&=\Cost{
			(\lambda \RenamePat{\alpha}{\PIn}.\Fusion{\PIn}{\alpha}{\emptyset})
		(\lambda \POut.\lintranspose{\affcontext{\overleftarrow{\Phi}},\PIn}{U})
		}
		\\
		&=\Cost{\lambda \RenamePat{\alpha}{\PIn}.\Fusion{\PIn}{\alpha}{\emptyset}}
		+\Cost{\lambda \POut.\lintranspose{\affcontext{\overleftarrow{\Phi}},\PIn}{U}}
		\\
		&=\Cost{\Fusion{\PIn}{\alpha}{\emptyset}}+\sum_{x:A\in\FV{\RenamePat{\alpha}{\PIn}}\setminus\FV{\Fusion{\PIn}{\alpha}{\emptyset}}} 	\CostType{A}
		+\Cost{\lambda \POut.\lintranspose{\affcontext{\overleftarrow{\Phi}},\PIn}{U}}
		\\
		&\hspace{-4mm}\overset{\text{Lemma}~\ref{lemma:cost_fusion_renaming}}{=}
		\Cost{\Dom{\alpha}\cap\Dom{\emptyset}\cap\FV{\PIn}}
		\\&\hspace{15mm} +\sum_{x:A\in\FV{\RenamePat{\alpha}{\PIn}}\setminus\FV{\Fusion{\PIn}{\alpha}{\emptyset}}} 	\CostType{A}
		+\Cost{\lambda \POut.\lintranspose{\affcontext{\overleftarrow{\Phi}},\PIn}{U}}
		\\
		&=
		\Cost{\emptyset}
		+\sum_{x:A\in \emptyset} \CostType{A}
		+\Cost{\lambda \POut.\lintranspose{\affcontext{\overleftarrow{\Phi}},\PIn}{U}}
		\\
		&= \Cost{\lambda \POut.\lintranspose{\affcontext{\overleftarrow{\Phi}},\PIn}{U}}
	\end{align*} 

	Summing up, in this case of the lemma it is enough to prove that 
	$$\Cost{\lambda \POut.\lintranspose{\affcontext{\overleftarrow{\Phi}},\PIn}{U}}
	+\CostType{L}
	+\Cost{\affinebang\overleftarrow{\Phi}^{\mathrm{in}}_{\lintranspose{\affcontext{\overleftarrow{\Phi}},\PIn}{U}}} \leq
	\Cost{\lambda {\PIn}.U}
	+\CostType{H}
	+\Cost{\affinebang\overleftarrow{\Phi}^{\mathrm{out}}_{\lintranspose{\affcontext{\overleftarrow{\Phi}},\PIn}{U}}}$$
	and we proceed by analyzing the cases in Figure~\ref{subfigure:transpose on lambdaLL_t}.  
    \begin{itemize}
    \item Case $U=\nTuple{U_1,U_2}$:\\
    In this case we have that $H=H_1\&H_2$ and $\POut=\nTuple{\POut_1,\POut_2}$. By hypothesis we have $\oc\Sigma, \affinebang\Phi, \PIn:L \vdash \nTuple{U_1,U_2}: H_1\&H_2$, so by $\&$- typing rule we have $\oc\Sigma, \affinebang\Phi, \PIn:L \vdash U_i: H_i$.

    Moreover, by item 1 of Lemma~\ref{lemma:typing_weak} we have $\oc\Sigma, \affinebang\Phi, \RenameTerm\Rename{\PIn}:L \vdash \RenameTerm\Rename{U_i}: H_i$.

    By inductive hypothesis on $U_i$ we have
    $$
    \Cost{\lambda \POut_i.\lintranspose{\affcontext{\overleftarrow{\Phi}},\PIn}{U_i}}
    +\CostType{L}
    +\Cost{\affinebang\overleftarrow{\Phi}^{\mathrm{in}}_{\lintranspose{\affinebang\Phi,\PIn}{U_i}}} \leq
    \Cost{\lambda {\RenameTerm\Rename{\PIn}}.\RenameTerm\Rename{U_i}}
    +\CostType{H_i}
    +\Cost{\affinebang\overleftarrow{\Phi}^{\mathrm{out}}_{\lintranspose{\affinebang\Phi,\PIn}{U_i}}}
    $$
    We have:
    \begin{align}
        &\Cost{\lambda \POut.\lintranspose{\affcontext{\overleftarrow{\Phi}},\PIn}{U}}
        =\Cost{\lambda \nTuple{\POut_1,\POut_2}.\lintranspose{\affcontext{\overleftarrow{\Phi}},\PIn}{\nTuple{U_1,U_2}}} \notag \\
        &=\label{eq:Wtransp_tuple_Wpout_1}
        \mathcal{W}\left(\lambda \nTuple{\POut_1,\POut_2}.\left(
            \begin{aligned}
                &\letterm{\nTuple{\RenamePat{\Rename_1}\PatAddA, \RenamePat{\Rename_2}\PatAddA}}
                {\nTuple{
                        \lintranspose{\affcontext{\overleftarrow{\Phi}},\RenameTerm{\Rename_1}{\PIn}}{\RenameTerm{\Rename_1}{U_1}},
                        \lintranspose{\affcontext{\overleftarrow{\Phi}},\RenameTerm{\Rename_2}\PIn}{\RenameTerm{\Rename_2}{U_2}}
                    }
                }{}\\
                & \Fusion{\PIn}{\Rename_1}{\Rename_2}    
            \end{aligned} 
        \right)\right)
        \\
        &\approx \label{eq:Wtransp_tuple_Wpout_2}
        \mathcal{W}\left(\lambda \nTuple{\POut_1,\POut_2}.\left(
            \begin{aligned}
                &\left(\lambda \nTuple{\RenamePat{\Rename_1}\PatAddA, \RenamePat{\Rename_2}\PatAddA}.\Fusion{\PIn}{\Rename_1}{\Rename_2} \right)
                \\
                &\hspace{15mm}\nTuple{
                        \lintranspose{\affcontext{\overleftarrow{\Phi}},\RenameTerm{\Rename_1}{\PIn}}{\RenameTerm{\Rename_1}{U_1}},
                        \lintranspose{\affcontext{\overleftarrow{\Phi}},\RenameTerm{\Rename_2}\PIn}{\RenameTerm{\Rename_2}{U_2}}
                }    
            \end{aligned} 
        \right)\right)
        \\
        &=\label{eq:Wtransp_tuple_Wpout_3}
        \mathcal{W}\left( 
            \begin{aligned}
                &\left(\lambda \nTuple{\RenamePat{\Rename_1}\PatAddA, \RenamePat{\Rename_2}\PatAddA}.\Fusion{\PIn}{\Rename_1}{\Rename_2} \right)
                \\
                &\hspace{15mm}\nTuple{
                        \lambda \POut_1.
                        \lintranspose{\affcontext{\overleftarrow{\Phi}},\RenameTerm{\Rename_1}{\PIn}}{\RenameTerm{\Rename_1}{U_1}},
                        \lambda \POut_2.
                        \lintranspose{\affcontext{\overleftarrow{\Phi}},\RenameTerm{\Rename_2}\PIn}{\RenameTerm{\Rename_2}{U_2}}
                }    
            \end{aligned}  
        \right)
        \\
        &=\label{eq:Wtransp_tuple_Wpout_4}
        \Cost{\lambda \nTuple{\RenamePat{\Rename_1}\PatAddA, \RenamePat{\Rename_2}\PatAddA}.\Fusion{\PIn}{\Rename_1}{\Rename_2}}
        \\
        &\hspace{14mm}\notag
        +\Cost{
            \nTuple{
                        \lambda \POut_1.
                        \lintranspose{\affcontext{\overleftarrow{\Phi}},\RenameTerm{\Rename_1}{\PIn}}{\RenameTerm{\Rename_1}{U_1}},
                        \lambda \POut_2.
                        \lintranspose{\affcontext{\overleftarrow{\Phi}},\RenameTerm{\Rename_2}\PIn}{\RenameTerm{\Rename_2}{U_2}}
            }
        }
        \\ 
         &=\label{eq:Wtransp_tuple_Wpout_5}
        \Cost{\lambda \nTuple{\RenamePat{\Rename_1}\PatAddA, \RenamePat{\Rename_2}\PatAddA}.\Fusion{\PIn}{\Rename_1}{\Rename_2}}
        +\sum^2_{i=1} \Cost{\lambda \POut_i.\lintranspose{\affcontext{\overleftarrow{\Phi}},\PIn}}
        \\ 
        &=\label{eq:Wtransp_tuple_Wpout_6}
        \Cost{\FV{U_1}\cap\FV{U_2}\cap\FV{\PIn}}
        +\sum^2_{i=1} \Cost{\lambda \POut_i.\lintranspose{\affcontext{\overleftarrow{\Phi}},\PIn}{U_i}} 
    \end{align}
    where:
    \begin{itemize}
        \item The passage from line~\eqref{eq:Wtransp_tuple_Wpout_1} to line~\eqref{eq:Wtransp_tuple_Wpout_2} is by syntactic.
        \item The passage from line~\eqref{eq:Wtransp_tuple_Wpout_2} to line~\eqref{eq:Wtransp_tuple_Wpout_3} is because by typing we know that $\FV{\POut_i}\cap\FV{\Fusion{\PIn}{\alpha_1}{\alpha_2}}=\emptyset$ with $i\in\{1,2\}$ and $\FV{\POut_{3-i}}\cap \FV{\lambda \POut_i.\lintranspose{\affcontext{\overleftarrow{\Phi}},\RenameTerm{\Rename_i}{\PIn}}{ \RenameTerm{\Rename_i}{U_i}}}=\emptyset$.
        \item The passage from line~\eqref{eq:Wtransp_tuple_Wpout_3} (resp.~\eqref{eq:Wtransp_tuple_Wpout_4}) to line~\eqref{eq:Wtransp_tuple_Wpout_4} (resp.~\eqref{eq:Wtransp_tuple_Wpout_5}) is by definition of workload.
        \item The passage from line~\eqref{eq:Wtransp_tuple_Wpout_2} to line~\eqref{eq:Wtransp_tuple_Wpout_3} is obtained by applying item 3 of Lemma~\ref{lemma:typing_weak} which states that $\Cost{\RenameTerm\Rename M}=\Cost{M}$.
        \item Recall that $\Dom{\Rename_i}=\FV{U_i}\cap\FV{\PIn}$, thus the passage from line~\eqref{eq:Wtransp_tuple_Wpout_5} to line~\eqref{eq:Wtransp_tuple_Wpout_6} is by Lemma~\ref{lemma:cost_fusion_renaming} obtaining that $\Cost{\Dom{\Rename_1}\cap\Dom{\Rename_2}\cap\FV{p}}=\Cost{\FV{U_1}\cap\FV{U_2}\cap\FV{\PIn}}$. More precisely, $\Cost{\FV{U_1}\cap\FV{U_2}\cap\FV{\PIn}}$ is the number of sums performed by $\Fusion \PIn{\Rename_1}{\Rename_2}$.
    \end{itemize} 
    We have also that
    \begin{align*}
        \Cost{\lambda \PIn.U}
        &= \Cost{\lambda \PIn.\nTuple{U_1,U_2}}\\
        &= \Cost{\FV{\PIn}\setminus(\FV{U_1}\cup\FV{U_2})}+\sum^2_{i=1} \Cost{U_i}
    \end{align*}
    where the last line is obtained by applying the definition of workload for the lambda abstraction.

    Finally, we show that 
    $$\Cost{\lambda \POut.\lintranspose{\affcontext{\overleftarrow{\Phi}},\PIn}{U}}
    +\CostType{L}
    +\Cost{\affinebang\overleftarrow{\Phi}^{\mathrm{in}}_{\lintranspose{\affinebang\Phi,\PIn}{U}}} \leq
    \Cost{\lambda {\PIn}.U}
    +\CostType{H}
    +\Cost{\affinebang\overleftarrow{\Phi}^{\mathrm{out}}_{\lintranspose{\affinebang\Phi,\PIn}{U}}}$$ 
    by using the following remark.
    \begin{remark}\label{rk:Wtranspose}
        
        We need to analyze the quantity $\Cost{\FV{U_1}\cap\FV{U_2}\cap\FV{\PIn}}
        +\Cost{\FV{\PIn}\setminus\FV{U_1}}
        +\Cost{\FV{\PIn}\setminus\FV{U_2}}$.

        \def\firstcircle{(-0.5,-1) circle (1cm)}
\def\secondcircle{(0,0) circle (1cm)}
\def\thirdcircle{(0.5,-1) circle (1cm)}

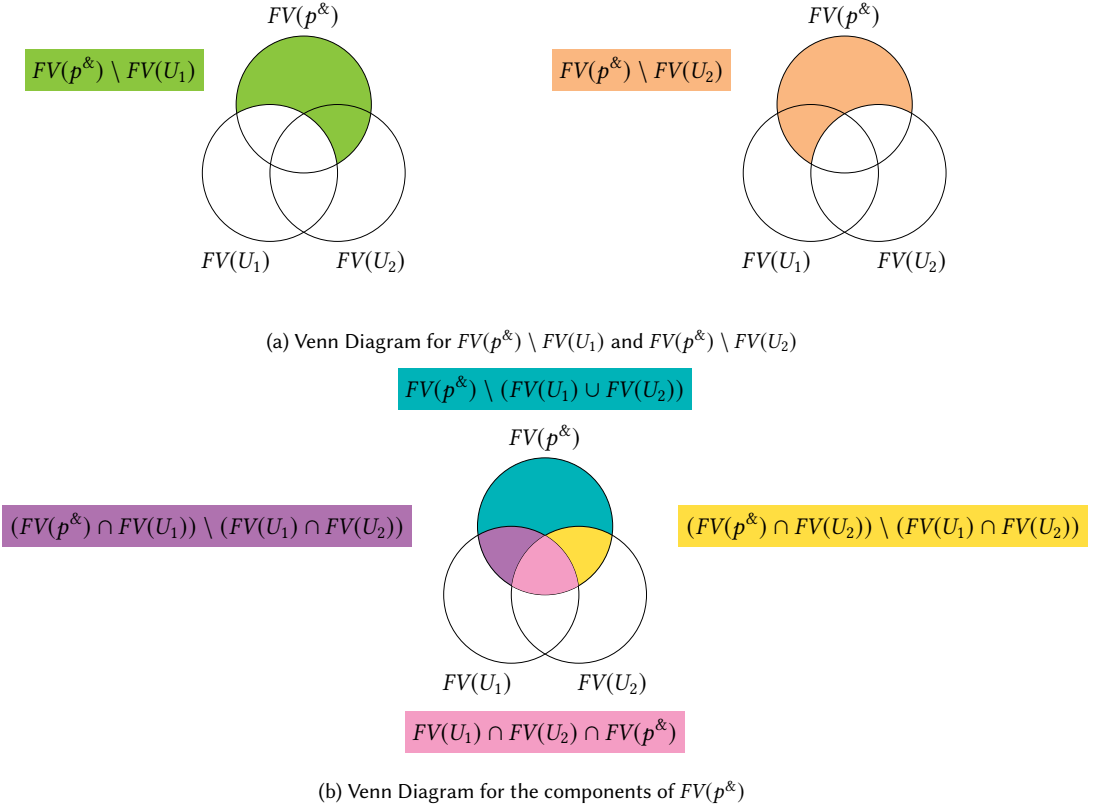
\begin{figure}[h!]
    \vspace{-15mm}
    \scalebox{.9}{\parbox{\textwidth+2cm}{
    \centering 
    \begin{subfigure}{1\linewidth}
        \begin{center}
            \begin{tikzpicture}     
                \node at (4.2,0.5) {$\myhighlight{LimeGreen}{\FV{\PIn}\setminus\FV{U_1}}$};
                \node at (12,0.5) {$\myhighlight{Apricot}{\FV{\PIn}\setminus\FV{U_2}}$};

                    \begin{scope}[shift={(7cm,0cm)}]
                        \begin{scope}[even odd rule] 
                            \clip \firstcircle (-3,-3) rectangle (3,3);
                        \fill[LimeGreen] \secondcircle;
                        \end{scope}
                        \draw \firstcircle node [black,yshift=-1.3cm,xshift=-0.5cm] {$\FV{U_1}$};
                        \draw \secondcircle node [black,yshift=1.3cm] {$\FV{\PIn}$};
                        \draw \thirdcircle node [black,yshift=-1.3cm,xshift=0.5cm] {$\FV{U_2}$};
                    \end{scope}

                \begin{scope}[shift={(15cm,0cm)}]
                    \begin{scope}[even odd rule] 
                        \clip \thirdcircle (-3,-3) rectangle (3,3);
                    \fill[Apricot] \secondcircle;
                    \end{scope}
                    \draw \firstcircle node [black,yshift=-1.3cm,xshift=-0.5cm] {$\FV{U_1}$};
                    \draw \secondcircle node [black,yshift=1.3cm] {$\FV{\PIn}$};
                    \draw \thirdcircle node [black,yshift=-1.3cm,xshift=0.5cm] {$\FV{U_2}$};
                \end{scope}
            
            \end{tikzpicture} 
        \end{center}
        
        \caption{Venn Diagram for $\FV{\PIn}\setminus\FV{U_1}$ and $\FV{\PIn}\setminus\FV{U_2}$}
         \label{subfig:vennDiagramWPT_1}
    \end{subfigure}

    \begin{subfigure}{1\linewidth}
    \begin{center}
        \begin{tikzpicture}
            \node at (0,2) {$\myhighlight{BlueGreen}{\FV{\PIn}\setminus(\FV{U_1}\cup\FV{U_2})}$};
            \node at (-5,0) {$\myhighlight{Orchid}{(\FV{\PIn}\cap\FV{U_1})\setminus(\FV{U_1}\cap\FV{U_2})}$};
            \node at (5,0) {$\myhighlight{Goldenrod}{(\FV{\PIn}\cap\FV{U_2})\setminus(\FV{U_1}\cap\FV{U_2})}$};
            \node at (0,-3) {$\myhighlight{Lavender}{\FV{U_1}\cap\FV{U_2}\cap\FV{\PIn}}$};
        
            \begin{scope}  
                    \fill[BlueGreen] \secondcircle;
            \end{scope}
        
            \begin{scope}
                \clip \thirdcircle;
                    \fill[Goldenrod] \secondcircle;
            \end{scope}
        
            \begin{scope}
                \begin{scope}
                    \clip \firstcircle;
                    \fill[Orchid] \secondcircle;
                \end{scope}
                \draw \firstcircle node [black,yshift=-1.3cm,xshift=-0.5cm] {$\FV{U_1}$};
                \draw \secondcircle node [black,yshift=1.3cm] {$\FV{\PIn}$};
                \draw \thirdcircle node [black,yshift=-1.3cm,xshift=0.5cm] {$\FV{U_2}$}; 
            \end{scope}
        
            \begin{scope}
                \clip \firstcircle;
                \clip \secondcircle;
                \fill[Lavender] \thirdcircle;
            \end{scope}
        
        \end{tikzpicture}
    \end{center}
    
        \caption{Venn Diagram for the components of $\FV{\PIn}$}
        \label{subfig:vennDiagramWPT_2}
    \end{subfigure}
    }}
    \caption{Venn Diagram for Work Preservation Transpose}
    \label{fig:vennDiagramWPT}
\end{figure}

        Let us consider the colours in Figure~\ref{fig:vennDiagramWPT}.

        The quantity we are analyzing is the following
        $$\Cost{\myhighlight{Lavender}{\FV{U_1}\cap\FV{U_2}\cap\FV{\PIn}}}
        +\Cost{\myhighlight{LimeGreen}{\FV{\PIn}\setminus\FV{U_1}}}
        +\Cost{\myhighlight{Apricot}{\FV{\PIn}\setminus\FV{U_2}}}.$$
        By observing the figure we have that:
        \begin{align}
            \label{eq:colorsTranspose_1}
            \myhighlight{LimeGreen}{\FV{\PIn}\setminus\FV{U_1}}&=
            \myhighlight{BlueGreen}{\FV{\PIn}\setminus(\FV{U_1}\cup\FV{U_2})}
            \\
            & \hspace{1.5cm}\cup
            \myhighlight{Goldenrod}{(\FV{\PIn}\cap\FV{U_2})\setminus(\FV{U_1}\cap\FV{U_2})}
            \notag\\[2mm]
            \label{eq:colorsTranspose_2}
            \myhighlight{Apricot}{\FV{\PIn}\setminus\FV{U_2}}&=
            \myhighlight{BlueGreen}{\FV{\PIn}\setminus(\FV{U_1}\cup\FV{U_2})}
            \\
            & \hspace{1.5cm}
            \cup
            \myhighlight{Orchid}{(\FV{\PIn}\cap\FV{U_1})\setminus(\FV{U_1}\cap\FV{U_2})}
            \notag \\[2mm]
            \FV{\PIn}&=
            \myhighlight{BlueGreen}{\FV{\PIn}\setminus(\FV{U_1}\cup\FV{U_2})}
            \cup
            \myhighlight{Lavender}{\FV{U_1}\cap\FV{U_2}\cap\FV{\PIn}}
            \cup \notag\\
            &\hspace{3mm}\cup
            \myhighlight{Orchid}{(\FV{\PIn}\cap\FV{U_1})\setminus(\FV{U_1}\cap\FV{U_2})}
            \label{eq:colorsTranspose_3}\\
            &\hspace{3mm}\cup
            \myhighlight{Goldenrod}{(\FV{\PIn}\cap\FV{U_2})\setminus(\FV{U_1}\cap\FV{U_2})}\notag
        \end{align}

        \medskip
        We have:
        \begin{align*}
        &\Cost{\myhighlight{Lavender}{\FV{U_1}\cap\FV{U_2}\cap\FV{\PIn}}}
        +\Cost{\myhighlight{LimeGreen}{\FV{\PIn}\setminus\FV{U_1}}}
        +\Cost{\myhighlight{Apricot}{\FV{\PIn}\setminus\FV{U_2}}}\\[3mm]
        &\hspace{5mm}\overset{\tiny \text{Eq.}~\ref{eq:colorsTranspose_1},~\ref{eq:colorsTranspose_2}}{=} 
        \Cost{\myhighlight{BlueGreen}{\FV{\PIn}\setminus(\FV{U_1}\cup\FV{U_2})}}+\\
        &\hspace{2cm}
        +\Cost{\myhighlight{BlueGreen}{\FV{\PIn}\setminus(\FV{U_1}\cup\FV{U_2})}}
        +\Cost{\myhighlight{Lavender}{\FV{U_1}\cap\FV{U_2}\cap\FV{\PIn}}}+\\
        &\hspace{2cm}
        +\Cost{\myhighlight{Orchid}{(\FV{\PIn}\cap\FV{U_1})\setminus(\FV{U_1}\cap\FV{U_2})}}+
        \\ &\hspace{2cm}
        +\Cost{\myhighlight{Goldenrod}{(\FV{\PIn}\cap\FV{U_2})\setminus(\FV{U_1}\cap\FV{U_2})}}\\[3mm]
        &\hspace{7mm}\overset{\tiny \text{Eq.}~\ref{eq:colorsTranspose_3}}{=} 
        \Cost{\myhighlight{BlueGreen}{\FV{\PIn}\setminus(\FV{U_1}\cup\FV{U_2})}}+\Cost{\FV{\PIn}}
        \end{align*}
        Moreover, if $\FV{\PIn}$ is of type $L$ then we can conclude that the analyzed quantity is equal to
        $$\Cost{\FV{\PIn}\setminus(\FV{U_1}\cup\FV{U_2})} + \Cost{L}$$
    \end{remark}
    More precisely, we proceed as follows
    \begin{align*}
        \Cost{&\lambda \POut.\lintranspose{\affcontext{\overleftarrow{\Phi}},\PIn}{U}}
        +\CostType{L}
        +\Cost{\affinebang\overleftarrow{\Phi}^{\mathrm{in}}_{\lintranspose{\affinebang\Phi,\PIn}{U}}}\\
        &=\Cost{\FV{U_1}\cap\FV{U_2}\cap\FV{\PIn}}
        +\sum^2_{i=1} \left( \Cost{\lambda \POut_i.\lintranspose{\affcontext{\overleftarrow{\Phi}},\PIn}{U_i}}\right) 
        \\
        &\hspace{2cm}
        +\CostType{L}
        +\Cost{\affinebang\overleftarrow{\Phi}^{\mathrm{in}}_{\lintranspose{\affinebang\Phi,\PIn}{U}}}\\
        &=\Cost{\FV{U_1}\cap\FV{U_2}\cap\FV{\PIn}}
        +\CostType{L}
        \\
        &\hspace{2cm}
        +\sum^2_{i=1} \Cost{\lambda \POut_i.\lintranspose{\affcontext{\overleftarrow{\Phi}},\PIn}{U_i}} 
        +\Cost{\affinebang\overleftarrow{\Phi}^{\mathrm{in}}_{\lintranspose{\affinebang\Phi,\PIn}{U_i}}}\\
        &\hspace{-1mm}\overset{\tiny \text{IHs}}{\leq} 
        \Cost{\FV{U_1}\cap\FV{U_2}\cap\FV{\PIn}}
        -\Cost{L}
        \\
        &\hspace{2cm}
        +\sum^2_{i=1} \Cost{\lambda {\RenameTerm\Rename{\PIn}}.\RenameTerm\Rename{U_i}}
        +\CostType{H_i}
        +\Cost{\affinebang\overleftarrow{\Phi}^{\mathrm{out}}_{\lintranspose{\affinebang\Phi,\PIn}{U_i}}} \\
        &\hspace{-5.2mm}\overset{\tiny \text{Lemma}~\ref{lemma:typing_weak}}{=} 
        \Cost{\FV{U_1}\cap\FV{U_2}\cap\FV{\PIn}}
        -\Cost{L}
        \\
        &\hspace{2cm}
        +\sum^2_{i=1} \Cost{\lambda {\PIn}.U_i}
        +\CostType{H_i}
        +\Cost{\affinebang\overleftarrow{\Phi}^{\mathrm{out}}_{\lintranspose{\affinebang\Phi,\PIn}{U_i}}}\\
        &=\Cost{\FV{U_1}\cap\FV{U_2}\cap\FV{\PIn}}\\
        &\hspace{2cm}-\Cost{L}
        +\sum^2_{i=1} \Cost{U_i}+\Cost{\FV{\PIn}\setminus\FV{U_i}}
        +\CostType{H_i}
        +\Cost{\affinebang\overleftarrow{\Phi}^{\mathrm{out}}_{\lintranspose{\affinebang\Phi,\PIn}{U_i}}}\\
        &=\Cost{\FV{U_1}\cap\FV{U_2}\cap\FV{\PIn}}
        +\Cost{\FV{\PIn}\setminus\FV{U_1}}
        +\Cost{\FV{\PIn}\setminus\FV{U_2}}\\
        &\hspace{2cm}
        -\Cost{L}
        +\sum^2_{i=1} \Cost{U_i}
        +\CostType{H_i}
        +\Cost{\affinebang\overleftarrow{\Phi}^{\mathrm{out}}_{\lintranspose{\affinebang\Phi,\PIn}{U_i}}}\\
        &\hspace{-4.2mm}\overset{\tiny \text{Remark}~\ref{rk:Wtranspose}}{=}
        \Cost{\FV{\PIn}\setminus(\FV{U_1}\cup\FV{U_2})} + \Cost{L} \\
        &\hspace{2cm}
        -\Cost{L}
        +\sum^2_{i=1} \Cost{U_i}
        +\CostType{H_i}
        +\Cost{\affinebang\overleftarrow{\Phi}^{\mathrm{out}}_{\lintranspose{\affinebang\Phi,\PIn}{U_i}}}\\
        &=\Cost{\FV{\PIn}\setminus(\FV{U_1}\cup\FV{U_2})} 
        +\sum^2_{i=1} \Cost{U_i}
        +\CostType{H_i}
        +\Cost{\affinebang\overleftarrow{\Phi}^{\mathrm{out}}_{\lintranspose{\affinebang\Phi,\PIn}{U_i}}}\\
        &=
        \Cost{\lambda \PIn.U} + \CostType{H}
        +\Cost{\affinebang\overleftarrow{\Phi}^{\mathrm{out}}_{\lintranspose{\affinebang\Phi,\PIn}{U}}}
    \end{align*}
    and so we can conclude.
    
    \item Case $U=FU'$:\\
    By hypothesis we have $\oc\Sigma, \affinebang\Phi, \PIn:L \vdash FU': H$, so by typing we have
    \begin{align*}
        \affinebang\Phi &\vdash F :L_0 \multimap H\\
        \oc\Sigma, \affinebang\Phi, \PIn:L &\vdash U': L_0
    \end{align*}
    By inductive hypothesis on $F$ (case 2 of the lemma) we have
    $$\Cost{\lintranspose{\affinebang\Phi}{F}}+\CostType{L}+\Cost{\affinebang\overleftarrow{\Phi}^{\mathrm{in}}_{\lintranspose{\affinebang\Phi}{F}}}
    \leq\Cost{F}+\CostType{H}+\Cost{\affinebang\overleftarrow{\Phi}^{\mathrm{out}}_{\lintranspose{\affinebang\Phi}{F}}}$$
    By inductive hypothesis on $U'$ (case 1 of the lemma) we have
    $$\Cost{\lambda \POut.\lintranspose{\affcontext{\overleftarrow{\Phi}},\PIn}{U'}}
        +\CostType{L}
        +\Cost{\affinebang\overleftarrow{\Phi}^{\mathrm{in}}_{\lintranspose{\affinebang\Phi,\PIn}{U'}}} \leq
        \Cost{\lambda {\PIn}.U'}
        +\CostType{L_0}
        +\Cost{\affinebang\overleftarrow{\Phi}^{\mathrm{out}}_{\lintranspose{\affinebang\Phi,\PIn}{U'}}}
    $$
    We have:
    \begin{align*}
        \Cost{\lambda \POut.\lintranspose{\affcontext{\overleftarrow{\Phi}},\PIn}{U}}
        &=\Cost{\lambda \POut.\lintranspose{\affcontext{\overleftarrow{\Phi}},\PIn}{FU'}}\\
        &=\Cost{(\lambda{\POut}'.\lintranspose{\affcontext{\overleftarrow{\Phi}},\PIn}{U'})
        (\lintranspose{\affcontext{\overleftarrow{\Phi}}}{F}\POut)}\\
        &=\Cost{\lambda{\POut}'.\lintranspose{\affcontext{\overleftarrow{\Phi}},\PIn}{U'}}
        +\Cost{\lintranspose{\affcontext{\overleftarrow{\Phi}}}{F}}
        \\[3mm]
        \Cost{\lambda {\PIn}.U}
        &=\Cost{\lambda {\PIn}.FU'}\\
        &=\Cost{F(\lambda {\PIn}.U')}=\Cost{F}+\Cost{\lambda {\PIn}.U'}
    \end{align*}
    where the last line is obtained by observing that by typing we have that $\PIn$ is free only in $U'$.
   
    We show that 
    $$\Cost{\lambda \POut.\lintranspose{\affcontext{\overleftarrow{\Phi}},\PIn}{U}}
    +\CostType{L}
    +\Cost{\affinebang\overleftarrow{\Phi}^{\mathrm{in}}_{\lintranspose{\affinebang\Phi,\PIn}{U}}} \leq
    \Cost{\lambda {\PIn}.U}
    +\CostType{H}
    +\Cost{\affinebang\overleftarrow{\Phi}^{\mathrm{out}}_{\lintranspose{\affinebang\Phi,\PIn}{U}}}$$ 
    as follows
    \begin{align*}
        \Cost{&\lambda \POut.\lintranspose{\affcontext{\overleftarrow{\Phi}},\PIn}{FU'}}
        +\CostType{L}
        +\Cost{\affinebang\overleftarrow{\Phi}^{\mathrm{in}}_{\lintranspose{\affinebang\Phi,\PIn}{U}}}\\
        &=\Cost{\lambda{\POut}'.\lintranspose{\affcontext{\overleftarrow{\Phi}},\PIn}{U'}}
        +\Cost{\lintranspose{\affcontext{\overleftarrow{\Phi}}}{F}}+\CostType{L}
        +\Cost{\affinebang\overleftarrow{\Phi}^{\mathrm{in}}_{\lintranspose{\affinebang\Phi,\PIn}{U'}}}+\Cost{\affinebang\overleftarrow{\Phi}^{\mathrm{in}}_{\lintranspose{\affinebang\Phi}{F}}}\\
        &\hspace{-4mm}\overset{\tiny \text{IH on U'}}{\leq}
        \Cost{\lambda {\PIn}.U'}
        +\CostType{L_0}
        +\Cost{\affinebang\overleftarrow{\Phi}^{\mathrm{out}}_{\lintranspose{\affinebang\Phi,\PIn}{U'}}}
        +\Cost{\lintranspose{\affcontext{\overleftarrow{\Phi}}}{F}}+\Cost{\affinebang\overleftarrow{\Phi}^{\mathrm{in}}_{\lintranspose{\affinebang\Phi}{F}}}\\
        &\hspace{-4mm}\overset{\tiny \text{IH on F}}{\leq}
        \Cost{\lambda {\PIn}.U'} 
        +\Cost{\affinebang\overleftarrow{\Phi}^{\mathrm{out}}_{\lintranspose{\affinebang\Phi,\PIn}{U'}}}
        +\Cost{F}+\CostType{H}+\Cost{\affinebang\overleftarrow{\Phi}^{\mathrm{out}}_{\lintranspose{\affinebang\Phi}{F}}}\\
        &=
        \Cost{\lambda {\PIn}.U'} 
        +\Cost{F}+\CostType{H}
        +\Cost{\affinebang\overleftarrow{\Phi}^{\mathrm{out}}_{\lintranspose{\affinebang\Phi,\PIn}{U}}}\\
        &=\Cost{\lambda {\PIn}.U}
        +\CostType{H}
        +\Cost{\affinebang\overleftarrow{\Phi}^{\mathrm{out}}_{\lintranspose{\affinebang\Phi,\PIn}{U}}}
    \end{align*}
    so we can conclude.
    \item Case $U=u$:\\
    Observe that by typing 
    \begin{equation}\label{eq:Wtransp_var}
        L=H, \quad 
        \Cost{\affinebang\overleftarrow{\Phi}^{\mathrm{in}}_{\lintranspose{\affinebang\Phi,\PIn}{u}}}=\Cost{\affinebang\overleftarrow{\Phi}^{\mathrm{out}}_{\lintranspose{\affinebang\Phi,\PIn}{u}}}=0 \quad \text{ and } \quad \POut=\PIn=u.
    \end{equation}
    We have: 
    \begin{align*}
        \Cost{&\lambda \POut.\lintranspose{\affcontext{\overleftarrow{\Phi}},\PIn}{u}}
        +\CostType{L}
        +\Cost{\affinebang\overleftarrow{\Phi}^{\mathrm{in}}_{\lintranspose{\affinebang\Phi,\PIn}{U}}}\\
        &=\Cost{\lambda \POut.u}
        +\CostType{L}
        +\Cost{\affinebang\overleftarrow{\Phi}^{\mathrm{in}}_{\lintranspose{\affinebang\Phi,\PIn}{U}}}\\
        &\hspace{-4mm}\overset{\tiny \text{Eq.}~\ref{eq:Wtransp_var}}{=}
        \Cost{\lambda u.u}
        +\CostType{L}\\
        &=\CostType{L}\\
        &\leq
        \Cost{\lambda {\PIn}.u}
        +\CostType{H}
        +\Cost{\affinebang\overleftarrow{\Phi}^{\mathrm{out}}_{\lintranspose{\affinebang\Phi,\PIn}{U}}}\\
        &\hspace{-2mm}\overset{\tiny \text{Eq.}~\ref{eq:Wtransp_var}}{=}
        \Cost{\lambda u.u}+\CostType{H}\\
        &=\CostType{H}
    \end{align*}
    and we can conclude because by typing we know that $L=H$.
    \item Case $U=\zeroterm{}$:\\
    By typing $H=\mathbb{R}$, so $\POut$ is of type $\mathbb{R}$.
    We have:
    \begin{align*}
        \Cost{\lambda \POut.\lintranspose{\affcontext{\overleftarrow{\Phi}},\PIn}{U}}
        &=\Cost{\lambda \POut.\lintranspose{\affcontext{\overleftarrow{\Phi}},\PIn}{\zeroterm{}}}
        =\Cost{\lambda \POut.\linemptytupleterm}=\Cost{\typR}=1\\[3mm]
        \Cost{\lambda {\PIn}.U}
        &=\Cost{\lambda {\PIn}.\zeroterm{}}=\Cost{\zeroterm{}}+\Cost{L}=\Cost{L}
    \end{align*}
    By definition we have that 
    \begin{equation}\label{eq:Wtransp_zero}
        \Cost{\affinebang\overleftarrow{\Phi}^{\mathrm{in}}_{\lintranspose{\affinebang\Phi,\PIn}{\zeroterm{}}}}
        =\Cost{\affinebang\overleftarrow{\Phi}^{\mathrm{out}}_{\lintranspose{\affinebang\Phi,\PIn}{\zeroterm{}}}}
        =0
    \end{equation}
    because $\lintranspose{\affinebang\Phi,\PIn}{\zeroterm{}}$ has only free variables of ground type.

    We have:
    \begin{align*}
        \Cost{&\lambda \POut.\lintranspose{\affcontext{\overleftarrow{\Phi}},\PIn}{\zeroterm{}}}
        +\CostType{L}
        +\Cost{\affinebang\overleftarrow{\Phi}^{\mathrm{in}}_{\lintranspose{\affinebang\Phi,\PIn}{\zeroterm{}}}}\\
        &=1+\CostType{L}+\Cost{\affinebang\overleftarrow{\Phi}^{\mathrm{in}}_{\lintranspose{\affinebang\Phi,\PIn}{\zeroterm{}}}}\\
        &\hspace{-4mm}\overset{\tiny \text{Eq.}~\ref{eq:Wtransp_zero}}{=}
        1+\CostType{L}\\
        &\leq 
        \Cost{\lambda {\PIn}.\zeroterm{}}
        +\CostType{H}
        +\Cost{\affinebang\overleftarrow{\Phi}^{\mathrm{out}}_{\lintranspose{\affinebang\Phi,\PIn}{\zeroterm{}}}}\\
        &=\Cost{L}
        +\CostType{H}
        +\Cost{\affinebang\overleftarrow{\Phi}^{\mathrm{out}}_{\lintranspose{\affinebang\Phi,\PIn}{\zeroterm{}}}}\\
        &\hspace{-4mm}\overset{\tiny \text{Eq.}~\ref{eq:Wtransp_zero}}{=}
        \Cost{L}
        +\CostType{H}\\
        &=\Cost{L}
        +\CostType{\typR}\\
        &=\Cost{L}+1
    \end{align*}

    \item Case $U=\linemptytupleterm$:\\
    By typing $H=\typtop$, so $\POut$ is of type $\typtop$.
    We have:
    \begin{align*}
        \Cost{\lambda \POut.\lintranspose{\affcontext{\overleftarrow{\Phi}},\PIn}{U}}
        &=\Cost{\lambda \POut.\lintranspose{\affcontext{\overleftarrow{\Phi}},\PIn}{\linemptytupleterm}}
        =\Cost{\lambda \POut.\linemptytupleterm}=\Cost{\typtop}=0\\[3mm]
        \Cost{\lambda {\PIn}.U}
        &=\Cost{\lambda {\PIn}.\linemptytupleterm}=\Cost{\linemptytupleterm}+\Cost{L}=\Cost{L}
    \end{align*}
    By definition we have that 
    \begin{equation}\label{eq:Wtransp_emptytuple}
        \Cost{\affinebang\overleftarrow{\Phi}^{\mathrm{in}}_{\lintranspose{\affinebang\Phi,\PIn}{\linemptytupleterm}}}
        =\Cost{\affinebang\overleftarrow{\Phi}^{\mathrm{out}}_{\lintranspose{\affinebang\Phi,\PIn}{\linemptytupleterm}}}
        =0
    \end{equation}
    because $\lintranspose{\affinebang\Phi,\PIn}{\linemptytupleterm}$ has only free variables of ground type.

    We have:
    \begin{align*}
        \Cost{&\lambda \POut.\lintranspose{\affcontext{\overleftarrow{\Phi}},\PIn}{\linemptytupleterm}}
        +\CostType{L}
        +\Cost{\affinebang\overleftarrow{\Phi}^{\mathrm{in}}_{\lintranspose{\affinebang\Phi,\PIn}{\linemptytupleterm}}}\\
        &=0+\CostType{L}+\Cost{\affinebang\overleftarrow{\Phi}^{\mathrm{in}}_{\lintranspose{\affinebang\Phi,\PIn}{\linemptytupleterm}}}\\
        &\hspace{-4mm}\overset{\tiny \text{Eq.}~\ref{eq:Wtransp_emptytuple}}{=}
        \CostType{L}\\
        &\leq 
        \Cost{\lambda {\PIn}.\linemptytupleterm}
        +\CostType{H}
        +\Cost{\affinebang\overleftarrow{\Phi}^{\mathrm{out}}_{\lintranspose{\affinebang\Phi,\PIn}{\linemptytupleterm}}}\\
        &=\Cost{L}
        +\CostType{H}
        +\Cost{\affinebang\overleftarrow{\Phi}^{\mathrm{out}}_{\lintranspose{\affinebang\Phi,\PIn}{\linemptytupleterm}}}\\
        &\hspace{-4mm}\overset{\tiny \text{Eq.}~\ref{eq:Wtransp_emptytuple}}{=}
        \Cost{L}
        +\CostType{H}\\
        &=\Cost{L}
        +\CostType{\typtop}\\
        &=\Cost{L} 
    \end{align*}
\end{itemize}
\end{sproof}

\begin{sproof}[Proof Claim 2: Cases of $\mathcal{T}$ on $\lambdaLL^{\mathtt f}$]
	By typing of $\lambdaLL^{\mathtt f}$ we have that a term $F\in \lambdaLL^{\mathtt f}$ is well-typed as: $\oc\Sigma, \affinebang\Phi \vdash F: L\multimap H$, so we are in the second case of the lemma and we want to prove that:
	$$
	\Cost{\lintranspose{\affcontext{\overleftarrow{\Phi}}}{F}}+\CostType{L}+\Cost{\affinebang\overleftarrow{\Phi}^{\mathrm{in}}_{\lintranspose{\affcontext{\overleftarrow{\Phi}}}{F}}}
	\leq\Cost{F}+\CostType{H}+\Cost{\affinebang\overleftarrow{\Phi}^{\mathrm{out}}_{\lintranspose{\affcontext{\overleftarrow{\Phi}}}{F}}}.
	$$
	We proceed by analyzing the cases in Figure~\ref{subfigure:transpose on lambdaLL_f}.
    \begin{itemize}
    \item Case $F=\lambda \PIn.U$:\\
    By hypothesis we have $\oc\Sigma, \affinebang\Phi \vdash \lambda \PIn.U: L\multimap H$, so by typing we have $\oc\Sigma, \affinebang\Phi, \PIn:L \vdash U: H$
    
    By inductive hypothesis on $U$ (case 1 of the lemma) we have
    \begin{multline*}
        \Cost{
            \lambda \POut.
			\letterm{\RenamePat{\alpha}{\PIn}}{\lintranspose{\affcontext{\overleftarrow{\Phi}},\PIn}{U}}
			{\Fusion{\PIn}{\alpha}{\emptyset}}
        }
        +\CostType{L}
        +\Cost{\affinebang\overleftarrow{\Phi}^{\mathrm{in}}_{\lintranspose{\affcontext{\overleftarrow{\Phi}},\PIn}{U}}} \\
		\leq
        \Cost{\lambda {\PIn}.U}
        +\CostType{H}
        +\Cost{\affinebang\overleftarrow{\Phi}^{\mathrm{out}}_{\lintranspose{\affcontext{\overleftarrow{\Phi}},\PIn}{U}}}
    \end{multline*}
    We have:
    \begin{align*}
        \Cost{\lintranspose{\affcontext{\overleftarrow{\Phi}}}{F}}
        &=\Cost{\lintranspose{\affcontext{\overleftarrow{\Phi}}}{\lambda \PIn.U}}
        =\Cost{
            \lambda \POut.
			\letterm{\RenamePat{\alpha}{\PIn}}{\lintranspose{\affcontext{\overleftarrow{\Phi}},\PIn}{U}}
			{\Fusion{\PIn}{\alpha}{\emptyset}}
        }\\[3mm]
        \Cost{F}&=\Cost{\lambda \PIn.U}
    \end{align*}

    We show that 
    $$
    \Cost{\lintranspose{\affcontext{\overleftarrow{\Phi}}}{\lambda \PIn.U}}+\CostType{L}+\Cost{\affinebang\overleftarrow{\Phi}^{\mathrm{in}}_{\lintranspose{\affcontext{\overleftarrow{\Phi}}}{\lambda \PIn.U}}}
    \leq\Cost{F}+\CostType{H}+\Cost{\affinebang\overleftarrow{\Phi}^{\mathrm{out}}_{\lintranspose{\affcontext{\overleftarrow{\Phi}}}{\lambda \PIn.U}}}
    $$
    as follows
    \begin{align*}
        \Cost{&\lintranspose{\affcontext{\overleftarrow{\Phi}}}{\lambda \PIn.U}}+\CostType{L}+\Cost{\affinebang\overleftarrow{\Phi}^{\mathrm{in}}_{\lintranspose{\affcontext{\overleftarrow{\Phi}}}{\lambda \PIn.U}}}\\
        &=\Cost{
            \lambda \POut.
			\letterm{\RenamePat{\alpha}{\PIn}}{\lintranspose{\affcontext{\overleftarrow{\Phi}},\PIn}{U}}
			{\Fusion{\PIn}{\alpha}{\emptyset}}
        }
        +\CostType{L}
        +\Cost{\affinebang\overleftarrow{\Phi}^{\mathrm{in}}_{\lintranspose{\affcontext{\overleftarrow{\Phi}},\PIn}{U}}}\\
        &\hspace{-4mm}\overset{\tiny \text{IH on U}}{\leq}
        \Cost{\lambda {\PIn}.U}
        +\CostType{H}
        +\Cost{\affinebang\overleftarrow{\Phi}^{\mathrm{out}}_{\lintranspose{\affcontext{\overleftarrow{\Phi}},\PIn}{U}}}
    \end{align*}
    and so we can conclude because in this case $\affinebang\overleftarrow{\Phi}^{\mathrm{out}}_{\lintranspose{\affcontext{\overleftarrow{\Phi}}}{\lambda \PIn.U}}=\affinebang\overleftarrow{\Phi}^{\mathrm{out}}_{\lintranspose{\affcontext{\overleftarrow{\Phi}},\PIn}{U}}$.
    
    \item Case $F=\letterm{\affinebang f^{\affinebang(L_0\multimap H_0)}}{\affinebang G_1}{G_2}$:\\
    By hypothesis we have $\oc\Sigma, \affinebang\Phi \vdash \letterm{\affinebang f^{\affinebang(L_0\multimap H_0)}}{\affinebang G_1}{G_2}: L\multimap H$, so by typing we have:

    \begin{align*}
        \oc\Sigma, \affinebang\Phi &\vdash  G_1: L_0\multimap H_0\\
        \oc\Sigma, \affinebang\Phi,f: L_0\multimap H_0 &\vdash  G_2: L\multimap H
    \end{align*}
    Observe that $f$ may not be free in $G_2$, so we have to analyze the following subcases:
    \begin{itemize}
        \item Subcase $f\in\FV{G_2}$:\\
        By inductive hypothesis on $G_1$ (case 2 of the lemma) we have
        $$
        \Cost{\lintranspose{\affcontext{\overleftarrow{\Phi}}}{G_1}}+\CostType{L_0}+\Cost{\affinebang\overleftarrow{\Phi}^{\mathrm{in}}_{\lintranspose{\affcontext{\overleftarrow{\Phi}}}{G_1}}}
    	\leq\Cost{F}+\CostType{H_0}+\Cost{\affinebang\overleftarrow{\Phi}^{\mathrm{out}}_{\lintranspose{\affcontext{\overleftarrow{\Phi}}}{G_1}}}.
    	$$
        By inductive hypothesis on $G_2$ (case 2 of the lemma) we have
        \begin{align*}
            \Cost{\lintranspose{\affcontext{\overleftarrow{\Phi}},\affinebang \overleftarrow{f}}{G_2}}+&\CostType{L}+\Cost{\affinebang\overleftarrow{\Phi}^{\mathrm{in}}_{\lintranspose{\affcontext{\overleftarrow{\Phi}},\affinebang \overleftarrow{f}}{G_2}}}+\myhighlight{Apricot}{\CostType{H_0}}
            \\
            &\leq\Cost{F}+\CostType{H}+\Cost{\affinebang\overleftarrow{\Phi}^{\mathrm{out}}_{\lintranspose{\affcontext{\overleftarrow{\Phi}},\affinebang \overleftarrow{f}}{G_2}}}+\myhighlight{Apricot}{\CostType{L_0}}\\
            &\hspace{-1cm}\text{\color{Apricot} \text{because}  $\overleftarrow f: \affinebang (H_0\multimap L_0) \in\FV{\lintranspose{\affcontext{\overleftarrow{\Phi}},\affinebang \overleftarrow{f}}{G_2}}$}
        \end{align*} 
        We have:
        \begin{align*}
            \Cost{\lintranspose{\affcontext{\overleftarrow{\Phi}}}{F}}
            &=\Cost{\lintranspose{\affcontext{\overleftarrow{\Phi}}}{\letterm{\affinebang f^{\affinebang(L_0\multimap H_0)}}{\affinebang G_1}{G_2}}}\\
            &=\Cost{\letterm{\overleftarrow{f}^{\affinebang(H_0\multimap L_0)}}{\affinebang \lintranspose{\affcontext{\overleftarrow{\Phi}},\affinebang \overleftarrow{f}}{G_2}}{\lintranspose{\affcontext{\overleftarrow{\Phi}}}{G_1}}}\\
            &\approx\Cost{(\lambda f.\lintranspose{\affcontext{\overleftarrow{\Phi}}}{G_1})\affinebang \lintranspose{\affcontext{\overleftarrow{\Phi}},\affinebang \overleftarrow{f}}{G_2}}\\
            &=\Cost{\lintranspose{\affcontext{\overleftarrow{\Phi}}}{G_1}}+\Cost{\affinebang \lintranspose{\affcontext{\overleftarrow{\Phi}},\affinebang \overleftarrow{f}}{G_2}}\\
            &\hspace{-5mm}\overset{\tiny \text{Lemma~\ref{lemma:cost_affine}}}{=}
            \Cost{\lintranspose{\affcontext{\overleftarrow{\Phi}}}{G_1}}+\Cost{\lintranspose{\affcontext{\overleftarrow{\Phi}},\affinebang \overleftarrow{f}}{G_2}}
            \\[3mm]
            \Cost{F}
            &=\Cost{\letterm{\affinebang f^{\affinebang(L_0\multimap H_0)}}{\affinebang G_1}{G_2}}\\
            &=\Cost{\affinebang G_1}+\Cost{G_2}\\
            &\hspace{-5mm}\overset{\tiny \text{Lemma~\ref{lemma:cost_affine}}}{=}
            \Cost{G_1}+\Cost{G_2}
        \end{align*} 

        We can conclude as follows
        \begin{align*}
            \Cost{\lintranspose{\affcontext{\overleftarrow{\Phi}}}{F}}&+\CostType{L}+\Cost{\affinebang\overleftarrow{\Phi}^{\mathrm{in}}_{\lintranspose{\affcontext{\overleftarrow{\Phi}}}{F}}}\\
            &=
            \Cost{\lintranspose{\affcontext{\overleftarrow{\Phi}}}{G_1}}
            +\Cost{\lintranspose{\affcontext{\overleftarrow{\Phi}},\affinebang \overleftarrow{f}}{G_2}}
            +\CostType{L}
            +\Cost{\affinebang\overleftarrow{\Phi}^{\mathrm{in}}_{\lintranspose{\affcontext{\overleftarrow{\Phi}}}{F}}}\\
            &=
            \Cost{\lintranspose{\affcontext{\overleftarrow{\Phi}}}{G_1}}
            +\underline{\Cost{\lintranspose{\affcontext{\overleftarrow{\Phi}},\affinebang \overleftarrow{f}}{G_2}}}
            +\underline{\CostType{L}}
            \\
            &\hspace{2cm}
            +\Cost{\affinebang\overleftarrow{\Phi}^{\mathrm{in}}_{\lintranspose{\affcontext{\overleftarrow{\Phi}}}{G_1}}}
            +\underline{\Cost{\affinebang\overleftarrow{\Phi}^{\mathrm{in}}_{\lintranspose{\affcontext{\overleftarrow{\Phi}},\affinebang \overleftarrow{f}}{G_2}}}}\\
            &\hspace{-4mm}\overset{\tiny \text{IH on $G_2$}}{\leq}
            \underline{\Cost{\lintranspose{\affcontext{\overleftarrow{\Phi}}}{G_1}}}
            +\underline{\Cost{\affinebang\overleftarrow{\Phi}^{\mathrm{in}}_{\lintranspose{\affcontext{\overleftarrow{\Phi}}}{G_1}}}}
            -\myhighlight{Apricot}{\CostType{H_0}}+
            \\
            &\hspace{2cm}
            +\Cost{G_2}+\CostType{H}+\Cost{\affinebang\overleftarrow{\Phi}^{\mathrm{out}}_{\lintranspose{\affcontext{\overleftarrow{\Phi}},\affinebang \overleftarrow{f}}{G_2}}}+\underline{\myhighlight{Apricot}{\CostType{L_0}}}\\
            &\hspace{-4mm}\overset{\tiny \text{IH on $G_1$}}{\leq}
            \Cost{G_1}+\CostType{H_0}+\Cost{\affinebang\overleftarrow{\Phi}^{\mathrm{out}}_{\lintranspose{\affcontext{\overleftarrow{\Phi}}}{G_1}}}
            -\myhighlight{Apricot}{\CostType{H_0}}
            \\
            &\hspace{2cm}
            +\Cost{G_2}+\CostType{H}+\Cost{\affinebang\overleftarrow{\Phi}^{\mathrm{out}}_{\lintranspose{\affcontext{\overleftarrow{\Phi}},\affinebang \overleftarrow{f}}{G_2}}} \\
            &=\Cost{G_1} +\Cost{\affinebang\overleftarrow{\Phi}^{\mathrm{out}}_{\lintranspose{\affcontext{\overleftarrow{\Phi}}}{G_1}}}
            +\Cost{G_2}+\CostType{H}+\Cost{\affinebang\overleftarrow{\Phi}^{\mathrm{out}}_{\lintranspose{\affcontext{\overleftarrow{\Phi}},\affinebang \overleftarrow{f}}{G_2}}} \\
            &=\Cost{G_1}+\Cost{G_2}+\Cost{H}+\Cost{\affinebang\overleftarrow{\Phi}^{\mathrm{out}}_{\lintranspose{\affcontext{\overleftarrow{\Phi}}}{F}}}\\
            &\leq\Cost{F}+\CostType{H}+\Cost{\affinebang\overleftarrow{\Phi}^{\mathrm{out}}_{\lintranspose{\affcontext{\overleftarrow{\Phi}}}{F}}}\\
            &=\Cost{G_1}+\Cost{G_2}+\Cost{H}+\Cost{\affinebang\overleftarrow{\Phi}^{\mathrm{out}}_{\lintranspose{\affcontext{\overleftarrow{\Phi}}}{F}}}\\
        \end{align*}

        \item Subcase $f\notin\FV{G_2}$:\\
        By typing we have $\oc\Sigma, \affinebang\Phi \vdash  G_2: L\multimap H$.

        By inductive hypothesis on $G_2$ (case 2 of the lemma) we have
        \begin{align*}
            \Cost{\lintranspose{\affcontext{\overleftarrow{\Phi}}}{G_2}}+\CostType{L}+\Cost{\affinebang\overleftarrow{\Phi}^{\mathrm{in}}_{\lintranspose{\affcontext{\overleftarrow{\Phi}}}{G_2}}}
    	    \leq\Cost{F}+\CostType{H}+\Cost{\affinebang\overleftarrow{\Phi}^{\mathrm{out}}_{\lintranspose{\affcontext{\overleftarrow{\Phi}}}{G_2}}}
        \end{align*}  

        We have:
        \begin{align*}
            \Cost{\lintranspose{\affcontext{\overleftarrow{\Phi}}}{F}}
            &=\Cost{\lintranspose{\affcontext{\overleftarrow{\Phi}}}{\letterm{\affinebang f^{\affinebang(L_0\multimap H_0)}}{\affinebang G_1}{G_2}}}
            = \Cost{\lintranspose{\affcontext{\overleftarrow{\Phi}}}{G_2}}
            \\[3mm]
            \Cost{F}
            &=\Cost{\letterm{\affinebang f^{\affinebang(L_0\multimap H_0)}}{\affinebang G_1}{G_2}}\\
            &\approx \Cost{(\lambda \affinebang f.G_2)\affinebang G_1}\\
            &=\Cost{\affinebang G_1}+\Cost{G_2}+\Cost{\FV{f}\setminus \FV{G_2}}\\
            &=\Cost{\affinebang G_1}+\Cost{G_2}+\Cost{f}\\
            &=\Cost{\affinebang G_1}+\Cost{G_2}+\Cost{\affinebang(L_0\multimap H_0)}\\
            &\approx\Cost{\affinebang G_1}+\Cost{G_2}+\Cost{\typone\&(L_0\multimap H_0)}\\
            &=\Cost{\affinebang G_1}+\Cost{G_2}+\Cost{\typone}+\Cost{L_0}+\Cost{H_0}\\
            &=\Cost{\affinebang G_1}+\Cost{G_2}+\Cost{L_0}+\Cost{H_0}\\
            &\hspace{-5mm}\overset{\tiny \text{Lemma~\ref{lemma:cost_affine}}}{=}
            \Cost{G_1}+\Cost{G_2}+\Cost{L_0}+\Cost{H_0}\\
        \end{align*} 

        Observe that in this case we have
        \begin{equation}\label{eq:WtranspF_let2}
            \affinebang\overleftarrow{\Phi}^{\mathrm{in}}_{\lintranspose{\affcontext{\overleftarrow{\Phi}}}{F}}
            =\affinebang\overleftarrow{\Phi}^{\mathrm{in}}_{\lintranspose{\affcontext{\overleftarrow{\Phi}}}{G_2}}
            \quad \text{and} \quad
            \affinebang\overleftarrow{\Phi}^{\mathrm{out}}_{\lintranspose{\affcontext{\overleftarrow{\Phi}}}{F}}=\affinebang\overleftarrow{\Phi}^{\mathrm{out}}_{\lintranspose{\affcontext{\overleftarrow{\Phi}}}{G_2}} 
        \end{equation}
        because $\lintranspose{\affcontext{\overleftarrow{\Phi}}}{F}=\lintranspose{\affcontext{\overleftarrow{\Phi}}}{G_2}$.

        We can conclude as follows
        \begin{align*}
            \Cost{\lintranspose{\affcontext{\overleftarrow{\Phi}}}{F}}&+\CostType{L}+\Cost{\affinebang\overleftarrow{\Phi}^{\mathrm{in}}_{\lintranspose{\affcontext{\overleftarrow{\Phi}}}{F}}}\\
            &=\Cost{\lintranspose{\affcontext{\overleftarrow{\Phi}}}{G_2}}+\CostType{L}+\Cost{\affinebang\overleftarrow{\Phi}^{\mathrm{in}}_{\lintranspose{\affcontext{\overleftarrow{\Phi}}}{F}}}\\
            &\hspace{-3mm}\overset{\tiny \text{Eq.}~\ref{eq:WtranspF_let2}}{=}
            \Cost{\lintranspose{\affcontext{\overleftarrow{\Phi}}}{G_2}}
            +\CostType{L}
            +\Cost{\affinebang\overleftarrow{\Phi}^{\mathrm{in}}_{\lintranspose{\affcontext{\overleftarrow{\Phi}}}{G_2}}}\\
            &\hspace{-4mm}\overset{\tiny \text{IH on $G_2$}}{\leq}
            \Cost{G_2}
            +\CostType{H}
            +\Cost{\affinebang\overleftarrow{\Phi}^{\mathrm{out}}_{\lintranspose{\affcontext{\overleftarrow{\Phi}}}{G_2}}}\\
            &\leq\Cost{F}+\CostType{H}+\Cost{\affinebang\overleftarrow{\Phi}^{\mathrm{out}}_{\lintranspose{\affcontext{\overleftarrow{\Phi}}}{F}}}\\
            &=\Cost{G_1}+\Cost{G_2}+\Cost{L_0}+\Cost{H_0}+\Cost{H}+\Cost{\affinebang\overleftarrow{\Phi}^{\mathrm{out}}_{\lintranspose{\affcontext{\overleftarrow{\Phi}}}{F}}}\\
            &\hspace{-3mm}\overset{\tiny \text{Eq.}~\ref{eq:WtranspF_let2}}{=}
            \Cost{G_1}+\Cost{G_2}+\Cost{L_0}+\Cost{H_0}+\Cost{H}+\Cost{\affinebang\overleftarrow{\Phi}^{\mathrm{out}}_{\lintranspose{\affcontext{\overleftarrow{\Phi}}}{G_2}}}
        \end{align*}
    \end{itemize}

\end{itemize}  
\end{sproof}

\begin{sproof}[Proof Claim 3: Cases of $\mathcal{T}$ on $\lambdaLL^{\mathtt A}$]
	By typing of $\lambdaLL^{\mathtt A}$ we have that a term $R\in \lambdaLL^{\mathtt A}$ is well-typed as: $\oc\Sigma, \affinebang\Phi \vdash R: \oc E \otimes \affinebang(L\multimap H)$, so we are in the third case of the lemma and we want to prove that:
	$$
	\Cost{\lintranspose{\affcontext{\overleftarrow{\Phi}}}{R}}+\CostType{L}+\Cost{\affinebang\overleftarrow{\Phi}^{\mathrm{in}}_{\lintranspose{\affcontext{\overleftarrow{\Phi}}}{R}}}
	\leq\Cost{R}+\CostType{H}+\Cost{\affinebang\overleftarrow{\Phi}^{\mathrm{out}}_{\lintranspose{\affcontext{\overleftarrow{\Phi}}}{R}}}
	$$
	We proceed by analyzing the cases in Figure~\ref{subfigure:transpose on lambdaLL_a}. 
    \begin{itemize}
    \item Case $R=\letterm{\affinebang f^{\affinebang(L_0\multimap H_0)}}{\affinebang F}{S}$:\\
    By hypothesis we have $\oc\Sigma, \affinebang\Phi \vdash \letterm{\affinebang f^{\affinebang(L_0\multimap H_0)}}{\affinebang F}{S}: \oc E \otimes \affinebang(L\multimap H)$ and by typing we have:
    \begin{align*}
        \oc\Sigma, \affinebang\Phi &\vdash F: L_0 \multimap H_0\\
        \oc\Sigma, \affinebang\Phi, f:\affinebang(L_0\multimap H_0) &\vdash S: \oc E \otimes \affinebang(L\multimap H)
    \end{align*}
    Observe that $f$ may not be free in $S$, so we have to analyze the following subcases (similar to what we did in the case of composition of $\lambdaLL^{\mathtt f}$):
    \begin{itemize}
        \item Subcase $f\in\FV{S}$:\\
        By inductive hypothesis on $F$ (case 2 of the lemma) we have:
        $$
        \Cost{\lintranspose{\affcontext{\overleftarrow{\Phi}}}{F}}+\CostType{L}+\Cost{\affinebang\overleftarrow{\Phi}^{\mathrm{in}}_{\lintranspose{\affcontext{\overleftarrow{\Phi}}}{F}}}
    	\leq\Cost{F}+\CostType{H}+\Cost{\affinebang\overleftarrow{\Phi}^{\mathrm{out}}_{\lintranspose{\affcontext{\overleftarrow{\Phi}}}{F}}}.
    	$$
        By inductive hypothesis on $S$ (case 3 of the lemma) we have:
        \begin{align*}
            \Cost{\lintranspose{\affcontext{\overleftarrow{\Phi}},\affinebang \overleftarrow{f}}{S}}+\CostType{L}+\Cost{\affinebang\overleftarrow{\Phi}^{\mathrm{in}}_{\lintranspose{\affcontext{\overleftarrow{\Phi}},\affinebang \overleftarrow{f}}{S}}}
            &+\myhighlight{Apricot}{\Cost{H_0}}
            \\
            &\hspace{-1cm}
    	    \leq\Cost{S}+\CostType{H}+\Cost{\affinebang\overleftarrow{\Phi}^{\mathrm{out}}_{\lintranspose{\affcontext{\overleftarrow{\Phi}},\affinebang \overleftarrow{f}}{S}}}
            +\myhighlight{Apricot}{\Cost{L_0}}\\
            &\hspace{-2cm}\text{\color{Apricot} because $\overleftarrow f: \affinebang (H_0\multimap L_0) \in\FV{\lintranspose{\affcontext{\overleftarrow{\Phi}},\affinebang \overleftarrow{f}}{S}}$}
        \end{align*}
        We have: 
        \begin{align*}
            \Cost{\lintranspose{\affcontext{\overleftarrow{\Phi}}}{R}}
            &=\Cost{\lintranspose{\affcontext{\overleftarrow{\Phi}}}{\letterm{\affinebang f^{\affinebang(L_0\multimap H_0)}}{\affinebang F}{S}}}\\
            &=\Cost{\letterm{\overleftarrow{f}^{\affinebang(H_0\multimap L_0)}}{\affinebang \lintranspose{\affcontext{\overleftarrow{\Phi}},\affinebang \overleftarrow{f}}{S}}{\lintranspose{\affcontext{\overleftarrow{\Phi}}}{F}}}\\
            &\approx\Cost{(\lambda f.\lintranspose{\affcontext{\overleftarrow{\Phi}}}{F})\affinebang \lintranspose{\affcontext{\overleftarrow{\Phi}},\affinebang \overleftarrow{f}}{S}}\\
            &=\Cost{\lintranspose{\affcontext{\overleftarrow{\Phi}}}{F}}+\Cost{\affinebang \lintranspose{\affcontext{\overleftarrow{\Phi}},\affinebang \overleftarrow{f}}{S}}\\
            &\hspace{-5mm}\overset{\tiny \text{Lemma~\ref{lemma:cost_affine}}}{=}
            \Cost{\lintranspose{\affcontext{\overleftarrow{\Phi}}}{F}}+\Cost{\lintranspose{\affcontext{\overleftarrow{\Phi}},\affinebang \overleftarrow{f}}{S}}
            \\[3mm]
            \Cost{R}
            &=\Cost{\letterm{\affinebang f^{\affinebang(L_0\multimap H_0)}}{\affinebang F}{S}}\\
            &=\Cost{\affinebang F}+\Cost{S}\\
            &\hspace{-5mm}\overset{\tiny \text{Lemma~\ref{lemma:cost_affine}}}{=}
            \Cost{F}+\Cost{S}
        \end{align*} 
        We can conclude as follows
        \begin{align*}
            \Cost{\lintranspose{\affcontext{\overleftarrow{\Phi}}}{R}}&+\CostType{L}+\Cost{\affinebang\overleftarrow{\Phi}^{\mathrm{in}}_{\lintranspose{\affcontext{\overleftarrow{\Phi}}}{R}}}\\
            &=
            \Cost{\lintranspose{\affcontext{\overleftarrow{\Phi}}}{F}}
            +\Cost{\lintranspose{\affcontext{\overleftarrow{\Phi}},\affinebang \overleftarrow{f}}{S}}
            +\CostType{L}
            +\Cost{\affinebang\overleftarrow{\Phi}^{\mathrm{in}}_{\lintranspose{\affcontext{\overleftarrow{\Phi}}}{R}}}\\
            &=
            \Cost{\lintranspose{\affcontext{\overleftarrow{\Phi}}}{F}}
            +\underline{\Cost{\lintranspose{\affcontext{\overleftarrow{\Phi}},\affinebang \overleftarrow{f}}{S}}}
            +\underline{\CostType{L}}
            \\
            &\hspace{2cm}
            +\Cost{\affinebang\overleftarrow{\Phi}^{\mathrm{in}}_{\lintranspose{\affcontext{\overleftarrow{\Phi}}}{F}}}
            +\underline{\Cost{\affinebang\overleftarrow{\Phi}^{\mathrm{in}}_{\lintranspose{\affcontext{\overleftarrow{\Phi}},\affinebang \overleftarrow{f}}{S}}}}\\
            &\hspace{-4mm}\overset{\tiny \text{IH on $S$}}{\leq}
            \underline{\Cost{\lintranspose{\affcontext{\overleftarrow{\Phi}}}{F}}}
            +\underline{\Cost{\affinebang\overleftarrow{\Phi}^{\mathrm{in}}_{\lintranspose{\affcontext{\overleftarrow{\Phi}}}{F}}}}
            -\myhighlight{Apricot}{\CostType{H_0}}+
            \\
            &\hspace{2cm}
            +\Cost{S}+\CostType{H}+\Cost{\affinebang\overleftarrow{\Phi}^{\mathrm{out}}_{\lintranspose{\affcontext{\overleftarrow{\Phi}},\affinebang \overleftarrow{f}}{S}}}+\underline{\myhighlight{Apricot}{\CostType{L_0}}}\\
            &\hspace{-4mm}\overset{\tiny \text{IH on $F$}}{\leq}
            \Cost{F}+\CostType{H_0}+\Cost{\affinebang\overleftarrow{\Phi}^{\mathrm{out}}_{\lintranspose{\affcontext{\overleftarrow{\Phi}}}{F}}}
            -\myhighlight{Apricot}{\CostType{H_0}}
            +\Cost{S}+\CostType{H}
            \\
            &\hspace{2cm}
            +\Cost{\affinebang\overleftarrow{\Phi}^{\mathrm{out}}_{\lintranspose{\affcontext{\overleftarrow{\Phi}},\affinebang \overleftarrow{f}}{S}}} \\
            &=\Cost{F} +\Cost{\affinebang\overleftarrow{\Phi}^{\mathrm{out}}_{\lintranspose{\affcontext{\overleftarrow{\Phi}}}{F}}}
            +\Cost{S}+\CostType{H}+\Cost{\affinebang\overleftarrow{\Phi}^{\mathrm{out}}_{\lintranspose{\affcontext{\overleftarrow{\Phi}},\affinebang \overleftarrow{f}}{S}}} \\
            &=\Cost{F}+\Cost{S}+\Cost{H}+\Cost{\affinebang\overleftarrow{\Phi}^{\mathrm{out}}_{\lintranspose{\affcontext{\overleftarrow{\Phi}}}{R}}}\\
            &\leq\Cost{R}+\CostType{H}+\Cost{\affinebang\overleftarrow{\Phi}^{\mathrm{out}}_{\lintranspose{\affcontext{\overleftarrow{\Phi}}}{R}}}\\
            &=\Cost{F}+\Cost{S}+\Cost{H}+\Cost{\affinebang\overleftarrow{\Phi}^{\mathrm{out}}_{\lintranspose{\affcontext{\overleftarrow{\Phi}}}{R}}}\\
        \end{align*}

        \item Subcase $f\notin\FV{S}$:\\
        By typing we have $\oc\Sigma, \affinebang\Phi \vdash  S: \oc E \otimes \affinebang(L\multimap H)$.

        By inductive hypothesis on $S$ (case 3 of the lemma) we have:
        $$
            \Cost{\lintranspose{\affcontext{\overleftarrow{\Phi}}}{S}}+\CostType{L}+\Cost{\affinebang\overleftarrow{\Phi}^{\mathrm{in}}_{\lintranspose{\affcontext{\overleftarrow{\Phi}}}{S}}}
    	    \leq\Cost{S}+\CostType{H}+\Cost{\affinebang\overleftarrow{\Phi}^{\mathrm{out}}_{\lintranspose{\affcontext{\overleftarrow{\Phi}}}{S}}}
        $$
        We have:
        \begin{align*}
            \Cost{\lintranspose{\affcontext{\overleftarrow{\Phi}}}{R}}
            &=\Cost{\lintranspose{\affcontext{\overleftarrow{\Phi}}}{\letterm{\affinebang f^{\affinebang(L_0\multimap H_0)}}{\affinebang F}{S}}}
            = \Cost{\lintranspose{\affcontext{\overleftarrow{\Phi}}}{S}}
            \\[3mm]
            \Cost{R}
            &=\Cost{\letterm{\affinebang f^{\affinebang(L_0\multimap H_0)}}{\affinebang F}{S}}\\
            &\approx \Cost{(\lambda \affinebang f. S)\affinebang F}\\
            &=\Cost{\affinebang F}+\Cost{S}+\Cost{\FV{f}\setminus \FV{S}}\\
            &=\Cost{\affinebang F}+\Cost{S}+\Cost{L_0\multimap H_0}\\
            &=\Cost{\affinebang F}+\Cost{S}+\Cost{L_0}+\Cost{H_0}\\ 
            &\hspace{-5mm}\overset{\tiny \text{Lemma~\ref{lemma:cost_affine}}}{=}
            \Cost{F}+\Cost{S}+\Cost{L_0}+\Cost{H_0}\\
        \end{align*} 

        Observe that in this case we have
        \begin{equation}\label{eq:WtranspR_letF}
            \affinebang\overleftarrow{\Phi}^{\mathrm{in}}_{\lintranspose{\affcontext{\overleftarrow{\Phi}}}{R}}
            =\affinebang\overleftarrow{\Phi}^{\mathrm{in}}_{\lintranspose{\affcontext{\overleftarrow{\Phi}}}{S}}
            \quad \text{and} \quad
            \affinebang\overleftarrow{\Phi}^{\mathrm{out}}_{\lintranspose{\affcontext{\overleftarrow{\Phi}}}{R}}=\affinebang\overleftarrow{\Phi}^{\mathrm{out}}_{\lintranspose{\affcontext{\overleftarrow{\Phi}}}{S}} 
        \end{equation}
        because $\lintranspose{\affcontext{\overleftarrow{\Phi}}}{R}=\lintranspose{\affcontext{\overleftarrow{\Phi}}}{S}$.

        We can conclude as follows
        \begin{align*}
            \Cost{\lintranspose{\affcontext{\overleftarrow{\Phi}}}{R}}&+\CostType{L}+\Cost{\affinebang\overleftarrow{\Phi}^{\mathrm{in}}_{\lintranspose{\affcontext{\overleftarrow{\Phi}}}{R}}}\\
            &=\Cost{\lintranspose{\affcontext{\overleftarrow{\Phi}}}{S}}+\CostType{L}+\Cost{\affinebang\overleftarrow{\Phi}^{\mathrm{in}}_{\lintranspose{\affcontext{\overleftarrow{\Phi}}}{R}}}\\
            &\hspace{-3mm}\overset{\tiny \text{Eq.}~\ref{eq:WtranspR_letF}}{=}
            \Cost{\lintranspose{\affcontext{\overleftarrow{\Phi}}}{S}}
            +\CostType{L}
            +\Cost{\affinebang\overleftarrow{\Phi}^{\mathrm{in}}_{\lintranspose{\affcontext{\overleftarrow{\Phi}}}{S}}}\\
            &\hspace{-1.4cm}\overset{\tiny \text{IH on $S$}}{\leq}
            \Cost{S}
            +\CostType{H}
            +\Cost{\affinebang\overleftarrow{\Phi}^{\mathrm{out}}_{\lintranspose{\affcontext{\overleftarrow{\Phi}}}{S}}}\\
            &\hspace{-1cm}
            \leq\Cost{R}+\CostType{H}+\Cost{\affinebang\overleftarrow{\Phi}^{\mathrm{out}}_{\lintranspose{\affcontext{\overleftarrow{\Phi}}}{R}}}\\
            &\hspace{-1cm}
            =\Cost{F}+\Cost{S}+\Cost{L_0}+\Cost{H_0}+\Cost{H}+\Cost{\affinebang\overleftarrow{\Phi}^{\mathrm{out}}_{\lintranspose{\affcontext{\overleftarrow{\Phi}}}{R}}}\\
            &\hspace{-1.3cm}\overset{\tiny \text{Eq.}~\ref{eq:WtranspR_letF}}{=}
            \Cost{F}+\Cost{S}+\Cost{L_0}+\Cost{H_0}+\Cost{H}+\Cost{\affinebang\overleftarrow{\Phi}^{\mathrm{out}}_{\lintranspose{\affcontext{\overleftarrow{\Phi}}}{S}}}
        \end{align*}
    \end{itemize}

    \item Case $R=\letterm{(\oc x,\affinebang f^{\affinebang(L_0\multimap H_0)})}{S_1}{S_2}$:\\
    By hypothesis we have $\oc\Sigma, \affinebang\Phi \vdash \letterm{(\oc x,\affinebang f^{\affinebang(L_0\multimap H_0)})}{S_1}{S_2}: \oc E \otimes \affinebang(L\multimap H)$ and by typing we have:
    \begin{align*}
        \oc\Sigma, \affinebang\Phi &\vdash S_1: \oc E_0 \otimes \affinebang(L_0 \multimap H_0)\\
        \oc\Sigma, \oc x:\oc E_0, \affinebang\Phi, f:\affinebang(L_0\multimap H_0) &\vdash S_2: \oc E \otimes \affinebang(L\multimap H)
    \end{align*}
    Observe that $f$ may not be free in $S_2$, so we have to analyze the following subcases:

    \begin{itemize}
        \item Subcase $f\in\FV{S_2}$:\\
        By inductive hypothesis on $S_1$ (case 3 of the lemma) we have:
        $$
            \Cost{\lintranspose{\affcontext{\overleftarrow{\Phi}}}{S_1}}+\CostType{L}+\Cost{\affinebang\overleftarrow{\Phi}^{\mathrm{in}}_{\lintranspose{\affcontext{\overleftarrow{\Phi}}}{S_1}}}
    	    \leq\Cost{S_1}+\CostType{H}+\Cost{\affinebang\overleftarrow{\Phi}^{\mathrm{out}}_{\lintranspose{\affcontext{\overleftarrow{\Phi}}}{S_1}}}
        $$
        By inductive hypothesis on $S_2$ (case 3 of the lemma) we have:
        \begin{align*}
            \Cost{\lintranspose{\affcontext{\overleftarrow{\Phi}},\affinebang \overleftarrow{f}}{S_2}}+\CostType{L}+\Cost{\affinebang\overleftarrow{\Phi}^{\mathrm{in}}_{\lintranspose{\affcontext{\overleftarrow{\Phi}},\affinebang\overleftarrow{f}}{S_2}}}
            &+\myhighlight{Apricot}{\Cost{H_0}}
            \\ 
            &\hspace{-2cm}\leq\Cost{S_2}+\CostType{H}+\Cost{\affinebang\overleftarrow{\Phi}^{\mathrm{out}}_{\lintranspose{\affcontext{\overleftarrow{\Phi}},\affinebang\overleftarrow{f}}{S_2}}}
            +\myhighlight{Apricot}{\Cost{L_0}}\\
            &\hspace{-3.5cm}\text{\color{Apricot} because $\overleftarrow f: \affinebang (H_0\multimap L_0) \in\FV{\lintranspose{\affcontext{\overleftarrow{\Phi}},\affinebang\overleftarrow{f}}{S_2}}$}
        \end{align*}
        We have: 
        \begin{align*}
            \Cost{\lintranspose{\affcontext{\overleftarrow{\Phi}}}{R}}
            &=\Cost{\lintranspose{\affcontext{\overleftarrow{\Phi}}}{\letterm{(\oc x,\affinebang f)}{S_1}{S_2}}}\\
            &=\Cost{\letterm{(\oc x,\affinebang\overleftarrow{f})}
            {\lintranspose{\affcontext{\overleftarrow{\Phi}}}{S_1}}
            {\lintranspose{\affcontext{\overleftarrow{\Phi}}, \affinebang\overleftarrow{f}}{S_2}}}\\
            &\approx
            \Cost{(\lambda f.\lintranspose{\affcontext{\overleftarrow{\Phi}}}{S_1}) \lintranspose{\affcontext{\overleftarrow{\Phi}}, \affinebang\overleftarrow{f}}{S_2}}\\
            &=\Cost{\lintranspose{\affcontext{\overleftarrow{\Phi}}}{S_1}}+\Cost{\lintranspose{\affcontext{\overleftarrow{\Phi}},\affinebang \overleftarrow{f}}{S_2}}
            \\[3mm]
            \Cost{R}
            &=\Cost{\letterm{(\oc x,\affinebang f)}{S_1}{S_2}}\\
            &=\Cost{S_1}+\Cost{S_2}\\
        \end{align*} 
        We can conclude as follows
        \begin{align*}
            \Cost{\lintranspose{\affcontext{\overleftarrow{\Phi}}}{R}}&+\CostType{L}+\Cost{\affinebang\overleftarrow{\Phi}^{\mathrm{in}}_{\lintranspose{\affcontext{\overleftarrow{\Phi}}}{R}}}\\
            &=
            \Cost{\lintranspose{\affcontext{\overleftarrow{\Phi}}}{S_1}}
            +\Cost{\lintranspose{\affcontext{\overleftarrow{\Phi}},\affinebang \overleftarrow{f}}{S_2}}
            +\CostType{L}
            +\Cost{\affinebang\overleftarrow{\Phi}^{\mathrm{in}}_{\lintranspose{\affcontext{\overleftarrow{\Phi}}}{R}}}\\
            &=
            \Cost{\lintranspose{\affcontext{\overleftarrow{\Phi}}}{S_1}}
            +\underline{\Cost{\lintranspose{\affcontext{\overleftarrow{\Phi}},\affinebang \overleftarrow{f}}{S_2}}}
            +\underline{\CostType{L}}
            \\
            &\hspace{2cm}+\Cost{\affinebang\overleftarrow{\Phi}^{\mathrm{in}}_{\lintranspose{\affcontext{\overleftarrow{\Phi}}}{S_1}}}
            +\underline{\Cost{\affinebang\overleftarrow{\Phi}^{\mathrm{in}}_{\lintranspose{\affcontext{\overleftarrow{\Phi}},\affinebang \overleftarrow{f}}{S_2}}}}\\
            &\hspace{-4mm}\overset{\tiny \text{IH on $S_2$}}{\leq}
            \underline{\Cost{\lintranspose{\affcontext{\overleftarrow{\Phi}}}{S_1}}}
            +\underline{\Cost{\affinebang\overleftarrow{\Phi}^{\mathrm{in}}_{\lintranspose{\affcontext{\overleftarrow{\Phi}}}{S_1}}}}
            -\myhighlight{Apricot}{\CostType{H_0}}+\\
            &\hspace{2cm}
            +\Cost{S_2}+\CostType{H}+\Cost{\affinebang\overleftarrow{\Phi}^{\mathrm{out}}_{\lintranspose{\affcontext{\overleftarrow{\Phi}},\affinebang \overleftarrow{f}}{S_2}}}+\underline{\myhighlight{Apricot}{\CostType{L_0}}}\\
            &\hspace{-4mm}\overset{\tiny \text{IH on $S_1$}}{\leq}
            \Cost{S_1}+\CostType{H_0}+\Cost{\affinebang\overleftarrow{\Phi}^{\mathrm{out}}_{\lintranspose{\affcontext{\overleftarrow{\Phi}}}{S_1}}}
            -\myhighlight{Apricot}{\CostType{H_0}}
            \\
            &\hspace{2cm}
            +\Cost{S_2}+\CostType{H}+\Cost{\affinebang\overleftarrow{\Phi}^{\mathrm{out}}_{\lintranspose{\affcontext{\overleftarrow{\Phi}},\affinebang \overleftarrow{f}}{S_2}}} \\
            &=\Cost{S_1} +\Cost{\affinebang\overleftarrow{\Phi}^{\mathrm{out}}_{\lintranspose{\affcontext{\overleftarrow{\Phi}}}{S_1}}}
            +\Cost{S_2}+\CostType{H}+\Cost{\affinebang\overleftarrow{\Phi}^{\mathrm{out}}_{\lintranspose{\affcontext{\overleftarrow{\Phi}},\affinebang \overleftarrow{f}}{S_2}}} \\
            &=\Cost{S_1}+\Cost{S_2}+\Cost{H}+\Cost{\affinebang\overleftarrow{\Phi}^{\mathrm{out}}_{\lintranspose{\affcontext{\overleftarrow{\Phi}}}{R}}}\\
            &\leq\Cost{R}+\CostType{H}+\Cost{\affinebang\overleftarrow{\Phi}^{\mathrm{out}}_{\lintranspose{\affcontext{\overleftarrow{\Phi}}}{R}}}\\
            &=\Cost{S_1}+\Cost{S_2}+\Cost{H}+\Cost{\affinebang\overleftarrow{\Phi}^{\mathrm{out}}_{\lintranspose{\affcontext{\overleftarrow{\Phi}}}{R}}}\\
        \end{align*}
        
        \item Subcase $f\notin\FV{S_2}$:\\
        By typing we have $\oc\Sigma, \oc x:\oc E_0, \affinebang\Phi \vdash S_2: \oc E \otimes \affinebang(L\multimap H)$.

        By inductive hypothesis on $S_2$ (case 3 of the lemma) we have:
        $$
            \Cost{\lintranspose{\affcontext{\overleftarrow{\Phi}}}{S_2}}+\CostType{L}+\Cost{\affinebang\overleftarrow{\Phi}^{\mathrm{in}}_{\lintranspose{\affcontext{\overleftarrow{\Phi}}}{S_2}}}
    	    \leq\Cost{S_2}+\CostType{H}+\Cost{\affinebang\overleftarrow{\Phi}^{\mathrm{out}}_{\lintranspose{\affcontext{\overleftarrow{\Phi}}}{S_2}}}
        $$
        We have:
        \begin{align*}
            \Cost{\lintranspose{\affcontext{\overleftarrow{\Phi}}}{R}}
            &=\Cost{\lintranspose{\affcontext{\overleftarrow{\Phi}}}{\letterm{(\oc x,\affinebang f)}{S_1}{S_2}}}\\
            &=\Cost{\letterm{\oc x}{\ExpContextA[1]{P_1}}{\lintranspose{\affcontext{\overleftarrow{\Phi}}}{S_2}}} \quad \text{ for } \unzippingPre{S_1}=(\ExpContextA[1]{}, P_1, F_1)\\
            &=\Cost{\ExpContextA[1]{P_1}}
            +\Cost{\lintranspose{\affcontext{\overleftarrow{\Phi}}}{S_2}}\\[3mm]
            \Cost{R}
            &=\Cost{\letterm{(\oc x,\affinebang f)}{S_1}{S_2}}\\
            &\approx\Cost{ (\lambda (\oc x,\affinebang f). S_2)S_1}\\
            &=\Cost{S_1}+\Cost{S_2}+\Cost{\FV{(\oc x,\affinebang f)}\setminus \FV{S_2}}\\
            &=\Cost{S_1}+\Cost{S_2}+\Cost{f}\\
            &=\Cost{S_1}+\Cost{S_2}+\Cost{L_0 \multimap H_0}\\
            &=\Cost{S_1}+\Cost{S_2}+\Cost{L_0}+\Cost{H_0}\\ 
        \end{align*}
        Observe that in this case we have that 
        $
        \Cost{\affinebang\overleftarrow{\Phi}^{\mathrm{in}}_{\lintranspose{\affcontext{\overleftarrow{\Phi}}}{R}}}=
        \Cost{\affinebang\overleftarrow{\Phi}^{\mathrm{in}}_{\lintranspose{\affcontext{\overleftarrow{\Phi}}}{S_2}}}
        +\Cost{\affinebang\overleftarrow{\Phi}^{\mathrm{in}}_{\oc\ExpContextA[1]{P_1}}}$ because $\lintranspose{\affcontext{\overleftarrow{\Phi}}}{R}=\letterm{\oc x}{\ExpContextA[1]{P_1}}{\lintranspose{\affcontext{\overleftarrow{\Phi}}}{S_2}}
        $. Moreover, $\ExpContextA[1]{P_1}\in \lambdaLL^{\mathtt{P}}$ and so $\Cost{\affinebang\overleftarrow{\Phi}^{\mathrm{in}}_{\ExpContextA[1]{P_1}}}=0$. The same reasoning can be applied to $\Cost{\affinebang\overleftarrow{\Phi}^{\mathrm{out}}_{\lintranspose{\affcontext{\overleftarrow{\Phi}}}{R}}}$.
        Summing up we have that:
        \begin{equation}\label{eq:WtranspR_letR}
            \Cost{\affinebang\overleftarrow{\Phi}^{\mathrm{in}}_{\lintranspose{\affcontext{\overleftarrow{\Phi}}}{R}}}=\Cost{\affinebang\overleftarrow{\Phi}^{\mathrm{in}}_{\lintranspose{\affcontext{\overleftarrow{\Phi}}}{S_2}}} 
            \quad \text{and} \quad
            \Cost{\affinebang\overleftarrow{\Phi}^{\mathrm{out}}_{\lintranspose{\affcontext{\overleftarrow{\Phi}}}{R}}}=\Cost{\affinebang\overleftarrow{\Phi}^{\mathrm{out}}_{\lintranspose{\affcontext{\overleftarrow{\Phi}}}{S_2}}} 
        \end{equation} 
        We can conclude as follows
        \begin{align*}
            \Cost{\lintranspose{\affcontext{\overleftarrow{\Phi}}}{R}}&+\CostType{L}+\Cost{\affinebang\overleftarrow{\Phi}^{\mathrm{in}}_{\lintranspose{\affcontext{\overleftarrow{\Phi}}}{R}}}\\
            &=\Cost{\ExpContextA[1]{P_1}}
            +\Cost{\lintranspose{\affcontext{\overleftarrow{\Phi}}}{S_2}}
            +\CostType{L}
            +\Cost{\affinebang\overleftarrow{\Phi}^{\mathrm{in}}_{\lintranspose{\affcontext{\overleftarrow{\Phi}}}{R}}}\\
            &\hspace{-4mm}\overset{\tiny \text{Eq.}~\ref{eq:WtranspR_letR}}{=}
            \Cost{\ExpContextA[1]{P_1}}
            +\underline{\Cost{\lintranspose{\affcontext{\overleftarrow{\Phi}}}{S_2}}}
            +\underline{\CostType{L}}
            +\underline{\Cost{\affinebang\overleftarrow{\Phi}^{\mathrm{in}}_{\lintranspose{\affcontext{\overleftarrow{\Phi}}}{S_2}}}}\\
            &\hspace{-4mm}\overset{\tiny \text{IH on $S_2$}}{\leq}
            \Cost{\ExpContextA[1]{P_1}}
            +\Cost{S_2}
            +\CostType{H}
            +\Cost{\affinebang\overleftarrow{\Phi}^{\mathrm{out}}_{\lintranspose{\affcontext{\overleftarrow{\Phi}}}{S_2}}}\\
            &\leq\Cost{R}+\CostType{H}+\Cost{\affinebang\overleftarrow{\Phi}^{\mathrm{out}}_{\lintranspose{\affcontext{\overleftarrow{\Phi}}}{R}}}\\
            &=\Cost{S_1}+\Cost{S_2}+\Cost{L_0}+\Cost{H_0}+\CostType{H}+\Cost{\affinebang\overleftarrow{\Phi}^{\mathrm{out}}_{\lintranspose{\affcontext{\overleftarrow{\Phi}}}{R}}}\\
            &\hspace{-4mm}\overset{\tiny \text{Eq.}~\ref{eq:WtranspR_letR}}{=}
            \Cost{S_1}+\Cost{S_2}+\Cost{L_0}+\Cost{H_0}+\CostType{H}+\Cost{\affinebang\overleftarrow{\Phi}^{\mathrm{out}}_{\lintranspose{\affcontext{\overleftarrow{\Phi}}}{S_2}}}
        \end{align*}
        because by Lemma \ref{lemma:cost_transp_UnzipPre} $\Cost{\ExpContextA[1]{P_1}}\leq \Cost{S_1}$.

    \end{itemize}
    
\end{itemize}
\end{sproof}

\subsection{Bridging the Gap between Mathematical Differentiation and AD}
We show that our transpose transformation on $U\in\lambdaLL^{\mathtt t}$ produces a term which is extensionally equivalent to $\overleftarrow{U}$ of Equation~\ref{eq:transpose_function} but satisfying the condition of Proposition~\ref{prop:safe_reduction_is_safe}. To do so we use the following lemma and corollary. 

\begin{lemma}
    Given a pattern $\PIn:L$ and $\judgment[]{\PIn:L}{U_i}{H_i}$ we have that
    $$
    \Fusion \PatAddA{\Rename_1}{\emptyset} 
    \{\sfrac{U_1}{\RenamePat{\Rename_1}{U_1}}\}
    \mbox{ } \dot{+}_L \mbox{ }
    \Fusion \PatAddA{\Rename_2}{\emptyset} 
    \{\sfrac{U_2}{\RenamePat{\Rename_2}{U_2}}\}
    \quad \sim_L \quad
    \Fusion \PatAddA{\Rename'_1}{\Rename'_2}
    \{\sfrac{\Rename'_i[U_i]}{\RenamePat{\Rename'_i}{\Rename'_i[U_i]}}\}^2_{i=1}
    $$
    where
    \begin{itemize}
        \item  $\Rename_1$ and $\Rename_2$ are two identity renamings such that $\Dom{\Rename_i}=\Codom{\Rename_i}=\FV{\PIn}\cap\FV{U_i}$.
        \item $\Rename'_1$ and $\Rename'_2$ are two renamings such that $\Dom{\Rename'_i}=\FV{\PIn}\cap\FV{U_i}$ and $\Codom{\Rename'_1}\cap\Codom{\Rename'_2}=\emptyset$.
    \end{itemize}
\end{lemma}

\begin{corollary}
    \label{cor:transpose_vs_function_on_U}
    Given a pattern $\PIn:L$ we have that
    $$ 
         \begin{aligned}
            &\letterm{\RenamePat{\Rename'_1}{\PatAddA}}
            {\lintranspose{\PatAddA}{U_1}}
            {\Fusion \PatAddA{\Rename'_1}{\emptyset}}
            \\
            &\hspace{20mm}\mbox{ } \dot{+}_L \mbox{ }
            \\
            &\letterm{\RenamePat{\Rename'_2}{\PatAddA}}
            {\lintranspose{\PatAddA}{U_2}}
            {\Fusion \PatAddA{\Rename'_2}{\emptyset}}
        \end{aligned}   
        \sim_{L}
        \begin{aligned}
            &\left(
                \lambda \nTuple{\RenamePat{\Rename'_1}\PatAddA, \RenamePat{\Rename'_2}\PatAddA}.
                \Fusion{\PIn}{\Rename'_1}{\Rename'_2}
            \right)  
            \\
            &\hspace{15mm}\nTuple{
                \RenameTerm{\Rename'_1}{\lintranspose{\PIn}{U_1}},
                \RenameTerm{\Rename'_2}{\lintranspose{\PIn}{U_2}}
            }
        \end{aligned}
    $$
\end{corollary}

\medskip

Finally, we prove Lemma~\ref{lemma:transpose_vs_function_on_U} using Corollary~\ref{cor:transpose_vs_function_on_U} in the tuple case as follows

\begin{proof}[Proof of Lemma~\ref{lemma:transpose_vs_function_on_U}] 
    By structural induction on $U$ we prove that for any $V\sim_{H}V'$, we have
    $$
    \overleftarrow{U}\{\sfrac{V}{\POut}\}
	\sim_{\POut:H\vdash L} 
	\left(\lambda \POut.\letterm{\RenamePat\Rename\PatAddA}{\lintranspose{\PatAddA}{U}}{\Fusion \PatAddA{\Rename}{\emptyset}}\right)V'
    $$
    where $\Rename$ is the identity renaming restricted to $\FV{\PIn}\cap \FV{U}$.

    Let us consider the case $U=\nTuple{U_1,U_2}$, then we have $H=H_1 \& H_2$ and $\POut=\nTuple{\POut_1,\POut_2}$. 
    
    By definition of value for a pattern we have $V=\nTuple{V_1,V_2}$. We proceed by analyzing $\overleftarrow{U}\{\sfrac{V}{\POut}\}$ as follows 
    \begin{align*}
        \overleftarrow{U}\{\sfrac{V}{\POut}\}
        &\overset{\text{Eq.~\ref{eq:transpose_function}}}{=}
        \left(\Osi[L]\left(\lambda {\PIn}.\Iso[H](\POut)U\right)\right)\{\sfrac{V}{\POut}\}
        \\
        &\hspace{4mm}=
        \left(\Osi[L]\left(\lambda {\PIn}.\Iso[H](\POut)U\right)\right)\{\sfrac{V}{\POut}\}
        \\
        & \hspace{-1mm}\overset{\text{Def. $\Osi[L]$}}{=}  
        \left(\left(\lambda f.\!\!\sum_{W\in\mathcal B_L}\!\! (f(W)) \dot*_L W\right)\left(\lambda {\PIn}.\Iso[H](\POut)U\right)\right)\{\sfrac{V}{\POut}\}
        \\
        & \hspace{-1mm}\overset{\text{Def. $\Osi[H]$}}{=}  
        \left(
            \left(\lambda f.\sum_{W\in\mathcal B_L} (f(W)) \dot*_L W \right)
            \left(\lambda {\PIn}.(\lambda h.\lambda h'.
            \mathcal I_{H_1\with H_2} (h,h'))(\POut)U \right)
        \right)\{\sfrac{V}{\POut}\}
        \\
        &\hspace{4mm}=
        \left(\lambda f.\sum_{W\in\mathcal B_L} (f(W)) \dot*_L W \right)
        \left(\lambda {\PIn}.(\lambda h.\lambda h'.
        \mathcal I_{H_1\with H_2} (h,h'))(V)U \right)
        \\
        &\hspace{4mm}=
        \left(\lambda f.\sum_{W\in\mathcal B_L} (f(W)) \dot*_L W \right)
        \left(\lambda {\PIn}.(\lambda h.\lambda h'.
        \mathcal I_{H_1\with H_2} (h,h'))(\nTuple{V_1,V_2})\nTuple{U_1,U_2} \right)
        \\
        &\hspace{2mm}\rightarrow^*
        \left(\lambda f.\sum_{W\in\mathcal B_L} (f(W)) \dot*_L W \right)
        \left(\lambda {\PIn}.
        \mathcal I_{H_1\with H_2} (\nTuple{V_1,V_2},\nTuple{U_1,U_2}) \right)
        \\
        &\hspace{2mm}\rightarrow
        \sum_{W\in\mathcal B_L} \left(\lambda {\PIn}.
        \mathcal I_{H_1\with H_2} (\nTuple{V_1,V_2},\nTuple{U_1,U_2}) \right)(W) \dot*_L W 
        \\
        &\hspace{2mm}\rightarrow
        \sum_{W\in\mathcal B_L}
        \mathcal I_{H_1\with H_2} (\nTuple{V_1,V_2},\nTuple{U_1\{\sfrac{W}{\PIn}\},U_2\{\sfrac{W}{\PIn}\}}) \dot*_L W 
        \\
        &\hspace{-6mm}\overset{\text{Def. $\mathcal I_H$ +$\beta$-red}}{\rightarrow^*}
        \sum_{W\in\mathcal B_L}
        \left( 
            \mathcal I_{H_1}(V_1,U_1\{\sfrac{W}{\PIn}\})
            \mbox{ } \dot{+} \mbox{ } 
            \mathcal I_{H_2}(V_2,U_2\{\sfrac{W}{\PIn}\})
        \right) \dot*_L W 
        \\
        &\overset{\text{assoc}}{\sim}
        \left( 
            \sum_{W\in\mathcal B_L}
            \mathcal I_{H_1}(V_1,U_1\{\sfrac{W}{\PIn}\})
            \dot{*} W 
        \right)
        \mbox{ } \dot{+}_L \mbox{ } 
        \left( 
            \sum_{W\in\mathcal B_L}
            \mathcal I_{H_2}(V_2,U_2\{\sfrac{W}{\PIn}\})
            \dot{*} W 
        \right)
        \\
        &\hspace{2mm}\sim
        \left( 
            \left(\Osi[L]\left(\lambda {\PIn}.\Iso[H_1](\POut)U_1\right)\right)\{\sfrac{V_1}{\POut_1}\}
        \right)
        \mbox{ } \dot{+}_L \mbox{ } 
        \left( 
            \left(\Osi[L]\left(\lambda {\PIn}.\Iso[H_2](\POut)U_2\right)\right)\{\sfrac{V_2}{\POut_2}\}
        \right)
        \\
        &\hspace{2mm}=
        \overleftarrow{U_1}\{\sfrac{V_1}{\POut_1}\}
        \mbox{ } \dot{+}_L \mbox{ } 
        \overleftarrow{U_2}\{\sfrac{V_2}{\POut_2}\}
    \end{align*}

    Summing up we have that
    \begin{equation}
        \label{eq:tuple_fun_transp_1}
        \overleftarrow{U}\{\sfrac{V}{\POut}\} 
        \sim 
        \overleftarrow{U_1}\{\sfrac{V_1}{\POut_1}\}
        \mbox{ } \dot{+}_L \mbox{ } 
        \overleftarrow{U_2}\{\sfrac{V_2}{\POut_2}\}
    \end{equation}

    By definition of value for a pattern we have $V'=\nTuple{V'_1,V'_2}$. We proceed by analyzing\\ $\left(\lambda\POut.\letterm{\RenamePat\Rename\PatAddA}{\lintranspose{\PatAddA}{U}}{\Fusion \PatAddA{\Rename}{\emptyset}}\right)V'$ where $\Rename$ is the identity renaming restricted to \\$\FV{\PIn}\cap(\FV{U_1}\cup \FV{U_2})$ as follows
    \begin{align*}
        &\left(\lambda\POut.\letterm{\RenamePat\Rename\PatAddA}{\lintranspose{\PatAddA}{U}}{\Fusion \PatAddA{\Rename}{\emptyset}}\right)V'
        \\
        &=
        \left(\lambda\POut.\letterm{\RenamePat\Rename\PatAddA}{\lintranspose{\PatAddA}{\nTuple{U_1,U_2}}}{\Fusion \PatAddA{\Rename}{\emptyset}}\right)V'
        \\
        &=\left(
        \begin{aligned}
            &\lambda\POut. \\
            &\left( 
                \begin{aligned}
                &\letterm{\RenamePat\Rename\PatAddA}{
                \left(
                \begin{aligned}
                    &\letterm{\nTuple{\RenamePat{\Rename'_1}\PatAddA, \RenamePat{\Rename'_2}\PatAddA}}
                    {\nTuple{\lintranspose{\RenameTerm{\Rename'_1}{\PIn}}{\RenameTerm{\Rename'_1}{U_1}},
                    \lintranspose{\RenameTerm{\Rename'_2}\PIn}{\RenameTerm{\Rename'_2}{U_2}}
                    }}{}\\
                    &{\Fusion{\PIn}{\Rename'_1}{\Rename'_2}}
                \end{aligned}
                \right)
                }{}\\
                &\Fusion \PatAddA{\Rename}{\emptyset}
                \end{aligned}
            \right)
        \end{aligned}
        \right) V'
        \\
        &\begin{aligned}
            &\text{where $\Rename'_1$ and $\Rename'_2$ are two renamings with distinct codomains such that $\Dom{\Rename'_i}=\FV{\PIn}\cap\FV{U_i}$}
        \end{aligned}
        \\
        &=
        \left(
        \begin{aligned}
            &\lambda\POut. \\
            &\left( 
                \begin{aligned}
                &\letterm{\RenamePat\Rename\PatAddA}{
                \left(
                \begin{aligned}
                    &\letterm{\nTuple{\RenamePat{\Rename'_1}\PatAddA, \RenamePat{\Rename'_2}\PatAddA}}
                    {\nTuple{\lintranspose{\RenameTerm{\Rename'_1}{\PIn}}{\RenameTerm{\Rename'_1}{U_1}},
                    \lintranspose{\RenameTerm{\Rename'_2}\PIn}{\RenameTerm{\Rename'_2}{U_2}}
                    }}{}\\
                    &{\Fusion{\PIn}{\Rename'_1}{\Rename'_2}}
                \end{aligned}
                \right)
                }{}\\
                &\RenamePat\Rename\PatAddA
                \end{aligned}
            \right)
        \end{aligned}
        \right) V'
        \\
        & \sim
       \left(
            \lambda\POut.
            \letterm{\nTuple{\RenamePat{\Rename'_1}\PatAddA, \RenamePat{\Rename'_2}\PatAddA}}
                {\nTuple{\lintranspose{\RenameTerm{\Rename'_1}{\PIn}}{\RenameTerm{\Rename'_1}{U_1}},
                \lintranspose{\RenameTerm{\Rename'_2}\PIn}{\RenameTerm{\Rename'_2}{U_2}}
            }}{\Fusion{\PIn}{\Rename'_1}{\Rename'_2}}
        \right) V'
        \\
        &\sim
        \left(\lambda\nTuple{\POut_1,\POut_2}.
        \left(
        \begin{aligned}
            &\letterm{\nTuple{\RenamePat{\Rename'_1}\PatAddA, \RenamePat{\Rename'_2}\PatAddA}}
                {\nTuple{\lintranspose{\RenameTerm{\Rename'_1}{\PIn}}{\RenameTerm{\Rename'_1}{U_1}},
                \lintranspose{\RenameTerm{\Rename'_2}\PIn}{\RenameTerm{\Rename'_2}{U_2}}
            }}{}
            \\
            &{\Fusion{\PIn}{\Rename'_1}{\Rename'_2}}
        \end{aligned}
        \right)\right)
        \nTuple{V'_1,V'_2}
        \\
        &\approx
        \left(
        \lambda\nTuple{\POut_1,\POut_2}.
        \left(
            \lambda \nTuple{\RenamePat{\Rename'_1}\PatAddA, \RenamePat{\Rename'_2}\PatAddA}.
            \Fusion{\PIn}{\Rename'_1}{\Rename'_2}
        \right)  
        \nTuple{
            \lintranspose{\RenameTerm{\Rename'_1}{\PIn}}{\RenameTerm{\Rename'_1}{U_1}},
            \lintranspose{\RenameTerm{\Rename'_2}\PIn}{\RenameTerm{\Rename'_2}{U_2}}
        }
        \right)\nTuple{V'_1,V'_2}
    \end{align*}

    Summing up we have that
    \begin{equation}\label{eq:tuple_fun_transp_2}
    \begin{aligned}
        &\left(\lambda\POut.\letterm{\RenamePat\Rename\PatAddA}{\lintranspose{\PatAddA}{U}}{\Fusion \PatAddA{\Rename}{\emptyset}}\right)V'
        \\ 
        &\sim
        \left(
        \lambda\nTuple{\POut_1,\POut_2}.
        \left(
            \lambda \nTuple{\RenamePat{\Rename'_1}\PatAddA, \RenamePat{\Rename'_2}\PatAddA}.
            \Fusion{\PIn}{\Rename'_1}{\Rename'_2}
        \right)  
        \nTuple{
            \lintranspose{\RenameTerm{\Rename'_1}{\PIn}}{\RenameTerm{\Rename'_1}{U_1}},
            \lintranspose{\RenameTerm{\Rename'_2}\PIn}{\RenameTerm{\Rename'_2}{U_2}}
        }
        \right)\nTuple{V'_1,V'_2}
    \end{aligned} 
    \end{equation}
    
    We can conclude that $Eq.~\ref{eq:tuple_fun_transp_1}\sim Eq.~\ref{eq:tuple_fun_transp_2}$ by applying induction hypotheses and Corollary~\ref{cor:transpose_vs_function_on_U} as follows
    \begin{align*}
        &\overleftarrow{U}\{\sfrac{V}{\POut}\} 
        \\
        &\overset{\text{Eq.~\ref{eq:tuple_fun_transp_1}}}{\sim} 
        \overleftarrow{U_1}\{\sfrac{V_1}{\POut_1}\}
        \mbox{ } \dot{+}_L \mbox{ } 
        \overleftarrow{U_2}\{\sfrac{V_2}{\POut_2}\}
        \\
        &\hspace{3mm}\overset{\text{IHs}}{\sim}
        \begin{aligned}
            &\left(\lambda\POut_1.\letterm{\RenamePat{\Rename'_1}{\PatAddA}}
            {\lintranspose{\PatAddA}{U_1}}
            {\Fusion \PatAddA{\Rename'_1}{\emptyset}}\right)V'_1
            \\
            &\hspace{35mm}\mbox{ } \dot{+}_L \mbox{ } 
            \\
            &\left(\lambda\POut_2.\letterm{\RenamePat{\Rename'_2}{\PatAddA}}
            {\lintranspose{\PatAddA}{U_2}}
            {\Fusion \PatAddA{\Rename'_2}{\emptyset}}\right)V'_2
        \end{aligned}
        \\
        &\hspace{6mm}\text{where $\Rename'_1$ and $\Rename'_2$ are two identity renamings such that $\Dom{\Rename_i}=\FV{\PIn}\cap\FV{U_i}$.}
        \\
        &
        \hspace{4mm}\sim
        \left(
        \lambda\nTuple{\POut_1,\POut_2}.  
        \left(
            \begin{aligned}
                &\letterm{\RenamePat{\Rename'_1}{\PatAddA}}
                {\lintranspose{\PatAddA}{U_1}}
                {\Fusion \PatAddA{\Rename'_1}{\emptyset}}
                \\
                &\hspace{20mm}\mbox{ } \dot{+}_L \mbox{ }
                \\
                &\letterm{\RenamePat{\Rename'_2}{\PatAddA}}
                {\lintranspose{\PatAddA}{U_2}}
                {\Fusion \PatAddA{\Rename'_2}{\emptyset}}
            \end{aligned}
        \right)
        \right)\nTuple{V'_1,V'_2}
        \\
        & \hspace{1mm} \overset{\text{Cor.~\ref{cor:transpose_vs_function_on_U}}}{\sim} 
       \left(
            \lambda\nTuple{\POut_1,\POut_2}.
            \left(
                \lambda \nTuple{\RenamePat{\Rename'_1}\PatAddA, \RenamePat{\Rename'_2}\PatAddA}.
                \Fusion{\PIn}{\Rename'_1}{\Rename'_2}
            \right)  
            \nTuple{
                \RenameTerm{\Rename'_1}{\lintranspose{\PIn}{U_1}},
                \RenameTerm{\Rename'_2}{\lintranspose{\PIn}{U_2}}
            }
        \right)\nTuple{V'_1,V'_2}
        \\
        &\hspace{4mm}\sim
        \left(
            \lambda\nTuple{\POut_1,\POut_2}.
            \left(
                \lambda \nTuple{\RenamePat{\Rename'_1}\PatAddA, \RenamePat{\Rename'_2}\PatAddA}.
                \Fusion{\PIn}{\Rename'_1}{\Rename'_2}
            \right)  
            \nTuple{
                \lintranspose{\RenameTerm{\Rename'_1}{\PIn}}{\RenameTerm{\Rename'_1}{U_1}},
                \lintranspose{\RenameTerm{\Rename'_2}\PIn}{\RenameTerm{\Rename'_2}{U_2}}
            }
        \right)\nTuple{V'_1,V'_2}
        \\
        &\hspace{2mm}\overset{\text{Eq.~\ref{eq:tuple_fun_transp_2}}}{\sim}
        \left(\lambda\POut.\letterm{\RenamePat\Rename\PatAddA}{\lintranspose{\PatAddA}{U}}{\Fusion \PatAddA{\Rename}{\emptyset}}\right)V' 
    \end{align*}
\end{proof}

\subsection{Example Modularity} \label{subsect:exMod}
\begin{figure}[h]
  \vspace{-0.2cm}
    \begin{subfigure}{1\textwidth}
        \centering
        \begin{tikzpicture}[black] 
            \draw[ thick,rounded corners](-6.8,1.7) rectangle (5.7,-0.3);
      
            \node  at (-0.5,-0.5) {\mysizehead{Autodiff}};
      
            \node  at (-5.8,1) {\mysizebody{$e$}};
            \node  at (-5.8,0.4) {\mysizebody{Linear B}};
            \node  at (-5.8,0.05) {\mysizebody{\eqref{linear_B:primal}}};

            \node  (FM) at (-3.3,1) {\mysizebody{$\mathcal{F}^{\mathtt{Jax}}(e)$}};
            \node  at (-3.3,0.3) {\mysizebody{\ref{eq:linearAsyntax}}};
      
            \node  (U) at (0,1) {\mysizebody{$\mathcal{U}^{\mathtt{Jax}}(\mathcal{F}^{\mathtt{Jax}}(e))$}};
            \node  at (0,0.3) {\mysizebody{\ref{linear_B}}};
            
            \node  (T) at (4.1,1) {\mysizebody{$\mathcal{T}^{\mathtt{Jax}}(\mathcal{U}^{\mathtt{Jax}}(\mathcal{F}^{\mathtt{Jax}}(e)))$}};
            \node  at (4.1,0.3) {\mysizebody{\ref{eq:linearAsyntax}}};
      
            \draw [->,decorate,decoration={snake,amplitude=.4mm,
            segment length=2mm,post length=1mm}]  (-5.5,1) -- (-3.9,1);
            \node  at (-4.6,1.3) {\small $\mathcal{F}^{\mathtt{Jax}}$};
      
            \draw [->,decorate,decoration={snake,amplitude=.4mm,
            segment length=2mm,post length=1mm}]  (-2.7,1) -- (-1,1);
            \node  at (-1.8,1.3) {\small $\mathcal{U}^{\mathtt{Jax}}$};
      
            \draw [->,decorate,decoration={snake,amplitude=.4mm,
            segment length=2mm,post length=1mm}] (1.1,1) -- (2.7,1);
            \node  at (2,1.3) {\small $\mathcal{T}^{\mathtt{Jax}}$};
          \end{tikzpicture}
        \caption{Autodiff Linear A.}
        \label{fig:ADJax}
    \end{subfigure}
    \hfill
    \begin{subfigure}{1\textwidth}
        \centering
        \begin{tikzpicture}
            \draw[thick,rounded corners](-4,1.7) rectangle (3.8,0);
    
            \node  at (-0.5,-0.3) {\mysizehead{AD System of $\mathbf{\lambda}$LL}};
    
            \node  at (-3.5,1) {\mysizebody{$P$}}; 
            \node  at (-3.5,0.5) {\mysizebody{\ref{eq:LLPrimalterms}}}; 
    
            \node  (FM) at (-0.5,1) {\mysizebody{$\mathcal{F}(P)$}};  
            \node  at (-0.5,0.5) {\mysizebody{\ref{eq:LLBTerms}}}; 
            
            \node  (T) at (3,1) {\mysizebody{$\mathcal{T}(\mathcal{F}(P))$}}; 
            \node  at (3,0.5) {\mysizebody{\ref{eq:LLBTerms}}}; 
    
            \draw [->,decorate,decoration={snake,amplitude=.4mm,
            segment length=2mm,post length=1mm}]  (-3.2,1) -- (-0.9,1);
            \node  at (-2,1.3) {\mysizebody{\small $\mathcal{F}$}}; 
    
            \draw [->,decorate,decoration={snake,amplitude=.4mm,
            segment length=2mm,post length=1mm}] (0,1) -- (2.4,1);
            \node  at (1.2,1.3) {\mysizebody{\small$\mathcal{T}$}};
        \end{tikzpicture}
      \caption{AD System of $\lambdaLL$ without Unzipping.}
      \label{fig:ADLambda}
    \end{subfigure}
    \caption{The two AD Systems.}
    \label{fig:ADSystems}
\end{figure}
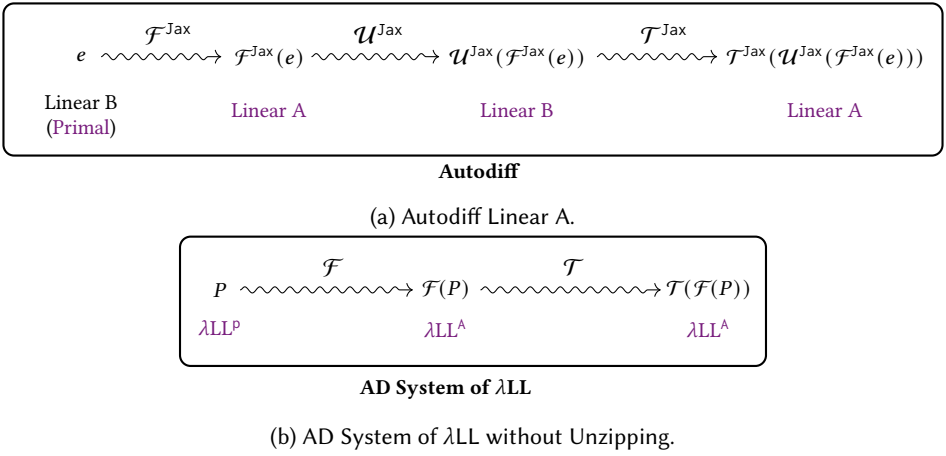

The implementation of the reverse mode as formalized in Linear A can obscure the parallel structure of a program due to the need for the unzipping transformation, which is not modular. By skipping unzipping as described above, we can preserve the program's inherent parallel structure. 
Let's illustrate this with an example. 
Consider the program $P = Q_1 * Q_2$ where $Q_1$ and $Q_2$ are two complex, independent subprograms of P, sharing only one input. Once a value for this latter is provided, $Q_1$ and $Q_2$ can be executed in parallel, needing to synchronise only at the end of their execution to perform the multiplication (seen as a numeric function, not the specialised $\dot{*}$).
We consider the two AD systems summarized in Figure~\ref{fig:ADSystems}. In order to keep the comparison between them more evident, we use $Q_1$ and $Q_2$ both as subexpressions in Linear A and as subterms in $\lambdaLL$, even if technically we should translate them into the two languages.

\paragraph{Autodiff Linear A.}
We start by applying the AD system of JAX, called Autodiff, to the program with some syntactic simplifications for the sake of clarity. Recall that Autodiff decomposes reverse mode AD into three different transformations as described in Figure~\ref{fig:ADJax}.

\begin{figure}[h!]
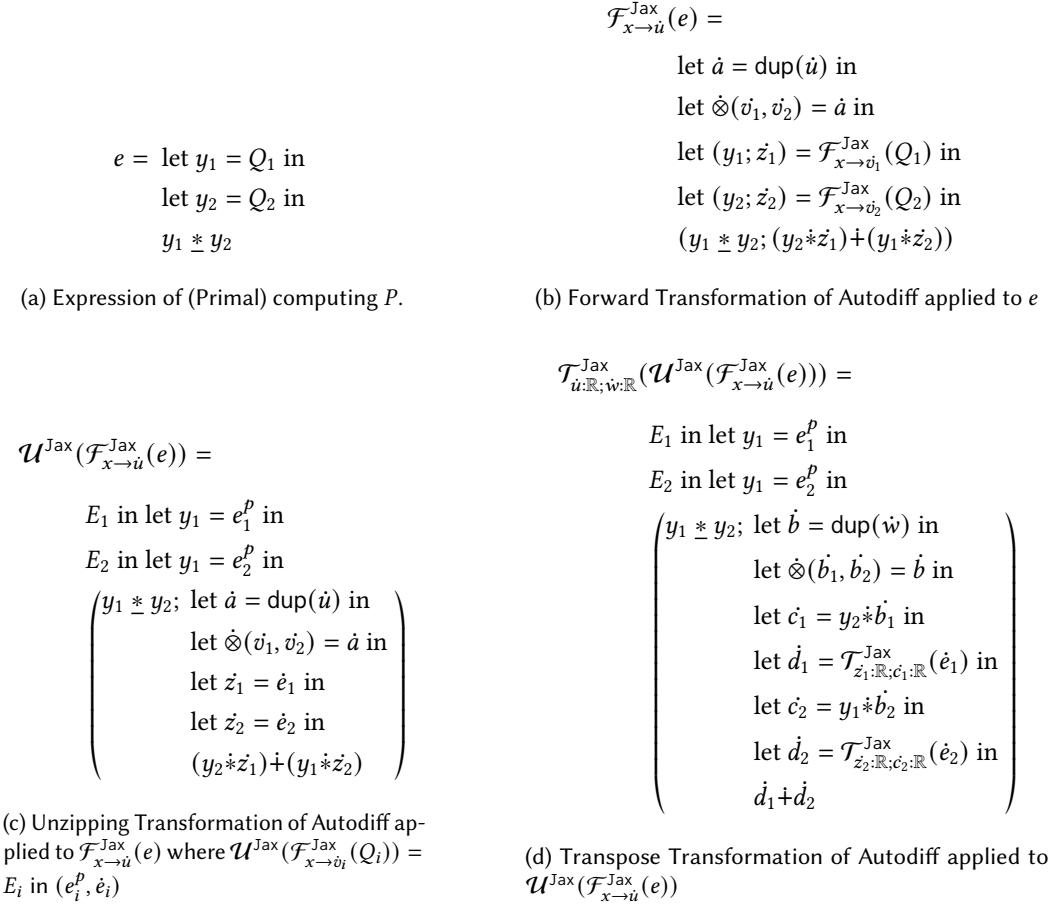
 
    \centering 
    \begin{subfigure}{0.4\textwidth}
        \begin{align*}
            e=\mbox{ }&\letJAX{y_1}{Q_1}{}\\
            &\letJAX{y_2}{Q_2}{}\\
            &y_1 \mbox{ }\underline{*}\mbox{ } y_2
        \end{align*}
        \caption{Expression of (Primal) computing $P$.}
        \label{fig:exexpLinA}
    \end{subfigure}
    \hfill
    \begin{subfigure}{0.5\textwidth} 
        \begin{align*}
            \forwardJAX{x\rightarrow \dot{u}}{&e}=\mbox{ }\\[1mm]
              &\letJAX{\dot{a}}{\mathtt{dup}(\dot{u})}{}\\
              &\letJAX{\lintupleJAX{v_1}{v_2}}{\dot{a}}{}\\
              &\letJAX{\retexpJAX{y_1}{\dot{z_1}}}{\forwardJAX{x\rightarrow \dot{v_1}}{Q_1}}{}\\
              &\letJAX{\retexpJAX{y_2}{\dot{z_2}}}{\forwardJAX{x\rightarrow \dot{v_2}}{Q_2}}{}\\
              &\retexpJAX{y_1 \mbox{ }\underline{*}\mbox{ } y_2}{(y_2\dot{*}\dot{z_1})\dot{+}(y_1\dot{*}\dot{z_2})}
        \end{align*} 
        \caption{Forward Transformation of Autodiff applied to $e$}
        \label{fig:exforwardLinA}
    \end{subfigure}
    \\ \vspace{0.3cm}
    \begin{subfigure}{0.4\textwidth} 
        \centering
        \begin{align*}
            \unzippingJAX{&\forwardJAX{x\rightarrow \dot{u}}{e}}=\\[2mm]
              &E_1 \text{ in } \letJAX{y_1}{e_1^p}{}\\
              &E_2 \text{ in } \letJAX{y_1}{e_2^p}{}\\
              &\left(
                \begin{aligned}
                  y_1 \mbox{ }\underline{*}\mbox{ } y_2;\mbox{ }
                  & \letJAX{\dot{a}}{\mathtt{dup}(\dot{u})}{}\\
                  & \letJAX{\lintupleJAX{v_1}{v_2}}{\dot{a}}{}\\
                  & \letJAX{\dot{z_1}}{\dot{e}_1}{}\\
                  & \letJAX{\dot{z_2}}{\dot{e}_2}{}\\
                  & (y_2\dot{*}\dot{z_1})\dot{+}(y_1\dot{*}\dot{z_2})\\
                \end{aligned}
              \right)
        \end{align*} 
        \caption{Unzipping Transformation of Autodiff applied to $\forwardJAX{x\rightarrow \dot{u}}{e}$ where $\mathcal{U}^{\mathtt{Jax}}(\forwardJAX{x\rightarrow \dot{v}_i}{Q_i})=E_i \text{ in } (e^p_i, \dot{e}_i)$}
        \label{fig:exunzipLinA}
    \end{subfigure}
    \hfill
    \begin{subfigure}{0.5\textwidth}
        \centering
        \begin{align*}
            \transpJAX{\dot{u}:\mathbb{R}}{\dot{w}:\mathbb{R}}{&\unzippingJAX{\forwardJAX{x\rightarrow \dot{u}}{e}}}=\\[2mm]
              &E_1 \text{ in } \letJAX{y_1}{e_1^p}{}\\
              &E_2 \text{ in } \letJAX{y_1}{e_2^p}{}\\
              & \left(
                \begin{aligned}
                  y_1 \mbox{ }\underline{*}\mbox{ } y_2;\mbox{ }
                  & \letJAX{\dot{b}}{\mathtt{dup}(\dot{w})}{}\\       
                  & \letJAX{\lintupleJAX{b_1}{b_2}}{\dot{b}}{}\\
                  & \letJAX{\dot{c_1}}{y_2\dot{*}\dot{b_1}}{}\\
                  & \letJAX{\dot{d_1}}{\transpJAX{\dot{z_1}:\mathbb{R}}{\dot{c_1}:\mathbb{R}}{\dot{e}_1}}{}\\
                  & \letJAX{\dot{c_2}}{y_1\dot{*}\dot{b_2}}{}\\
                  & \letJAX{\dot{d_2}}{\transpJAX{\dot{z_2}:\mathbb{R}}{\dot{c_2}:\mathbb{R}}{\dot{e}_2}}{}\\
                  & \dot{d_1}\dot{+}\dot{d_2}\\
                \end{aligned}
              \right)
          \end{align*}
        \caption{Transpose Transformation of Autodiff applied to $\unzippingJAX{\forwardJAX{x\rightarrow \dot{u}}{e}}$}
        \label{fig:extranposeLinA}
    \end{subfigure} 
    \caption{Autodiff Linear A applied to the program $P$ by skipping Unzipping.}
    \label{fig:exLinA}
\end{figure}

We translate the program $P$ into a purely primal expression in Linear B and we obtain the expression $e$ in Figure~\ref{fig:exexpLinA}. In order to keep the comparison more evident we use the subprograms of $P$, namely $Q_1$ and $Q_2$, as subexpressions of $e$ without translating them.

We proceed by following the steps described in Figure~\ref{fig:ADJax}. 
First, we apply the transformation $\mathcal{F}^{\mathtt{Jax}}$ (defined in Figure~\ref{fig:FMJAX}) to the expression $e$ and we obtain the expression in Figure~\ref{fig:exforwardLinA}, where $\letJAX{\dot{a}}{\mathtt{dup}(\dot{u})}{e'}$ is syntactic sugar for $\letJAX{(b;\dot{a})}{\mathtt{dup}(\dot{u})}{\letJAX{{\emptytupleJAX}}{b}{e'}}$.

Note that we can not directly apply the transpose transformation to the expression $\forwardJAX{x\rightarrow \dot{u}}{e}$ in Figure~\ref{fig:exforwardLinA} because it is an expression in Linear A not in Linear B and the transpose transformation is defined on Linear B. We proceed by applying the unzipping transformation $\mathcal{U}^{\mathtt{Jax}}$ (given in Figure~\ref{fig:UnzipJAX}), assuming $\mathcal{U}^{\mathtt{Jax}}(\forwardJAX{x\rightarrow \dot{v}_i}{Q_i})=E_i \text{ in } (e^p_i, \dot{e}_i)$. After unzipping, we obtain an expression in Linear B, the latter is described in Figure~\ref{fig:exunzipLinA} by using some simplifications.

Finally, we can apply the transpose transformation $\mathcal{T}^{\mathtt{Jax}}$ (defined in Figure~\ref{fig:TlinJAX} and Figure~\ref{fig:TJAX}) to the expression in Figure~\ref{fig:exunzipLinA}. We obtain the Linear A expression defined in Figure~\ref{fig:extranposeLinA} which computes the gradient of the initial program $P$ backward.

\paragraph{AD System of $\lambdaLL$ without Unzipping.}
We apply the AD system of $\lambdaLL$ as described in Figure~\ref{fig:ADLambda}. 
We start by translating the program $P$ into a purely primal term in $\PrimalLL$ and we obtain the term $M$ in Figure~\ref{fig:exexpLambda}. As we did for JAX, we use the subprograms of $P$, namely $Q_1$ and $Q_2$, as subterms of $M$ without translating them.

Observe that by notational conventions defined in Subsection~\ref{subsect:syntax} we use the $\mathtt{let}$ notation for the application to an abstraction and the multiplication is a binary numeric function of type $\oc\typR \otimes \oc\typR \multimap \oc\typR$, so we have that $M\approx 
(\lambda \oc y_1.(\lambda \oc y_2.\underline{*}(\oc y_1, \oc y_2))Q_2)Q_1$. 

We proceed by following the steps described in Figure~\ref{fig:ADLambda}. 
First, we apply the transformation $\mathcal{F}$ (given in Section~\ref{sect:forward} and in Figure~\ref{fig:FMlambda}) to the term $M$ and we obtain, after some $\beta$-steps and simplifications, the term in Figure~\ref{fig:exforwardLambda}. Moreover, for the sake of readability, we reduce $\forward{x:\mathbb{!R}}{M}$ into the term $N$ in Figure~\ref{fig:exNLambda} via $\beta$-reduction (defined in Figure~\ref{fig:beta_rules}). Then, we apply the transpose  transformation $\mathcal{T}$ (given in Section~\ref{sect:transpose} and in Figure~\ref{fig:def_transpose}) to the term $N$ and we obtain a term which is logical equivalent to the term in Figure~\ref{fig:extransposeLambda}.

\begin{figure}[h]
  \centering
  \begin{subfigure}{0.3\textwidth}
      \begin{align*}
          M=\mbox{ }& \letterm{\oc y_1}{Q_1}{}\\
          &\letterm{\oc y_2}{Q_2}{}\\
          &\oc y_1 \mbox{ }\underline{*}\mbox{ } \oc y_2 
        \end{align*}
      \caption{Term of $\PrimalLL$ computing $P$.}
      \label{fig:exexpLambda}
  \end{subfigure} 
  \hfill
  \begin{subfigure}{0.5\textwidth} 
      \begin{align*}
          \forward{x:\mathbb{!R}}{&M}=\\
          &\letterm{(\oc y_1, \affinebang f_1)}{\forward{x:\mathbb{!R}}{Q_1}}{}\\
          &\letterm{(\oc y_2,\affinebang f_2)}{\forward{x:\mathbb{!R}}{Q_2}}{}\\
          &\letterm{\oc z_1}{y_2}{\letterm{\oc z_2}{y_1}}\\
          &\letterm{\affinebang g}{
          \affbangterm{(\lambda \lintupleterm{u_1}{u_2}.
          (z_1\dot{*}u_1) \dot{+} (z_2\dot{*}u_2))}}{}\\
          &(\oc y_1 \mbox{ }\underline{*}\mbox{ } \oc y_2,\affbangterm{(\lambda u^\mathbb{R}.
          \letterm{\nTuple{u_1,u_2}}{\nTuple{u,u}}
          \affderterm{g}\lintupleterm{\affderterm{f_1}\mbox{}u_1}{\affderterm{f_2}\mbox{}u_2})})
      \end{align*}
      \caption{Forward Transformation in $\LinearBLL$ applied to $M$}
      \label{fig:exforwardLambda}
  \end{subfigure} 
  \\ \vspace{0.2cm}
  \begin{subfigure}{0.7\textwidth} 
      \begin{align*}
          N=\mbox{ }&
          \letterm{(\oc y_1,\affinebang f_1)}{\forward{x:\mathbb{!R}}{Q_1}}{}\\
          &\letterm{(\oc y_2,\affinebang f_2)}{\forward{x:\mathbb{!R}}{Q_2}}{}\\ 
          &(\oc y_1 \mbox{ }\underline{*}\mbox{ } \oc y_2,\affbangterm{(\lambda u^\mathbb{R}.
          \letterm{\nTuple{u_1,u_2}}{\nTuple{u,u}} 
              ({y_2 \dot{*}(\affderterm{f_1}\mbox{}u_1)})
              \mbox{ }\dot{+}\mbox{ }
              ({y_1\dot{*}(\affderterm{f_2}\mbox{}u_2)})
          )})
        \end{align*}
      \caption{Term of $\LinearBLL$ s.t. $\forward{x:\mathbb{!R}}{M} \rightarrow^* N$.}
      \label{fig:exNLambda}
  \end{subfigure} 
  \\ \vspace{0.3cm}
  \begin{subfigure}{0.7\textwidth} 
      \begin{align*} 
        \lintranspose{\affcontext{\overleftarrow{\Phi}}}{N}\sim\mbox{ }&
          \letterm{(y_1,\transptyp{{f_1}})}{\lintranspose{\affcontext{\overleftarrow{\Phi}}}{\forward{x:\mathbb{!R}}{Q_1}}}{}\\
          &\letterm{(y_2,\transptyp{{f_2}})}{ \lintranspose{\affcontext{\overleftarrow{\Phi},\transptyp{{f_1}}}}{\forward{x:\mathbb{!R}}{Q_2}}}{}\\ 
          &(\oc y_1 \mbox{ }\underline{*}\mbox{ } \oc y_2,
          \affbangterm{(\lambda h^\mathbb{R}.
          \letterm{\nTuple{w_1,w_2}}{\nTuple{h,h}} 
          (y_2 \dot{*} (\affderterm{\transptyp{{f_1}}}w_1)) 
          \mbox{ }\dot{+}\mbox{ }
          (y_1\dot{*} \affderterm{\transptyp{{f_2}}}w_2))})
      \end{align*}     
      \caption{Transpose Transformation in $\LinearBLL$ applied to $N$.}
      \label{fig:extransposeLambda}
  \end{subfigure} 
  \caption{AD System of $\lambdaLL$ without unzipping applied to the program $P$.}
  \label{fig:exLambda}
\end{figure}

\paragraph{Comparison.}
An efficient approach to compute the gradient of $P$ would be to alternate the forward and transpose transformations by computing the independent computations related to $Q_1$ and $Q_2$ in parallel and only at the end compose the results.

Consider the programs in Figure~\ref{fig:extranposeLinA} and in  Figure~\ref{fig:extransposeLambda}.
The expression obtained by applying the reverse mode as formalized in Linear A requires to perform the two transformations sequentially. We can observe that by looking at the expression given in Figure~\ref{fig:extranposeLinA} where the forward transformation, computing the primals $y_1$ and $y_2$, and the transpose, computing the co-tangents $d_1$ and $d_2$, are executed in a strict order. This requires performing the forward for both $Q_1$ and $Q_2$ before executing the transpose for both. However, since the subroutines $Q_1$ and $Q_2$ are independent, we should, in principle, be able to mix the transformations of $Q_1$ and $Q_2$ in various ways, resulting in an equivalent program.
In our framework, we indeed have this flexibility. We can either apply unzipping to get a program similar to the one described above, or we can apply the transpose transformation without unzipping, resulting in the term described Figure~\ref{fig:extransposeLambda}. In this case, the syntax of the term highlights that the two phases of $\mathcal{T}(\mathcal{F}(Q_1))$ are independent from those of $\mathcal{T}(\mathcal{F}(Q_2))$ and can be computed in parallel.

The two programs are extensionally equivalent  as a consequence of Corollary~\ref{cor:skipping_unzipping}, which exemplifies the kind of results our encoding enables. However, they can be implemented differently, with the parallel structure of the latter being more explicit than in the former. 

\end{document}